\documentclass[11pt,a4paper]{article}

\usepackage{graphicx}
\usepackage{afterpage}
\usepackage{epsfig,cite}
\usepackage{amssymb}
\usepackage{amsmath}
\usepackage{dsfont}
\usepackage{multirow}
\usepackage{url,hyperref}

\usepackage{url}
\usepackage{hyperref}

\newcommand{\be}{\begin{equation}}
\newcommand{\ee}{\end{equation}}
\newcommand{\bea}{\begin{eqnarray}}
\newcommand{\eea}{\end{eqnarray}}
\newcommand{\bi}{\begin{itemize}}
\newcommand{\ei}{\end{itemize}}
\newcommand{\ben}{\begin{enumerate}}
\newcommand{\een}{\end{enumerate}}
\newcommand{\la}{\left\langle}
\newcommand{\ra}{\right\rangle}
\newcommand{\lc}{\left[}
\newcommand{\rc}{\right]}
\newcommand{\lp}{\left(}
\newcommand{\rp}{\right)}

\def\frac#1#2{{{#1}\over {#2}}}
\def\gsim{\mathrel{\rlap{\lower4pt\hbox{\hskip1pt$\sim$}}
    \raise1pt\hbox{$>$}}}         
\def\lsim{\mathrel{\rlap{\lower4pt\hbox{\hskip1pt$\sim$}}
    \raise1pt\hbox{$<$}}}         

\newcommand{\dat}{\mathrm{dat}}

\newcommand{\net}{\mathrm{net}}

\newcommand{\tot}{\mathrm{tot}}

\newcommand{\draft}[1]{}

\def\beq{\begin{equation}}  
\def\eeq{\end{equation}}  

\def \n0{N_j^{(0)}}

\def\lapprox{\lower .7ex\hbox{$\;\stackrel{\textstyle <}{\sim}\;$}}
\def\gapprox{\lower .7ex\hbox{$\;\stackrel{\textstyle >}{\sim}\;$}}

\begin{document}
\begin{figure}[h]
\epsfig{width=0.32\textwidth,figure=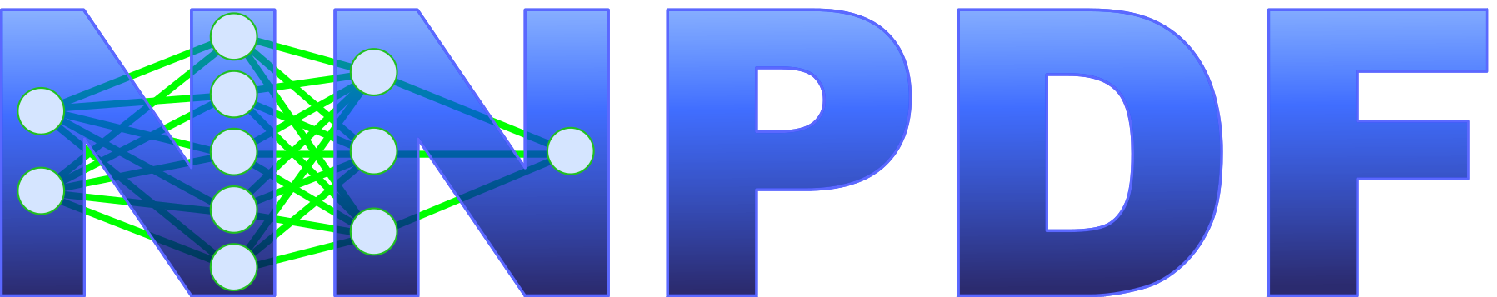}
\end{figure}
\vspace{-2.0cm}
\begin{flushright}
Edinburgh 2012/08\\
IFUM-FT-997\\
FR-PHENO-2012-014\\
RWTH TTK-12-25\\
CERN-PH-TH/2012-037\\
SFB/CPP-12-47\\
\end{flushright}

\begin{center}
{\Large \bf Parton distributions with LHC data}
\vspace{.7cm}

{\bf  The NNPDF Collaboration:}\\

Richard~D.~Ball$^{1}$, Valerio~Bertone$^{2,6}$,
Stefano~Carrazza$^{4}$,\\ Christopher~S.~Deans$^1$,
Luigi~Del~Debbio$^1$, Stefano~Forte$^4$, Alberto~Guffanti$^{5}$,\\ 
Nathan~P.~Hartland$^1$, Jos\'e~I.~Latorre$^3$, Juan~Rojo$^6$ 
and Maria~Ubiali$^7$.

\vspace{.3cm}
{\it ~$^1$ Tait Institute, University of Edinburgh,\\
JCMB, KB, Mayfield Rd, Edinburgh EH9 3JZ, Scotland\\
~$^2$  Physikalisches Institut, Albert-Ludwigs-Universit\"at Freiburg,\\ 
Hermann-Herder-Stra\ss e 3, D-79104 Freiburg i. B., Germany  \\
~$^3$ Departament d'Estructura i Constituents de la Mat\`eria, 
Universitat de Barcelona,\\ Diagonal 647, E-08028 Barcelona, Spain\\
~$^4$ Dipartimento di Fisica, Universit\`a di Milano and
INFN, Sezione di Milano,\\ Via Celoria 16, I-20133 Milano, Italy\\
~$^5$ The Niels Bohr International Academy and Discovery Center, \\
The Niels Bohr Institute, Blegdamsvej 17, DK-2100 Copenhagen, Denmark\\
~$^6$ PH Department, TH Unit, CERN, CH-1211 Geneva 23, Switzerland \\
~$^7$ Institut f\"ur Theoretische Teilchenphysik und Kosmologie, RWTH 
Aachen University,\\ D-52056 Aachen, Germany\\}
\end{center}

\vspace{0.1cm}

\begin{center}
{\bf \large Abstract}

\end{center}

We present the first  determination of parton distributions of
the nucleon at NLO and NNLO based on a global data set which includes 
LHC data: NNPDF2.3. 
Our data set includes, besides the deep inelastic,
Drell-Yan, gauge boson production and jet data already used in
previous global PDF determinations, all the relevant LHC data for which 
experimental systematic uncertainties are currently available: 
ATLAS and LHCb W and Z  rapidity
distributions from the 2010 run, CMS W electron asymmetry data from
the 2011 run, and ATLAS inclusive jet cross-sections from the 2010 run.
We introduce an improved implementation of the FastKernel method
which allows us to fit to this extended data set, and also to 
adopt a more effective minimization methodology. We present the NNPDF2.3
PDF sets, and compare them to the NNPDF2.1 sets to assess the impact of 
the LHC data. We find that all the LHC data are broadly consistent with 
each other and with all the older data sets included in the fit. We 
 present predictions for various
standard candle cross-sections, and compare them to those obtained 
previously using NNPDF2.1, and specifically
discuss the impact of ATLAS electroweak data on the determination of the
strangeness fraction of the proton. We also present  collider
 PDF sets, constructed using only 
data from HERA, Tevatron and LHC, but find that this data set is neither 
precise nor complete enough for a competitive PDF determination.

\clearpage

\tableofcontents

\clearpage

\section{Introduction}

\label{sec-intro}

The most accurate available information on the parton distribution 
functions (PDFs) of the nucleon, an essential ingredient for hadron
collider phenomenology~\cite{Forte:2010dt,Alekhin:2011sk,Watt:2011kp}, 
comes from global fits to
extended sets of data obtained from a variety of different electro- and
hadroproduction  processes, in particular, deep inelastic scattering (DIS),
Drell-Yan (DY) and gauge boson production, and jet production. The
combination of the information from all these processes allows one to 
determine six independent light quark and antiquark distributions
and the gluon. Data from the LHC are likely to offer
very significant improvements in the accuracy of these determinations, 
because of their greater precision and kinematic coverage, 
and also because of the greater range of precise cross-sections available,
including processes such as gauge boson production in 
association with jets and charm which  
hitherto have not been used for PDF determinations.

In fact, data for processes relevant for PDF determination
collected at the LHC during the first run
in 2010 already reached an accuracy comparable to pre-existing
data. In Ref.~\cite{Ball:2011gg,Ubiali:2011sb} we presented
a first study of the impact on PDFs of LHC data, specifically the 
$36~{\rm pb}^{-1}$ W-lepton asymmetry data from ATLAS and CMS, then 
available without full information on correlated systematics. 
We constructed a PDF set, NNPDF2.2, by reweighting~\cite{Ball:2010gb} 
the NNPDF2.1 NLO and NNLO PDF sets~\cite{Ball:2011mu,Ball:2011uy}.
Even with the modest amount of new
experimental information then available, small improvements
in the accuracy and changes in the shape of light-quark distributions in
the medium and small $x$ region were found.

Meanwhile, several LHC data sets with full information on correlated
systematics have been published, in particular gauge boson 
production data from ATLAS, CMS and LHCb, and inclusive jet and 
dijet data from ATLAS. Preliminary studies~\cite{leshouches} with some of these
data have shown that, thanks to the information on correlated
systematics,  their impact on PDFs is
significant: if included by reweighting the NNPDF2.1 set, one 
has to start with a large number of replicas in order to obtain
accurate results. A new set of PDFs
including this considerable amount of new information is thus needed, 
with the new data included in the fit, rather 
than added at a second stage via reweighting.

It is the purpose of this paper to present such a PDF determination, 
using the methodology developed by the NNPDF
collaboration~\cite{Forte:2002fg,DelDebbio:2004qj,DelDebbio:2007ee,Ball:2008by,Rojo:2008ke,Ball:2009mk,Ball:2009qv,Ball:2010de},
and used to produce the NNPDF2.1 LO, NLO, and NNLO PDF
sets~\cite{Ball:2011mu,Ball:2011uy} which are now part of the PDF4LHC
prescription~\cite{Botje:2011sn} for reliable determination of PDF
uncertainties in LHC processes. 
The new PDF set presented here, NNPDF2.3, is the most accurate 
determination to date from the NNPDF family, and it
supersedes previous existing sets. It differs from the NNPDF2.1 set
because of the inclusion of LHC data, and also because of some
improvements in fitting methodology, specifically in the genetic
algorithm which is used to determine the fit to the replicas.

We will determine NNPDF2.3 PDFs both at NLO and NNLO for a wide range
of values of $\alpha_s$, so that the user can select their own preferred 
value. While the default NNPDF sets use a 
variable-flavor number general-mass scheme based on the
FONLL~\cite{Cacciari:1998it,Forte:2010ta,LHhq} 
method for the inclusion of
heavy quark masses, in which the number of active flavor increases at
each quark threshold, we also provide PDF sets in which the maximum
number of active flavors is fixed at $n_f=4$ or $n_f=5$, 
which may be useful for specific applications (see for example 
Refs.\cite{Maltoni:2012pa,Cacciari:2011hy}).
We will also provide NNPDF2.3 NLO and NNLO sets based on reduced data sets: an
NNPDF2.3 noLHC set, which uses exactly the same data set as NNPDF2.1,
and differs from it only because of the improved methodology; and an 
NNPDF2.3 collider set, which only uses HERA inclusive and charm DIS data,
Tevatron gauge boson production and jet data, and LHC data, and excludes 
all fixed target data. The former
set is useful for the precise assessment of the impact
of LHC data, and also if one for some reason  
wishes to avoid using LHC data, for example for unbiased new physics 
searches at LHC. The latter set is interesting because the
fixed-target data, both for DIS and DY,
are problematic: they are generally at lower scales (hence many of them are
potentially subject to indeterminate higher-twist corrections), some 
carry no information on correlated systematics, and
finally many of them (such as NMC and BCDMS and all neutrino DIS data, 
and Tevatron DY data) are partly or fully obtained using nuclear 
targets which may be subject to significant but unknown nuclear
corrections. Jet  data, which are affected by larger theoretical
uncertainties, play at present a relatively minor role, so
a collider-only  fit is at present definitely
theoretically rather more reliable than the global fit: 
unfortunately we find that
it is not yet competitive in terms of statistical precision.

The paper is organized as follows. In Sec.~\ref{sec-expdata} we
will summarize the general features of the LHC data which are being
added to the data set, and specifically the choices of kinematic cuts.
In Sec.~\ref{sec-method} we will discuss the methodological
improvements which are being introduced in the NNPDF2.3
determination. In order to cope with the
non-negligible widening of the data set ---  the number of data for
hadron collider processes, which are computationally the most intensive 
since they depend quadratically on the PDFs, 
is considerably increased in comparison to NNPDF2.1 --- we introduce here
(Sec.~\ref{sec-fk}--\ref{sec-fkhad}) a
new, more efficient implementation of our previous FastKernel
method~\cite{Ball:2010de} for the computation of hadronic
observables. The ensuing considerable improvement in computational
efficiency allows us to switch to a new choice of settings for the
genetic algorithm which is used for minimization, which is
computationally rather more intensive but leads to more precise 
results (Sec.~\ref{sec:iga}).  In Sec.~\ref{sec-results} we
discuss the NNPDF2.3 PDF set: after summarizing the statistical
features of the fit, we present the PDFs, and compare them 
to our previous NNPDF2.1 set, specifically by separating the
effect of the LHC data and that of the improved methodology, through a
comparison which also involves the NNPDF2.3 noLHC set. The analysis
of the impact of LHC data is expanded upon in
Sec.~\ref{sec-collider}. First, we determine quantitatively through
the reweighting technique the amount of information introduced in the
fit by LHC data and their degree of consistency with the rest of the
data set, and examine how the description 
of the LHC data improves when they are included in the fit. Then, we
study in detail the
compatibility of collider data (including those from the LHC) with
fixed target data. Finally, we specifically address the issue of the
amount of
strangeness
 in the nucleon, which has
attracted some attention recently~\cite{Aad:2012sb}, and in particular
show that even though ATLAS data seem to favour a somewhat
larger strange fraction,
once uncertainties are properly accounted for there is no
incompatibility between ATLAS and fixed target data. Finally, in
Sec.~\ref{sec-pheno}, after briefly discussing NNPDF2.3 parton luminosities,
we compare several standard candle cross-sections obtained with them
to those obtained with NNPDF2.1.

\section{Experimental data}

\label{sec-expdata}

In this section we describe the data set used in the NNPDF2.3
analysis. The non-LHC data is the same as in the corresponding
NNPDF2.1 NLO and NNLO fits: this includes
NMC~\cite{Arneodo:1996kd,Arneodo:1996qe}, BCDMS~\cite{bcdms1,bcdms2}
and SLAC~\cite{Whitlow:1991uw} deep inelastic scattering (DIS) fixed
target data; the combined HERA-I DIS data set~\cite{H1:2009wt}, HERA
$F_L$~\cite{h1fl} and $F_2^c$ structure function
data~\cite{Breitweg:1999ad,Chekanov:2003rb,Chekanov:2008yd,Chekanov:2009kj,Adloff:2001zj,
  Collaboration:2009jy,H1F2c10:2009ut}, ZEUS HERA-II DIS
cross-sections~\cite{Chekanov:2009gm,Chekanov:2008aa},
CHORUS inclusive neutrino DIS~\cite{Onengut:2005kv}, and
NuTeV dimuon production data~\cite{Goncharov:2001qe,MasonPhD};
fixed-target E605~\cite{Moreno:1990sf} and
E866~\cite{Webb:2003ps,Webb:2003bj,Towell:2001nh} Drell-Yan production
data; CDF W asymmetry~\cite{Aaltonen:2009ta} and
CDF~\cite{Aaltonen:2010zza} and D0~\cite{Abazov:2007jy} Z rapidity
distributions; and CDF~\cite{Aaltonen:2008eq} and D0~\cite{D0:2008hua}
Run-II one-jet inclusive cross-sections. The kinematical cuts are unchanged 
from NNPDF2.1, so we will not review these data sets here. Instead we
will focus on the features of the LHC data included in
NNPDF2.3.  First we discuss electroweak gauge boson production and
then inclusive jets. In each case we describe the NLO and NNLO
codes and the corresponding settings used to
compute the theoretical predictions for each of these data sets, while
in the next section we describe the practical implementation of these
calculations into the FastKernel method used in the NNPDF fitting code.

The LHC has already provided an impressive set of measurements which
are sensitive to parton distributions, mostly from the 2010 run based
on a total integrated luminosity of $36~{\rm pb}^{-1}$:
the inclusive jet
and dijet
data~\cite{CMS:2011mea,Chatrchyan:2011qta,Collaboration:2011fc},
electroweak vector boson
production~\cite{Aad:2011dm,Aad:2011yn,cmsweasy,
  Chatrchyan:2011jz,Chatrchyan:2011wt,Aaij:2012vn}, both inclusive and
in association with heavy quarks~\cite{CMSWc}, and direct photon
production, both inclusive and associated with
jets~\cite{Aad:2011tw,Chatrchyan:2011ue,ATLAS:2012ar}.  Several other
measurements from the 2011 and 2012 runs, which are very relevant for
PDF fits, will be available in the next months, like the CMS and LHCb low
mass Drell-Yan differential distributions~\cite{CMSdy,lhcbdy} and the
inclusive jets and dijets from ATLAS and CMS~\cite{CMS2011jets}.

Precise determination of PDFs, adequate to current needs,  requires the use of experimental data 
which come with a full covariance matrix. In
NNPDF2.3 we include all currently available LHC data for which 
the experimental covariance
matrix has been provided: the ATLAS W and Z lepton rapidity distributions
from the 2010 data set, the CMS W electron asymmetry from the 2011
data set and the LHCb W lepton rapidity distributions from
the 2010 data set~\cite{Aaij:2012vn}, together with the ATLAS inclusive
jet cross-sections from the 2010
run~\cite{Collaboration:2011fc}. 

Tevatron Run II lepton asymmetry data
from W production were included in Ref.~\cite{Ball:2010gb} by
reweighting the NNPDF2.0 PDF set. Some of these data had issues of
compatibility with the rest of the NNPDF2.0-NNPDF2.1 dataset (these
two PDF sets only differ in the treatment of heavy quark mass terms),
and they only had a moderate impact on PDF uncertainty and essentially no
effect on the PDF shape. We prefer therefore not to include these data
in the NNPDF2.3 set, and concentrate on the impact of LHC data.

Also, for the time being we do not increase the set of physical
processes which are being used for PDF determination. In particular,
we choose to use the inclusive jet rather than the dijet cross-sections from
ATLAS: to use both would be double counting because they share the 
same underlying raw data. In principle, dijets
cross-sections carry
more detailed information on the underlying parton kinematics;
however, they are subject to significant scale uncertainties (see
for example Ref.~\cite{Thorne:2011kq} and references therein). 
In the future, as more data with full systematics becomes
available, it is likely that the inclusion of new processes in PDF
determination will be advantageous. This will require the
development of suitable fast interfaces for these processes. An
example is prompt photon production, whose impact on the gluon
determination was studied recently in Ref.~\cite{d'Enterria:2012yj}.

\begin{figure}[t]
\begin{center}
\epsfig{width=\textwidth,figure=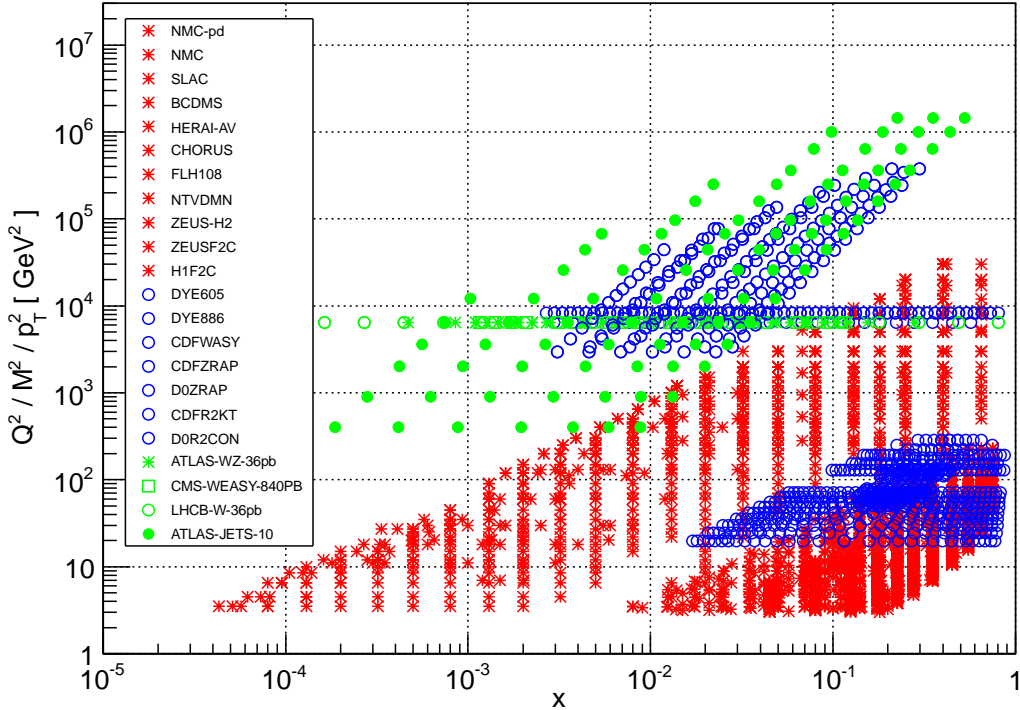}
\caption{ \small The kinematical coverage of the experimental data
used in the NNPDF2.3 PDF determination.
\label{fig:dataplottot}} 
\end{center}
\end{figure}

The kinematical coverage of the LHC data sets included
in the NNPDF2.3 analysis with the
corresponding average experimental uncertainties for each data set are
summarized in Tab.~\ref{tab:exp-sets-errors}.\footnote{For the one-jet
  inclusive cross-sections we consider here, the parton kinematics is
  not fixed even at leading-order. Therefore,  we
  plot only the minimum value of $x$  of the parton with smallest $x$,
  which is given by
  $x=\frac{p_t }{\sqrt{s}}e^{-|\eta|}$ in terms of the transverse
  momentum $p_T$ and rapidity $\eta$ of the jet and the center-of mass
  energy $\sqrt{s}$ of the hadronic collision}.  A 
scatter plot of the kinematical plane for all experimental
data from NNPDF2.3 is shown in
Fig.~\ref{fig:dataplottot}.  The LHC electroweak data
span a larger range in Bjorken-$x$ than
the Tevatron data thanks to the extended rapidity coverage
(up to $\eta=4.5$), while the inclusive
jets span a much wider kinematical range both
in $x$ and $Q^2$ than the one accessible at
the Tevatron. 

In Tab.~\ref{tab:sets-numpts} we also give the total number of
data points used for PDF fitting, both for the NLO and the NNLO global
sets, and for the various other PDF sets discussed in
Sec.~\ref{sec-results}. Note that the NLO and NNLO noLHC data sets
differ from those of the NNPDF2.1 NLO and NNLO fits of
Refs.~\cite{Ball:2011mu,Ball:2011uy} because of the inclusion in the
NNPDF2.3 data set of three  NMC data points
which were inadvertently neglected in the NNPDF2.1 fits.

  \begin{table}[h]                                                     
 \footnotesize
 \centering
 \begin{tabular}{|c||c|c|c||c|c|c|}
\hline
 Data Set & Ref.  & $N_{\rm dat}$ &  
$\lc\eta_{\rm min},\eta_{\rm max}\rc$ &
$\la \sigma_{\rm stat}\ra$ (\%) &
   $\la \sigma_{\rm sys}\ra$  (\%) & $\la \sigma_{\rm norm}\ra$  (\%)
 \\ \hline
\hline
CMS W electron asy $840~{\rm pb}^{-1}$       &   \cite{cmsweasy}  & 11 & $\lc 0,2.4\rc$&   2.1  & 4.7  
      & 0   \\  
\hline
ATLAS ${\rm W}^+$ $36~{\rm pb}^{-1}$      & \cite{Aad:2011dm} & 11 & $\lc 0,2.4\rc$ & 1.4 
 & 1.3    & 3.4      \\  
ATLAS ${\rm W}^-$ $36~{\rm pb}^{-1}$      & \cite{Aad:2011dm} & 11 & $\lc 0,2.4\rc$  & 1.6  & 1.4 & 3.4       \\  
ATLAS Z $36~{\rm pb}^{-1}$       & \cite{Aad:2011dm}  & 8 & $\lc 0,3.2\rc$ & 2.8
  &  2.4   & 3.4      \\  
\hline
LHCb ${\rm W}^+$ $36~{\rm pb}^{-1}$      & \cite{Aaij:2012vn} & 5  & $\lc 2,4.5\rc$ & 4.7  &   11.1  &  3.4   \\  
LHCb ${\rm W}^-$ $36~{\rm pb}^{-1}$      & \cite{Aaij:2012vn} & 5  & $\lc 2,4.5\rc$ & 3.4   &   7.8  &  3.4   \\  
\hline
\hline
ATLAS Inclusive Jets 36 pb$^{-1}$       & \cite{Collaboration:2011fc}  & 90 &
$\lc 0,4.5\rc$ & 10.2  & 23.4   & 3.4  \\  
\hline        
 \end{tabular}
\caption{\small \label{tab:exp-sets-errors} The number of data points,
kinematical coverage and average
statistical, systematic and normalization percentage uncertainties for
each of the experimental LHC data sets considered for
the NNPDF2.3 analysis. ATLAS inclusive jets refers to the
$R=0.4$ data set. There are $146$ LHC data points altogether.
}
 \end{table}

\begin{table}[h]                                                     
 \footnotesize
 \centering
 \begin{tabular}{|c|cc|}
\hline
    Fit & NLO  &   NNLO    \\
\hline
    NNPDF2.3 noLHC & 3341 &   3360 \\
    NNPDF2.3 Collider only &1212  &   1231 \\
    NNPDF2.3 & 3482 &  3501\\
\hline 
\end{tabular}
\caption{\small \label{tab:sets-numpts} Total 
number of data points for the various
global sets used for PDF fitting.
}
 \end{table}

\subsection{Electroweak boson production}

ATLAS has measured the W lepton and Z rapidity
distributions from $36~{\rm pb}^{-1}$, and provides the full experimental
covariance matrix~\cite{Aad:2011dm}. This measurement supersedes the
original muon asymmetry measurement from W decays~\cite{Aad:2011yn},
for which the covariance matrix was not available, and also adds 
the Z rapidity distributions, which are closely tied to the W
lepton distributions by the cross-correlated systematic
uncertainties.  The CMS
collaboration has presented a measurement of the electron asymmetry
with $840 {\rm pb}^{-1}$~\cite{cmsweasy} which supersedes the $36~{\rm pb}^{-1}$
data~\cite{Chatrchyan:2011jz} and also provides the experimental
covariance matrix. CMS has presented a measurement of the normalized
Z rapidity distribution with $36~{\rm pb}^{-1}$~\cite{Chatrchyan:2011wt},
but the covariance matrix is not available.  Finally, the LHCb
Collaboration has presented results for the W and Z lepton
rapidity distribution, from the 2010 data set~\cite{Aaij:2012vn}, again
with the experimental covariance matrix; however the $Z$ data are not
included in our determination  because they are in the process of
being reanalyzed~\cite{ronanprivate}.\footnote{Note also that in
  Ref.~\cite{Aaij:2012vn} there is a typo in the $W^+$ 
data   point for the $3.5-4$ rapidity bin, which reads
$125\pm5^{+19}_{-4}$ while it ought to be
$125\pm5^{+10}_{-4}$~\cite{ronanprivate}.} 

The theoretical predictions for LHC electroweak boson production have
been computed at NLO with {\tt MCFM}~\cite{MCFMurl,Campbell:2004ch}
interfaced with the {\tt APPLgrid} library for fast NLO
calculations~\cite{Carli:2010rw}.  As discussed in
Ref.~\cite{Ball:2011uy} (see in particular Sec.~3.2) NNLO
predictions are obtained by means of local $K$-factors. These have
been computed using the {\tt DYNNLO}
code~\cite{Catani:2010en}, and are found to be quite small, 
of order
2\% at most, and slowly varying with the lepton rapidity.

The calculation of NLO cross-sections requires the implementation 
of cuts on the lepton kinematics.
For the ATLAS data, these are the following:
\begin{itemize}
\item cuts for the W lepton rapidity  distributions
\begin{equation*}
\centering
p_T^{l} \ge 20~ {\rm GeV},\qquad  p_T^{\nu} \ge 25~ {\rm GeV}, \qquad m_T>40~ {\rm GeV} ,\qquad |\eta_l| \le 2.5;
\end{equation*}
\item cuts for the Z rapidity distribution
\begin{align*}
\centering
& p_T^{l} \ge 20~ {\rm GeV},\qquad 66~{\rm GeV} \le m_{l^+l^-} \le 116~{\rm GeV} ,\qquad \eta_{l^+,l^-} \le 4.9.
\end{align*}
\end{itemize}

ATLAS measures separately the rapidity
distributions in both the electron and muon channels, and then
combines them into a common data set. The above kinematical cuts 
correspond to the combination of electrons and muons, but differ
from the cuts applied in individual leptonic
channels. For Z rapidity distributions
 we have explicitly verified that results are
unchanged if the cut on the
rapidity of the leptons from the Z decay is removed.

For the CMS W electron asymmetry, the only cut is 
 \begin{equation*} 
 \centering 
p_T^{e} \ge 35~ {\rm GeV},
\end{equation*}
with the same binning in electron
rapidity as in Ref.~\cite{cmsweasy}. 

Finally, for the LHCb W data the
kinematical cuts are
  \begin{equation*} 
    \centering 
    p_T^{\mu} \ge 20~ {\rm GeV},\qquad 2.0 \le \eta^{\mu} \le 4.5.
  \end{equation*} 

For all three data sets, we performed extensive cross-checks at NLO
using two different codes, {\tt DYNNLO} and {\tt MCFM}: we
checked that, once common
settings are adopted, 
the results of the {\tt MCFM} and {\tt DYNNLO} runs agree
to better than  1\% for all the data bins. An even more accurate
agreement could be reached in principle~\cite{giancarlopriv}, 
but it is computationally
very costly and unnecessary for our purposes. In the particular 
case of the ATLAS W and Z
distributions, we also found good agreement with the {\tt APPLgrid}
tables used in the recent {\tt HERAfitter} analysis of ATLAS
data~\cite{Aad:2012sb}.

\subsection{Inclusive jet production}
        
The Tevatron jet data
play an important role in constraining the gluon distribution.
The kinematics coverage of this constraint is extended considerably by 
the LHC jet data.
From the 2010 $36 {\rm pb}^{-1}$ data set inclusive 
jet and dijet production have been
measured by CMS~\cite{CMS:2011mea,Chatrchyan:2011qta} and
ATLAS~\cite{Collaboration:2011fc}, however only ATLAS give the
full experimental covariance matrix. The covariance matrix is particularly 
important for these data because they are highly correlated.

The theoretical calculation of jet production cross-sections in hadron
collisions can be carried out by exclusive parton level Monte Carlo
codes such as {\tt NLOjet++} \cite{Nagy:2003tz} and  {\tt EKS}-{\tt MEKS}
\cite{Ellis:1990ek,Gao:2012he}. More recently, the NLO calculation matched to
parton showers in the context of the {\tt POWHEG} framework has also become
available \cite{Alioli:2010xa}, allowing direct hadron level
comparisons between theory and data. On the other hand, the full NNLO
corrections to the inclusive jet production are unknown, and only the
threshold corrections to the inclusive jet $p_T$ distribution are
available \cite{Kidonakis:2000gi}, thus the inclusion of jet data into
an NNLO analysis is necessarily approximate. Hadron collider jet
production data can be consistently included at NLO within a global
PDF analysis framework using fast NLO grid codes such as {\tt FastNLO}
\cite{Kluge:2006xs,Wobisch:2011ij} or {\tt
  APPLgrid}~\cite{Carli:2010rw}.

We compute inclusive jet cross-sections  using 
{\tt NLOjet++} interfaced to {\tt APPLgrid}. The jet
reconstruction parameters are identical to those used in the
experimental analysis~\cite{Aad:2010ad}. 
The NLO calculation uses the anti-$k_T$ algorithm~\cite{Cacciari:2008gp}, 
and 
the factorization and renormalization scales
are set to be $p_T^{\rm max}$, the
transverse momentum of the hardest jet in each event. We choose to
include in the analysis the data with $R=0.4$. These data are less
sensitive to nonperturbative corrections from the underlying event
and pileup as compared to the $R=0.6$ 
data~\cite{Dasgupta:2007wa,Cacciari:2008gd}, and
though they are a bit more sensitive to hadronization effects, all
in all the nonperturbative parton to hadron correction factors are
smaller for $R=0.4$ than for $R=0.6$. We have checked that the results are
essentially unchanged, both in terms of impact on PDFs and at the
level of the $\chi^2$ description if the $R=0.6$ data is
used instead of the $R=0.4$ data. 

On top of the 86 sources of fully correlated systematic errors, the
ATLAS jet spectra have an additional source of uncertainty due to the
theoretical uncertainty in the computation of the hadron to parton 
nonperturbative correction factors. We take these nonperturbative
corrections and their associated uncertainties from the ATLAS
analysis, where they are obtained from the variations of different
leading order Monte Carlo programs. It is clear from
Ref.~\cite{Aad:2010ad} that for a given Monte Carlo model
the nonperturbative correction is strongly correlated between data
bins, and thus conservatively we treat it as an additional source of
fully correlated systematic uncertainty, to be added to the covariance
matrix.

Because NNLO corrections to jet cross-sections are not available,  
hadron collider jet data can only be included in a
NNLO fit within some approximations. As in NNPDF2.1, NNLO 
theoretical predictions for
CDF and D0 inclusive jet data are obtained using the approximate NNLO
matrix element obtained from threshold resummation \cite{Kidonakis:2000gi} 
as implemented in the {\tt FastNLO} framework
\cite{Kluge:2006xs,Wobisch:2011ij}. For ATLAS data the threshold
approximation is expected to be worse because of the higher
centre-of-mass energy, and thus we simply used the NLO matrix element
with NNLO PDFs and $\alpha_s$. It was checked in
Ref.~\cite{Ball:2011uy} that, for Tevatron data, the difference between fits with
approximate NNLO jet matrix elements, and fits with purely NLO matrix
elements is significantly smaller than the difference between fits
with and without jet data.

\section{Methodological improvements}
\label{sec-method}

We discuss some methodological improvements introduced
in the current NNPDF release. First, we present a new
efficient method which has allowed us to 
speed up considerably the computation of hadronic observables
while maintaining full NLO accuracy for all experimental
data. As explained
below, it is a refinement of the FastKernel method introduced in
 Ref.~\cite{Ball:2010de}.
The rest of the section describes some improvements  of the
minimization algorithm, which allow for a more extensive
exploration of the space of parameters, and thus a more accurate
minimization. These new settings are computationally more intensive,
and are made possible by the new implementation of the FastKernel
method.

\subsection{The FastKernel method revisited: deep inelastic scattering}
\label{sec-fk}
We first briefly review the FastKernel method in the simplest case of
deep inelastic scattering,
closely following the original description of the algorithm in
Ref.~\cite{Ball:2010de}. 
In the FastKernel method the PDFs at the initial scale $Q_0^2$
are transformed into the evolution basis~\cite{Ball:2008by}, 
in which all nonsinglet combinations decouple, and only the
singlet and gluon evolve, coupled to each other. The basis of parametrized
PDFs is trivially related to 
the evolution basis through  a linear transformation.
The PDFs in the evolution basis at the initial scale are
denoted by $N^0_i(x)$. The index $i$ ranges from 1 to
$N_{\rm pdf}=13$, though only the light PDFs are independently
parametrized, 
heavy quarks being generated dynamically during the 
perturbative evolution. Observables are denoted
by $\sigma_I$, where $I$ is an index that runs over the number of data points
included in the fit. Each data point $\sigma_I$ is
characterized by a set of kinematic variables. For a DIS observable
the kinematic variables characterizing the data points are $(x_I,
Q_I^2)$, and the observable itself can be written as  
\begin{equation}
  \label{eq:DISobs}
  \sigma_I = \sum_{i=1}^{N_\mathrm{pdf}} K^I_{i} \otimes N^0_i\, ,
\end{equation}
where $K^I_{i}$ is a kernel obtained by convoluting the coefficient
functions $C^I_{j}$ for $\sigma_I$ with the DGLAP evolution kernels
$\Gamma_{ji}$, so 
\begin{equation}
  \label{eq:Kkernel}
  K^I_{i}(x_I,\alpha_s(Q^2),\alpha_s(Q_0^2)) = 
  \sum_{j=1}^{N_{\rm pdf}} C^I_{j} \otimes
  \Gamma_{ji}(x_I,\alpha_s(Q^2),\alpha_s(Q_0^2)) \, .
\end{equation}

The idea behind the FastKernel method is to approximate the PDFs at
the initial scale by a linear combination of
interpolating functions
\begin{equation}
  \label{eq:Nint}
  N^0_i(x) = \sum_{\alpha=1}^{N_x} 
  N^0_{\alpha i}\mathcal{I}^{(\alpha)}(x)\, .
\end{equation}
The coefficients of the linear combination are given by the value of
the function $N^0_j(x)$ computed on a grid of points,
$N^0_{\alpha i} \equiv N^0_i(x_\alpha)$. The index $\alpha$ runs from 1 to $N_x$,
the number of points in the $x$-grid.
 The same grid of $x$ values is used for all
data points $I$. More details on the choice of the grid, and of the interpolating
functions can be found in Ref.~\cite{Ball:2010de}, where
the choices that
guarantee an accurate interpolation at a sensible computational
cost are described. The idea of 
expanding PDFs in terms of a basis of interpolating polynomials
in order to perform PDF evolution more efficiently is the same
as that used in the evolution programmes {\tt HOPPET}~\cite{Salam:2008qg} and 
{\tt QCDNUM}~\cite{Botje:2010ay}. 
Using the interpolated PDFs, and writing explicitly the convolution in
Eq.~(\ref{eq:DISobs}) yields
\begin{equation}
  \label{eq:DISobs2}
  \sigma_I = \sum_{i=1}^{N_\mathrm{pdf}} \sum_{\alpha=1}^{N_x}
   \Big(\int_{x_I}^1 \frac{dy}{y} K^I_{i} \lp \frac{x_I}{y} \rp
  \mathcal I^{(\alpha)}(y)\Big)\, N^0_{\alpha  i}.
\end{equation}

The key observation now is that the integral is independent of the value
of the $N^0_{\alpha i}$, and can thus be precomputed. Denoting
the integral in Eq.~(\ref{eq:DISobs2}) by
$\Sigma^I_{\alpha i}$, we thus obtain
\begin{equation}
  \label{eq:DISobs3}
  \sigma_I = \sum_{\alpha,i}  \Sigma^I_{\alpha i}N^0_{\alpha i}
\equiv \Sigma_I\cdot N^0.
\end{equation}
This expression is a simple scalar product, with the dot indicating that 
the suppressed indices $(\alpha,i)$ have been contracted. The coefficients
$N^0_{\alpha i}$ are stored as an array of real numbers. In an actual
fit, this array is updated every time the PDFs at the initial scale
are changed. For each data point $I$, the coefficients
$\Sigma^I_{\alpha i}$ do not change when the PDFs are updated, so they can
be precomputed offline and stored.
Note that, for any given choice of $I$ and $\alpha$, the integral in
Eq.~(\ref{eq:DISobs2}) vanishes if $\mathcal I^{(\alpha)}(y)=0$ in the
interval $\left[x_I,1\right]$. As a consequence, the
$\Sigma^I_{\alpha i}$ array contains many zeroes. The computation
of the observables is optimized by including only the non-vanishing
terms in the scalar product in Eq.~(\ref{eq:DISobs3}), and can thus 
be evaluated very rapidly.

It is worth noting that, within this framework, all
the theoretical inputs are encoded in the arrays $\Sigma^I_{\alpha i}$, called 
the {\tt FK} tables in the following discussion. Any variation of  
the parameters
(e.g.~$\alpha_s$, CKM matrix elements, EW parameters, mass
thresholds), renormalization scheme, or renormalization or 
factorization scales,
is implemented by generating a new set of 
$\Sigma^I_{\alpha  i}$. Each data set then has several {\tt FK} tables 
associated with it, each table corresponding to one particular 
choice of theoretical inputs. 

\subsection{The {\tt FK}  method for hadronic collider data}
\label{sec-fkhad}

As described in Ref.~\cite{Ball:2010de}, the FastKernel
method can also be applied to hadronic data, and
indeed from NNPDF2.0 onwards all the fixed target Drell-Yan
data and the Tevatron W and Z data were included
at full NLO accuracy using this technique.
A similar approach to the FastKernel method has been applied by various
other groups to the NLO calculation of 
hadron collider observables. For
example the {\tt FastNLO}~\cite{Kluge:2006xs,Wobisch:2011ij} and
{\tt APPLgrid}~\cite{Carli:2010rw} collaborations provide software
tools which
are capable of performing efficient NLO QCD calculations for
a variety of hard scattering processes, like jet production
and electroweak boson production.

In all these frameworks,
Monte Carlo weights from an appropriate event generator, such
as {\tt NLOjet++}~\cite{Nagy:2003tz} and {\tt MCFM}~\cite{Campbell:2000bg}
 are stored,
partonic subprocess by subprocess, on an interpolating grid in $x$
and $Q$-space. With this grid the calculation of the observable is
reduced to simple products and sums; the parton distributions 
may then be varied without incurring a large computational overhead.

As an illustration, we review the procedure implemented in 
the {\tt APPLgrid} approach.\footnote{
The corresponding formalism for the FastKernel and 
 {\tt FastNLO} methods are very similar, with some technical
differences.} The
calculation of a collider observable is performed as 
\begin{equation}
\label{eq:applconv}
\sigma_I = \sum_{l=0}^{N_{\mathrm{sub}}} \sum_{\alpha,\beta}^{N_x} \sum_{\tau}^{N_{Q}}
W_{\alpha\beta\tau}^{I(l)} \, 
F^{(l)}\left(x_{\alpha}, x_{\beta},  Q^2_{\tau}\right)\, ,
\end{equation}
where the indices $\alpha,\beta$ run over points in the $x$-space
grid, $\tau$ runs over points in $Q^2$, and $l$ denotes the 
specific parton level subprocess. The W table contains 
the values of the Monte Carlo weights for a particular subprocess, 
and the $F^{(l)}$ are the
incoming subprocess parton luminosities constructed as a bilinear 
combination of PDFs appropriate for the subprocess in consideration.

Methods such as the one described above, and exemplified by
Eq.~(\ref{eq:applconv}), provide already fast and efficient NLO QCD
calculations of hadronic observables, and were adopted by NNPDF
starting with the NNPDF2.0 global fit~\cite{Ball:2010de} . However, 
it is still possible to 
reduce considerably the number of floating-point operations required
in a fit by combining this procedure with the evolution of the
PDFs. Such a combined approach for hadronic processes, which we call the
{\tt FK} approach, is extremely fast: it allows the
precomputation of all the $Q^2$ dependence in the calculation, and reduces
the final computational task to scalar products similar to that described
above for DIS.

The first step of the  {\tt FK} method for hadronic data is to
 construct a table of evolution
coefficients by taking the convolution of the DGLAP evolution kernels
with a suitable interpolation basis. The procedure is identical to the
case of DIS FastKernel tables summarized above, albeit without the
additional convolution with coefficient functions.  Using the same
interpolating functions, we introduce first an evolution matrix
\begin{equation}
  \label{eq:Edef}
  E^\tau_{\alpha\beta ij}
  = \int_{x_\alpha}^1 \frac{dy}{y}\Gamma_{ij}\left(
    \frac{x_\alpha}{y},\alpha_s(Q_\tau^2),\alpha_s(Q_0^2) \right)
  \mathcal{I}^{(\beta)}(y)\, . 
\end{equation}
Evolution from the initial scale in the evolution basis is then once
again reduced to a scalar product:
\begin{equation}
  \label{eq:Emat}
  N_i(x_\alpha,Q_\tau^2) = \sum_{\beta, j} E^\tau_{\alpha\beta ij}
  N^0_{\beta j} \equiv E^\tau_{\alpha i} \cdot N^0 \, ,
\end{equation}
where the dot again denotes an implicit sum, now over $(\beta,j)$. 
Introducing a suitable matrix $R$ to rotate to the flavor basis, 
more suitable for constructing the parton luminosities 
required in Eq.~(\ref{eq:applconv}), we can write a PDF $f_n$, 
$n=1,\dots N_{\rm pdf}$ in the flavor basis at the scale $Q_\tau^2$ in the form
\begin{align}
  f_n(x_{\alpha},Q^2_\tau) &=
  \sum_i^{N_{\rm pdf}} R_{ni} N_i(x_{\alpha},Q_\tau^2) =
  \sum_{\beta}^{N_x} 
  \sum_{i,j}^{N_{\rm pdf}}
  R_{ni} E^\tau_{\alpha\beta ij} N^0_{\beta j} \nonumber\\
  &=  \sum_{\beta}^{N_x} \sum_{j}^{N_{\mathrm{pdf}}}A^\tau_{\alpha\beta nj}N^0_{\beta j}
  \equiv A^\tau_{\alpha n} \cdot N^0, \label{eq:RxEdef}
\end{align}
where
\begin{equation}
  \label{eq:Adef}
  A^\tau_{\alpha\beta n j} = \sum_i^{N_{\rm pdf}} R_{ni} E^\tau_{\alpha\beta ij}\, 
\end{equation}
is now the rotated evolution matrix, and again the dot denotes an implicit 
sum over $(\beta,j)$. 

Having factorized the DGLAP evolution of the incoming parton
distributions, it is now simple to construct the required subprocess
luminosities for the observable. Firstly,  
\begin{equation}
  \label{eq:subproc1}
  F^{(l)}\left(x_{\alpha}, x_{\beta}, Q^2_{\tau}\right) =
  \sum_{m,n}^{N_{\rm pdf}} D^{(l)}_{mn} \,
    f_m(x_{\alpha},Q^2_\tau) f_n(x_{\beta},Q^2_\tau) \, ,  
\end{equation}
where the $D^{(l)}_{mn}$ are coefficients giving the
nonzero contributions of the flavor combination $f_mf_n$ to the
subprocess in question. Substituting Eq.~(\ref{eq:RxEdef}) into 
Eq.~(\ref{eq:subproc1}) then gives 
\begin{equation}
  \label{eq:subproc3}
  F^{(l)}\left(x_{\alpha}, x_{\beta}, Q^2_{\tau}\right) = 
  \sum_{m,n}^{N_{\rm pdf}} D^{(l)}_{mn}
  \left( A^\tau_{\alpha m} \cdot N^0\right) \left(  A^\tau_{\beta n}
    \cdot  N^0\right)\, , 
\end{equation}
and thus upon substitution into Eq.~(\ref{eq:applconv}),
\begin{equation}
\label{eq:applconv2}
\sigma_I = \sum_{l=0}^{N_{\mathrm{sub}}} \sum_{\gamma,\delta}^{N_x} \sum_{\tau}^{N_{Q}}
\sum_{m,n}^{N_{\rm pdf}} W_{\gamma\delta\tau}^{I(l)}D^{(l)}_{mn}
  \left( A^\tau_{\gamma m} \cdot N^0\right) \left(  A^\tau_{\delta n}
    \cdot  N^0\right) \, .
\end{equation}

The PDF evolution, now made explicit, may be absorbed into the Monte
Carlo weight grid allowing for a great deal more of the calculation of
Eq.~(\ref{eq:applconv}) to be precomputed: if we define a  {\tt FK} grid
\begin{equation}
  \label{eq:newwgt}
  \Sigma^I_{\alpha\beta i j} = 
  \sum_{l=0}^{N_{\mathrm{sub}}}
  \sum_{\tau}^{N_{Q}}
  \sum_{m,n}^{N_{\rm pdf}}
  \sum_{\gamma,\delta}^{N_x} 
  W_{\gamma\delta\tau}^{I(l)} \, 
  D^{(l)}_{mn} A^\tau_{\gamma\alpha m i} A^\tau_{\delta\beta n j}
  \, ,
\end{equation}
the hadronic observable can be evaluated as
\begin{equation}
\label{eq:FKsigma1} 
  \sigma_I = \sum_{i,j}^{N_{\mathrm{pdf}}} \sum_{\alpha,\beta}^{N_x} 
  \Sigma^I_{\alpha\beta i j} N_{\alpha i}^0N_{\beta j}^0 
  \equiv N^0 \cdot \Sigma_I \cdot N^0 \, . 
\end{equation}
This compact expression shows that the computation of hadronic
observables is now reduced to a sum of bilinear products over a grid in
$x$-space, and the basis of input PDFs, in complete analogy with the 
DIS case presented above. 

The coefficients $\Sigma_I$ are the {\tt FK} tables for hadronic
collider processes. As discussed above, they encode all the
theoretical inputs introduced in the calculation of a given
observables. Any variation in these inputs can be included in
the fit by generating a new {\tt FK} table, while the rest of the code is
left unchanged. 

From the practical point of view, to obtain the {\tt FK} 
tables as in Eq.~(\ref{eq:FKsigma1})  we first need to obtain
the partonic weights  as in  Eq.~(\ref{eq:applconv}) for
a given experimental data set, and then combine these partonic
grids with the interpolated PDF evolution coefficients using 
Eq.~(\ref{eq:newwgt}).
Note that in previous NNPDF fits the evolution and the coefficient functions
were not combined together in this way.
For the data sets considered in this paper, we have used
the following codes for the partonic weights, 
as discussed in Sec.~\ref{sec-expdata}:

\begin{itemize}

\item For the fixed target Drell-Yan data and the Tevatron
W and Z data we use the FastKernel tables from~\cite{Ball:2010de}.
These observables are now calculated
using Eq.~(\ref{eq:FKsigma1}), giving exactly the same result as previously,
but in a fraction of the time.

\item For the Tevatron Run II CDF and D0 inclusive jet
production we use the tables provides by {\tt FastNLO}.
Again these observables are calculated
using Eq.~(\ref{eq:FKsigma1}), giving the same result as 
in previous NNPDF fits but in a fraction of the time. 

\item For the ATLAS 2010 inclusive jet data we use the
tabulated partonic cross-sections from the {\tt APPLgrid} program.

\item For the LHC electroweak vector boson production data,
we have computed new {\tt APPLgrid} partonic cross-section tables,
using the built-in interface to the {\tt MCFM} program.

\end{itemize}

\begin{table}[t!]
\begin{center}
\footnotesize
\begin{tabular}{c|c|c|c||c|c|c|}
\cline{2-7}
   & \multicolumn{3}{|c||}{${\rm W}^{+}$ distribution [pb]} & \multicolumn{3}{|c|}{${\rm W}^{-}$ distribution [pb]}  \\
\hline
\multicolumn{1}{|c||}{$| \eta_{l} |$} & {\tt FK}	 & {\tt APPLgrid} &  $\epsilon_{\rm rel}$ & {\tt FK}  &  {\tt APPLgrid} &  $\epsilon_{\rm rel}$\\
\hline
\multicolumn{1}{|c||}{0.00--0.21} & 617.287 & 617.345 & 0.01\% & 456.540 & 456.819 & 0.06\% \\
\multicolumn{1}{|c||}{0.21--0.42} & 616.988 & 617.062 & 0.01\% & 453.045 & 453.315 & 0.06\%\\
\multicolumn{1}{|c||}{0.42--0.63} & 620.237 & 620.290 & 0.01\% & 448.902 & 449.172 & 0.06\%\\
\multicolumn{1}{|c||}{0.63--0.84} & 624.192 & 624.235 & 0.01\% & 441.789 & 442.045 & 0.06\%\\
\multicolumn{1}{|c||}{0.84--1.05} & 630.235 & 630.286 & 0.01\% & 432.206 & 432.435 & 0.05\%\\
\multicolumn{1}{|c||}{1.05--1.37} & 636.835 & 636.886 & 0.01\% & 419.027 & 419.222 & 0.05\%\\
\multicolumn{1}{|c||}{1.37--1.52} & 642.800 & 642.861 & 0.01\% & 403.908 & 404.084 & 0.04\%\\
\multicolumn{1}{|c||}{1.52--1.74} & 642.499 & 642.569 & 0.01\% & 390.564 & 390.724 & 0.04\%\\
\multicolumn{1}{|c||}{1.74--1.95} & 642.351 & 642.437 & 0.01\% & 377.328 & 377.473 & 0.04\%\\ 
\multicolumn{1}{|c||}{1.95--2.18} & 628.592 & 628.693 & 0.02\% & 359.373 & 359.498 & 0.03\%\\
\multicolumn{1}{|c||}{2.18--2.50} & 590.961 & 591.079 & 0.02\% & 337.255 & 337.366 & 0.03\% \\
\hline
 \end{tabular}
 \begin{tabular}{c|c|c|c|}
 \cline{2-4}
   & \multicolumn{3}{|c|}{Z distribution [pb]}  \\
    \hline
\multicolumn{1}{|c||}{$|y|$} & {\tt FK}	  &    {\tt APPLgrid}  &   $\epsilon_{\rm rel}$\\
\hline
\multicolumn{1}{|c||}{0.0--0.4} & 124.634 & 124.633 & 0.001\%\\
\multicolumn{1}{|c||}{0.4--0.8} & 123.478 & 123.488 & 0.01\%\\
\multicolumn{1}{|c||}{0.8--1.2} & 121.079 & 121.108 & 0.02\%\\
\multicolumn{1}{|c||}{1.2--1.6} & 118.057 & 118.108 & 0.04\%\\
\multicolumn{1}{|c||}{1.6--2.0} & 113.512 & 113.549 & 0.03\%\\
\multicolumn{1}{|c||}{2.0--2.4} & 106.552 & 106.562 & 0.01\%\\
\multicolumn{1}{|c||}{2.4--2.8} & 93.7637 & 937.838 & 0.02\%\\
\multicolumn{1}{|c||}{2.8--3.6} & 55.8421 & 558.538 & 0.02\%\\
\hline
 \end{tabular}
\begin{tabular}{c|c|c|c|}
\cline{2-4}
   & \multicolumn{3}{|c|}{ATLAS 2010 jets [pb]}  \\
    \hline
\multicolumn{1}{|c||}{$p_T$ (GeV)} & {\tt FK}	  &    {\tt APPLgrid}  &    $\epsilon_{\rm rel}$ \\
\hline
\multicolumn{1}{|c||}{20--30}   & $6.1078 \times 10^6$ & $6.1090 \times 10^6$ &	0.02\% \\
\multicolumn{1}{|c||}{30--45}   & 986285 	                     & 98654      & 0.03\% \\
\multicolumn{1}{|c||}{45--60}   & 190487 	                     & 190556    & 0.04\% \\
\multicolumn{1}{|c||}{60--80}   & 48008.7 	                     & 48029.7   & 0.04\% \\
\multicolumn{1}{|c||}{80--110} & 10706.6 	                     & 10710.4   & 0.03\% \\
\multicolumn{1}{|c||}{110--160} & 1822.62 			  & 1822.87   & 0.01\% \\
\multicolumn{1}{|c||}{160--210} & 303.34 			  & 303.443   & 0.03\% \\
\multicolumn{1}{|c||}{210--260} & 76.1127 			  & 76.1338   & 0.03\% \\
\hline
 \end{tabular}

 \end{center}
\caption{\small The {\tt FK} results for some of the LHC data
included in the NNPDF2.3 analysis compared with the
original {\tt APPLgrid} interfaced to LHAPDF, for the same
PDF set. The tables show the comparison for the  ATLAS differential cross-sections for ${\rm W}^{\pm}$ production, where the average relative discrepancy
over the whole W/Z data set is 0.03\% and the maximum relative discrepancy is 0.06\%. They also show the corresponding results for some selected
bins for the theoretical predictions for the ATLAS 2010 jet data
 in the first rapidity bin $|y|<0.3$, where in this case the average relative discrepancy
over the whole data set is 0.03\% and
the maximum relative discrepancy is 0.2\%.}
\label{tab:fkbenchatlasjet}
\end{table}

Given the crucial role played by the {\tt FK} tables in our current fitting
procedure, a careful benchmark of their accuracy is mandatory when
they are generated. It is clear from Eqs.~(\ref{eq:newwgt}--\ref{eq:FKsigma1}) 
that our prescription for computing the observables
is identical to the original  formula, Eq.~(\ref{eq:applconv}), 
except that we have changed
the order of the sums in order to precompute all the terms that do not
depend on the PDFs at the initial scale. Benchmarking the tables is
then straightforward, since the observables computed with the {\tt FK}
tables must agree with those computed using {\tt APPLgrid}/{\tt FastNLO}, 
provided the two procedures use the same PDFs as an input.

For the hadronic observables that were already in previous NNPDF fits,
namely fixed target Drell-Yan and Tevatron electroweak boson
production,
the {\tt FK} tables have been computed on exactly the same grid points in $x$,
and therefore the agreement between the old and the new computations
is at the level of the machine precision.
For the new LHC jet and electroweak observables 
computed using the {\tt APPLgrid} interface, 
the grid of
$x$ points used in the NNPDF analysis is different from the original grids
used by the other packages. Therefore the comparison is only accurate
to the precision of the interpolating functions. Results for the ATLAS 2010
jets double differential cross-section~\cite{Collaboration:2011fc}, and for the
ATLAS W/Z 
differential cross-section~\cite{Aad:2011dm}, computed at a sample of
kinematical points, are compared in Tab.~\ref{tab:fkbenchatlasjet}: is clear
that the numerical accuracy is more than satisfactory for the requirements
of precision phenomenology.

\subsection{Improved minimization}
\label{sec:iga}

We have introduced some  new settings for the minimization,
which 
allow for a more extensive exploration of the space of parameters and
thus a more accurate result. Some of these new settings are 
computationally more intensive, and thus only made possible by the
implementation of the {\tt FK}  method described in the previous section.

In the NNPDF2.1 fits, different parameters were chosen for the genetic
algorithm at
NLO and NNLO: specifically, at NNLO the number of mutations and
mutants were increased, in order to cope with the more complex shape of
NNLO coefficient functions. For  NNPDF2.3 
we use  the same parameters
(number of mutations and mutation rates) at NLO and NNLO, and we
choose them to be 
the same as in NNPDF2.1 NNLO, and summarized in 
Tab.~\ref{tab:gapars}. We refer to Sec.~4.2 of
Ref.~\cite{Ball:2011uy}, Sec.~4.3 of Ref.~\cite{Ball:2010de} 
and Sec.~4.2 of Ref.~\cite{Ball:2008by} for a more detailed
discussion of the genetic algorithm and the definition of these
parameters.  As mentioned, this choice corresponds to an increased
number of mutants and mutations, and thus a more detailed exploration
of parameter space.
These new  parameters of the genetic algorithm for 
the NNPDF2.3 fits are collected in Tab.~\ref{tab:gapars}.

Also, we have modified the 
criterion for dynamical stopping by making it a little more
stringent, which means that stopping happens on average at a somewhat
later stage: we take $r_v-1=4\times 10^{-4}$ at NLO and
$r_v-1=3\times 10^{-4}$ at NNLO, to be compared with the respective values 
 $r_v-1=3\times 10^{-4}$ and  $r_v-1=2\times 10^{-4}$ of the NNPDF2.1
fits, discussed in Sec.~4.6 of Ref.~\cite{Ball:2010de} (NNPDF2.0 NLO,
unchanged in NNPDF2.1) and Sec.~4.2 of Ref.~\cite{Ball:2011uy}
(NNPDF2.1 NNLO). We have also increased the maximum number 
of genetic algorithm generations at which the minimization stops if 
the stopping criterion is not satisfied, from $N_{\rm
  gen}^{\rm max}=3\times 10^4$ of NNPDF2.1 to $N_{\rm
  gen}^{\rm max}=5\times 10^4$. These new values of the  stopping
parameters have been determined, as discussed in
Ref.~\cite{Ball:2010de} (see in particular Sect.~4.6 and Fig.~8) by
inspection of the minimization profiles for individual replicas, with
the new dataset and minimization parameters used now.
Clearly both the increased number of mutants and mutations, and the
increase by almost a factor two of the maximum training length are
computationally quite demanding. 

We have also introduced two more small improvements of the minimization
procedure. First, we now discard outlier replicas such that their
value of $\chi^{2(k)}$ is more than four sigma larger than the mean
value evaluated over the replica sample (see Sec.~\ref{sec-stat} and
Tab.~\ref{tab:est}): such replicas are exceedingly unlikely in
the $N_{\rm rep}=100$ replica samples that we consider here, and
their inclusion would bias results. 

Second,  
for experiments with a small number of data points we
include all data in the training set, rather than equally dividing
them between training and validation sets.  Indeed
fit results are quite stable upon changes of
the value of the training fraction, provided the sample of training
and validation data are large enough to be representative of the full
dataset, as shown in 
Ref.~\cite{Ball:2008by},   where 
the size of the training fraction was varied by a factor two 
with essentially unchanged results. Experiments with a small number of
data points have little or no impact on the fulfillment of the
stopping criteria, and it is then advantageous to include all their
points in the training sample in order to maximise the information
which is extracted from these data.   
In practice, we 
include in the training set
all the data points for all experiments with up to 30 data points: 
H1 $F_L$, CDF W asymmetry, CDF and
D0 Z rapidity distributions, ATLAS W and Z data, CMS
W electron asymmetry and LHCb W and Z data.

\begin{table}
\begin{center}
  \begin{tabular}{|c||c|c|c|c|c|c|}
    \hline 
 &   $N_{\rm gen}^{\rm wt}$ & $N_{\rm gen}^{\rm mut}$
&   $N_{\rm gen}^{\rm max}$ & $E^{\mathrm{sw}}$ & $N_{\rm mut}^a$ 
&  $N_{\rm mut}^b $\\
    \hline
2.1 NLO &    $10000$ & 2500 & 30000 & 2.6 & 80 & 10\\
    \hline
2.1 NNLO &    $10000$ & 2500 & 30000 & 2.3 & 80 & 30\\
    \hline
    \hline
2.3 NLO &    $10000$ & 2500 & 50000 & 2.3 & 80 & 30\\
    \hline
2.3 NNLO &    $10000$ & 2500 & 50000 & 2.3 & 80 & 30\\
    \hline
  \end{tabular}\\

\bigskip

  \begin{tabular}{|c||c|c|c|c|}
\hline
&  \multicolumn{2}{|c|}{2.1 NLO} &
   \multicolumn{2}{|c|}{2.1 NNLO and 2.3}    \\
    \hline 
PDF &   $N_{\rm mut}$ &  $\eta^{\rm k}$ &  $N_{\rm mut}$ &  $\eta^{k}$  \\
    \hline
\hline 
$\Sigma(x)$   &2 & 10, 1 & 2 & 10, 1 \\
$g(x)$  & 2& 10, 1& 3 & 10, 3, 0.4 \\
$T_3(x)$  &2 & 1, 0.1 & 2 &  1, 0.1 \\
$V(x)$  &2 & 1, 0.1 & 3 &  8, 1, 0.1\\
$\Delta_S(x)$  &2 & 1, 0.1 &3 & 5, 1, 0.1 \\
$s^+(x)$  & 2&  5, 0.5 & 2 & 5, 0.5 \\
$s^-(x)$  & 2&  1, 0.1& 2 & 1, 0.1\\
\hline 
  \end{tabular}
  \end{center}
  \caption{\small Parameter values for the genetic 
algorithm for NNPDF2.3 fits, compared
to NNPDF2.1 NLO and  NNLO (top).
 The number of mutations and values of the mutation rates for each
 individual PDF are also  given (bottom). }
  \label{tab:gapars}
\end{table}

\label{sec-ga}

\section{Results}

\label{sec-results}

The main results of this paper are the NNPDF2.3 NLO and NNLO parton
distributions. 
In this section we will discuss the statistical
features of the corresponding fits, then present the NNPDF2.3 PDFs and compare
them to other available NLO and NNLO sets. 
We have produced NNPDF2.3 PDFs for all values of $\alpha_s$  
from $0.114$ to $0.124$ in  steps of $0.001$. 
We have also produced PDF sets in which the number of active flavors
does not increase beyond $N_f=4$ or $N_f=5$, which are loosely
referred to as ``fixed flavor number'' (even though, strictly
speaking they are ``maximum flavor number'' sets).  These are
provided for all values of $\alpha_s$  
from $0.116$ to $0.120$ in steps of $0.001$. A more extensive
selection of plots than that presented here is available from the NNPDF web
site, \url{http://nnpdf.hepforge.org/},

In order to ease the comparison with NNPDF2.1 and gauge the impact of LHC
data, we have also produced NLO and NNLO sets based on exactly the
same data set used for NNPDF2.1, but including various methodological
improvements (such as those discussed in Sec.~\ref{sec:iga}), which
we call  NNPDF2.3 noLHC: these are provided for values of $\alpha_s$  
from $0.116$ to $0.120$ in steps of $0.001$. Finally, in order to 
better elucidate the
compatibility of collider (including LHC) data with low-energy data,
we have also produced NLO and NNLO sets only based on collider data,
which we call NNPDF2.3 collider, also provided for all values of $\alpha_s$  
from $0.116$ to $0.120$ varied in steps of $0.001$. 
These will be discussed in more detail
in Sec.~\ref{sec-collider}.  In order to check the consistency of the LHC
data with the rest of the data included in the fit, we have also 
produced extended sets of NNPDF2.3 noLHC  NLO and
NNLO PDFs including 500 replicas, which can be used in order to
perform PDF determinations in which LHC data are included by
reweighting an existing set, according to the methodology of
Refs.~\cite{Ball:2010gb,Ball:2011gg}, to be discussed in Sec.~\ref{sec-diff}.

\begin{table}\label{tab:est}
\centering
\begin{tabular}{c|c|c}
\hline
\multicolumn{3}{c}{NNPDF2.3}  \\
\hline 
\hline 
 & NLO & NNLO  \\
\hline 
$\chi^{2}_{\tot}$   & 1.122   &   1.139    \\
$\la E \ra \pm \sigma_{E} $   &   2.17 $\pm$       0.05 & 2.19 $\pm$       0.07      \\
$\la E_{\rm tr} \ra \pm \sigma_{E_{\rm tr}}$&   2.15 $\pm$       0.07  &   2.17 $\pm$       0.08  \\
$\la E_{\rm val} \ra \pm \sigma_{E_{\rm val}}$&   2.20 $\pm$       0.07   &  2.24 $\pm$       0.10     \\
$\la{\rm TL} \ra \pm \sigma_{\rm TL}$   &  (24 $\pm$   16) $10^3$  & 
 (22 $\pm$   15) $10^3$ \\
\hline
$\la \chi^{2(k)} \ra \pm \sigma_{\chi^{2}} $  &    1.18 $\pm$       0.03   &
 1.21 $\pm$       0.04 \\
\hline
 $\la \sigma^{(\exp)}
\ra_{\dat}$(\%) &  12.1  &  12.2 \\
 $\la \sigma^{(\net)}
\ra_{\dat} $(\%)&   3.0  &  3.0\\
\hline
 $\la \rho^{(\exp)}
\ra_{\dat}$ &    0.18  &  0.18\\
 $\la \rho^{(\net)}
\ra_{\dat}$&  0.40  &  0.49\\
\hline
\end{tabular}
\caption{\small \label{tab:estfit1} Table of statistical estimators
  for the NNPDF2.3  NLO and NNLO fits with $N_{\rm rep}=
100$ replicas.  }
\end{table}

All tables and plots in this section will be produced using the PDF sets
which correspond to the value $\alpha_s(M_{\rm Z})=0.119$ (consistent with the
current PDG~\cite{PDG2012} value $\alpha_s(M_{\rm Z})=0.1184\pm0.0007$) 
for ease of comparison with previous NNPDF papers. A determination of
$\alpha_s$ based on the NNPDF2.1 parton set was performed both at
NLO~\cite{Lionetti:2011pw} and NNLO~\cite{Ball:2011us}.
However, the quality of the PDF fit is quite good for all values of
$\alpha_s$ considered here, and we see no indication of pathological
behaviour of PDFs for any value of $\alpha_s$. As a consequence, the
user can utilize the PDF set corresponding to the value of $\alpha_s$
of their choice. Combined PDF+$\alpha_s$ uncertainties may be determined
by combining replicas from sets corresponding to different values of
$\alpha_s$, as discussed in Sec.~3.2 of
Ref.~\cite{Demartin:2010er}. As a default, the current PDG value and
uncertainty may be used unless there are reasons to do otherwise: this
value is obtained by combining determinations of $\alpha_s$ at various
perturbative orders, and it is thus meant to be appropriate both at
NLO and NNLO.   

\subsection{Statistical features}
\label{sec-stat}

{
\begin{table}
\centering
\scriptsize
\begin{tabular}{c||c|c||c|c||c|c||c|c||c|c}
\hline 
& \multicolumn{2}{c||}{\bf NNPDF2.1} & \multicolumn{8}{|c}{\bf NNPDF2.3}  \\
\hline 
&  \multicolumn{2}{c||}{Global} &  \multicolumn{2}{|c||}{Global Fit} &
 \multicolumn{2}{|c||}{Global RW} &  \multicolumn{2}{|c||}{noLHC} &
 \multicolumn{2}{|c}{Collider}   \\
\hline 
\hline 
Experiment  & NLO & NNLO  & NLO  & NNLO  & NLO  & NNLO  & NLO  & NNLO  &
NLO  & NNLO   \\
\hline 
Total  &   1.145&   1.162 &    1.101 &    1.139 &    1.105 &    1.139 &    1.101 &    1.142 &    0.971 &    0.993 \\  
 \hline 
NMC-pd              & $          0.97      $ & $          0.93      $  &  $          0.95      $  &  $          0.95      $  &  $         0.93      $ &  $         0.93      $ &  $          0.93      $  &  $          0.94      $  &  $  \lc     5.33 \rc  $  &  $  \lc     5.13 \rc  $  \\  
NMC                 & $          1.68      $ & $          1.58      $  &  $          1.61      $  &  $          1.59      $  &  $         1.62      $ &  $         1.57      $ &  $          1.59      $  &  $          1.56      $  &  $  \lc     1.89 \rc  $  &  $  \lc     1.83 \rc  $  \\  
SLAC                & $          1.34      $ & $          1.04      $  &  $          1.24      $  &  $          1.00      $  &  $         1.27      $ &  $         1.01      $ &  $          1.28      $  &  $          1.04      $  &  $  \lc     1.72 \rc  $  &  $  \lc     1.41 \rc  $  \\  
BCDMS               & $          1.21      $ & $          1.29      $  &  $          1.20      $  &  $          1.28      $  &  $         1.20      $ &  $         1.28      $ &  $          1.20      $  &  $          1.28      $  &  $  \lc     1.85 \rc  $  &  $  \lc     2.15 \rc  $  \\  
CHORUS              & $          1.10      $ & $          1.08      $  &  $          1.10      $  &  $          1.07      $  &  $         1.10      $ &  $         1.06      $ &  $          1.09      $  &  $          1.07      $  &  $  \lc     1.73 \rc  $  &  $  \lc     1.70 \rc  $  \\  
NTVDMN              & $          0.70      $ & $          0.50      $  &  $          0.43      $  &  $          0.56      $  &  $         0.42      $ &  $         0.51      $ &  $          0.42      $  &  $          0.48      $  &  $  \lc    26.69 \rc  $  &  $  \lc    21.13 \rc  $  \\  
 \hline
HERAI-AV            & $          1.04      $ & $          1.04      $  &  $          1.00      $  &  $          1.01      $  &  $         1.00      $ &  $         1.02      $ &  $          1.01      $  &  $          1.03      $  &  $          0.97      $  &  $          0.99      $  \\  
FLH108              & $          1.34      $ & $          1.23      $  &  $          1.29      $  &  $          1.20      $  &  $         1.29      $ &  $         1.20      $ &  $          1.29      $  &  $          1.21      $  &  $          1.35      $  &  $          1.25      $  \\  
ZEUS-H2             & $          1.21      $ & $          1.21      $  &  $          1.20      $  &  $          1.22      $  &  $         1.20      $ &  $         1.22      $ &  $          1.20      $  &  $          1.22      $  &  $          1.29      $  &  $          1.32      $  \\  
ZEUS $F_2^c$        & $          0.75      $ & $          0.81      $  &  $          0.82      $  &  $          0.90      $  &  $         0.80      $ &  $         0.90      $ &  $          0.81      $  &  $          0.86      $  &  $          0.71      $  &  $          0.77      $  \\  
H1 $F_2^c$          & $          1.50      $ & $          1.44      $  &  $          1.59      $  &  $          1.53      $  &  $         1.57      $ &  $         1.52      $ &  $          1.58      $  &  $          1.49      $  &  $          1.33      $  &  $          1.30      $  \\  
 \hline
DYE605              & $          0.94      $ & $          1.08      $  &  $          0.86      $  &  $          1.04      $  &  $         0.88      $ &  $         1.04      $ &  $          0.85      $  &  $          1.06      $  &  $  \lc     3.58 \rc  $  &  $  \lc     1.02 \rc  $  \\  
DYE886              & $          1.42      $ & $          1.69      $  &  $          1.27      $  &  $          1.58      $  &  $         1.27      $ &  $         1.55      $ &  $          1.24      $  &  $          1.55      $  &  $  \lc     5.65 \rc  $  &  $  \lc     5.14 \rc  $  \\  
 \hline
CDF $W$ asy         & $          1.88      $ & $          1.63      $  &  $          1.57      $  &  $          1.64      $  &  $         1.57      $ &  $         1.72      $ &  $          1.45      $  &  $          1.67      $  &  $          1.05      $  &  $          1.21      $  \\  
CDF $Z$ rap         & $          1.77      $ & $          2.38      $  &  $          1.80      $  &  $          2.03      $  &  $         1.77      $ &  $         2.17      $ &  $          1.76      $  &  $          2.13      $  &  $          1.32      $  &  $          1.37      $  \\  
D0 $Z$ rap          & $          0.57      $ & $          0.67      $  &  $          0.56      $  &  $          0.61      $  &  $         0.57      $ &  $         0.63      $ &  $          0.57      $  &  $          0.63      $  &  $          0.56      $  &  $          0.58      $  \\  
ATLAS $W,Z$         & $  \lc     1.57 \rc  $ & $  \lc     2.21 \rc  $  &  $          1.26      $  &  $          1.43      $  &  $         1.31      $ &  $         1.65      $ &  $  \lc     1.37 \rc  $  &  $  \lc     1.94 \rc  $  &  $          1.02      $  &  $          1.05      $  \\  
CMS $W$ el asy      & $  \lc     2.02 \rc  $ & $  \lc     1.27 \rc  $  &  $          0.82      $  &  $          0.81      $  &  $         1.09      $ &  $         0.99      $ &  $  \lc     1.32 \rc  $  &  $  \lc     1.20 \rc  $  &  $          0.87      $  &  $          0.85      $  \\  
LHCb $W$            & $  \lc     0.89 \rc  $ & $  \lc     1.13 \rc  $  &  $          0.67      $  &  $          0.83      $  &  $         0.77      $ &  $         0.98      $ &  $  \lc     0.76 \rc  $  &  $  \lc     1.03 \rc  $  &  $          0.74      $  &  $          0.72      $  \\  
 \hline
CDF RII $k_T$       & $          0.68      $ & $          0.65      $  &  $          0.60      $  &  $          0.68      $  &  $         0.61      $ &  $         0.67      $ &  $          0.60      $  &  $          0.67      $  &  $          0.60      $  &  $          0.59      $  \\  
D0 RII cone         & $          0.90      $ & $          0.98      $  &  $          0.84      $  &  $          0.94      $  &  $         0.84      $ &  $         0.93      $ &  $          0.84      $  &  $          0.94      $  &  $          0.85      $  &  $          0.92      $  \\  
ATLAS jets          & $  \lc     1.06 \rc  $ & $  \lc     0.95 \rc  $  &  $          1.00      $  &  $          0.94      $  &  $         1.00      $ &  $         0.92      $ &  $  \lc     1.01 \rc  $  &  $  \lc     0.94 \rc  $  &  $          0.98      $  &  $          0.93      $  \\  

\hline
\end{tabular}
\caption{\small \label{tab:estfit2dataset} The $\chi^2$ values
per data point for individual experiments computed using 
the default NNPDF2.1 NLO and NNLO PDF and
 various   NNPDF2.3 PDF sets. All  $\chi^2$ values have been
obtained using  $N_{\rm rep}$=100 replicas with
  $\alpha_s(M_{\rm Z})=0.119$. Normalization uncertainties have been
  included using the experimental covariance matrix
(note that in Tab.~\ref{tab:est} the $t_0$~\cite{Ball:2009qv}
covariance matrix for normalization uncertainties is used). Values in square 
brackets correspond to experiments not included in the
corresponding fit: these are not included in the total $\chi^2$. 
}

\end{table}
}

In Tab.~\ref{tab:estfit1} we summarize the statistical estimators for
the NNPDF2.3 NLO and NNLO fits with $N_{\rm rep}= 100$ replicas.
A detailed discussion of statistical indicators and their meaning can
be found in
Refs.~\cite{Ball:2010de,Ball:2011mu,phystat,Ball:2009qv}; here we
merely recall that  $\chi^{2}_{\tot}$ is computed by comparing
the central (average) NNPDF2.3 fit to the original experimental data,
$\la \chi^{2(k)} \ra$ is computed by comparing each NNPDF2.3 replica
to the data and averaging over replicas, while $\la E \ra$ is the
quantity that is actually minimized, i.e.~it coincides with the
$\chi^{2}$ computed by comparing each NNPDF2.3 replica to the data
replica it is fitted to, with the three values given corresponding to
the total, training, and validation data sets. All these estimators, 
including in particular the $\chi^2$, are normalized to the relevant 
number of data points. When comparing two
different fits we will also show distances between central values and
uncertainties, computed using
sets of $N_\mathrm{rep}=100$ replicas: in this respect, recall that 
two points extracted from distributions that differ by one standard deviation have an average  distance
$d=10$, while two points extracted from the same distribution have an
average distance $d=1$ (the difference being due to the fact that
 the standard deviation of the mean is by a factor $\sqrt{N_{\rm rep}}$
   smaller than  the standard deviation of the distribution).

In Tab.~\ref{tab:estfit2dataset} we compare the $\chi^2_\tot$ for
NNPDF2.1 and NNPDF2.3 PDF sets both at NLO and at NNLO. The $\chi^2$
of each data set is also shown. For the NNPDF2.3 fit,
we also show the values for the noLHC and the collider only fits, and
the $\chi^2$ obtained by reweighting the NNPDF2.3
noLHC PDF sets with the new LHC data, using $500$
replicas.  The noLHC sets
are discussed in detail in Sec.~\ref{sec-comp} and also
Sec.~\ref{sec-diff} where they are also used to construct reweighted
sets,
while the collider
only PDF sets are discussed in Sec.~\ref{sec-reduced}.

It should be noticed that all the the statistical estimators of
Tab.~\ref{tab:estfit1}, and specifically the $\chi^2$ are determined
using the $t_0$ method for the treatment of normalization
uncertainties~\cite{Ball:2009qv}, while all $\chi^2$ values in 
Tab.~\ref{tab:estfit2dataset} are
computed using the experimental covariance matrix: the former is
needed for unbiased minimization, while the latter yields a measure of
the goodness-of-fit. The $t_0$ method for fitting normalization uncertainties
leads to statistically unbiased results, and it does not require the
ad-hoc use of quartic penalties in the treatment of normalization
uncertainties as often required when these are treated by means of
an offset method (see e.g. Ref.~\cite{Watt:2012tq}). However, recent
benchmarking exercises suggest that in the context of the existing
global fits  results obtained the $t_0$ and offset
method  are very close to each other~\cite{pavelpriv}.

We have checked that the difference between the 
two values ($t_0$ and experimental
covariance matrix) is of the order of the expected statistical fluctuation of
the total $\chi^2$ (i.e.~of order of $1/\sqrt{N_{\rm dat}}$ for the
$\chi^2$ per data point as given here). It should be noticed
that the $\chi^2$ values for the NNPDF2.1 NLO and NNLO fits
differ from those given in Tab.~4
of~\cite{Ball:2011mu} and Tab.~6 of~\cite{Ball:2011uy} respectively,
both because the latter were given using the $t_0$ definition of the
covariance matrix, and also because they were computed using a set
of 1000 replicas, while only 100 replica sets are used for all sets
of Tab.~\ref{tab:estfit2dataset}. Finally, the $\chi^2$ value
reported in Refs.~\cite{Ball:2009qv,Ball:2011uy} for  NTVDMN
(NuTeV dimuons) was affected by an error, to be discussed in
Eq.~(\ref{eq:bug}) below. 

{
\begin{table}
\centering
\footnotesize
\begin{tabular}{c|c||c|c||c|c||c|c||c|c}
\hline 
& & \multicolumn{2}{c||}{\bf NNPDF2.1} & \multicolumn{6}{|c}{\bf NNPDF2.3}  \\
\hline 
& & \multicolumn{2}{c||}{Global} &  \multicolumn{2}{|c||}{Global Fit}  &  \multicolumn{2}{|c||}{noLHC} &
 \multicolumn{2}{|c}{Collider}   \\
\hline 
\hline 
Set  & Data & NLO & NNLO  & NLO  & NNLO  & NLO  & NNLO  &
NLO  & NNLO   \\
\hline 
NMC-pd              &    1.9& $          0.6      $ & $          0.5      $  &  $          0.5      $  &  $          0.5      $  &  $          0.5      $  &  $          0.5      $  &  $  \lc     3.5 \rc  $  &  $  \lc     2.9 \rc  $  \\  
NMC                 &    5.1& $          1.6      $ & $          1.3      $  &  $          1.1      $  &  $          1.2      $  &  $          1.2      $  &  $          1.2      $  &  $  \lc     1.9 \rc  $  &  $  \lc     1.9 \rc  $  \\  
SLACp               &    4.6& $          1.6      $ & $          1.3      $  &  $          1.2      $  &  $          1.3      $  &  $          1.2      $  &  $          1.3      $  &  $  \lc     2.5 \rc  $  &  $  \lc     2.5 \rc  $  \\  
SLACd               &    4.0& $          1.7      $ & $          1.5      $  &  $          1.2      $  &  $          1.4      $  &  $          1.3      $  &  $          1.4      $  &  $  \lc     4.2 \rc  $  &  $  \lc     3.9 \rc  $  \\  
BCDMSp              &    5.2& $          2.7      $ & $          2.1      $  &  $          1.6      $  &  $          1.9      $  &  $          1.8      $  &  $          1.8      $  &  $  \lc     4.6 \rc  $  &  $  \lc     4.5 \rc  $  \\  
BCDMSd              &    6.4& $          3.3      $ & $          2.4      $  &  $          1.8      $  &  $          2.3      $  &  $          2.2      $  &  $          2.1      $  &  $  \lc     6.7 \rc  $  &  $  \lc     6.2 \rc  $  \\  
CHORUSnu            &   10.6& $          2.6      $ & $          2.1      $  &  $          1.9      $  &  $          2.0      $  &  $          2.0      $  &  $          2.1      $  &  $  \lc     6.9 \rc  $  &  $  \lc     6.7 \rc  $  \\  
CHORUSnb            &   19.6& $         13.6      $ & $          5.1      $  &  $          5.3      $  &  $          4.5      $  &  $          5.4      $  &  $          5.1      $  &  $  \lc    36.3 \rc  $  &  $  \lc    22.8 \rc  $  \\  
NTVnuDMN            &   16.4& $         12.8      $ & $         14.1      $  &  $         12.1      $  &  $         12.2      $  &  $         13.9      $  &  $         12.9      $  &  $  \lc    40.0 \rc  $  &  $  \lc    35.4 \rc  $  \\  
NTVnbDMN            &   25.8& $         16.6      $ & $         16.9      $  &  $         13.8      $  &  $         15.9      $  &  $         16.2      $  &  $         14.3      $  &  $  \lc    43.6 \rc  $  &  $  \lc    39.2 \rc  $  \\  
 \hline
HERA1-NCep          &    4.6& $          1.3      $ & $          1.2      $  &  $          0.9      $  &  $          1.0      $  &  $          1.0      $  &  $          1.1      $  &  $          1.6      $  &  $          1.6      $  \\  
HERA1-NCem          &   11.3& $          1.2      $ & $          1.0      $  &  $          0.8      $  &  $          0.9      $  &  $          0.9      $  &  $          1.0      $  &  $          1.5      $  &  $          1.6      $  \\  
HERA1-CCep          &   12.6& $          2.0      $ & $          2.0      $  &  $          1.6      $  &  $          1.7      $  &  $          1.8      $  &  $          1.9      $  &  $          3.9      $  &  $          3.7      $  \\  
HERA1-CCem          &   21.1& $          1.2      $ & $          1.1      $  &  $          0.9      $  &  $          1.0      $  &  $          1.0      $  &  $          1.1      $  &  $          2.0      $  &  $          1.8      $  \\  
FLH108              &   67.9& $          4.9      $ & $          5.1      $  &  $          3.4      $  &  $          4.0      $  &  $          4.1      $  &  $          4.0      $  &  $          6.8      $  &  $          6.6      $  \\  
Z06NC               &    6.7& $          1.1      $ & $          0.9      $  &  $          0.8      $  &  $          0.9      $  &  $          0.8      $  &  $          0.9      $  &  $          1.3      $  &  $          1.4      $  \\  
Z06CC               &   32.0& $          1.9      $ & $          1.7      $  &  $          1.2      $  &  $          1.5      $  &  $          1.4      $  &  $          1.5      $  &  $          2.9      $  &  $          2.9      $  \\  
ZEUS $F_2^c$ 99     &   21.2& $          3.2      $ & $          3.2      $  &  $          2.0      $  &  $          2.5      $  &  $          2.3      $  &  $          2.7      $  &  $          2.4      $  &  $          2.8      $  \\  
ZEUS $F_2^c$ 03     &   20.4& $          3.2      $ & $          3.2      $  &  $          2.1      $  &  $          2.4      $  &  $          2.4      $  &  $          2.6      $  &  $          2.6      $  &  $          2.8      $  \\  
ZEUS $F_2^c$ 08     &   29.9& $          2.9      $ & $          2.6      $  &  $          1.9      $  &  $          2.0      $  &  $          2.1      $  &  $          2.3      $  &  $          2.4      $  &  $          2.4      $  \\  
ZEUS $F_2^c$ 09     &   27.6& $          2.8      $ & $          2.6      $  &  $          1.8      $  &  $          2.2      $  &  $          2.0      $  &  $          2.5      $  &  $          2.3      $  &  $          2.6      $  \\  
H1 $F_2^c$ 01       &   25.2& $          2.8      $ & $          3.2      $  &  $          1.9      $  &  $          2.4      $  &  $          2.1      $  &  $          2.5      $  &  $          2.1      $  &  $          2.7      $  \\  
H1 $F_2^c$ 09       &   21.3& $          2.6      $ & $          2.5      $  &  $          1.8      $  &  $          2.1      $  &  $          1.9      $  &  $          2.3      $  &  $          2.2      $  &  $          2.4      $  \\  
H1 $F_2^c$ 10       &   14.0& $          2.8      $ & $          2.5      $  &  $          2.0      $  &  $          2.0      $  &  $          2.1      $  &  $          2.1      $  &  $          2.5      $  &  $          2.5      $  \\  
 \hline
DYE605              &   22.9& $          8.6      $ & $          7.4      $  &  $          7.0      $  &  $          7.6      $  &  $          6.4      $  &  $          6.8      $  &  $  \lc   134.6 \rc  $  &  $  \lc    85.5 \rc  $  \\  
DYE886p             &   21.8& $         10.3      $ & $          9.7      $  &  $          8.7      $  &  $          8.6      $  &  $          9.1      $  &  $          9.2      $  &  $  \lc   107.0 \rc  $  &  $  \lc    64.1 \rc  $  \\  
DYE886r             &    4.3& $          2.9      $ & $          3.1      $  &  $          2.6      $  &  $          2.8      $  &  $          2.8      $  &  $          2.9      $  &  $  \lc   101.5 \rc  $  &  $  \lc    50.2 \rc  $  \\  
 \hline
CDF $W$ asy         &    5.9& $          4.2      $ & $          4.0      $  &  $          3.3      $  &  $          3.2      $  &  $          3.6      $  &  $          3.7      $  &  $          4.8      $  &  $          4.7      $  \\  
CDF $Z$ rap         &   11.5& $          3.7      $ & $          3.1      $  &  $          2.3      $  &  $          2.9      $  &  $          2.7      $  &  $          3.0      $  &  $          4.5      $  &  $          3.9      $  \\  
D0 $Z$ rap          &   10.1& $          3.0      $ & $          2.5      $  &  $          1.9      $  &  $          2.3      $  &  $          2.1      $  &  $          2.5      $  &  $          3.6      $  &  $          3.2      $  \\  
ATLAS $W,Z$         &    4.4& $  \lc     2.4 \rc  $ & $  \lc     2.0 \rc  $  &  $          1.4      $  &  $          1.5      $  &  $  \lc     2.1 \rc  $  &  $  \lc     2.2 \rc  $  &  $          1.9      $  &  $          1.9      $  \\  
CMS $W$ el asy      &    5.2& $  \lc     4.6 \rc  $ & $  \lc     5.5 \rc  $  &  $          2.4      $  &  $          3.1      $  &  $  \lc     7.0 \rc  $  &  $  \lc     5.9 \rc  $  &  $          3.1      $  &  $          3.0      $  \\  
LHCb $W$            &   11.6& $  \lc     4.4 \rc  $ & $  \lc     2.9 \rc  $  &  $          2.4      $  &  $          2.5      $  &  $  \lc     2.7 \rc  $  &  $  \lc     2.9 \rc  $  &  $          6.0      $  &  $          5.2      $  \\  
 \hline
CDF RII $k_T$       &   21.7& $          5.6      $ & $          4.7      $  &  $          4.1      $  &  $          4.2      $  &  $          4.6      $  &  $          4.6      $  &  $          5.7      $  &  $          5.3      $  \\  
D0 RII cone         &   16.9& $          6.4      $ & $          5.3      $  &  $          4.8      $  &  $          4.9      $  &  $          5.5      $  &  $          5.2      $  &  $          6.7      $  &  $          6.2      $  \\  
ATLAS jets          &   27.4& $  \lc     3.7 \rc  $ & $  \lc     3.0 \rc  $  &  $          2.7      $  &  $          2.7      $  &  $  \lc     3.0 \rc  $  &  $  \lc     3.0 \rc  $  &  $          4.1      $  &  $          3.6      $  \\  

\hline
\end{tabular}
\caption{\small \label{tab:estfitsigdataset} The average percentage
value of the experimental uncertainty  $\la \sigma^{(\exp)}
\ra_{\dat}$ and of the PDF uncertainty  $\la \sigma^{(\net)}
\ra_{\dat}$ for each data set, for all the NNPDF2.1
and NNPDF2.3 NLO and NNLO PDF sets.  All the $\la \sigma^{(\exp)}
\ra_{\dat}$ values have been
obtained including normalization uncertainties
 using the experimental covariance matrix. Values in square brackets  
correspond to experiments not included in the
corresponding fit.
}

\end{table}
}

In Tab.~\ref{tab:estfitsigdataset} we compare the average uncertainties 
$\la \sigma^{(\exp)}\ra_{\dat}$ on the experimental data, for each separate 
data set, to the average uncertainties $\la \sigma^{(\net)}\ra_{\dat}$ on 
the predictions for those data due to PDF uncertainties, obtained using 
each of the various PDF sets. Clearly these are rather smaller than the 
experimental uncertainties, due to the extra information coming from 
the other data sets, but it is interesting to see how they compare 
between NLO and NNLO, and between different fits. 

\begin{figure}[t]
\begin{center}
\epsfig{width=0.49\textwidth,figure=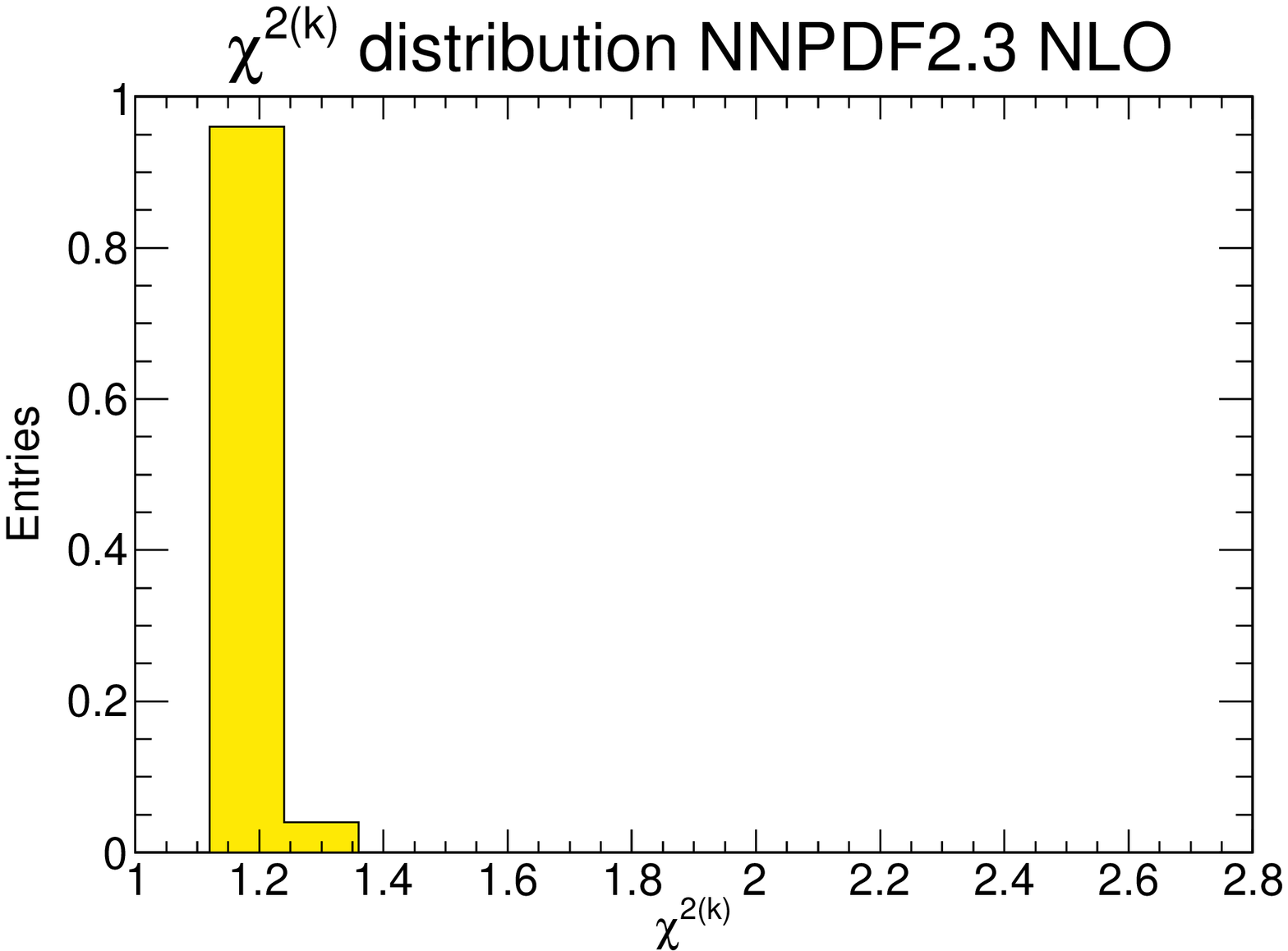}
\epsfig{width=0.49\textwidth,figure=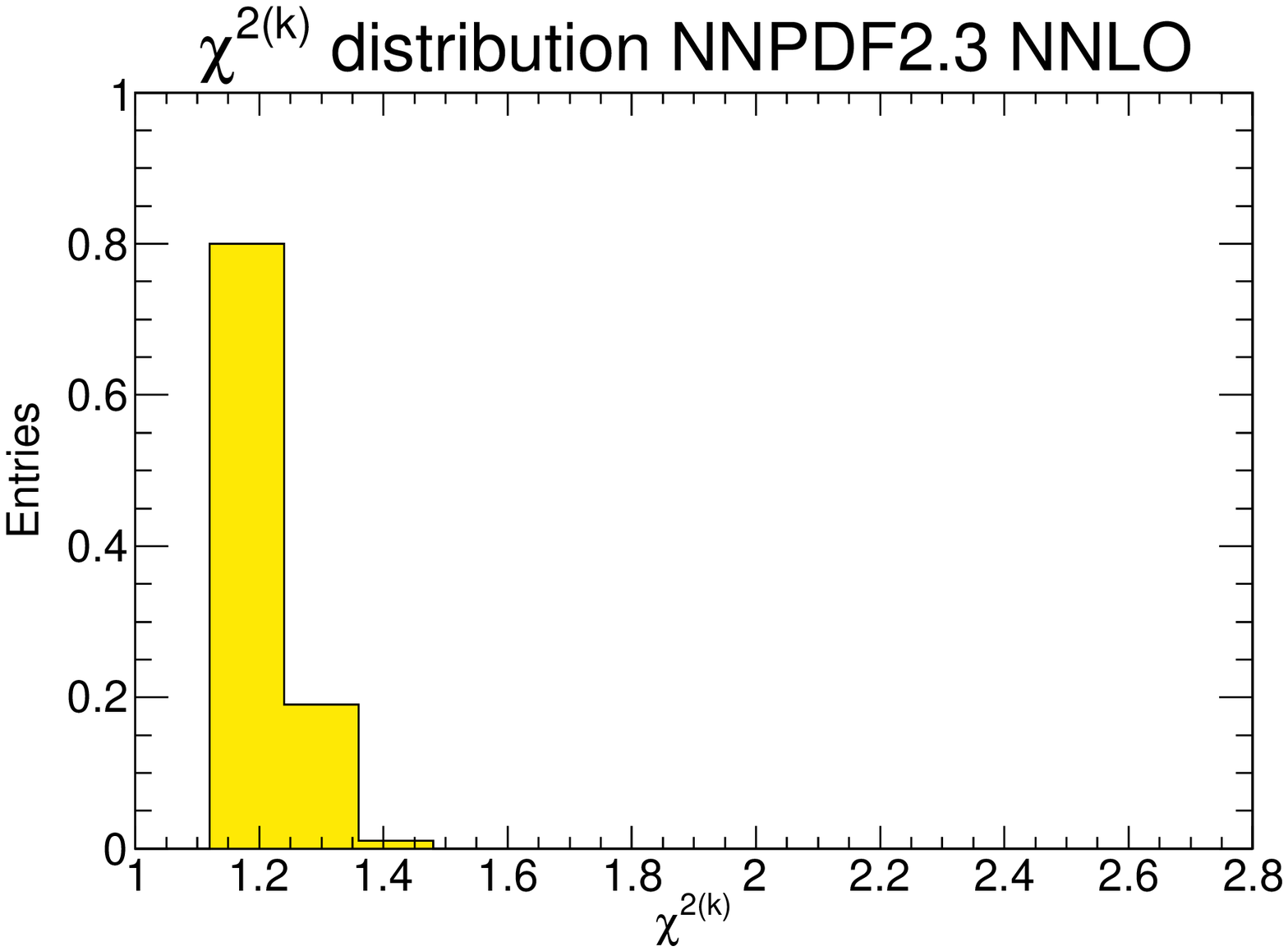}
\epsfig{width=0.49\textwidth,figure=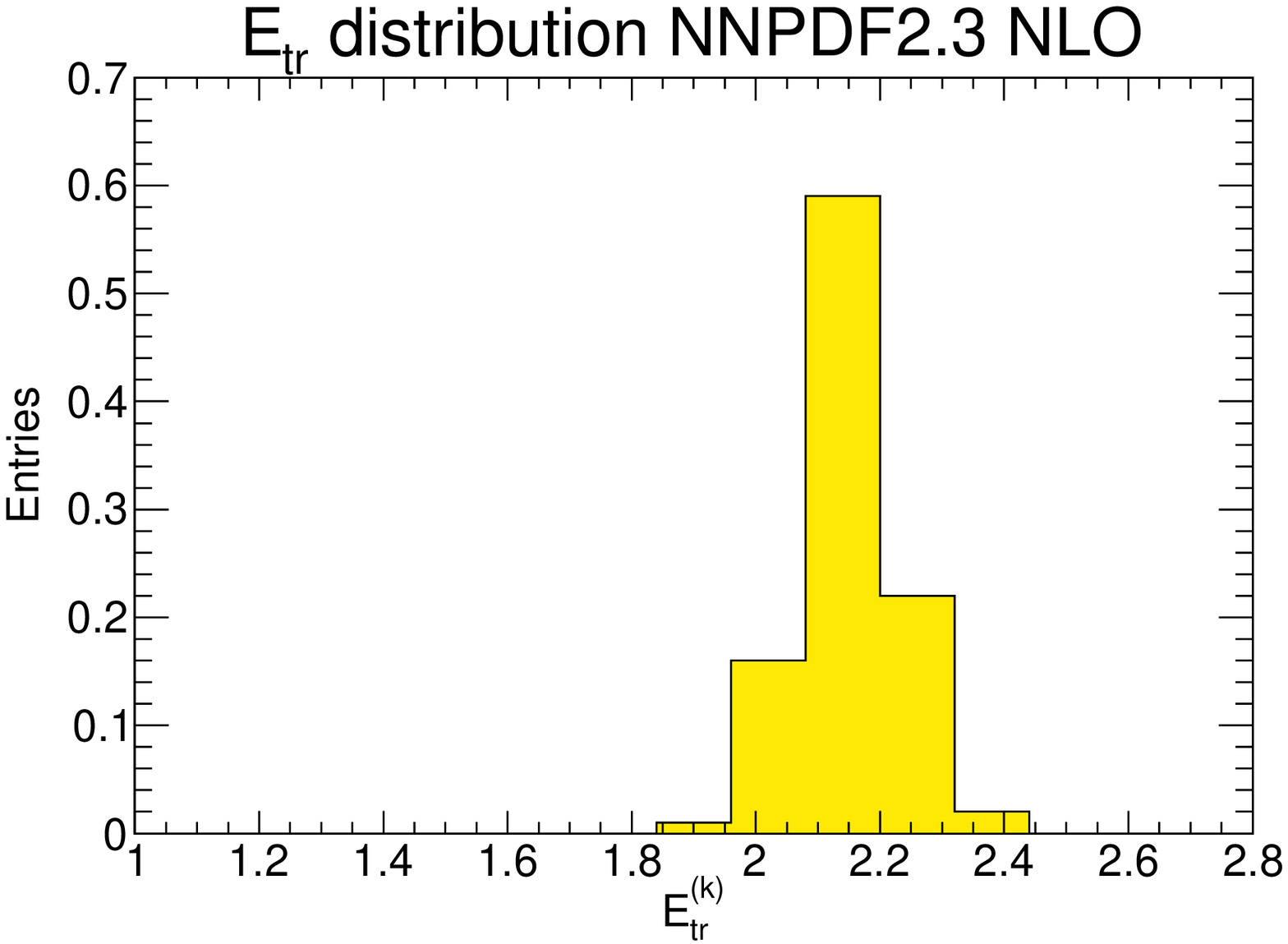}
\epsfig{width=0.49\textwidth,figure=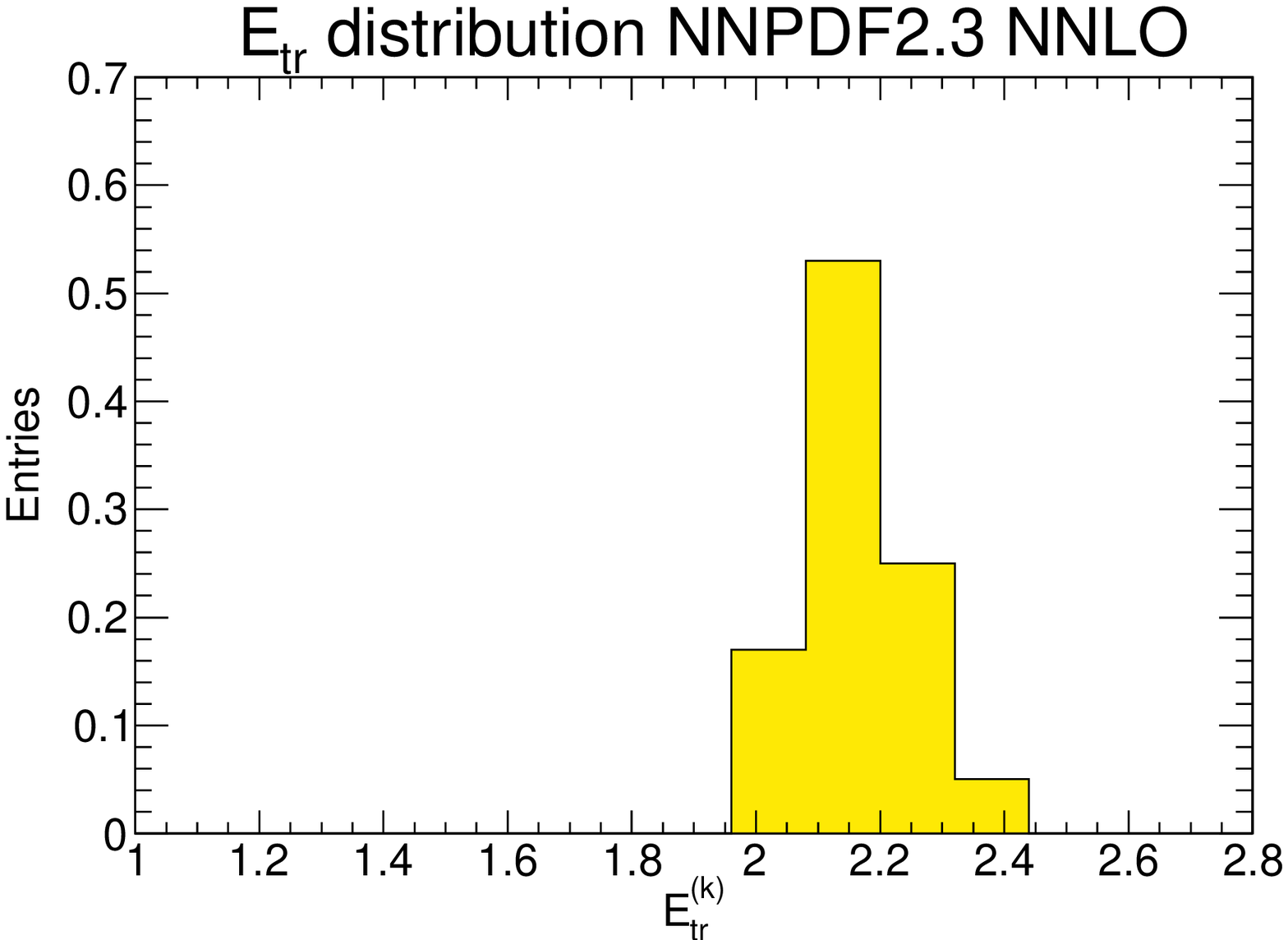}
\caption{\small Distribution of $\chi^{2(k)}$ (upper plots) and  $E^{(k)}_{\rm tr}$ (lower plots),
over the sample of $N_\mathrm{rep}=100$ replicas, for the NNPDF2.3 NLO
(left plots) and NNLO (right plots) PDF sets. \label{chi2histoplots}} 
\end{center}
\end{figure}

\begin{figure}[ht!]
\begin{center}
\epsfig{width=0.49\textwidth,figure=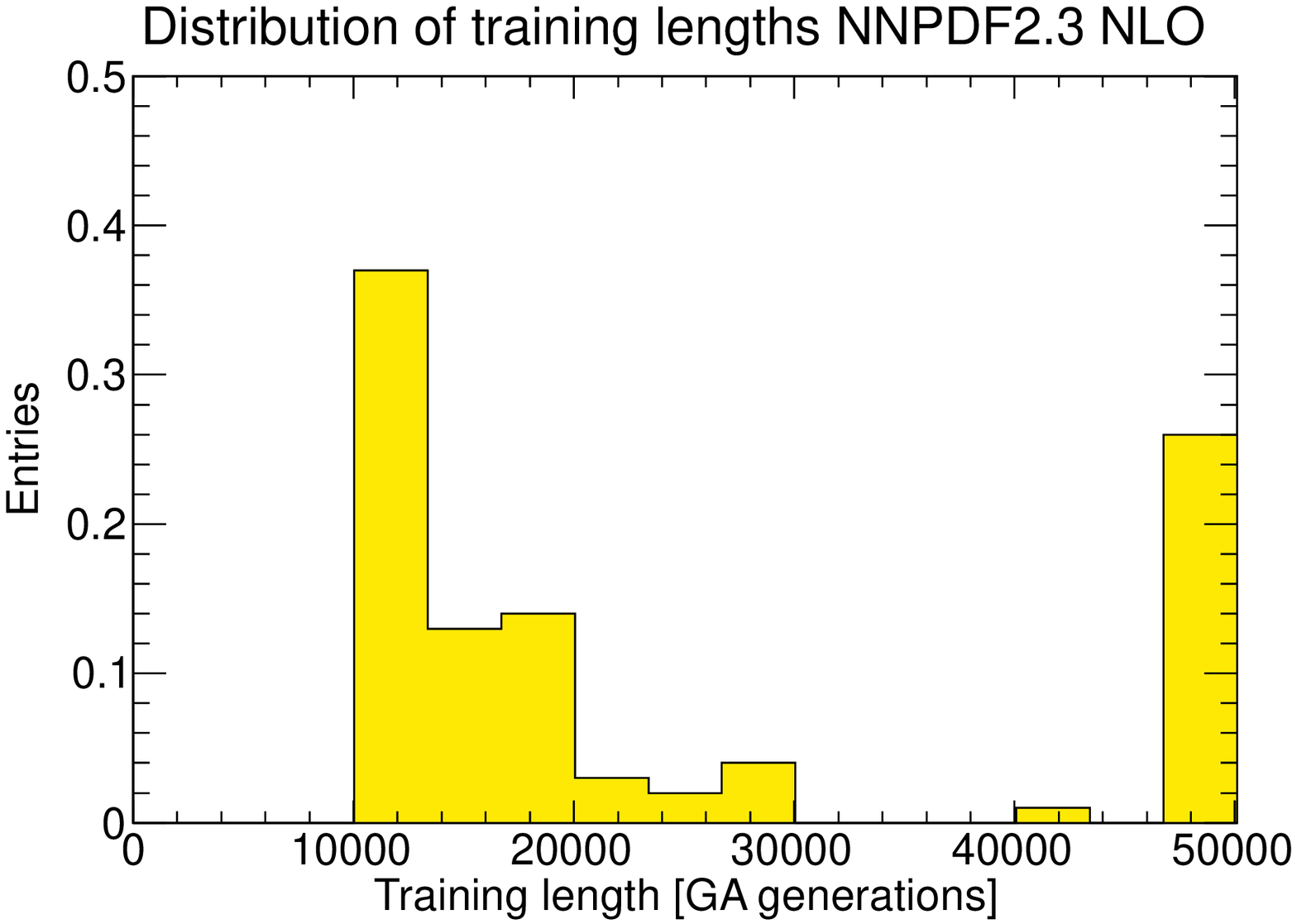}
\epsfig{width=0.49\textwidth,figure=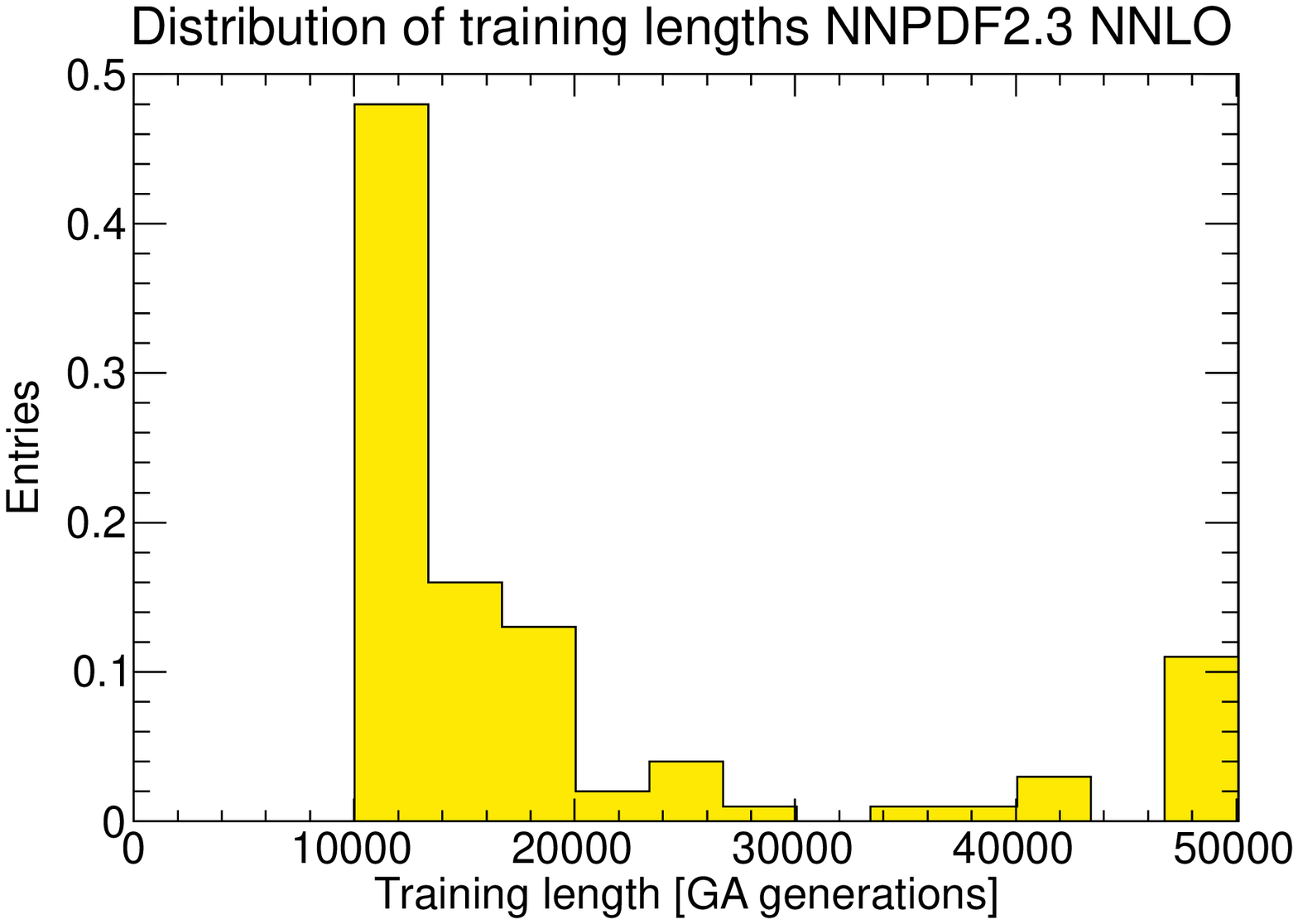}
\caption{\small Distribution of training lengths over the sample of
  $N_{\mathrm rep}=100$ replicas for the NNPDF2.3 NLO
(left plot) and NNPDF2.3 NNLO (right plot) PDF sets.  
\label{fig:tl}} 
\end{center}
\end{figure}

The distribution of $\chi^{2(k)}$, $E^{(k)}_{\rm tr}$, and training
lengths among the $100$ NNPDF2.3 NLO and NNLO replicas are
shown in Fig.~\ref{chi2histoplots} and Fig.~\ref{fig:tl} respectively.
While most of the replicas fulfil the stopping criterion, a fraction
($\sim 20\%$) of them stops at the maximum training length $N_{\rm
  gen}^{\rm max}$ which has been introduced in order to avoid
unacceptably long fits. As in previous PDF determinations,  we have 
explicitly verified that if we were to  
discard all replicas that do not stop dynamically,  PDFs change
by an amount which is smaller than a typical statistical
fluctuation.

\subsection{NNPDF2.3 parton distributions}
\label{sec-pdf}

\begin{figure}[t]
\begin{center}
\epsfig{width=0.49\textwidth,figure=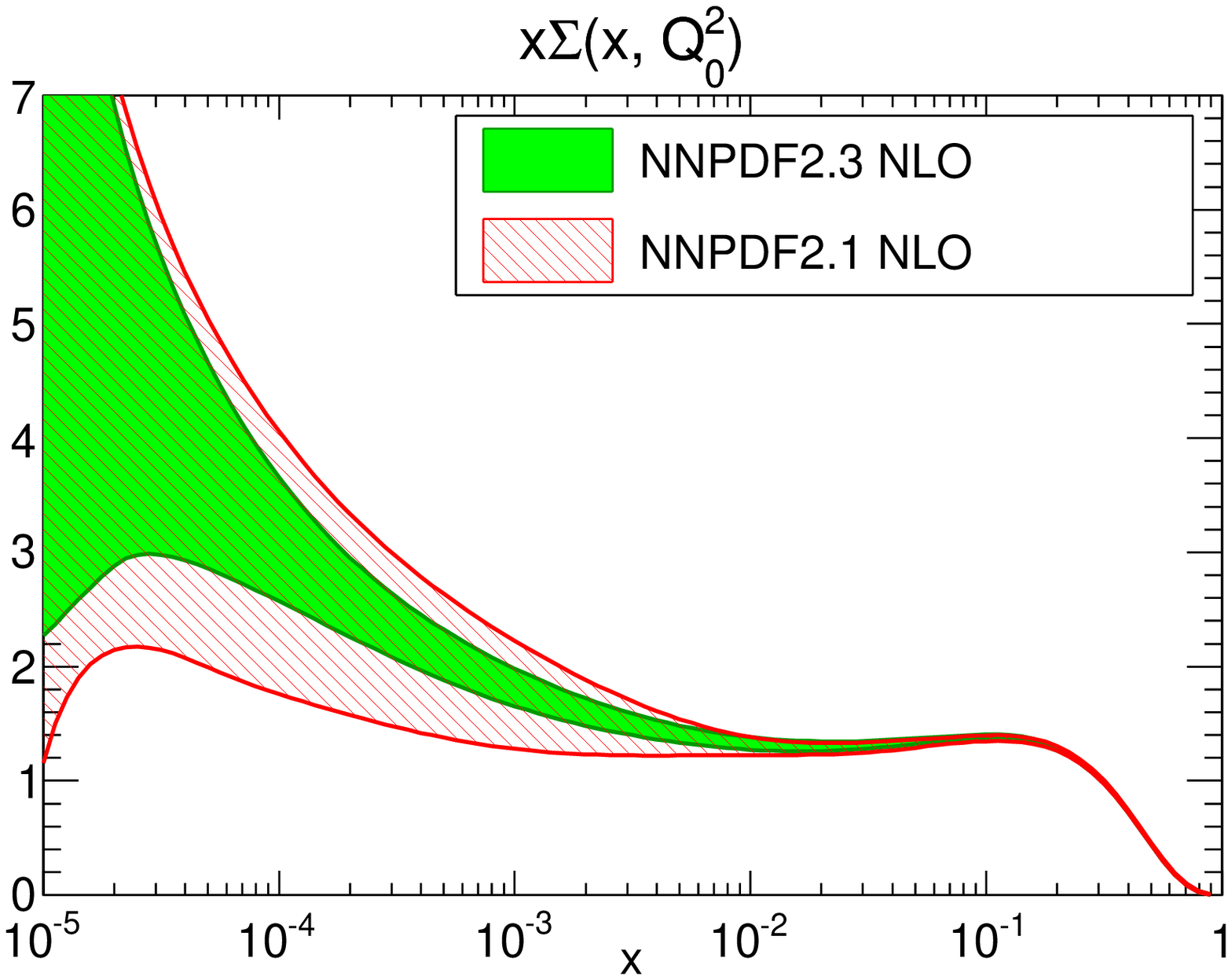}
\epsfig{width=0.49\textwidth,figure=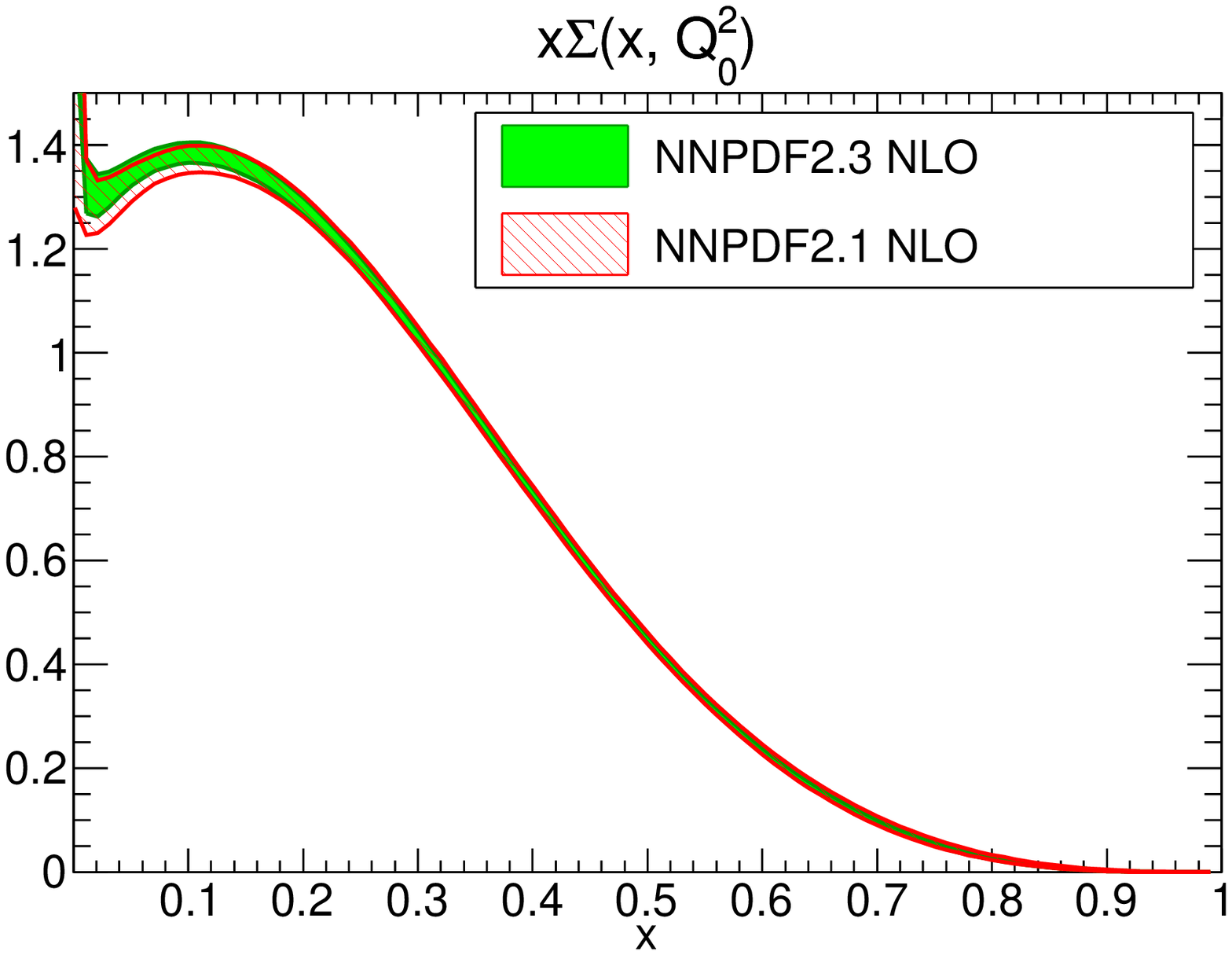}
\epsfig{width=0.49\textwidth,figure=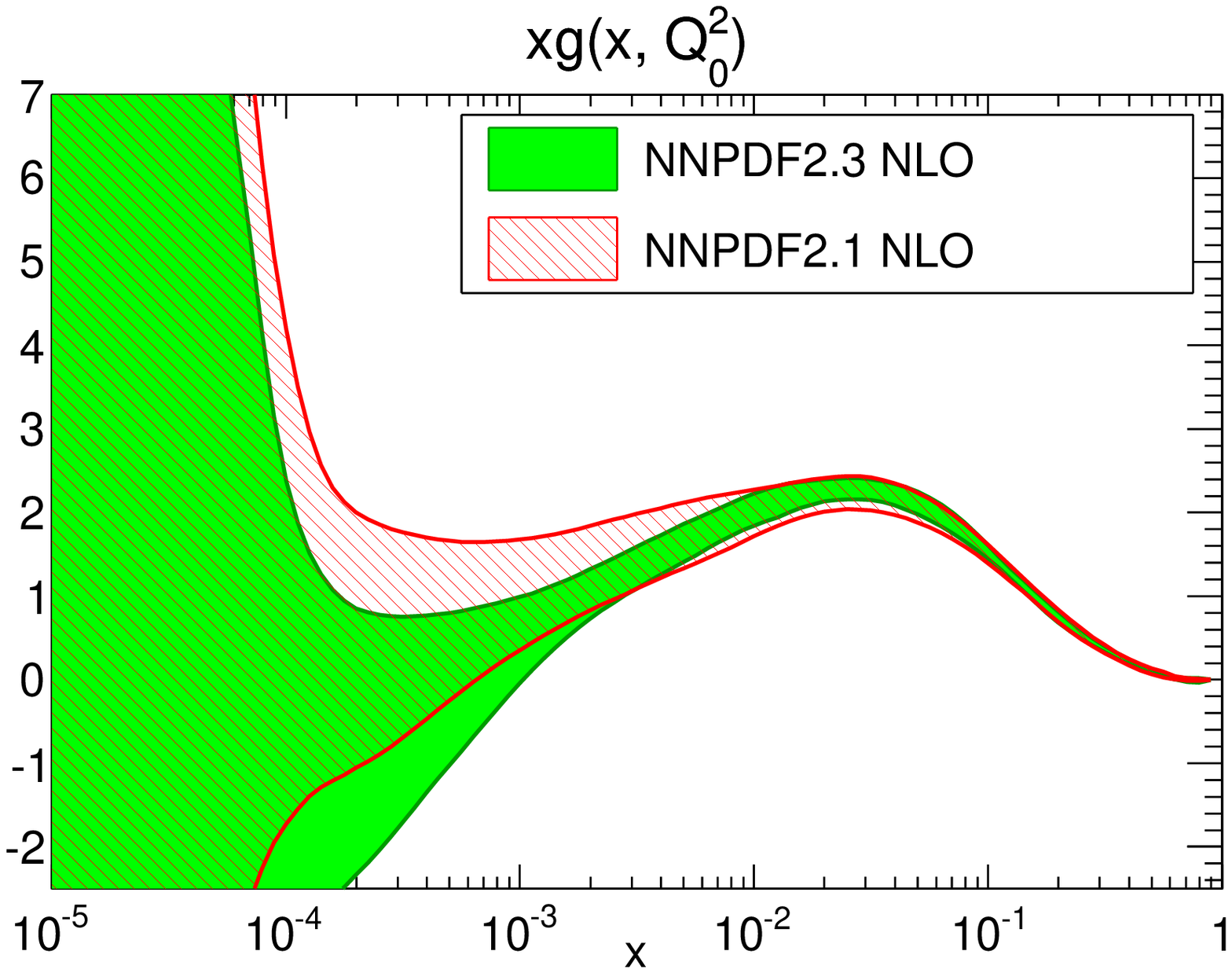}
\epsfig{width=0.49\textwidth,figure=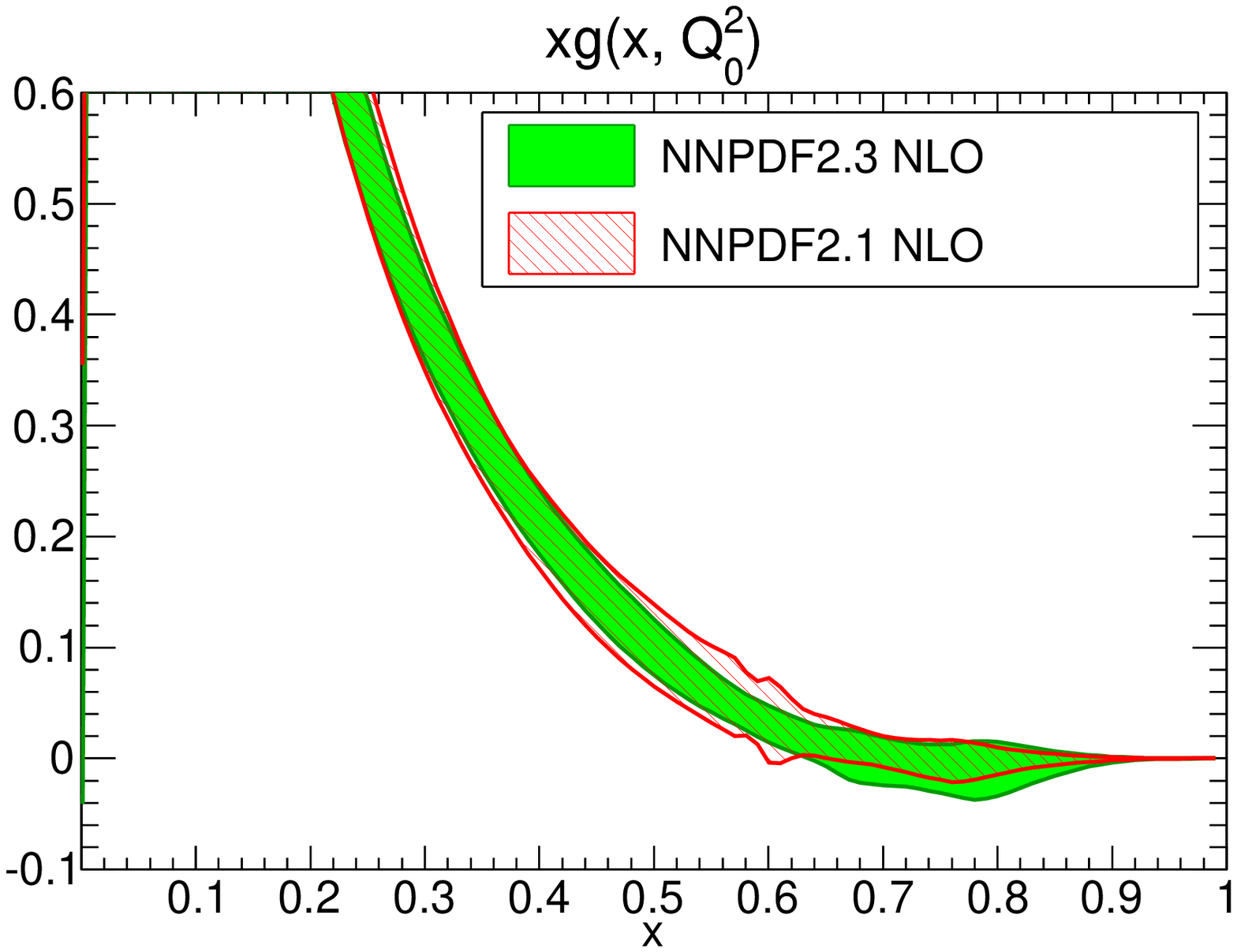}
\caption{\small NLO NNPDF2.3   singlet sector PDFs at $Q^2=2$ GeV$^2$, 
compared to their
  NNPDF2.1 counterparts, computed using $N_{\rm rep}=100$ replicas from both sets. All
  error bands shown correspond to a one sigma interval.
 \label{fig:singletPDFs}} 
\end{center}
\vskip-0.5cm
\end{figure}

\begin{figure}[t!]
\begin{center}
\epsfig{width=0.49\textwidth,figure=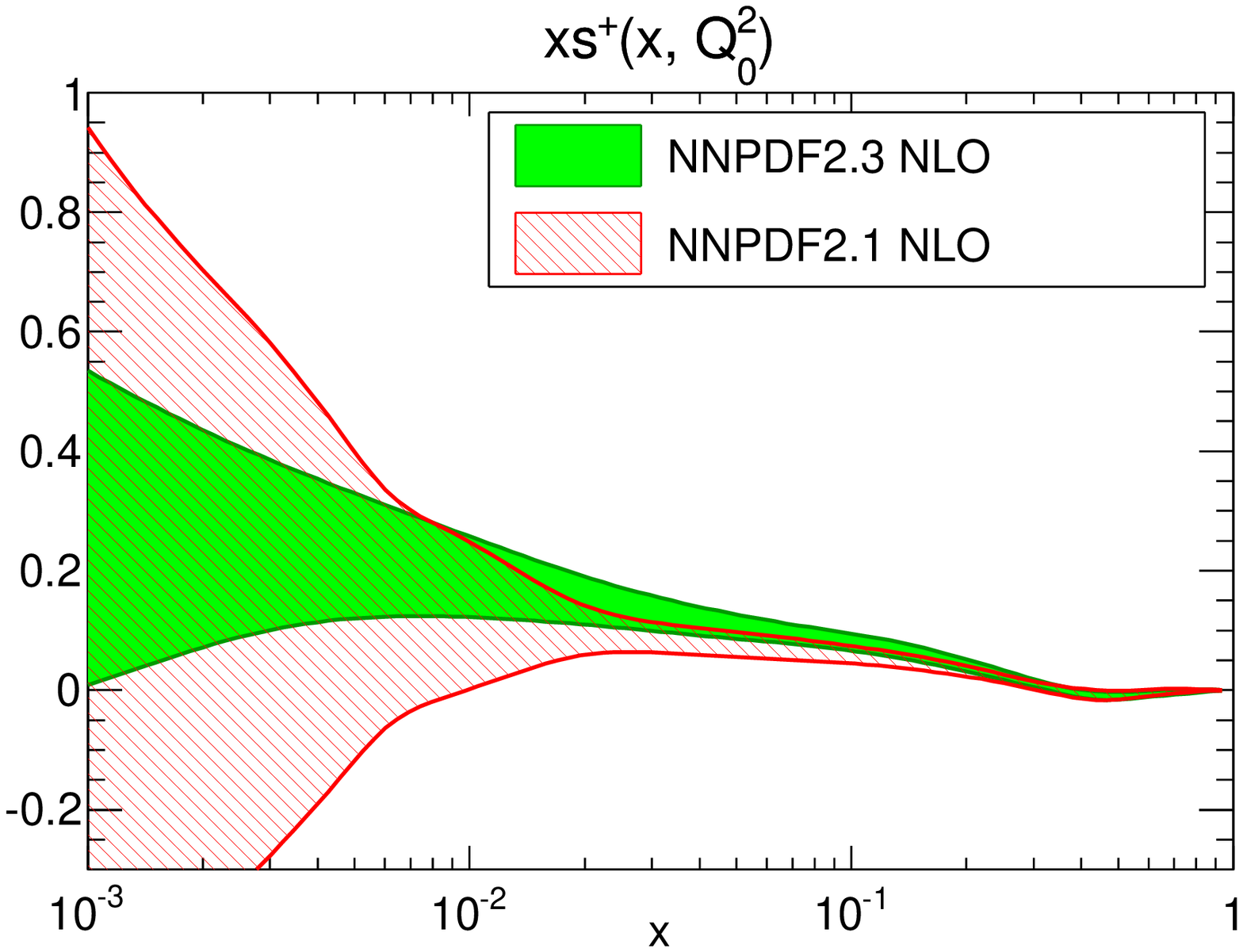}
\epsfig{width=0.49\textwidth,figure=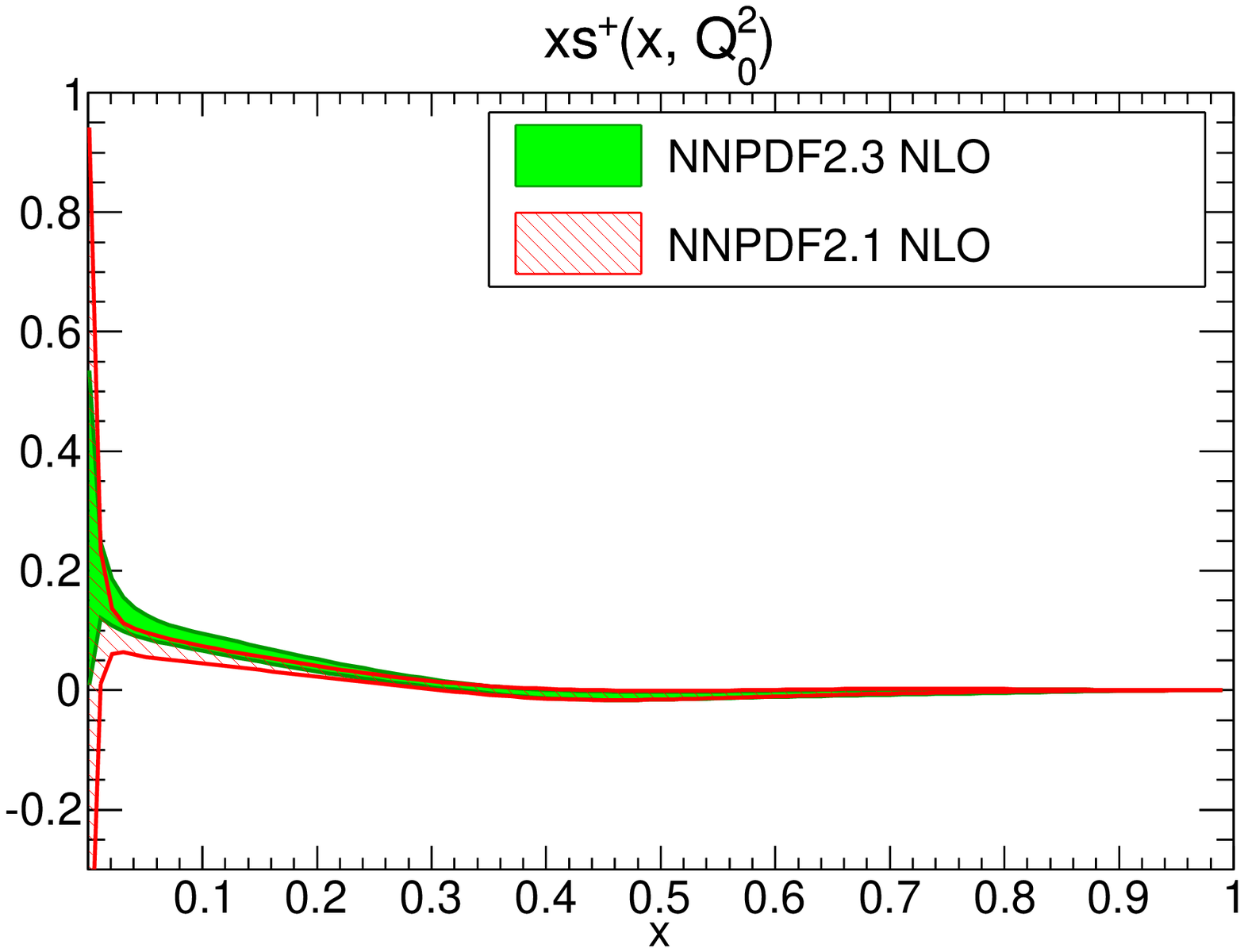}
\epsfig{width=0.49\textwidth,figure=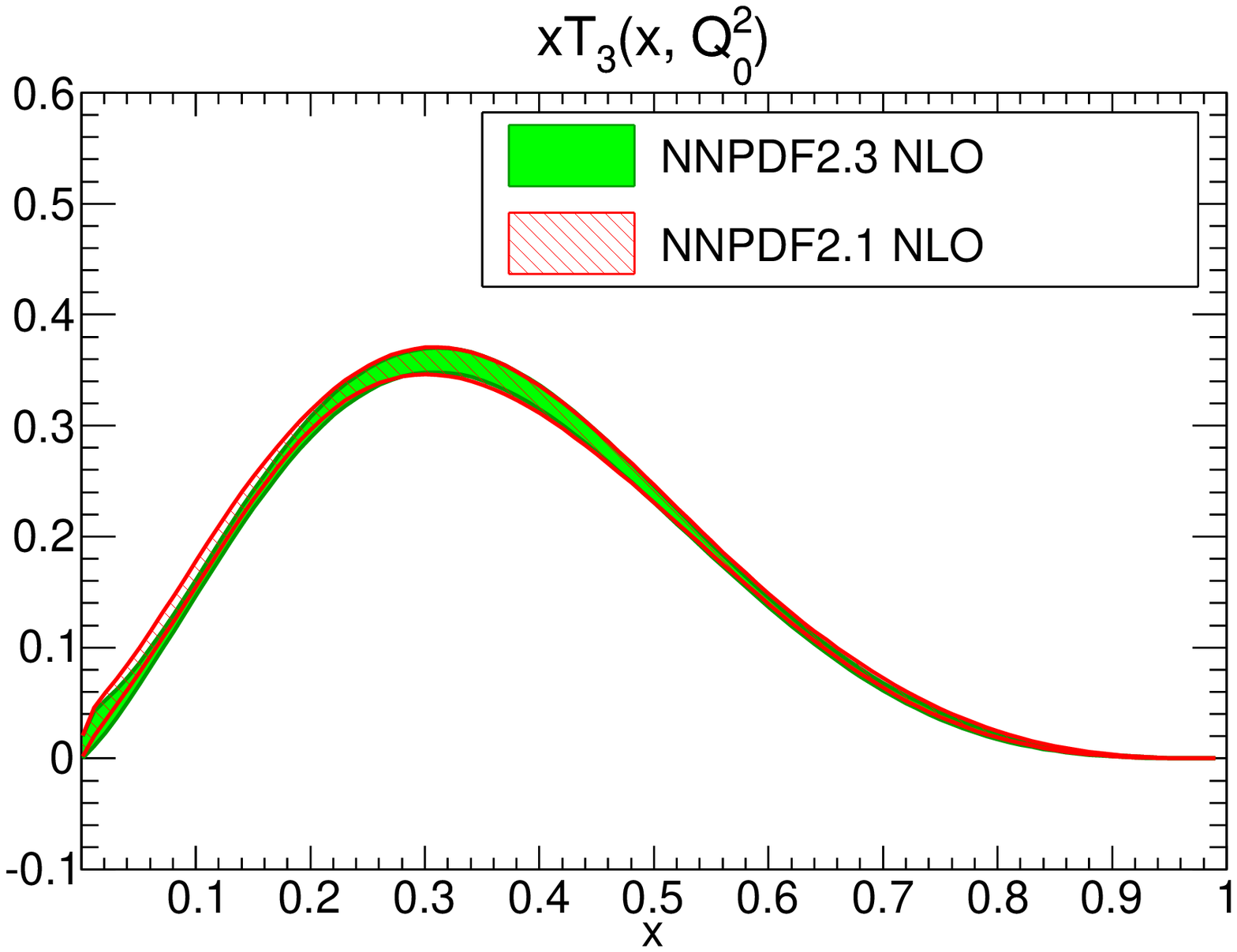}
\epsfig{width=0.49\textwidth,figure=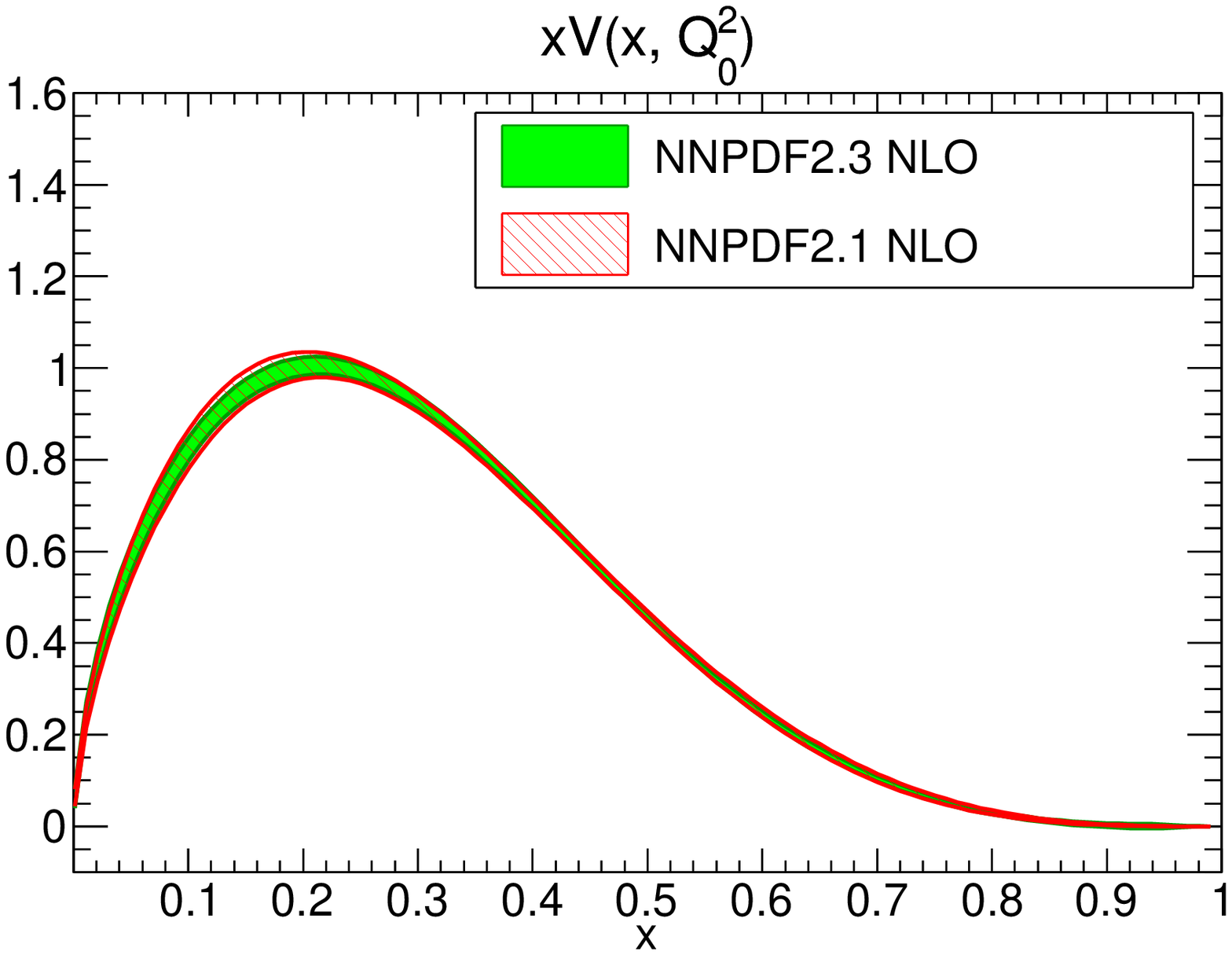}
\epsfig{width=0.49\textwidth,figure=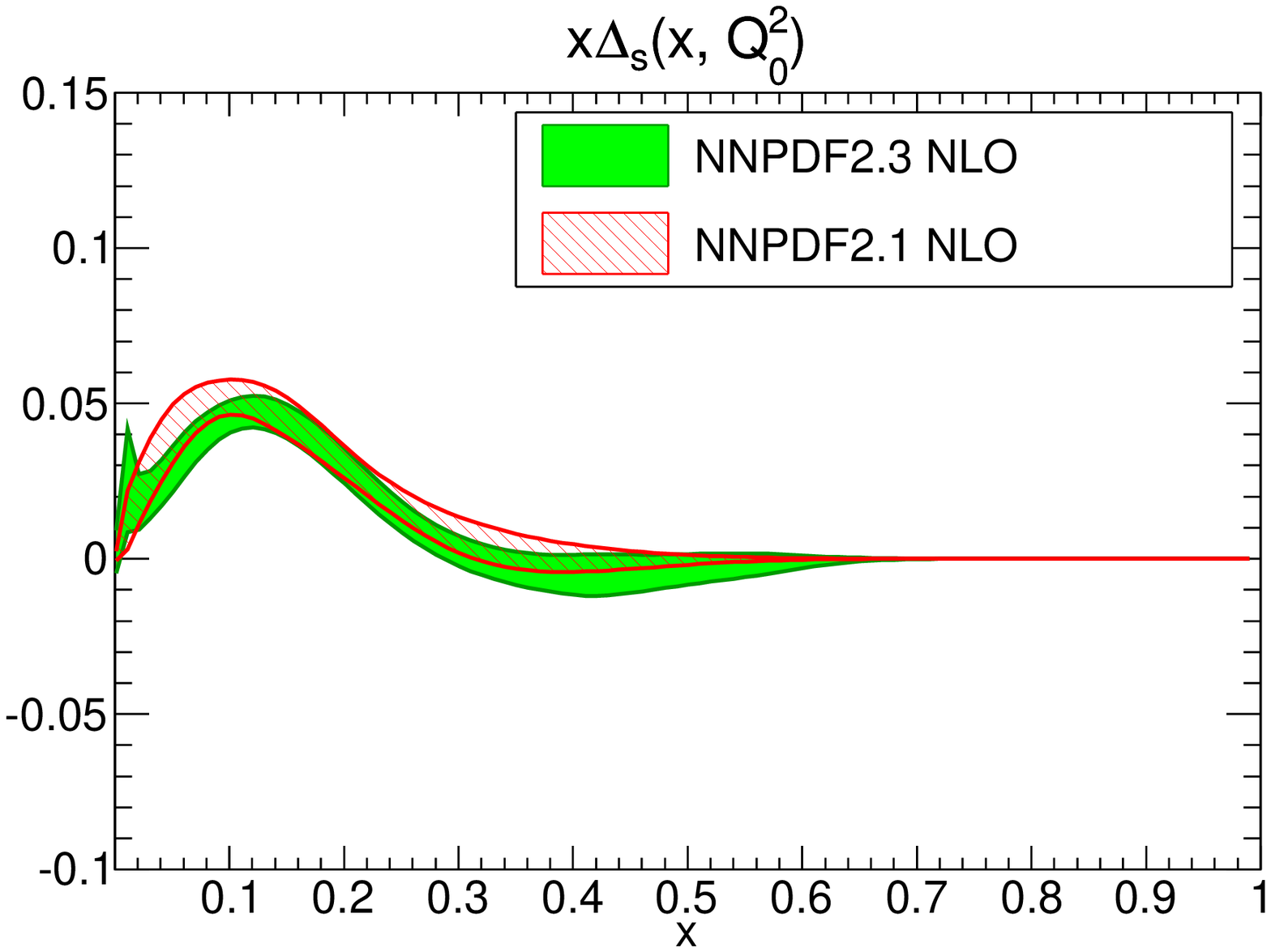}
\epsfig{width=0.49\textwidth,figure=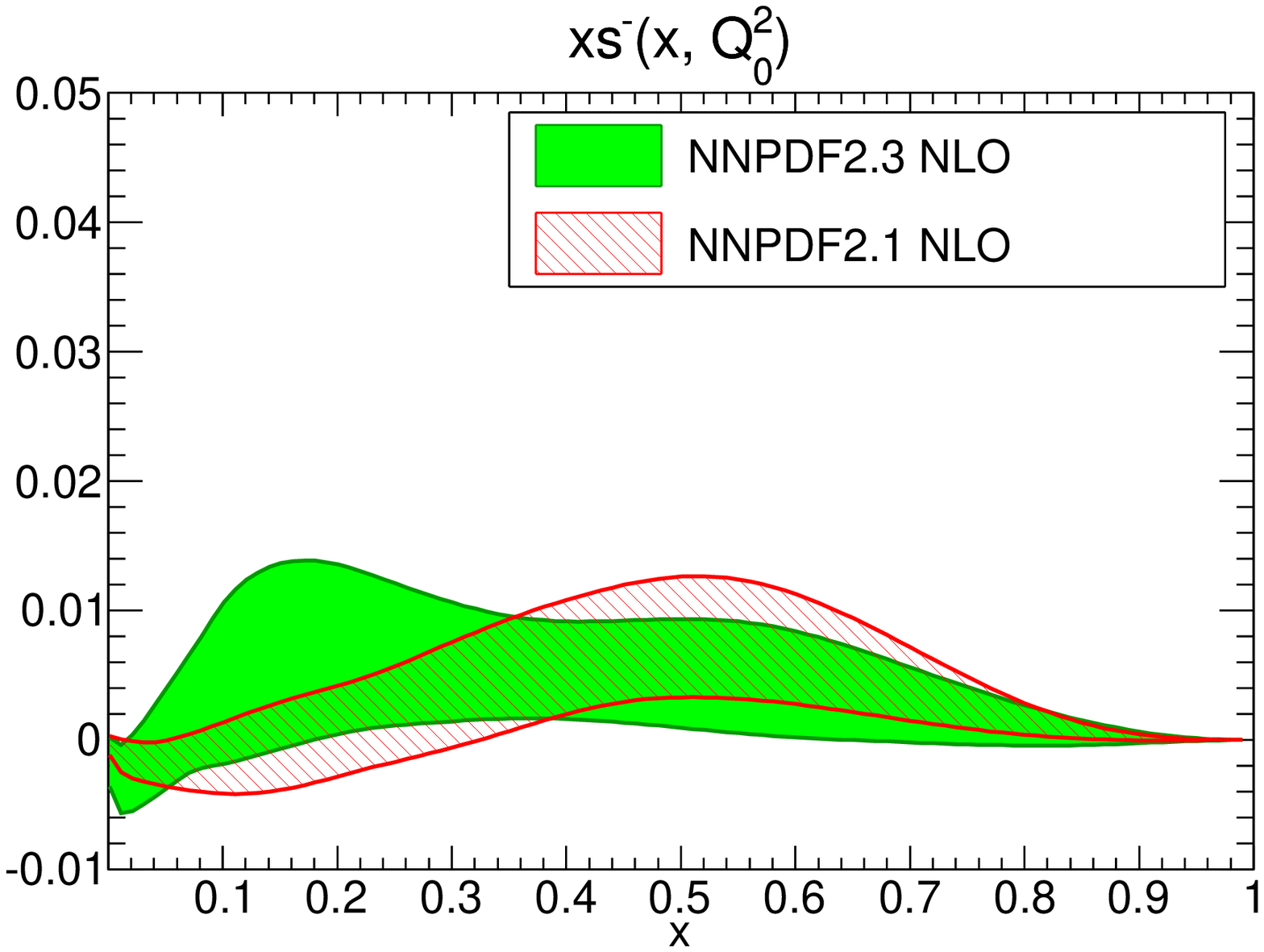}
\caption{\small Same as Fig.~\ref{fig:singletPDFs}
for the nonsinglet sector NLO PDFs.
 \label{fig:valencePDFs}} 
\end{center}
\vskip-0.5cm
\end{figure}

\begin{figure}[t]
\begin{center}
\epsfig{width=0.49\textwidth,figure=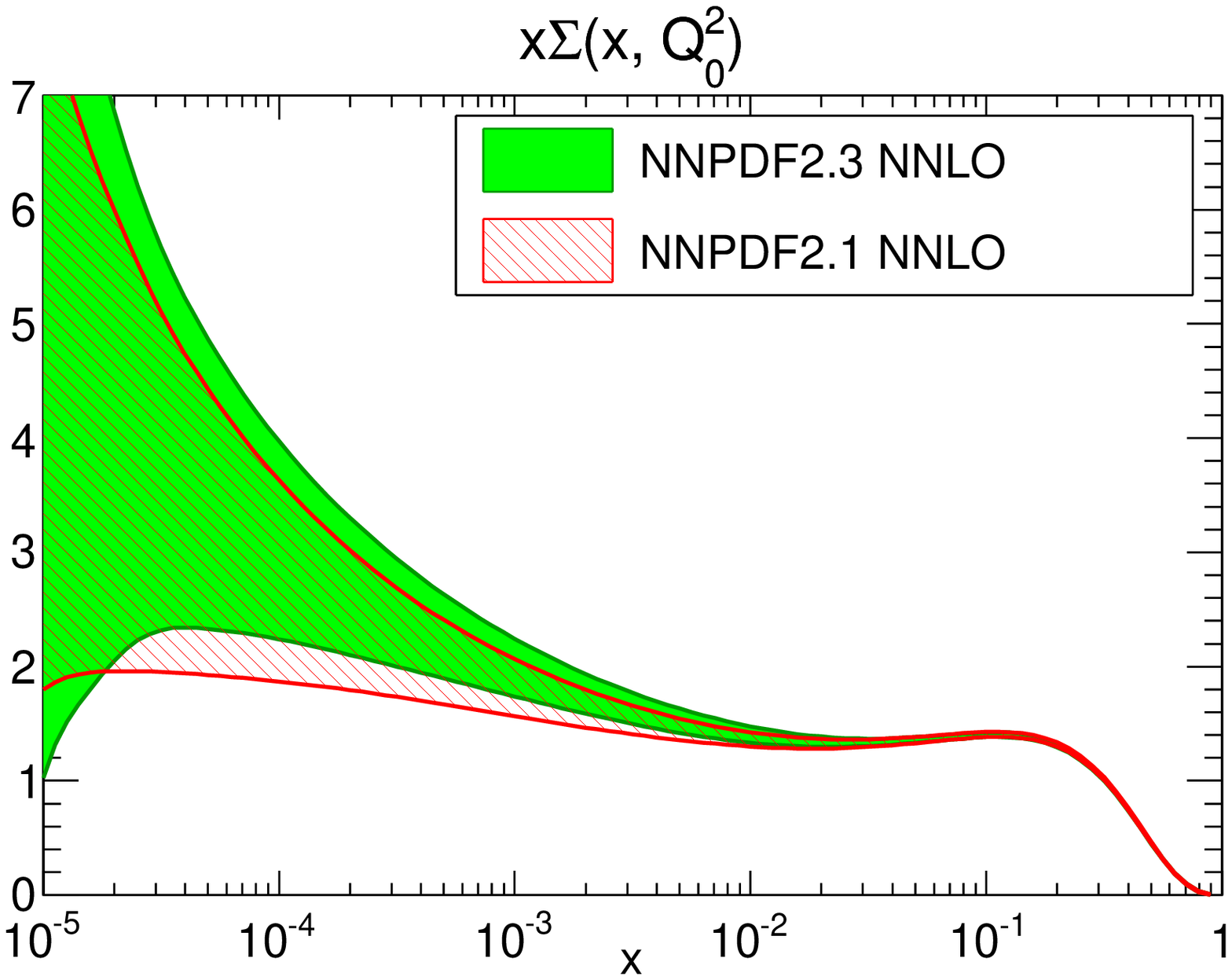}
\epsfig{width=0.49\textwidth,figure=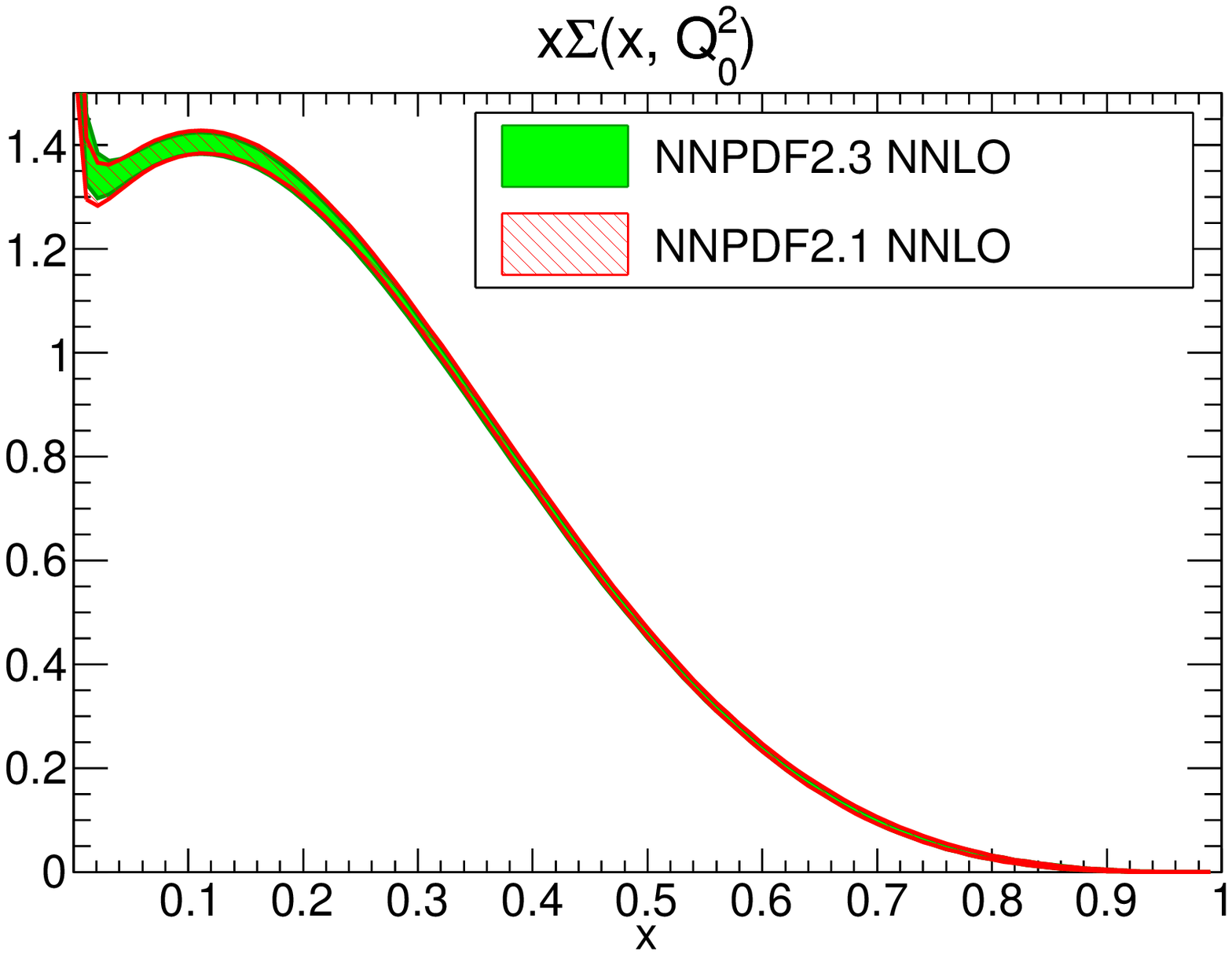}
\epsfig{width=0.49\textwidth,figure=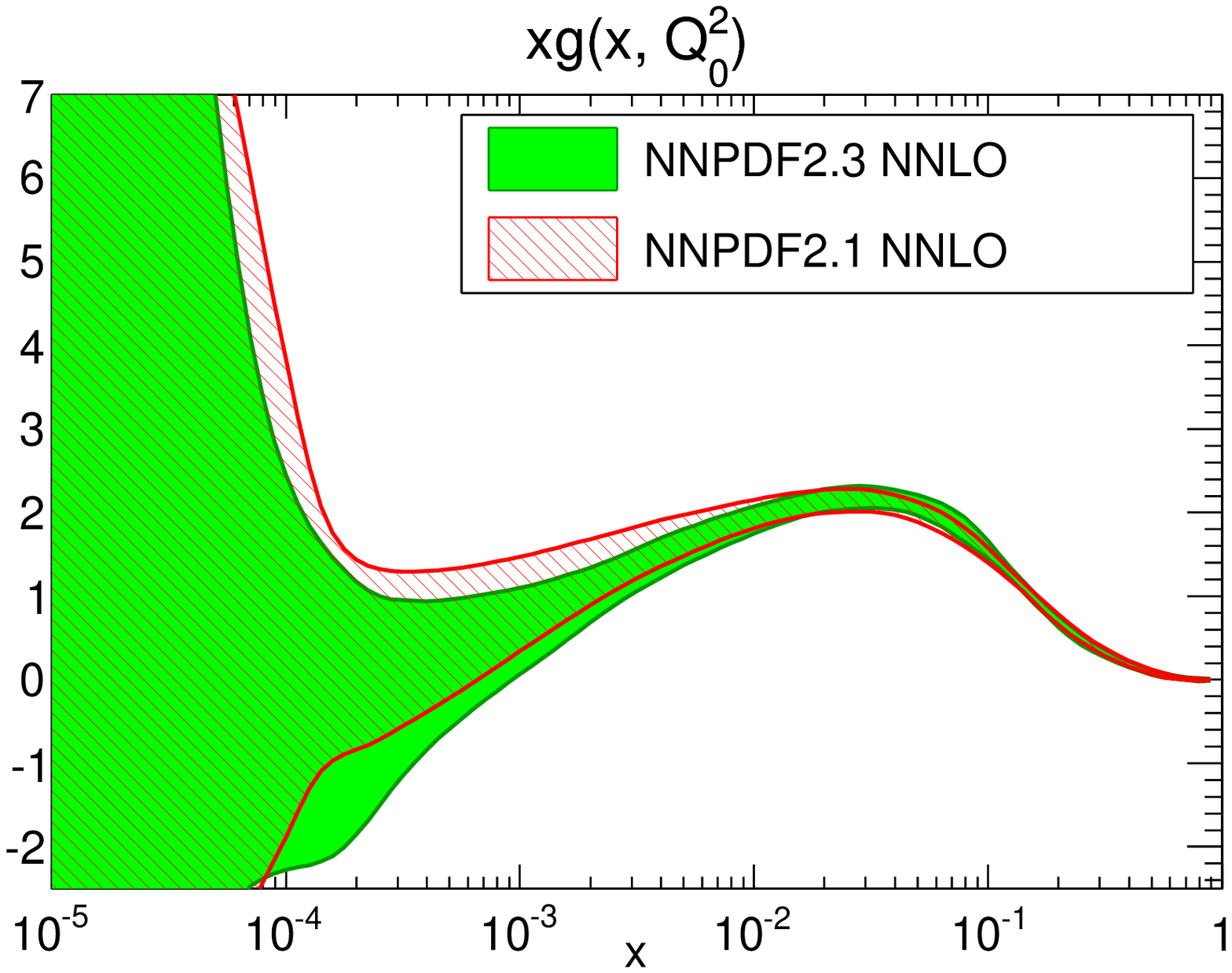}
\epsfig{width=0.49\textwidth,figure=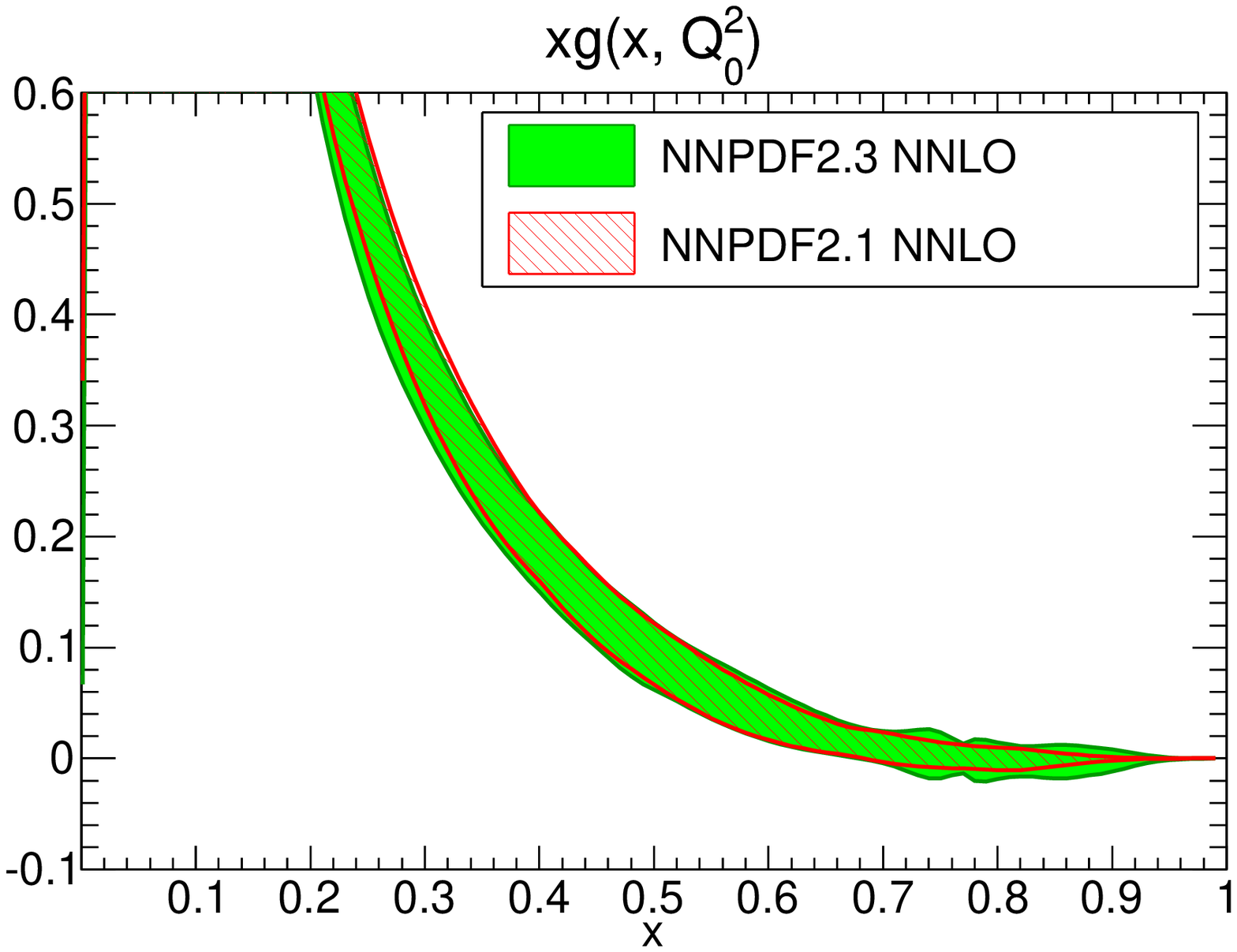}
\caption{\small Same as Fig.~\ref{fig:singletPDFs} but at NNLO.
 \label{fig:singletPDFsnn}} 
\end{center}
\vskip-0.5cm
\end{figure}

\begin{figure}[t!]
\begin{center}
\epsfig{width=0.49\textwidth,figure=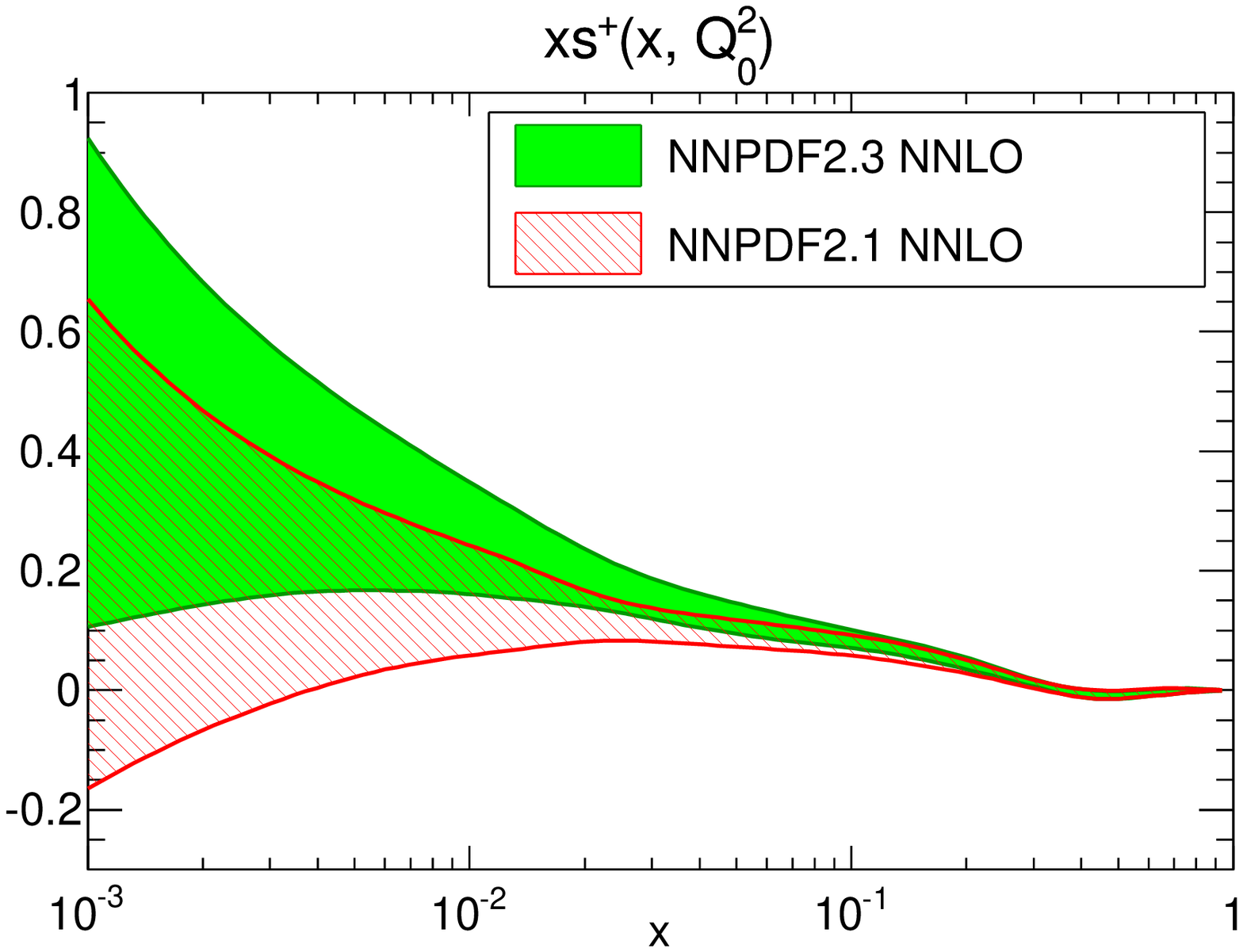}
\epsfig{width=0.49\textwidth,figure=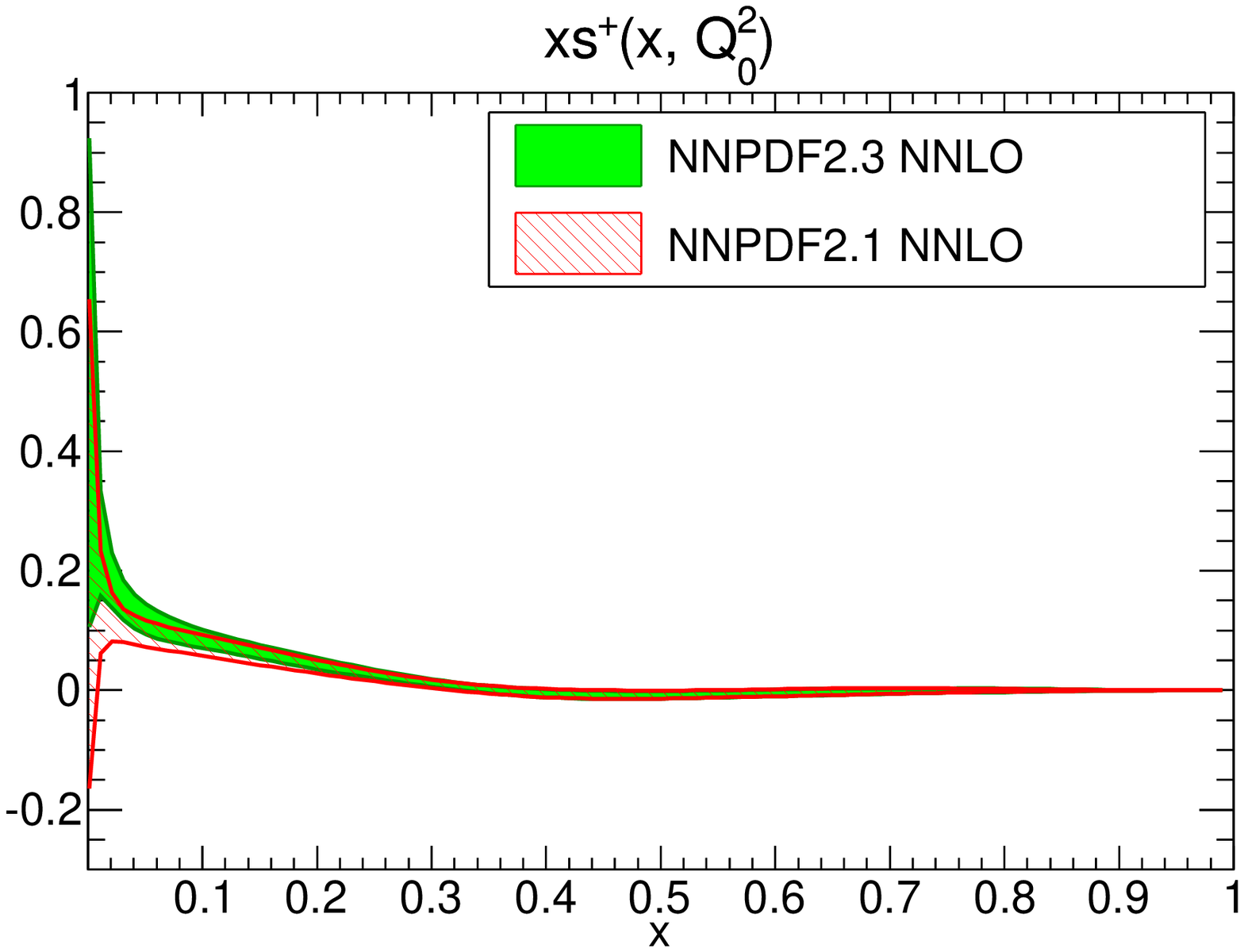}
\epsfig{width=0.49\textwidth,figure=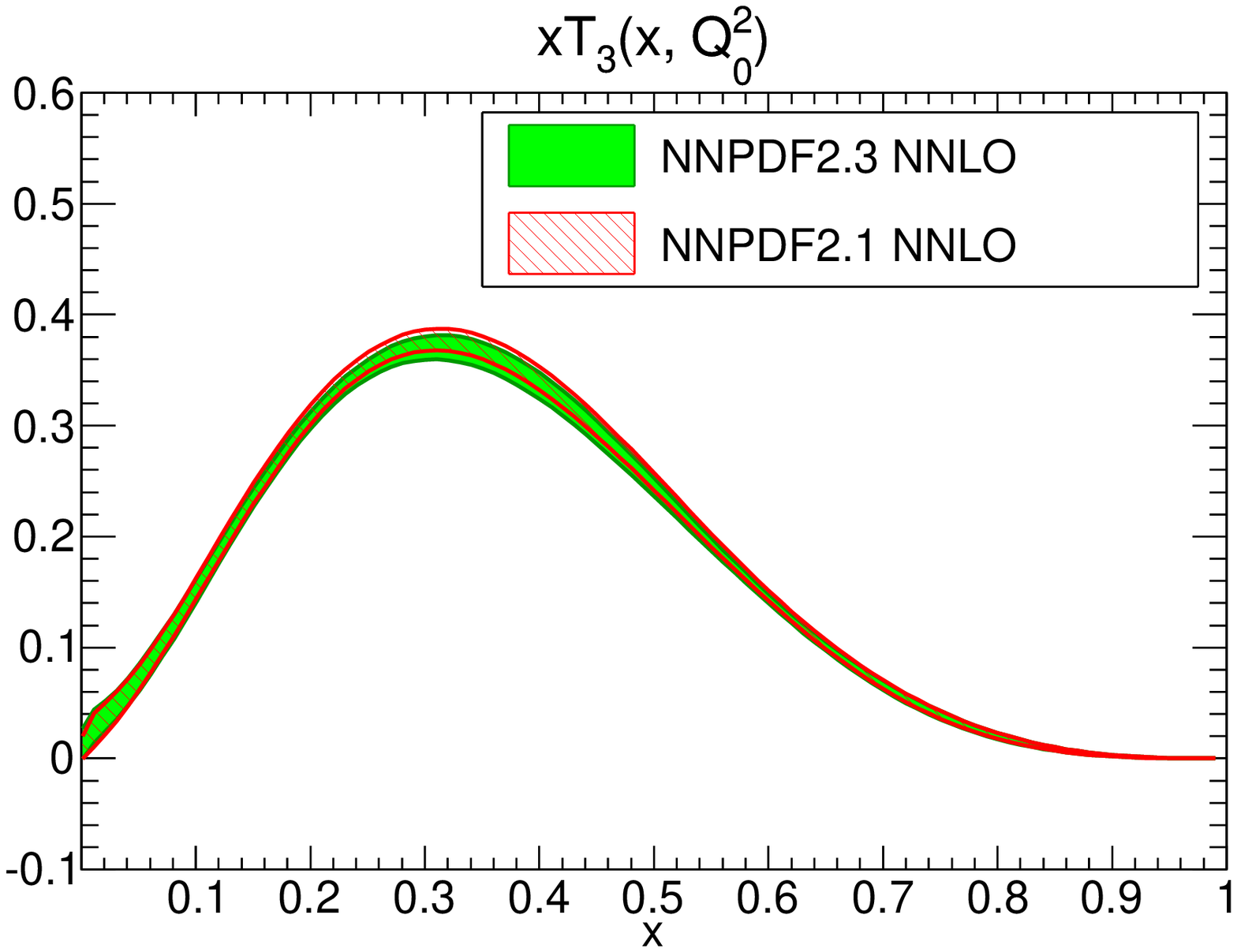}
\epsfig{width=0.49\textwidth,figure=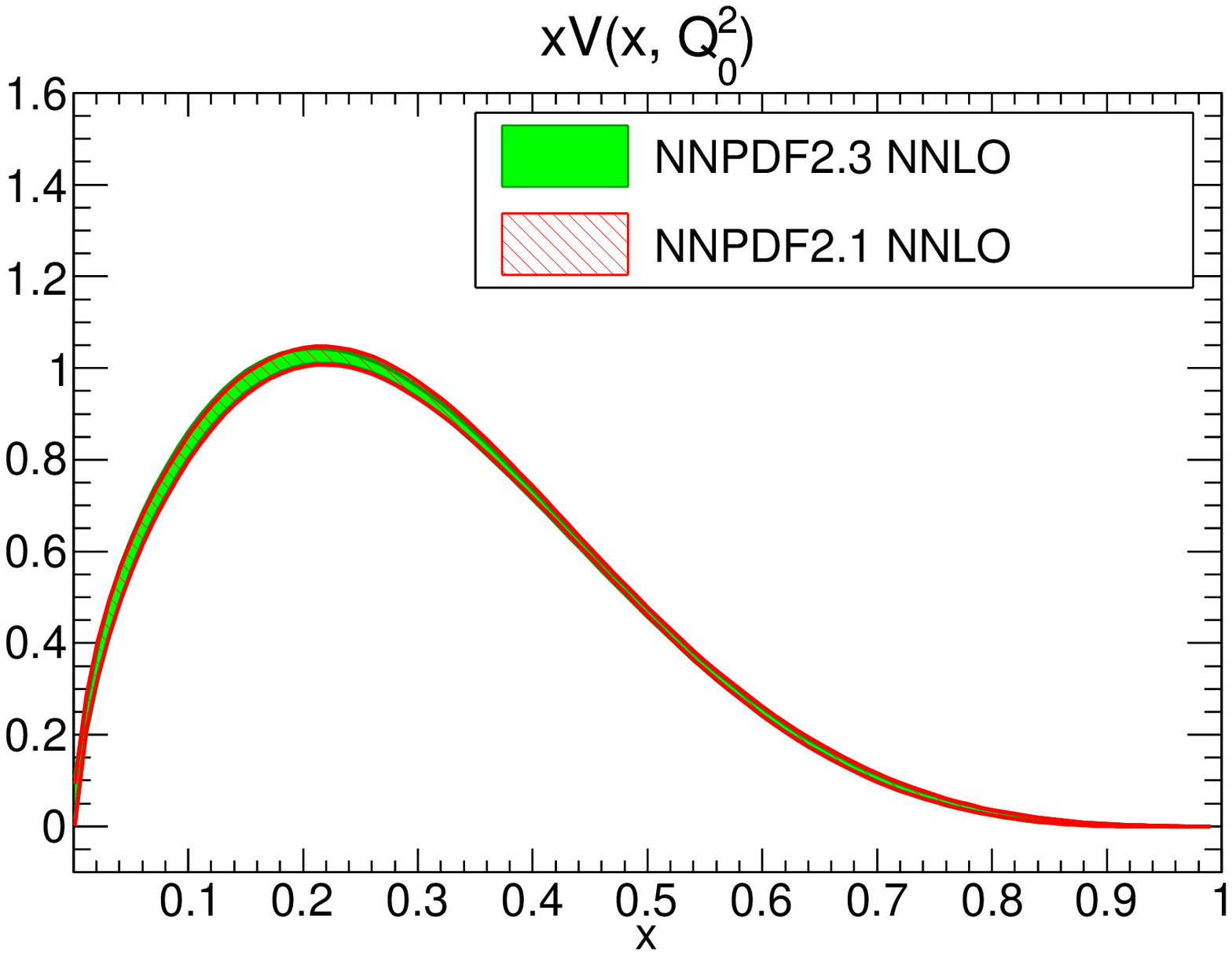}
\epsfig{width=0.49\textwidth,figure=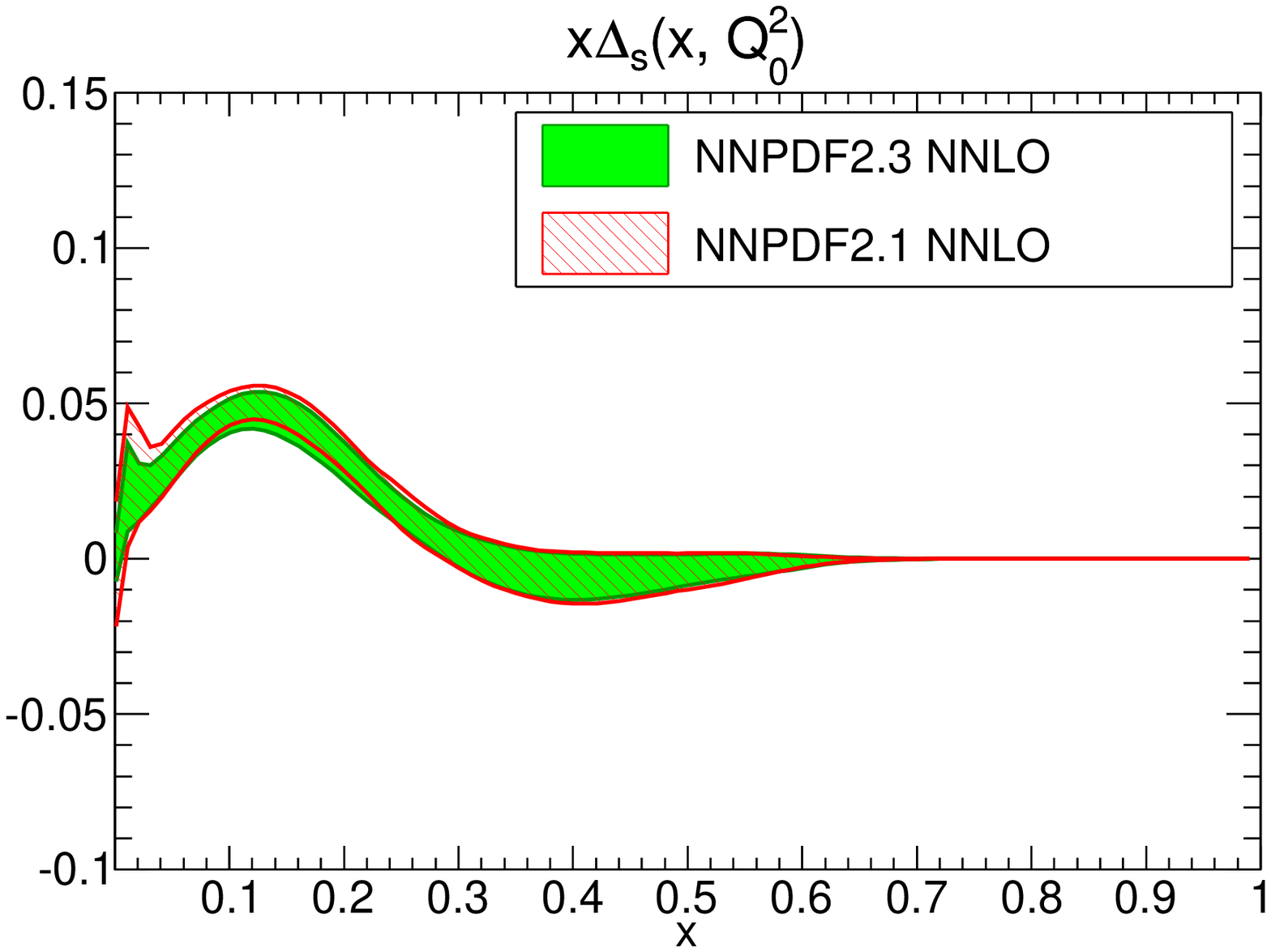}
\epsfig{width=0.49\textwidth,figure=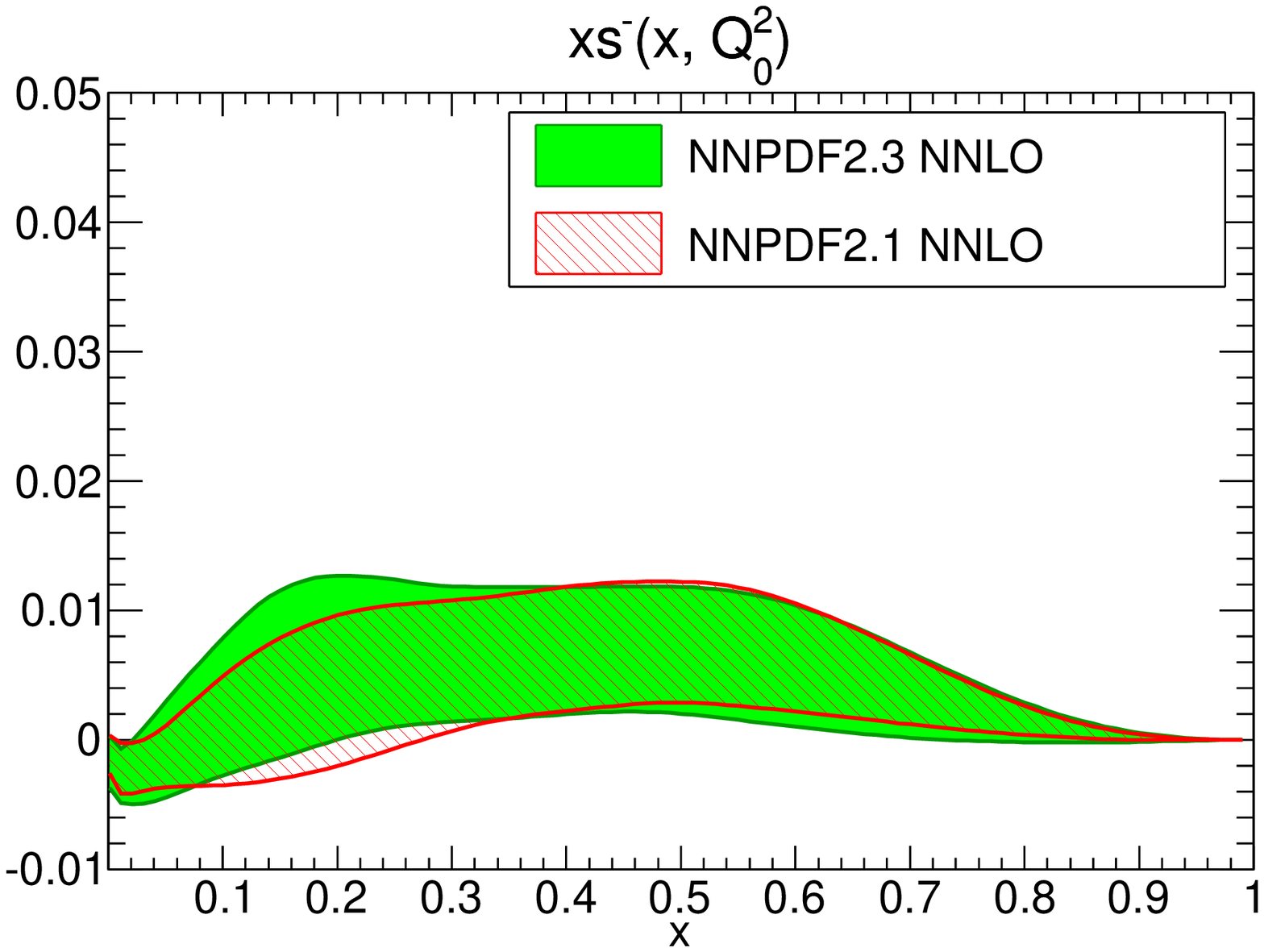}
\caption{\small Same as Fig.~\ref{fig:valencePDFs} but at NNLO.
 \label{fig:valencePDFsnn}} 
\end{center}
\vskip-0.5cm
\end{figure}

The NNPDF2.3 NLO and NNLO PDFs are shown, along with 
the corresponding
PDFs from NNPDF2.1, in
Figs.~\ref{fig:singletPDFs} \& \ref{fig:valencePDFs} (NLO) and 
Figs.~\ref{fig:singletPDFsnn} \& \ref{fig:valencePDFsnn} (NNLO). 
It is clear that all PDFs from the two sets differ by less, and
usually much less, than one sigma, with differences being generally
smaller at NNLO.

\begin{figure}[t]
\begin{center}
\epsfig{width=0.93\textwidth,figure=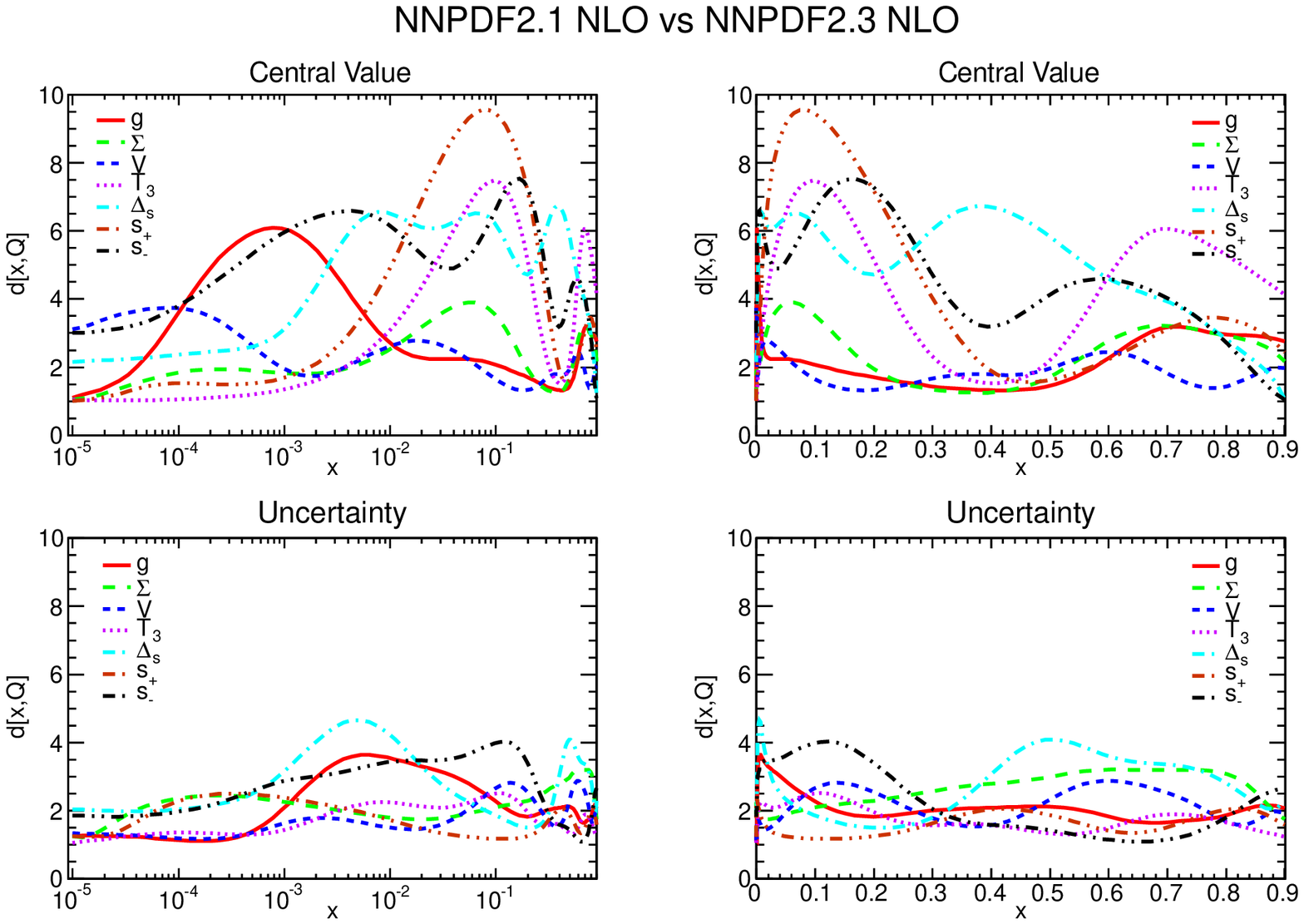}
\caption{ \small Distances between NNPDF2.1 and NNPDF2.3 NLO.
\label{fig:distances-21-vs-23-nlo.eps}} 
\end{center}
\end{figure}

\begin{figure}[t]
\begin{center}
\epsfig{width=0.93\textwidth,figure=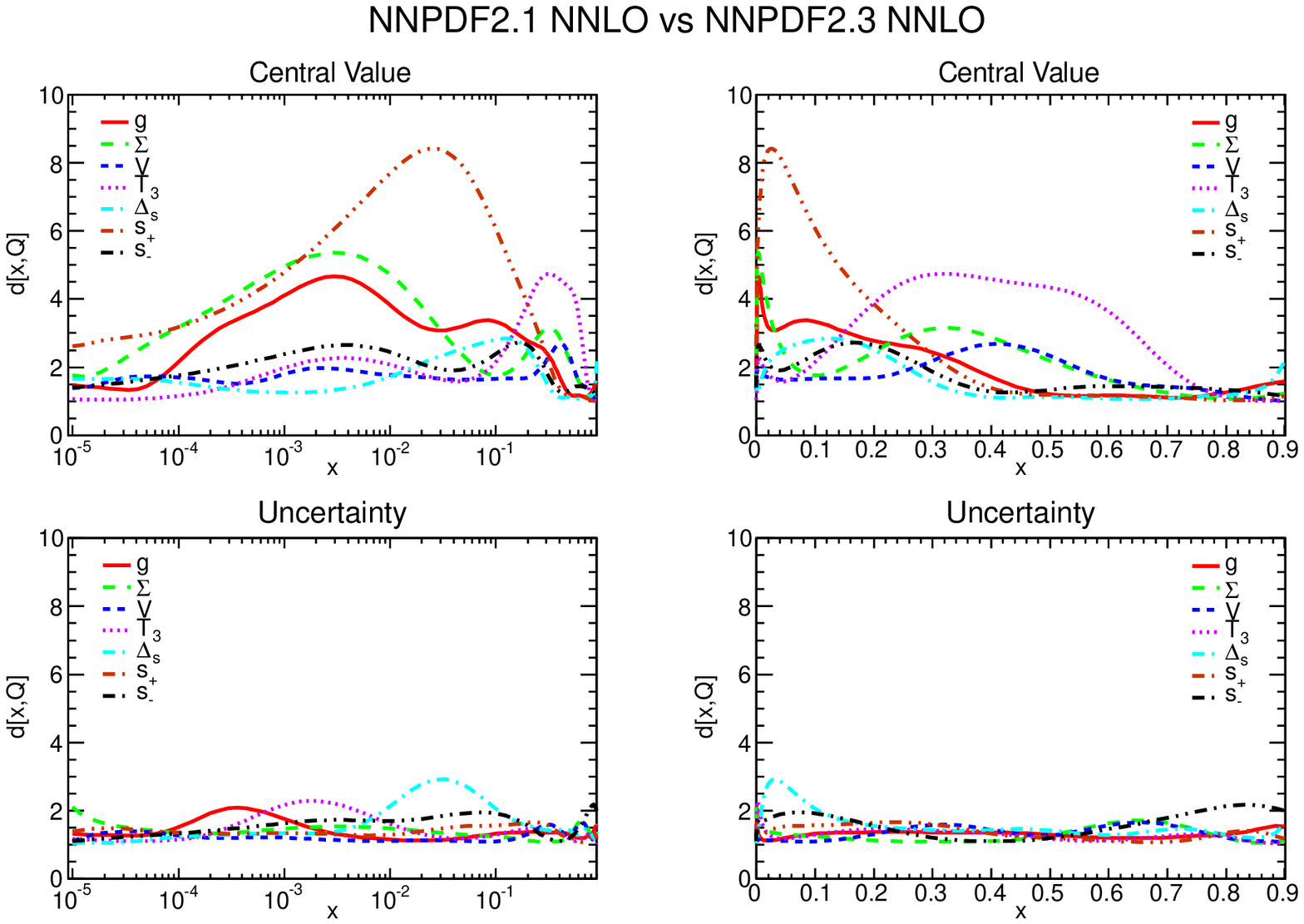}
\caption{ \small Distances between NNPDF2.1 and NNPDF2.3 NNLO.
\label{fig:distances-21-vs-23-nnlo.eps}} 
\end{center}
\end{figure}

\begin{figure}[t]
\begin{center}
\epsfig{width=0.47\textwidth,figure=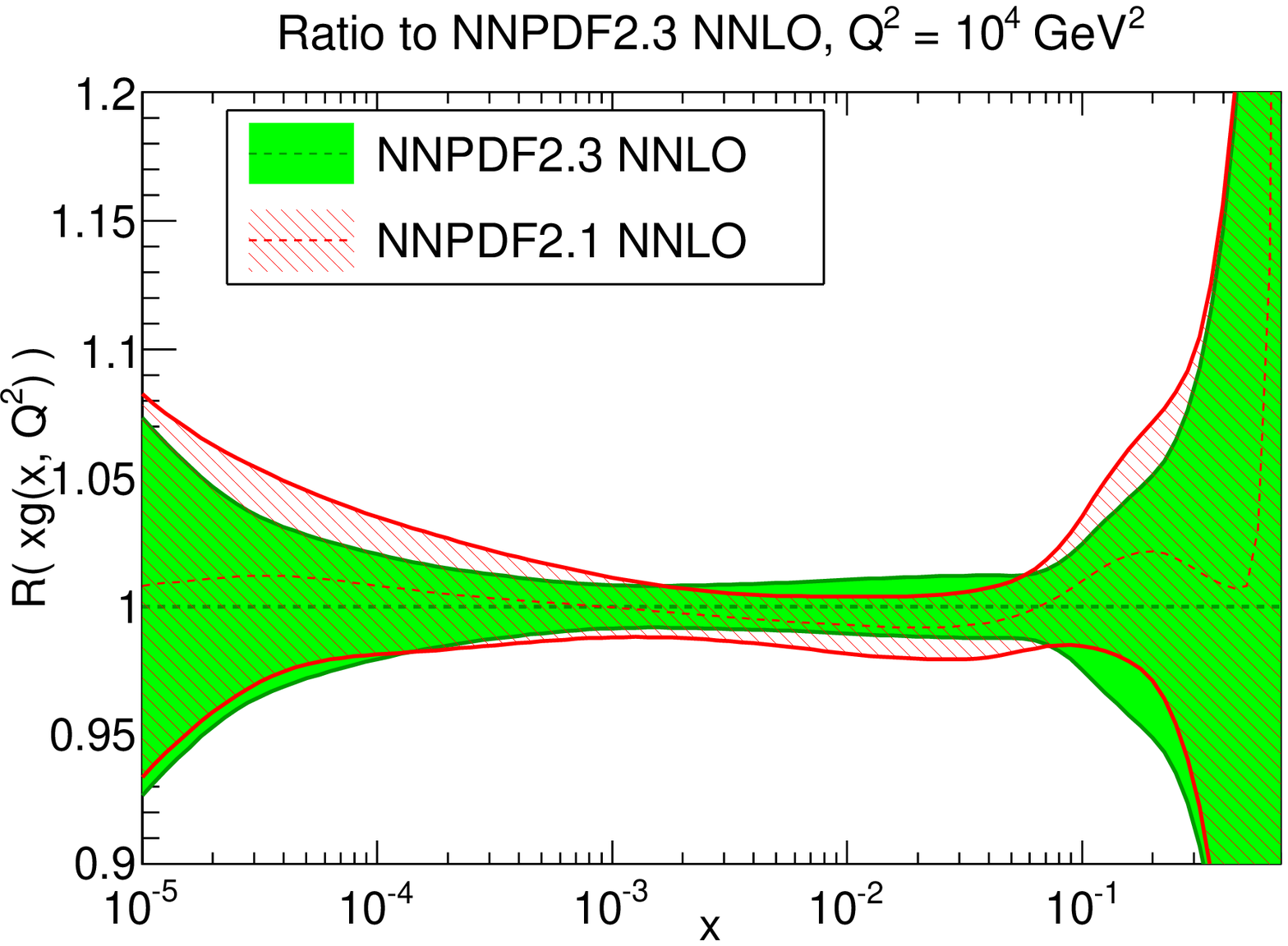}
\epsfig{width=0.47\textwidth,figure=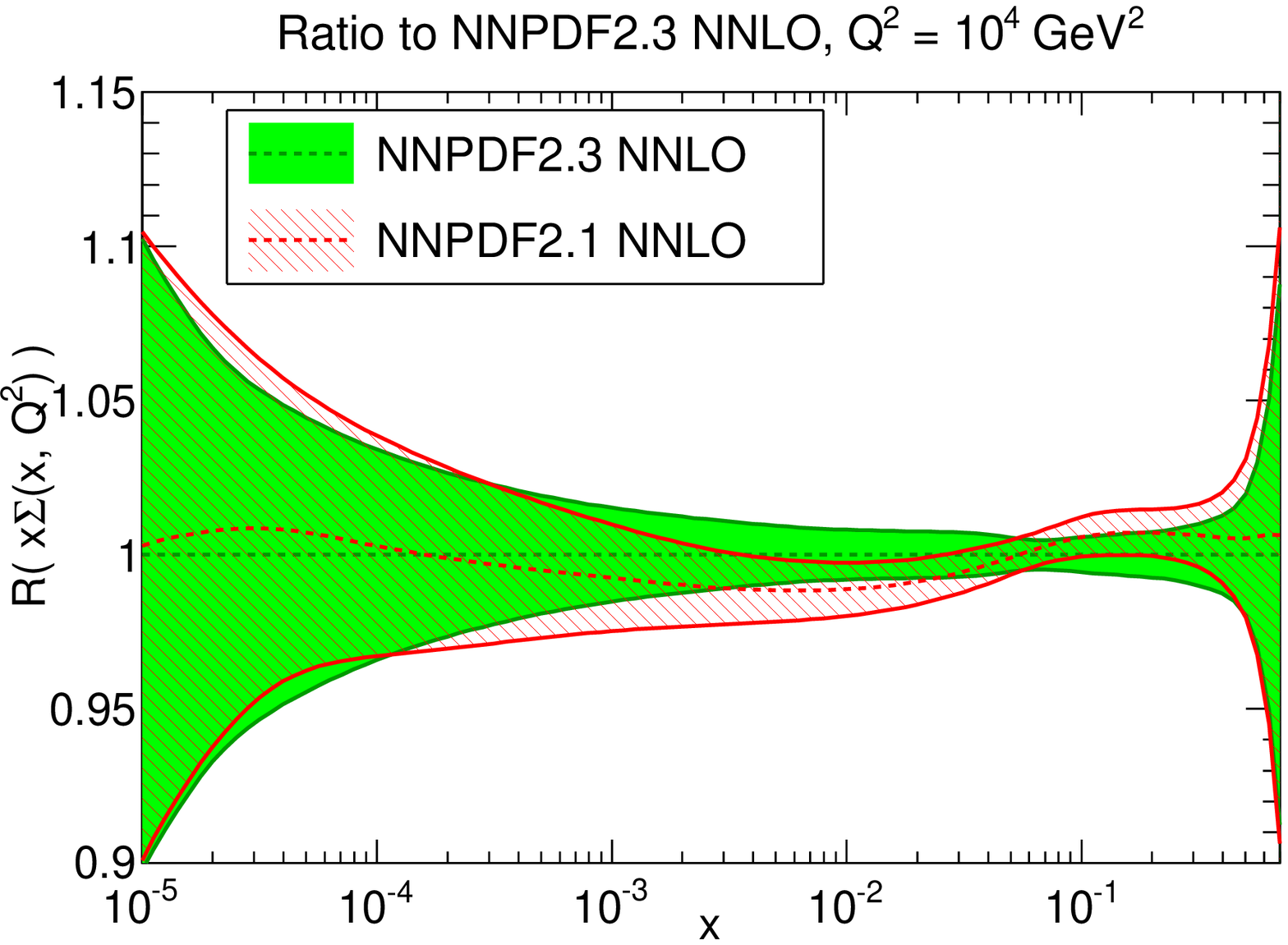}
\epsfig{width=0.47\textwidth,figure=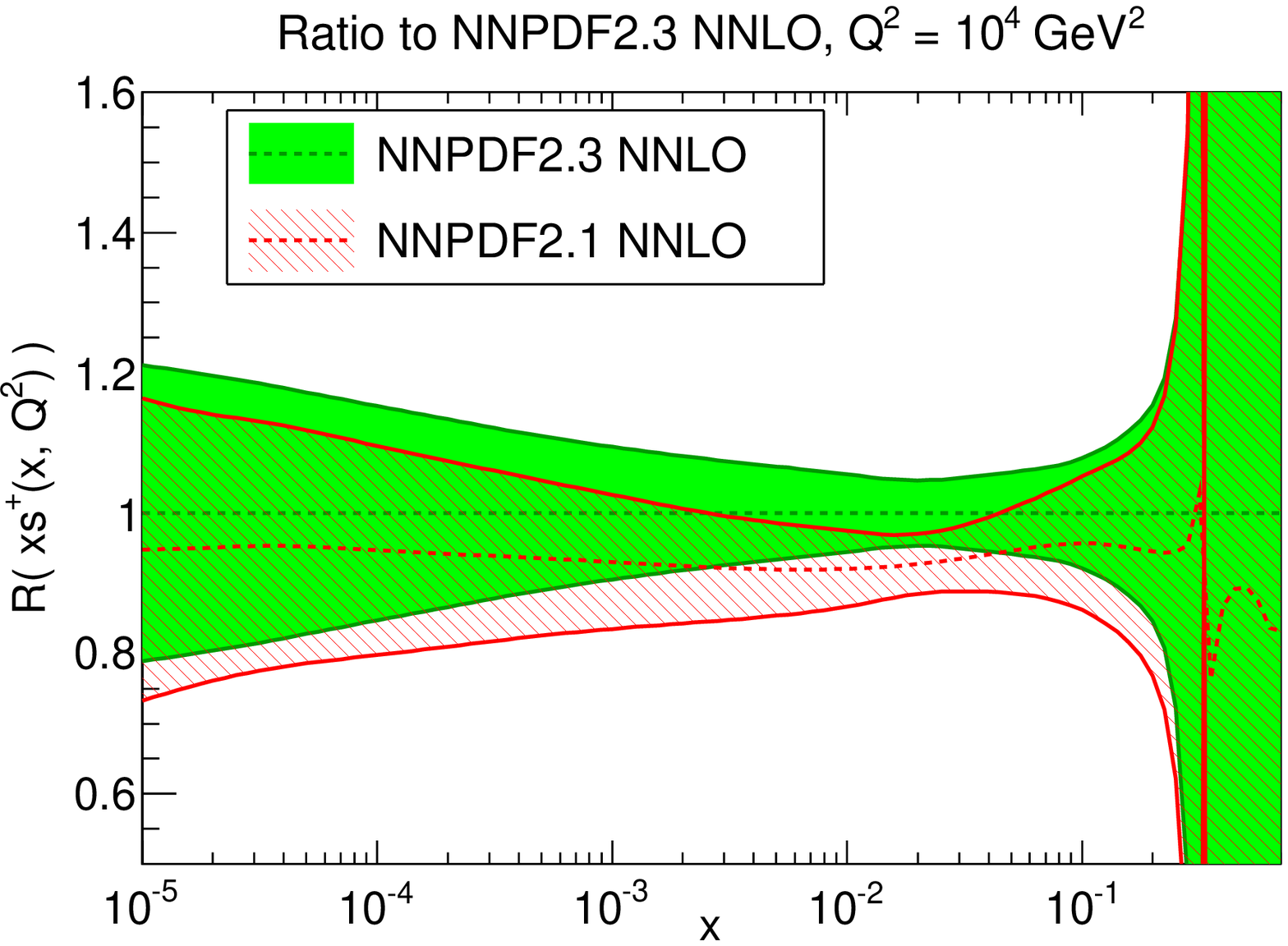}
\epsfig{width=0.47\textwidth,figure=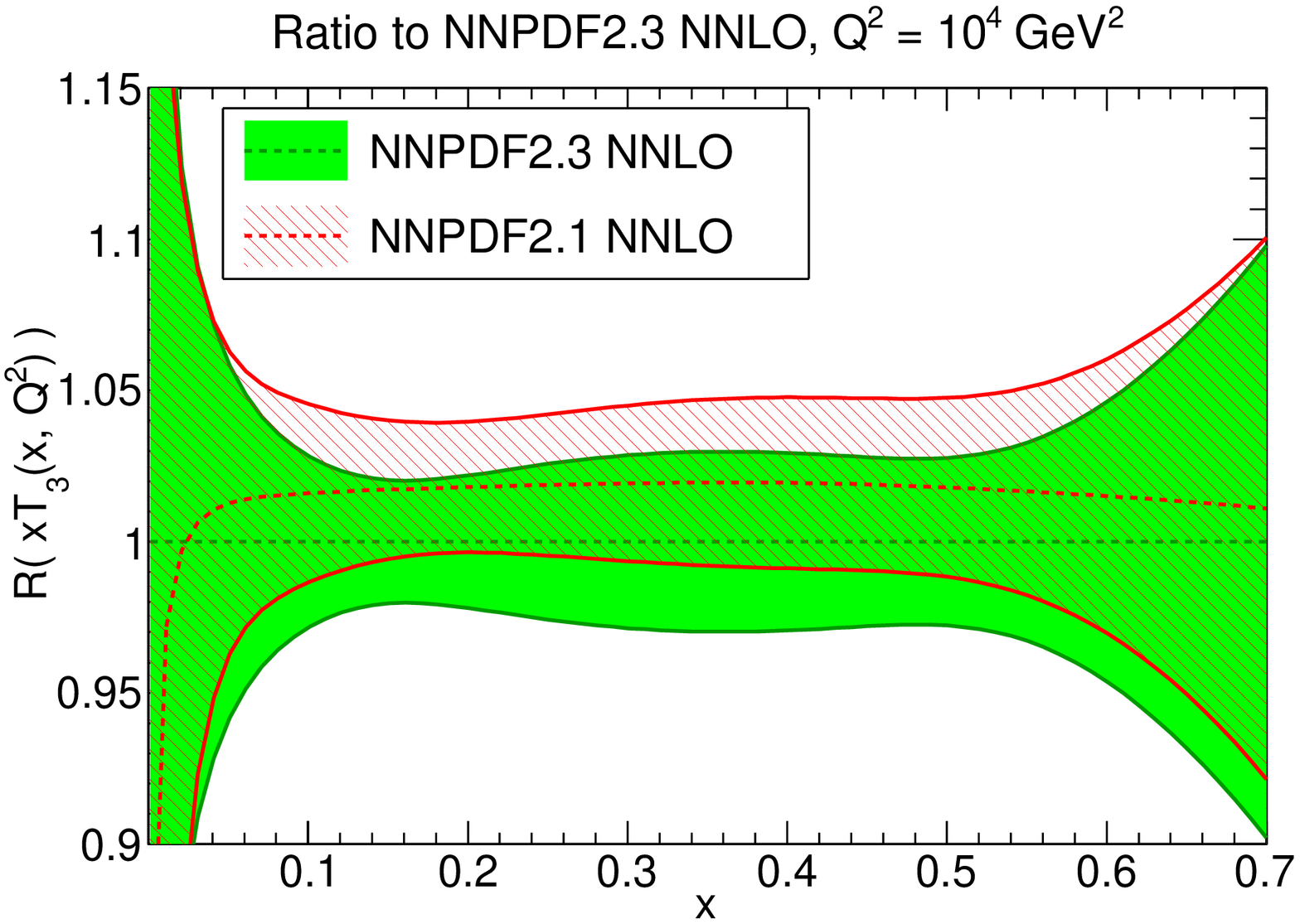}
\caption{ \small The ratio 
of NNLO NNPDF2.1 to NNPDF2.3  gluon, singlet, (top) total strangeness and
  triplet (bottom) PDFs at $10^4$ GeV$^2$.
\label{fig:pdfs-23-vs-21-nnlo-10000.eps}} 
\end{center}
\end{figure}

A more accurate assessment of the difference between the NNPDF2.1 and
NNPDF2.3 sets can be obtained by looking at the distance between the
two sets, shown in
Figs.~\ref{fig:distances-21-vs-23-nlo.eps} \& \ref{fig:distances-21-vs-23-nnlo.eps}. The
largest changes are in the gluon and in the valence flavor
decomposition (i.e.~in the sea asymmetry, triplet and strangeness), 
where we would indeed expect jet and gauge boson production to
have some impact. The ratio of the NNPDF2.1 to NNPDF2.3 NNLO PDFs 
at $Q^2=10^4$~GeV$^2$ is shown for the gluon, singlet, triplet ans
strangeness
in 
Fig.~\ref{fig:pdfs-23-vs-21-nnlo-10000.eps}.
The origin of these differences is
addressed in detail in Sec.~\ref{sec-comp}

\subsection{Detailed comparison to NNPDF2.1}
\label{sec-comp}

The NNPDF2.3 sets differ from the NNPDF2.1 ones not only because of the
addition of LHC data, but also due to the 
improvements in the neural network training  procedure presented in
Sec.~\ref{sec-method}, and finally because of the correction of an
error in Eq.~(33) of Ref.~\cite{Ball:2011mu}\footnote{The correct
  equation should read
\begin{eqnarray}
  \label{eq:bug}
   &&\tilde{\sigma}^{\nu (\bar{\nu}),c}(x,y,Q^2)=\frac{G_F^2M_N}{2\pi(1+Q^2/M_{\rm W}^2)^2}
  \Bigg[
    \left( \lp Y_+ - \frac{2M^2_Nx^2y^2}{Q^2} -y^2\rp +y^2\right)
    F_{2,c}^{\nu(\bar{\nu})}(x,Q^2) \nonumber\\ 
    &&\qquad\qquad\qquad\qquad\qquad\qquad\qquad
    -y^2F_{L,c}^{\nu(\bar{\nu})}(x,Q^2)\pm 
    \,Y_-\,xF_{3,c}^{\nu(\bar{\nu})}(x,Q^2)
    \Bigg]. \,
\end{eqnarray}
In Eq.~(33) of Ref.~\cite{Ball:2011mu} there is a spurious factor 
of $\lp 1+\frac{m^2_c}{Q^2}\rp$. This is the same expression as Eq.~(1)
of Ref.~\cite{Ball:2009mk}: in that reference, a so-called improved zero-mass
variable-flavor number scheme is used, and this factor  provides the
desired improvement. But in Ref.~\cite{Ball:2011mu} a general mass
scheme is used, in which this factor is unnecessary and thus spurious
given that the charm mass is treated exactly.}.
This error only affects the NuTeV dimuon cross-sections, which in turn
only have a significant effect on the strange distribution.

In order to isolate the effect of each of these changes, 
we have performed an NNPDF2.3 fit without LHC data,
i.e.~with the same data set as NNPDF2.1, but with the
methodological improvements discussed in the previous section,
supplemented by the correction of the error in the dimuon cross-section.

\begin{figure}[t]
\begin{center}
\epsfig{width=0.93\textwidth,figure=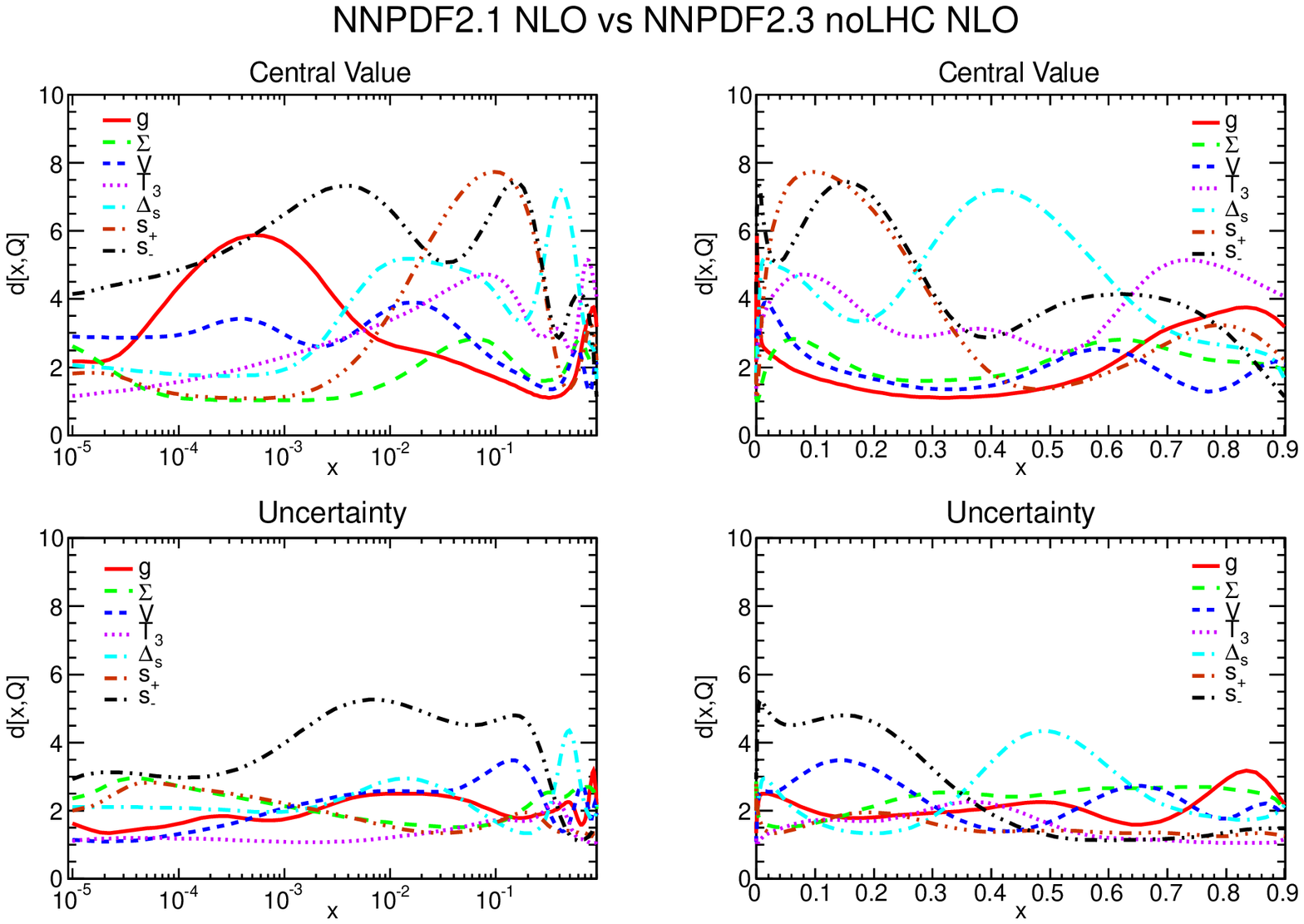}
\caption{ \small Distances between NNPDF2.3 noLHC and NNPDF2.1 NLO.
\label{fig:distances-23noLHC-vs-21-nlo.eps}} 
\end{center}
\end{figure}

\begin{figure}[t]
\begin{center}
\epsfig{width=0.93\textwidth,figure=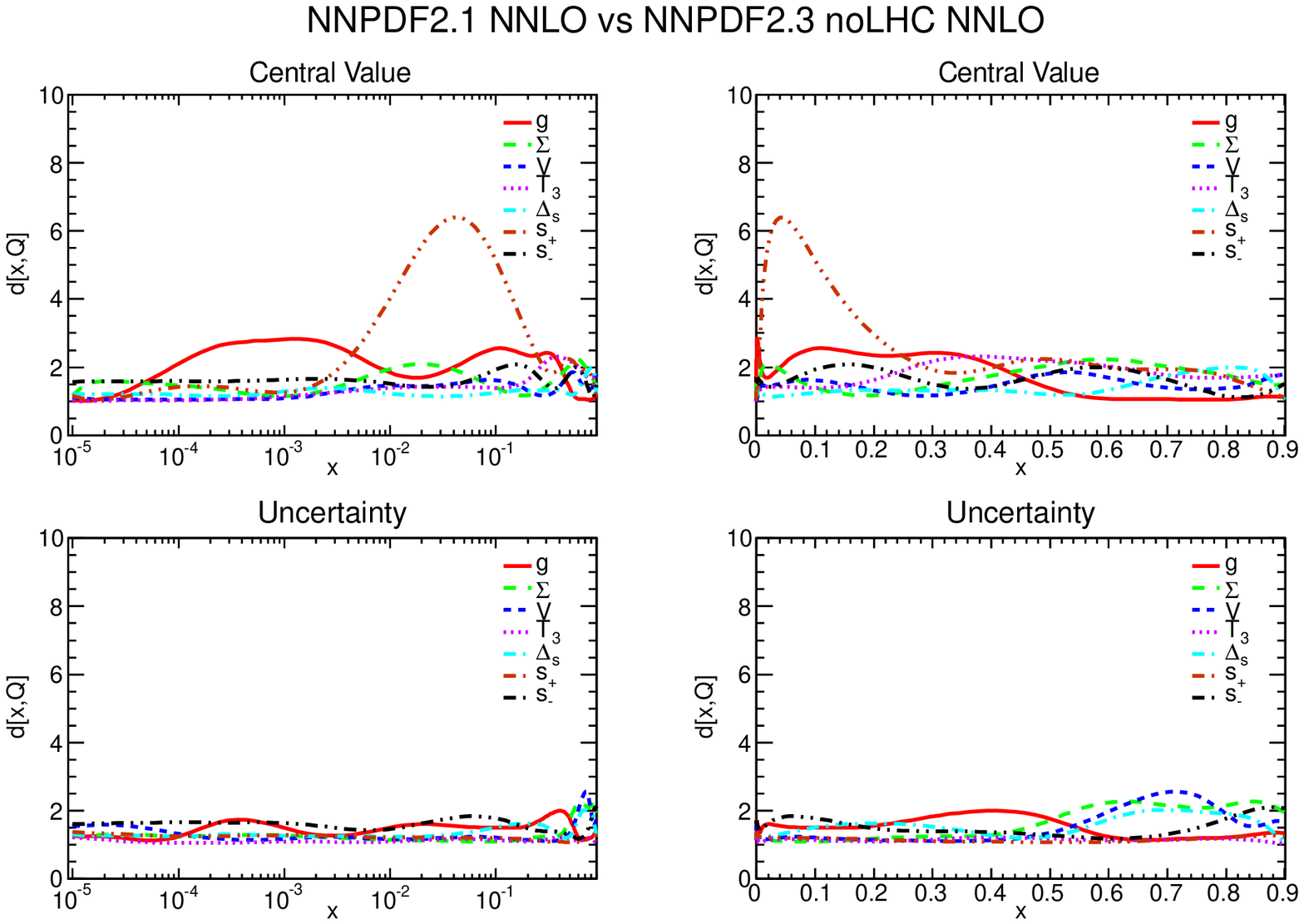}
\caption{ \small Distances between NNPDF2.3 noLHC and NNPDF2.1 NNLO.
\label{fig:distances-23noLHC-vs-21-nnlo.eps}} 
\end{center}
\end{figure}

\begin{figure}[t]
\begin{center}
\epsfig{width=0.44\textwidth,figure=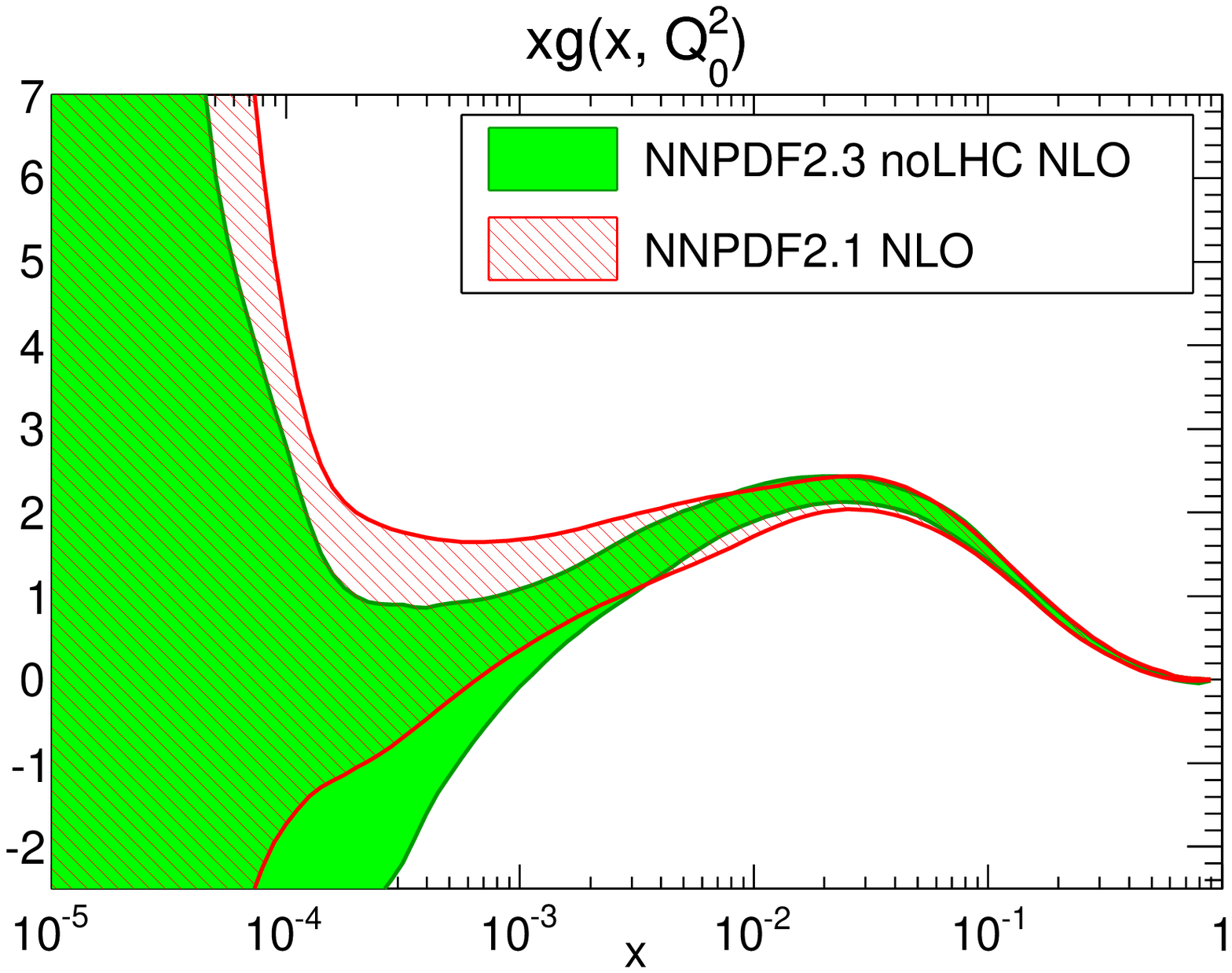}
\epsfig{width=0.44\textwidth,figure=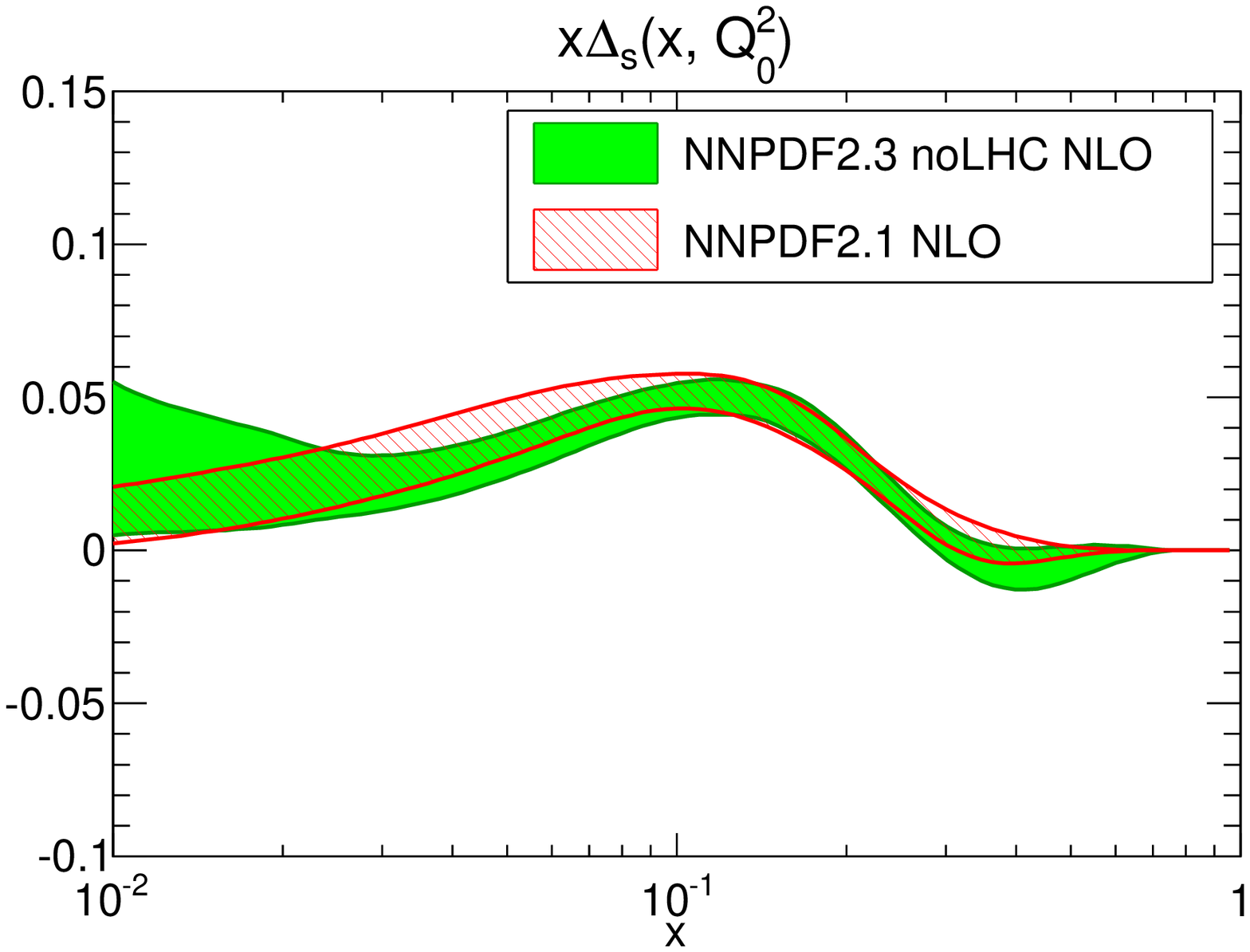}\\
\epsfig{width=0.44\textwidth,figure=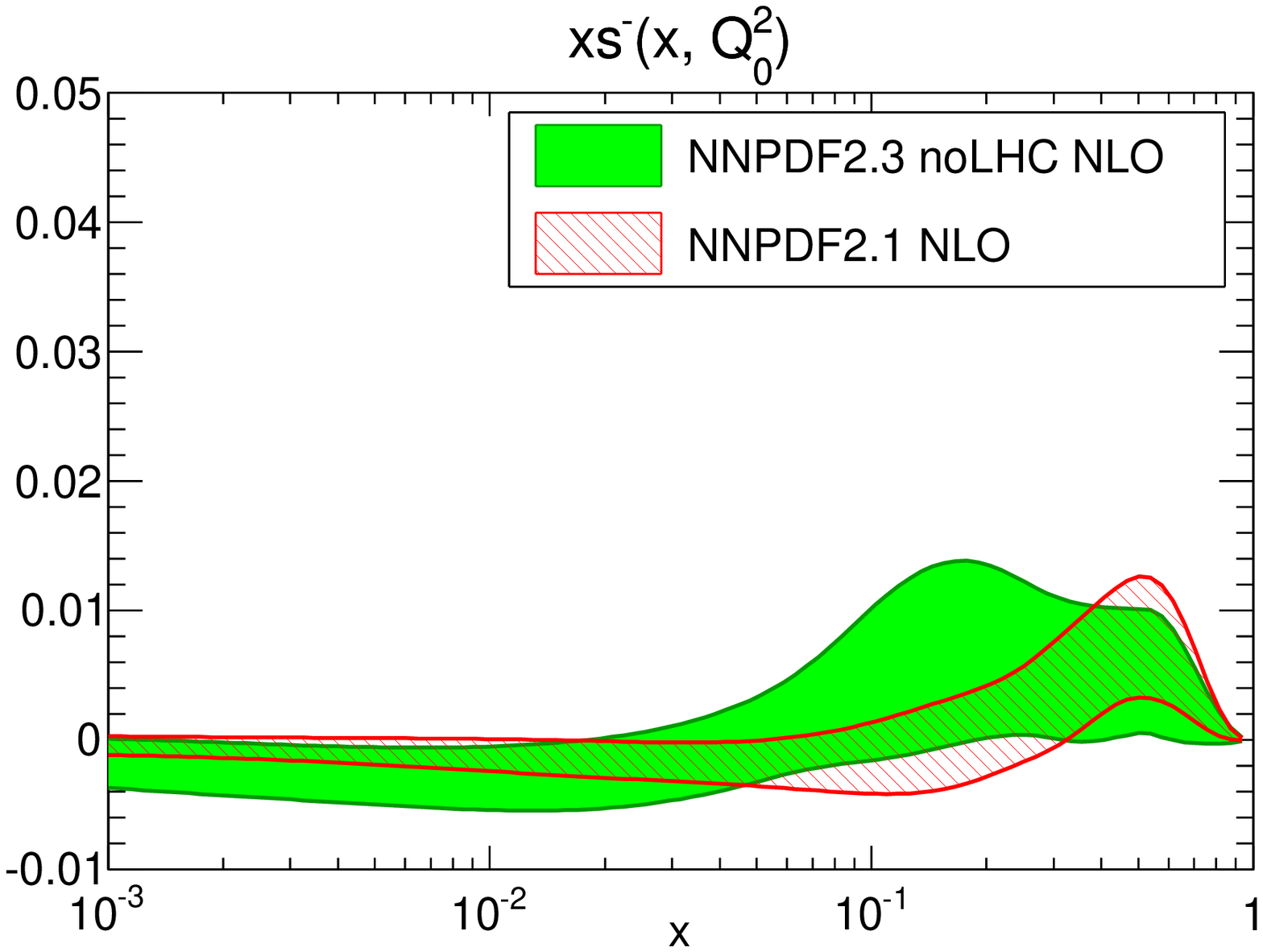}
\epsfig{width=0.44\textwidth,figure=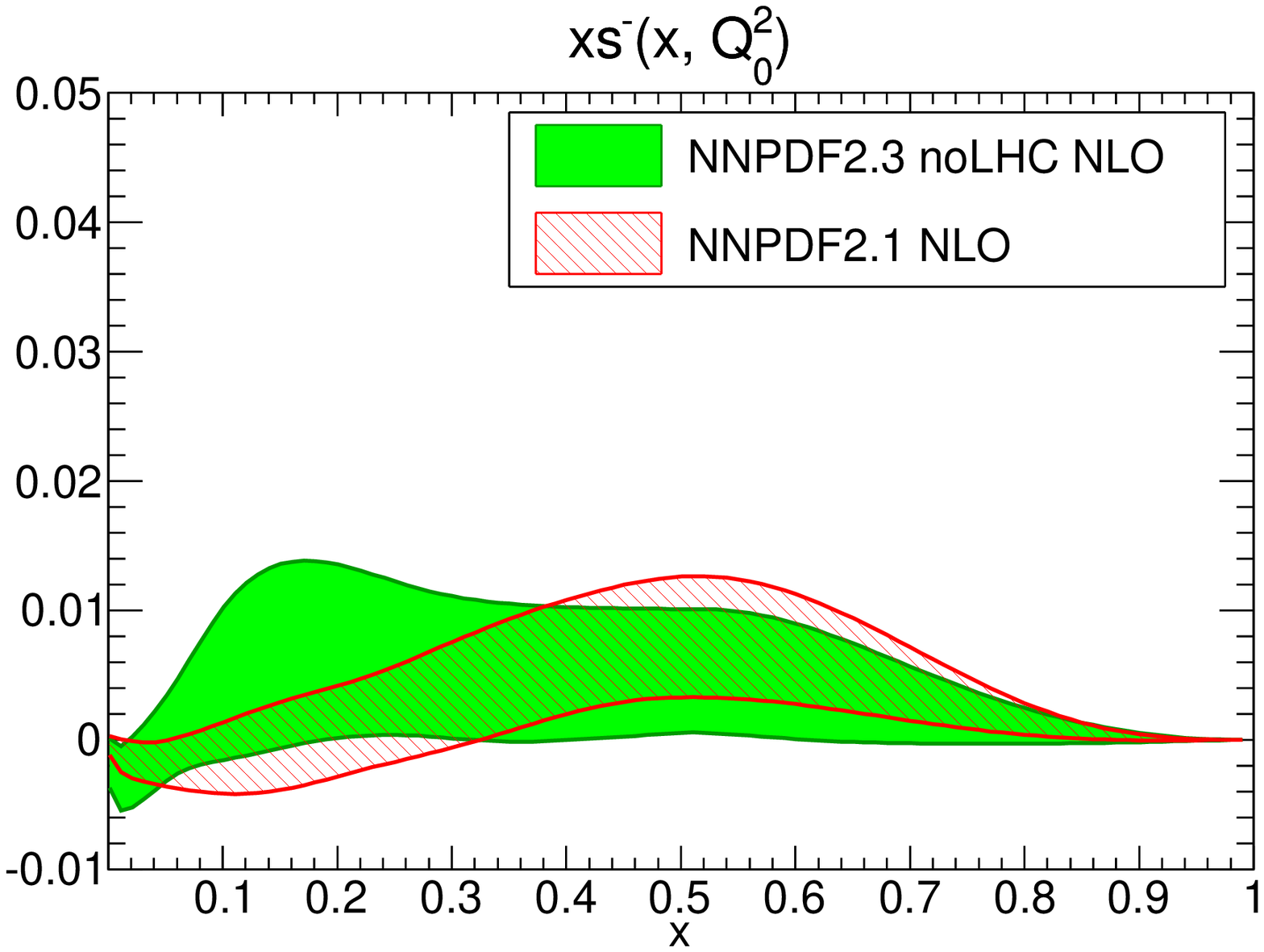}\\
\epsfig{width=0.44\textwidth,figure=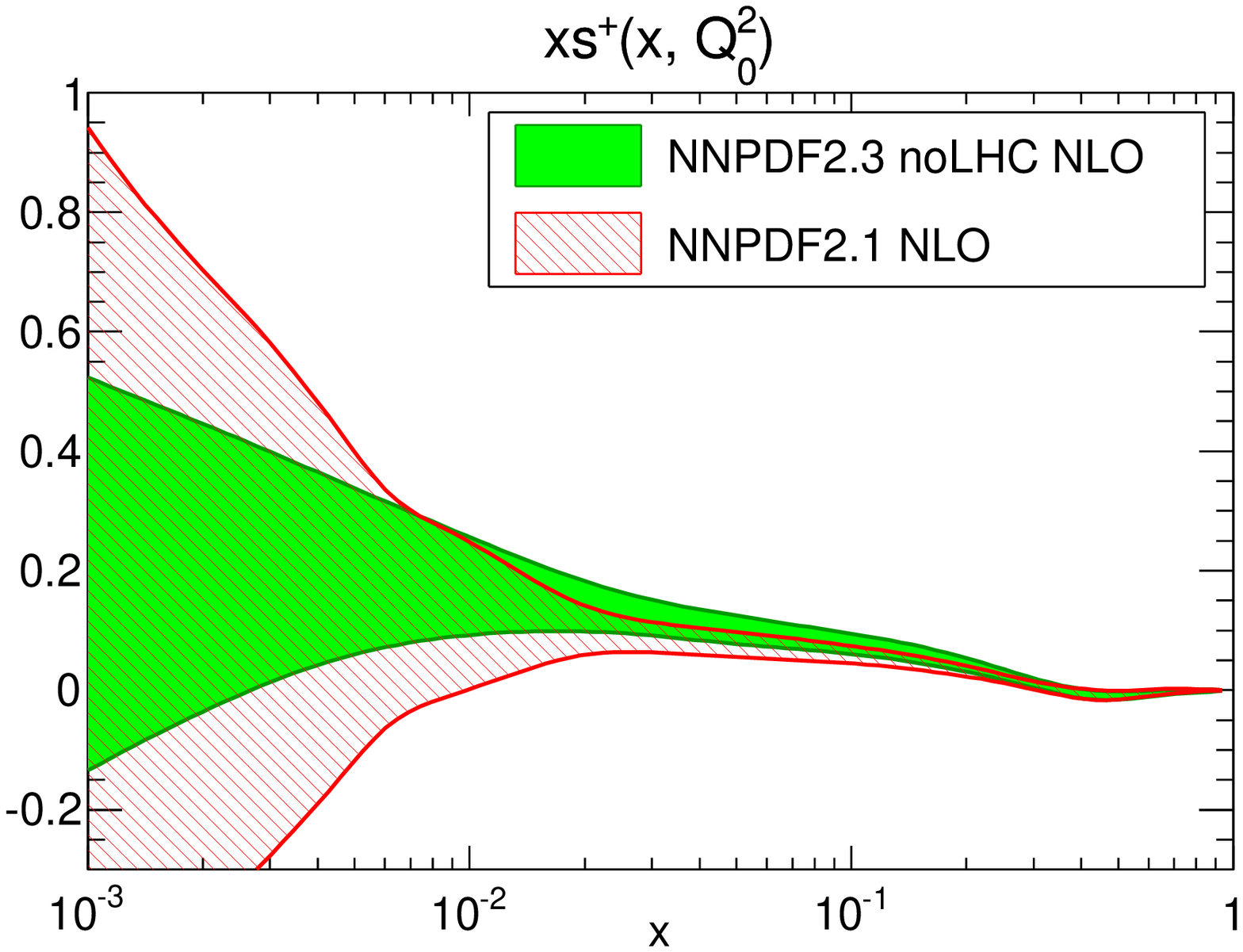}
\epsfig{width=0.44\textwidth,figure=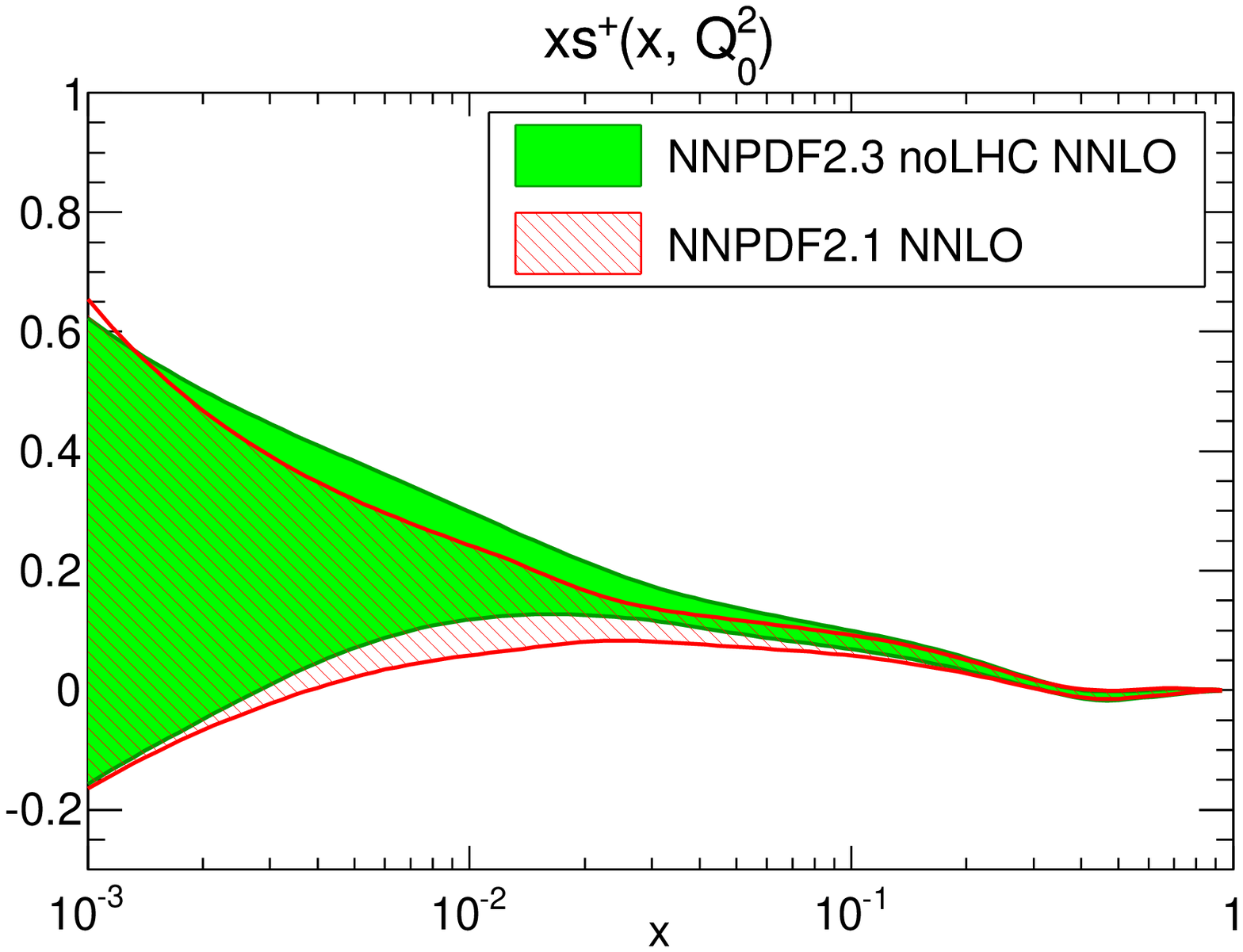}\\
\caption{ \small Comparison of some PDFs from the NNPDF2.1  and
NNPDF2.3 noLHC  sets: top: NLO small $x$ gluon (left),  $\bar d-\bar u$  
 (right);
middle: NLO
$s^-=s-\bar s$ small $x$ (left), large $x$  (right)
at $Q^2=2$ GeV$^2$;
bottom  small $x$ $s^+=s+\bar s$ at NLO (left) and NNLO (right).
\label{fig:xg_Q_2_log-21-vs-23noLHC.eps}} 
\end{center}
\end{figure}

We  first determine distances between NNPDF2.1 and NNPDF2.3 noLHC: this
measures the effect of  the various 
improvements, with fixed data set. The distances are shown in
Fig.~\ref{fig:distances-23noLHC-vs-21-nlo.eps} (at NLO)
and~\ref{fig:distances-23noLHC-vs-21-nnlo.eps} (at NNLO).  The largest
distances are observed between the NLO strange, gluon, and sea
asymmetry PDFs, for which a direct comparison between NNPDF2.1 and
NNPDF2.3 noLHC is shown in Fig.~\ref{fig:xg_Q_2_log-21-vs-23noLHC.eps}.

At NNLO the distances for essentially all PDFs except  total
strangeness 
are compatible with purely statistical fluctuations ($d\sim 1$ corresponds
to statistically equivalent fits). Strangeness (also shown in 
Fig.~\ref{fig:xg_Q_2_log-21-vs-23noLHC.eps}) changes in a
statistically significant way, though at most by
about half sigma, in the $10^{-2}\lsim x\lsim 10^{-1}$ range. We have
checked that if the error in Eq.~(\ref{eq:bug}) is corrected with
everything else left unchanged, then the distance in strangeness
is somewhat smaller, but roughly of the same order: hence the change
in strangeness is mostly due to the correction of this error. It is
interesting to observe that, despite the fact that the change in each
individual PDF is statistically insignificant, the improvement quality of the
global fit is still significant: the  $\chi^2$ per data point decreases from
$1.167$ to $1.147$, corresponding to an decrease by about
$70$ units of the total $\chi^2$ of the fit (which corresponds to a
decrease of the $\chi^2$ by slightly more than one sigma).
Note that this decrease
is not due  to the NuTeV dimuon data, whose $\chi^2$ only decreases by
a couple units, and thus it must 
be attributed to the improved minimization. Because at NNLO this only
amounts to  more
stringent stopping and higher maximum number of iterations, we must
conclude that the higher $\chi^2$ value in the NNPDF2.1 NNLO fit was
due to slight underlearning.

The improvement in fit quality is
even more marked at NLO, where essentially 
all PDFs undergo changes at the half sigma
level, and the $\chi^2$ decreases by about 150 units (i.e.~about two
sigma). 
The fact that more significant changes are observed in PDF
shapes at NLO can be understood as a consequence of having increased
the number of mutations and mutants in this case.


\begin{figure}[t]
\begin{center}
\epsfig{width=0.91\textwidth,figure=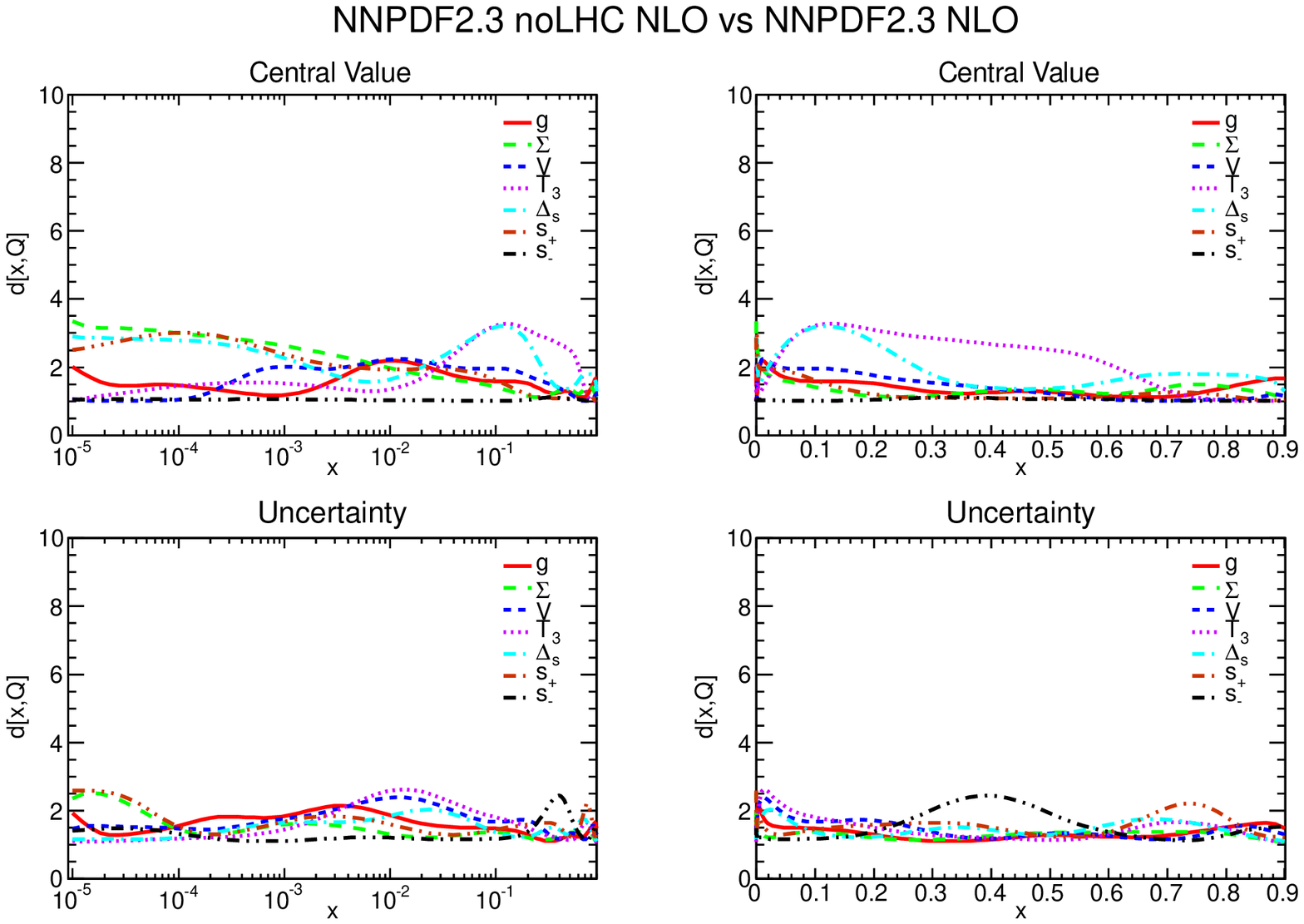}
\caption{ \small Distances between NNPDF2.3 noLHC and NNPDF2.3 NLO.
\label{fig:distances-23nolhc-vs-23-nlo.eps}} 
\end{center}
\end{figure}

\begin{figure}[t]
\begin{center}
\epsfig{width=0.93\textwidth,figure=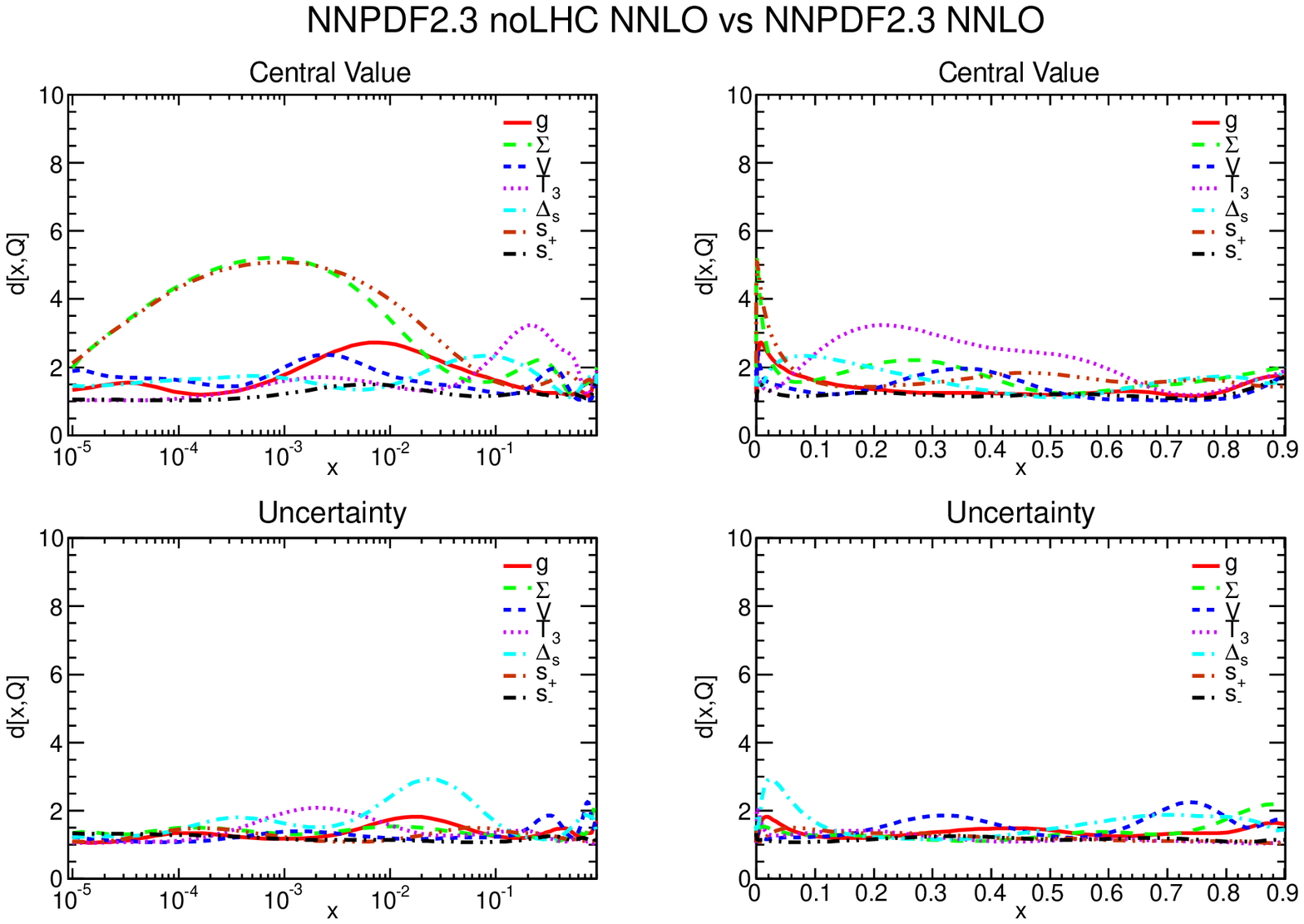}
\caption{ \small Distances between NNPDF2.3 noLHC and NNPDF2.3 NNLO.
\label{fig:distances-23nolhc-vs-23-nnlo.eps}} 
\end{center}
\end{figure}

\begin{figure}[t]
\begin{center}
\epsfig{width=0.49\textwidth,figure=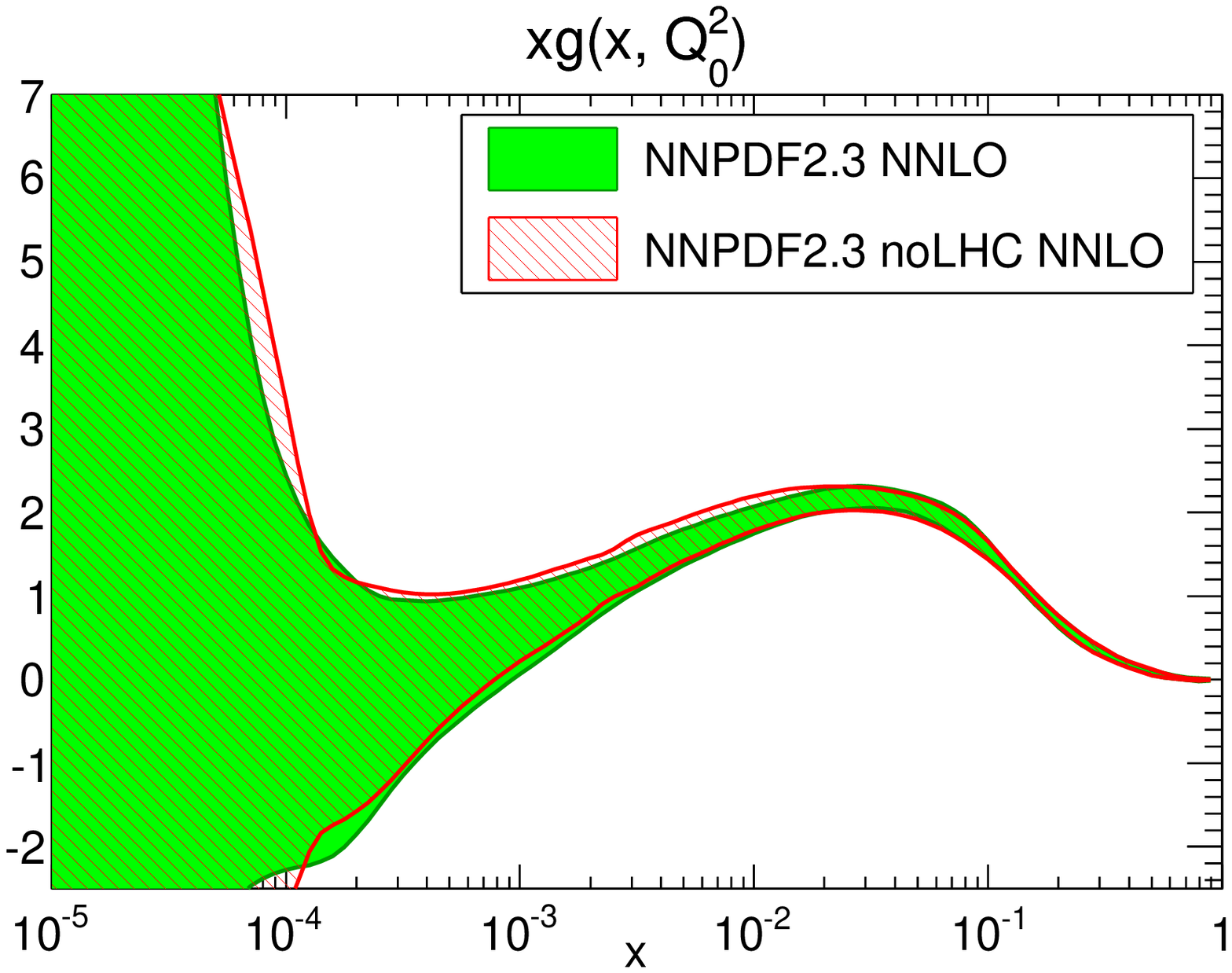 }
\epsfig{width=0.49\textwidth,figure=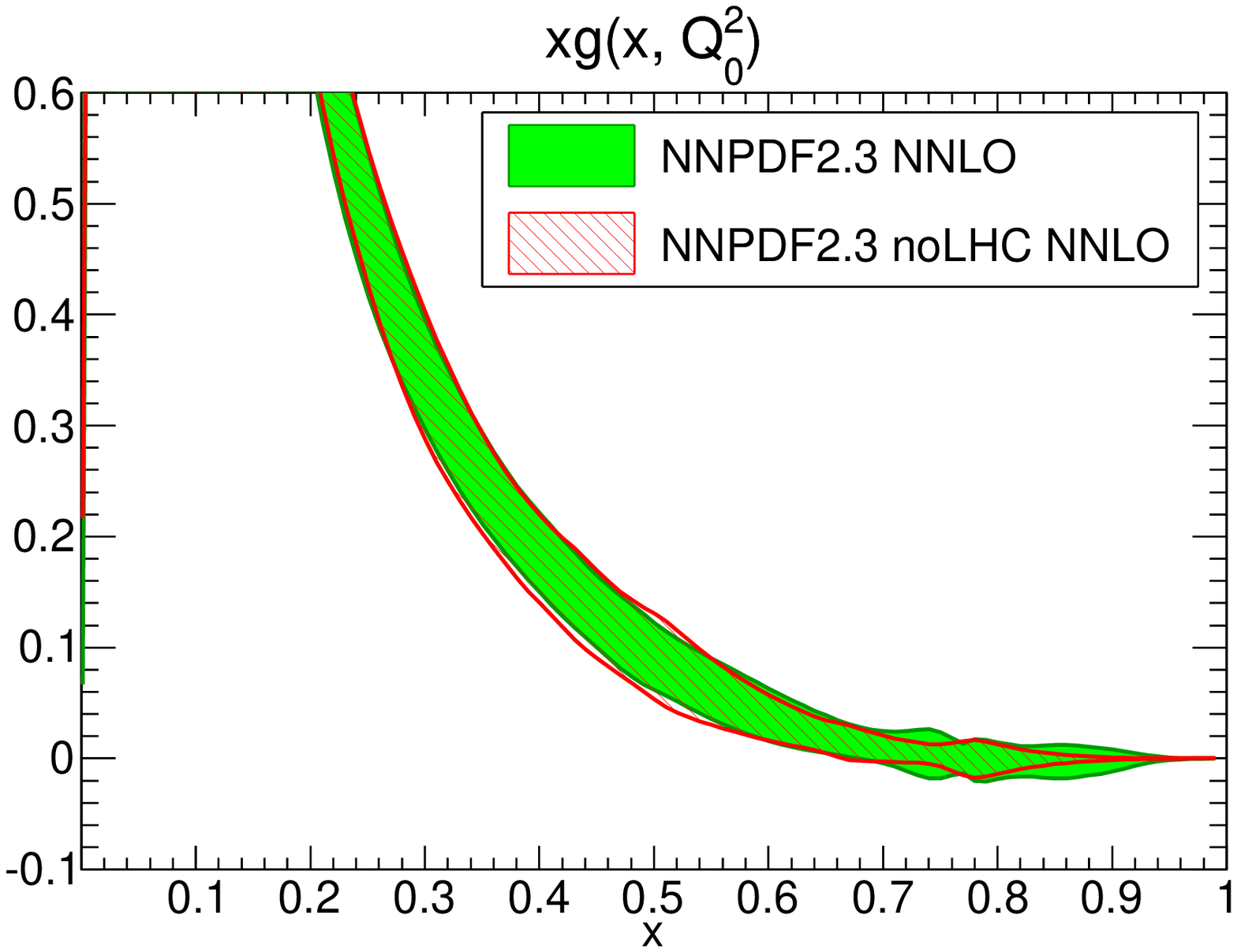 }\\
\epsfig{width=0.49\textwidth,figure=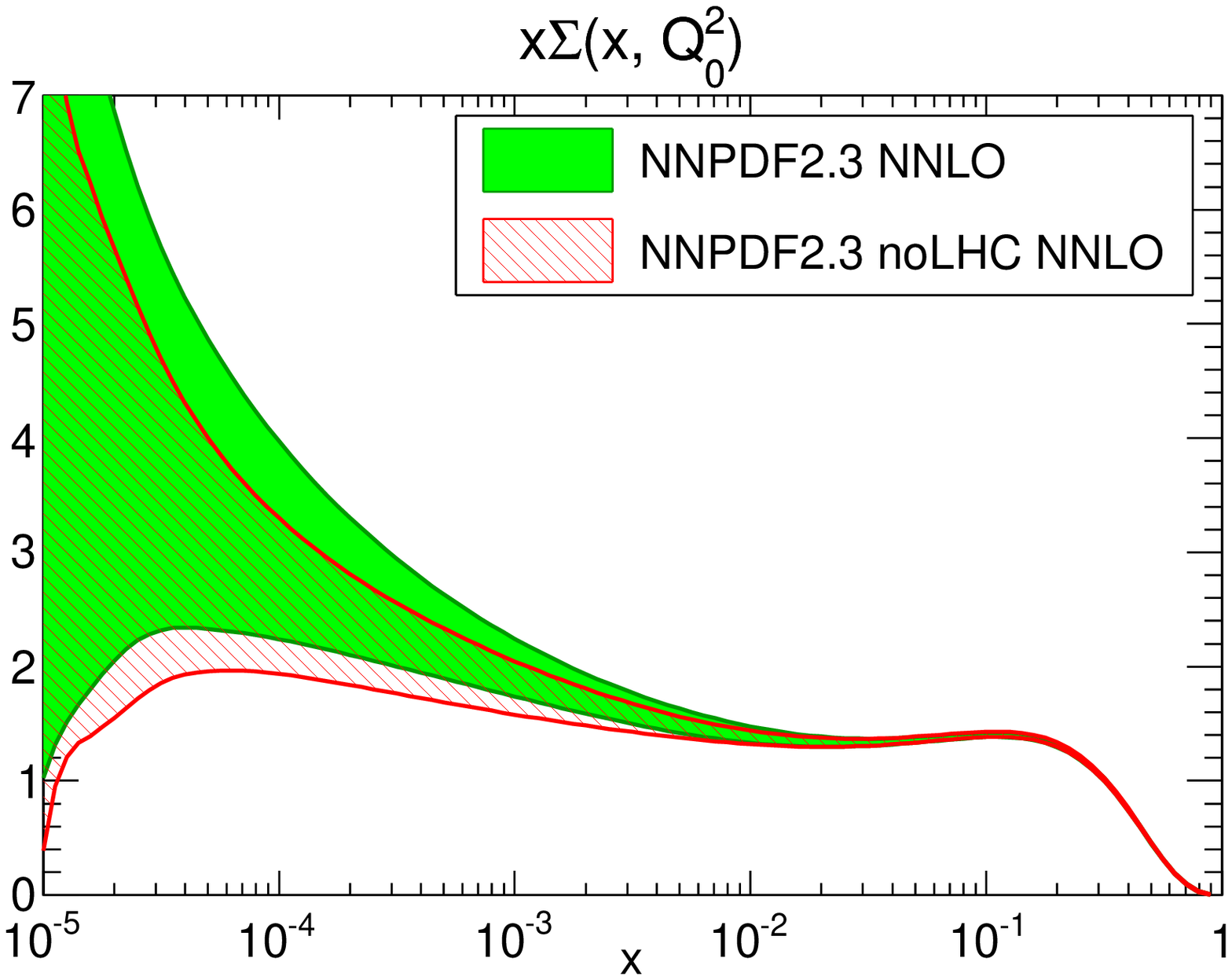 }
\epsfig{width=0.49\textwidth,figure=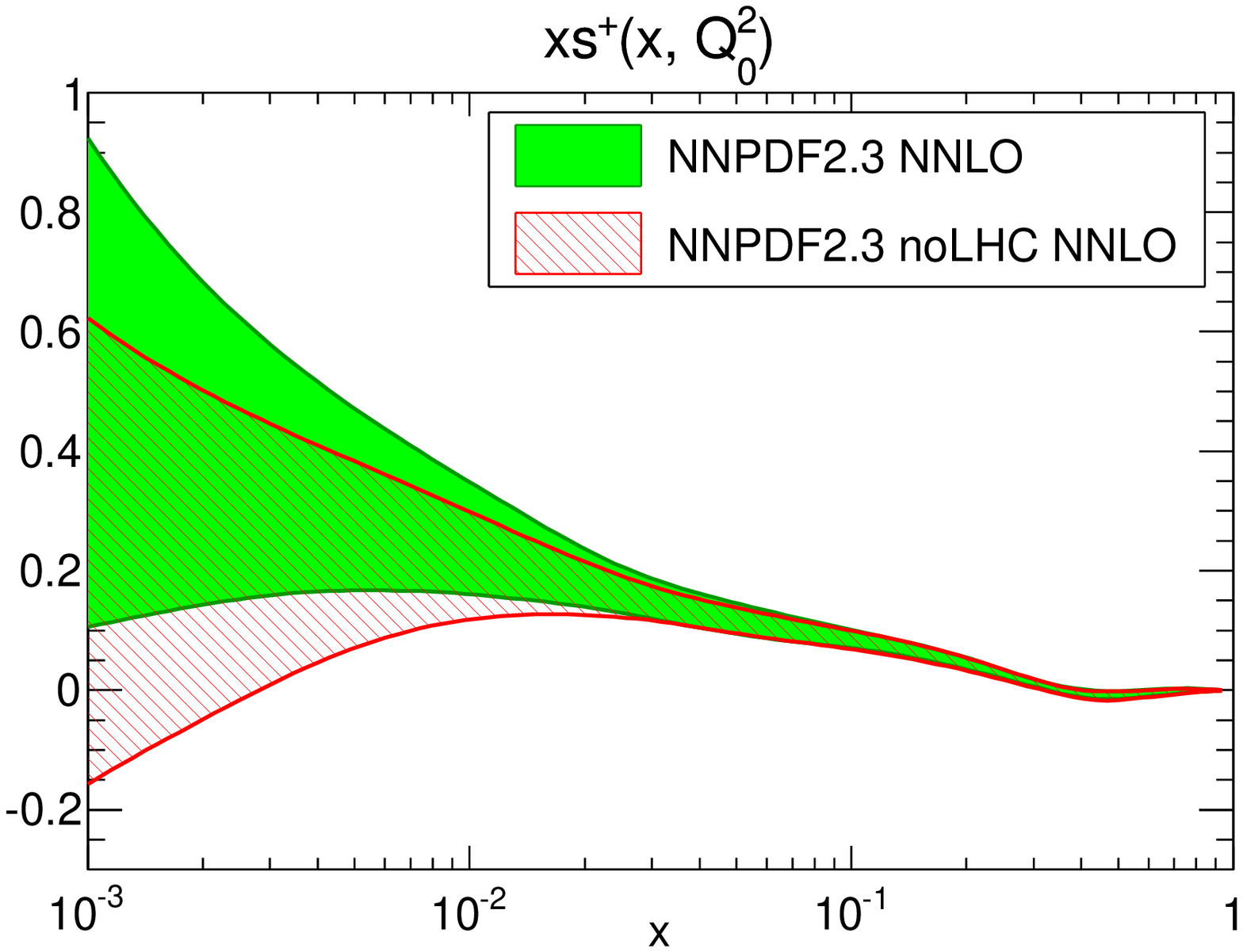}
\caption{ \small 
Comparison of some PDFs from the  NNPDF2.3 noLHC and
NNPDF2.3 NNLO sets. Top: gluon at  small $x$ (left) and large $x$ (right);
bottom:  small $x$  total singlet (left) and total strangeness (right).  
\label{fig:xg_Q_2_lin-23-vs-23noLHC-nnlo.eps}} 
\end{center}
\end{figure}

Next we compare the NNPDF2.3 noLHC and the NNPDF2.3 fits at NLO and
NNLO, in order to gauge the genuine impact of LHC data. The distances
between these two fits are shown in
Fig.~\ref{fig:distances-23nolhc-vs-23-nlo.eps}
and~\ref{fig:distances-23nolhc-vs-23-nnlo.eps}.
The largest
distances in central values
are observed between the NNLO total strangeness and quark singlet at
small $x$
PDFs, and to a lesser extent the gluon
for all of which a direct comparison between NNPDF2.1 and
NNPDF2.3 noLHC is shown in Fig.~\ref{fig:xg_Q_2_lin-23-vs-23noLHC-nnlo.eps}.

The impact of LHC data is clearly moderate, with no distance larger than
four at NLO, which means that the fitted PDFs differ by less than half sigma. 
At NNLO the effect seems to be a bit larger. This
confirms the consistency of the PDFs extracted from lower energy
experiments with the PDFs extracted from LHC data. The description of
the LHC experiments is of course better in NNPDF2.3 than in NNPDF2.3
noLHC, although the starting agreement was already very reasonable, as
is clearly seen in the $\chi^2$ comparison shown in
Tab.~\ref{tab:estfit2dataset}. 

While all changes are moderate, the main effect of the LHC data is to
lead to somewhat
smaller gluon uncertainties, thanks to the
inclusion of the ATLAS jet data, and especially a more accurate 
light quark flavor decomposition thanks to the LHC electroweak vector boson
production data. This in turn leads to an improvement in the
accuracy of standard candle cross-sections, both dependent on the
gluon (top production) and the quark (gauge boson production), as we will see in
Sec.~\ref{sec-tot}. 
 
\section{The LHC data}
\label{sec-collider}

We will now address in more detail the impact of LHC data on
NNPDF2.3 parton distributions. First, we extend the discussion of
Sec.~\ref{sec-comp}, where NNPDF2.3 fits with and without LHC data
were compared, by comparing a fit in which LHC data are included in
the full fitting data set with a fit in which the LHC data are
included by reweighting a fit which does not include them: this
enables us to assess more accurately both the impact and the
consistency of the LHC data in the global fit.  This analysis is
supplemented by an explicit assessment of the quality of the fit to
LHC data before and after their inclusion.
We then consider a PDF determination which is based on collider data
only. This enables us to discuss the consistency between collider
data and the lower-energy fixed-target data, several of which are
affected by nuclear corrections. The LHC data, by expanding the set of
available collider data, sheds some further light on this comparison.
Finally, we present a dedicated analysis of the strangeness fraction
of the nucleon, and the impact of LHC data on its determination.

\subsection{Impact and consistency of the LHC data in the global fit}
\label{sec-diff}

New data can be included in an existing PDF fit either by simply
redoing the fit, or by reweighting the replicas of an existing
fit\footnote{The possibility of fitting PDFs to new data via
  reweighting was originally suggested in
  Refs.~\cite{Giele:1998gw,Giele:2001mr}. However, the expression for
  the weights given in these references was incorrect due to an
  argument, explained in Sect.~2.3 of Ref.~\cite{Ball:2010gb},  
related to the so-called Borel-Kolmogorov paradox of probability
theory~\cite{jaynes}. Use of the formula for the weight of
Ref.~\cite{Giele:1998gw} generally leads to all the normalized weights
but one to be vanishingly small (i.e. only one replica survives
reweighting). The reweighting method of
Ref.~\cite{Ball:2010gb,Ball:2011gg} was recently succesfully used
in Ref.~\cite{Watt:2012tq} in the context of the MSTW PDF fits.}~\cite{Ball:2010gb,Ball:2011gg}. Besides providing a strong
consistency check of the fitting procedure, the inclusion of the data
by reweighting allows one to test for the impact and consistency
of the new data. For example, one can determine the probability
distribution $P(\alpha)$ for the rescaling of uncertainties of the new
data set by a factor $\alpha$: for fully consistent data the
mean value of $\alpha$ should be $\langle\alpha\rangle\sim
1$. Furthermore, one may determine the  effective number of replicas
$N_{\rm eff}$ left after reweighting the initial set of $N_{\rm rep}$
replicas: if $N_{\rm eff}\ll N_{\rm rep}$ the new data either bring in
considerable new information or they are very
inconsistent with the pre-existing data, and conversely. 

\begin{figure}[hb!]
    \begin{center}
      \includegraphics[width=0.4\textwidth]{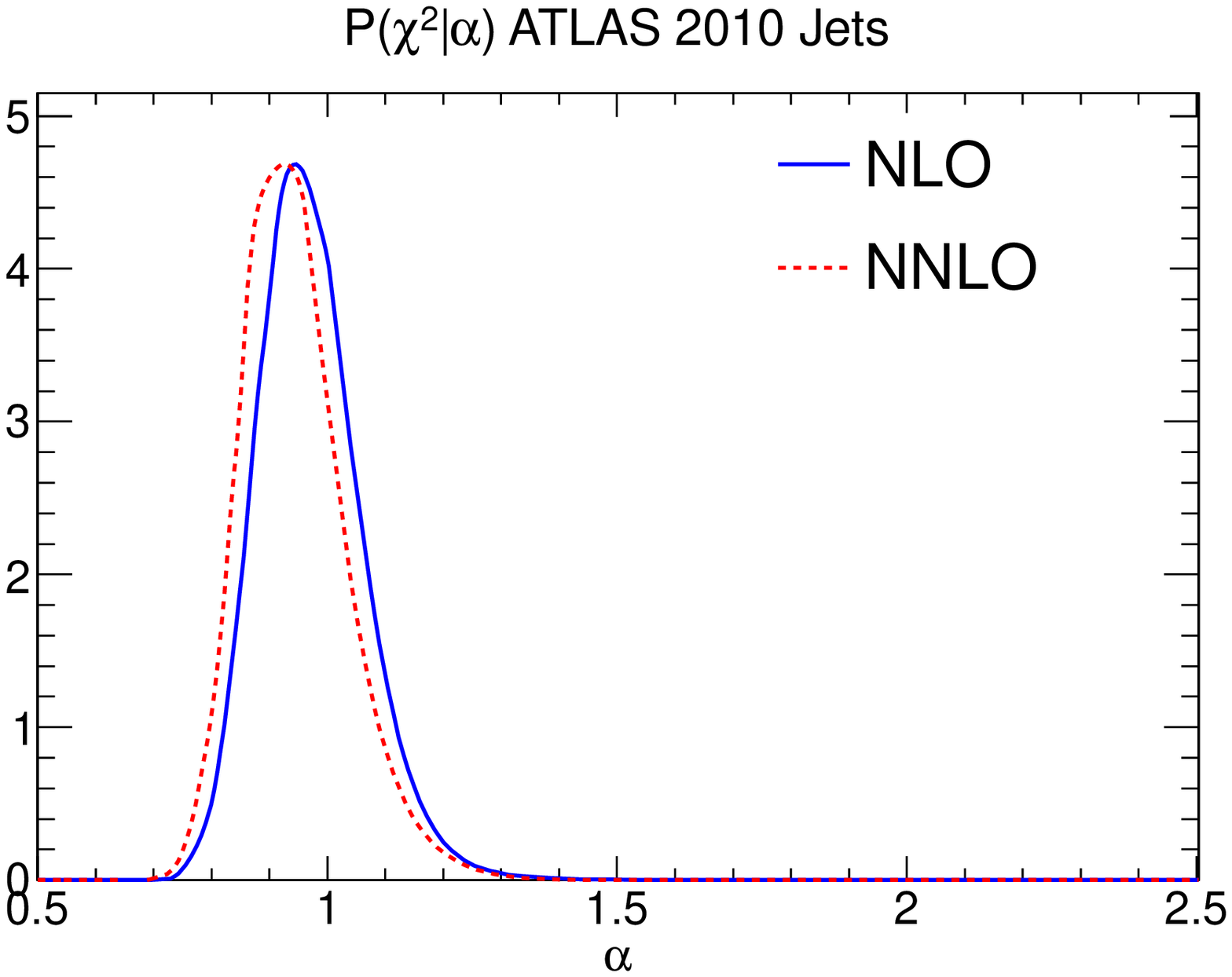}
      \includegraphics[width=0.4\textwidth]{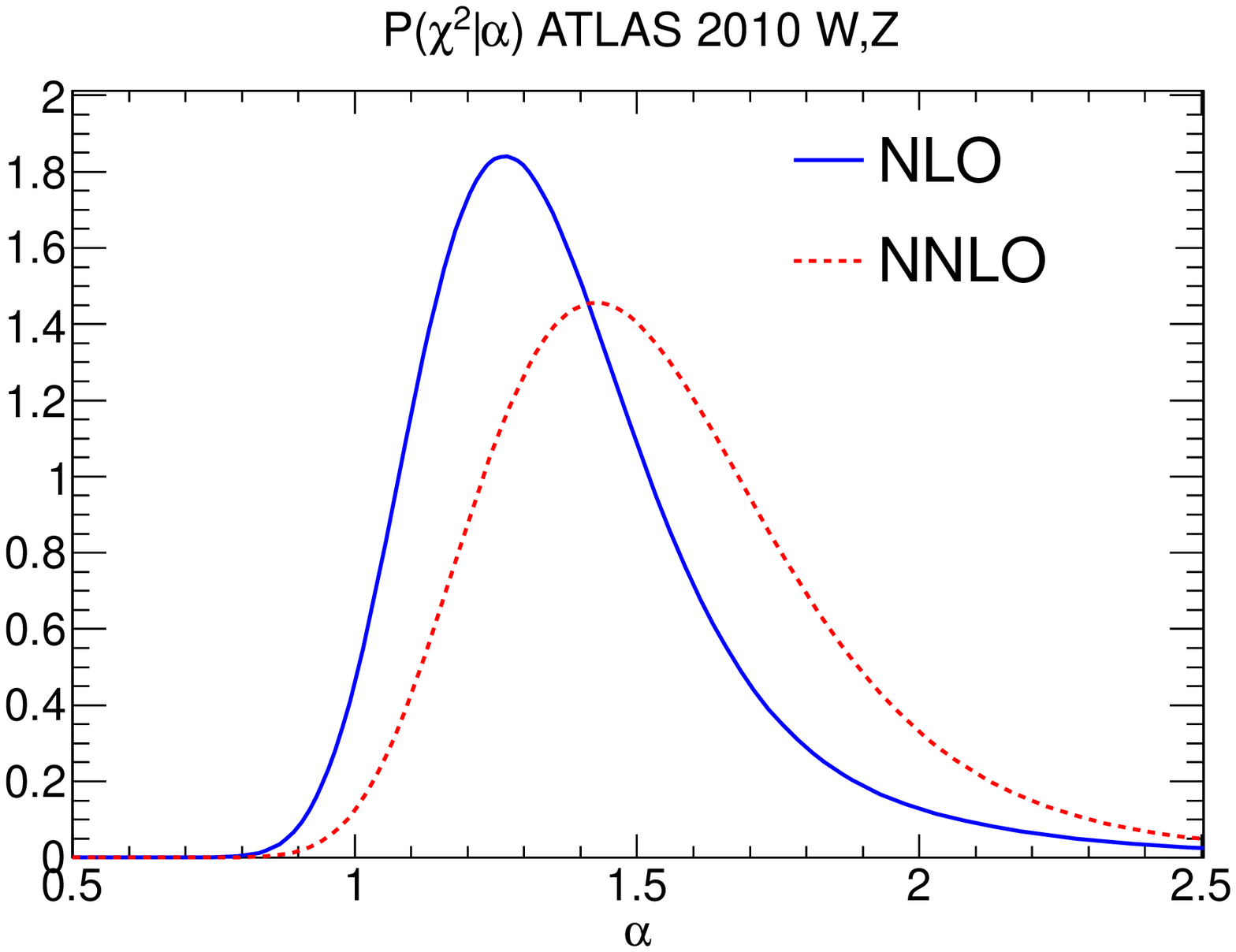}
      \includegraphics[width=0.4\textwidth]{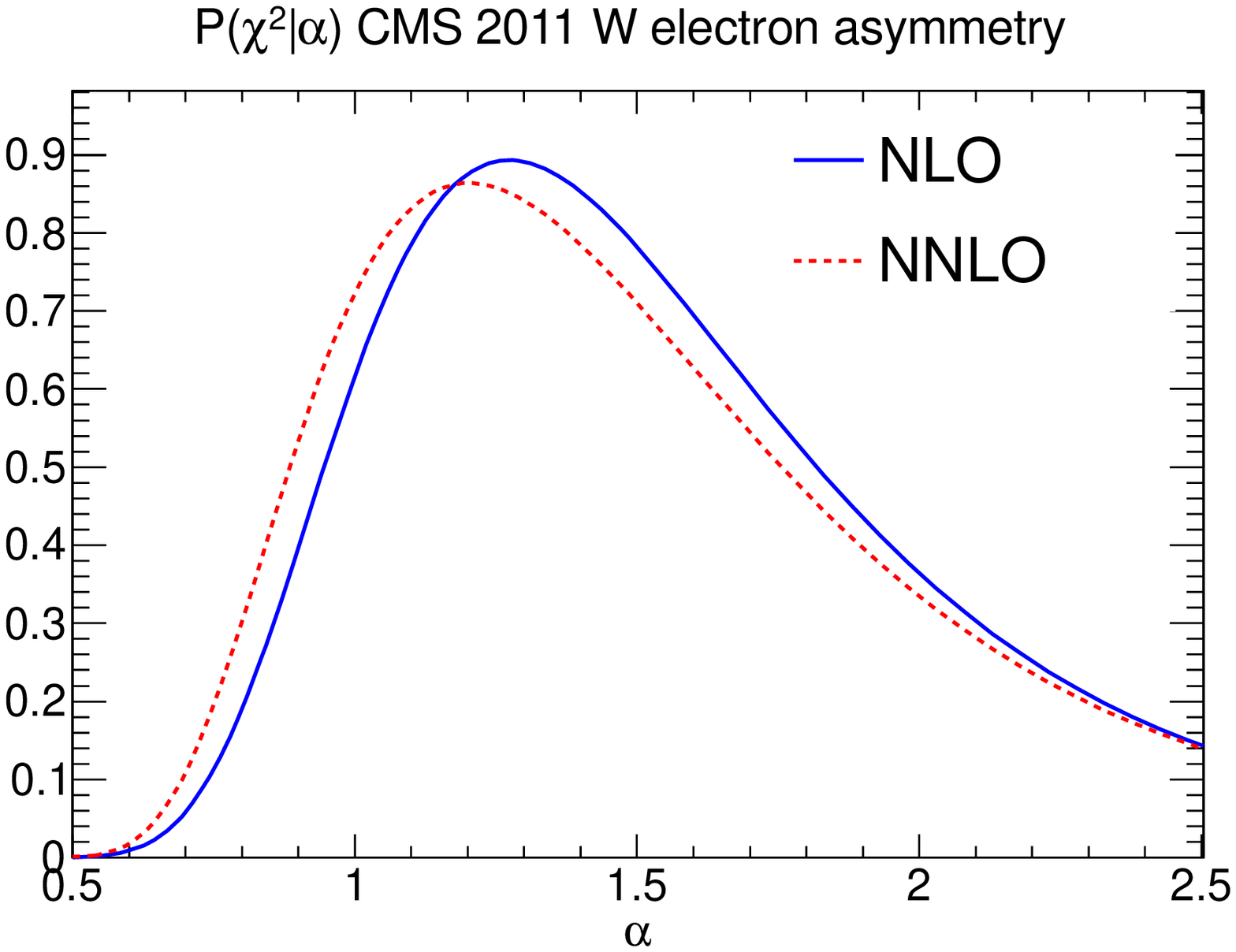}
      \includegraphics[width=0.4\textwidth]{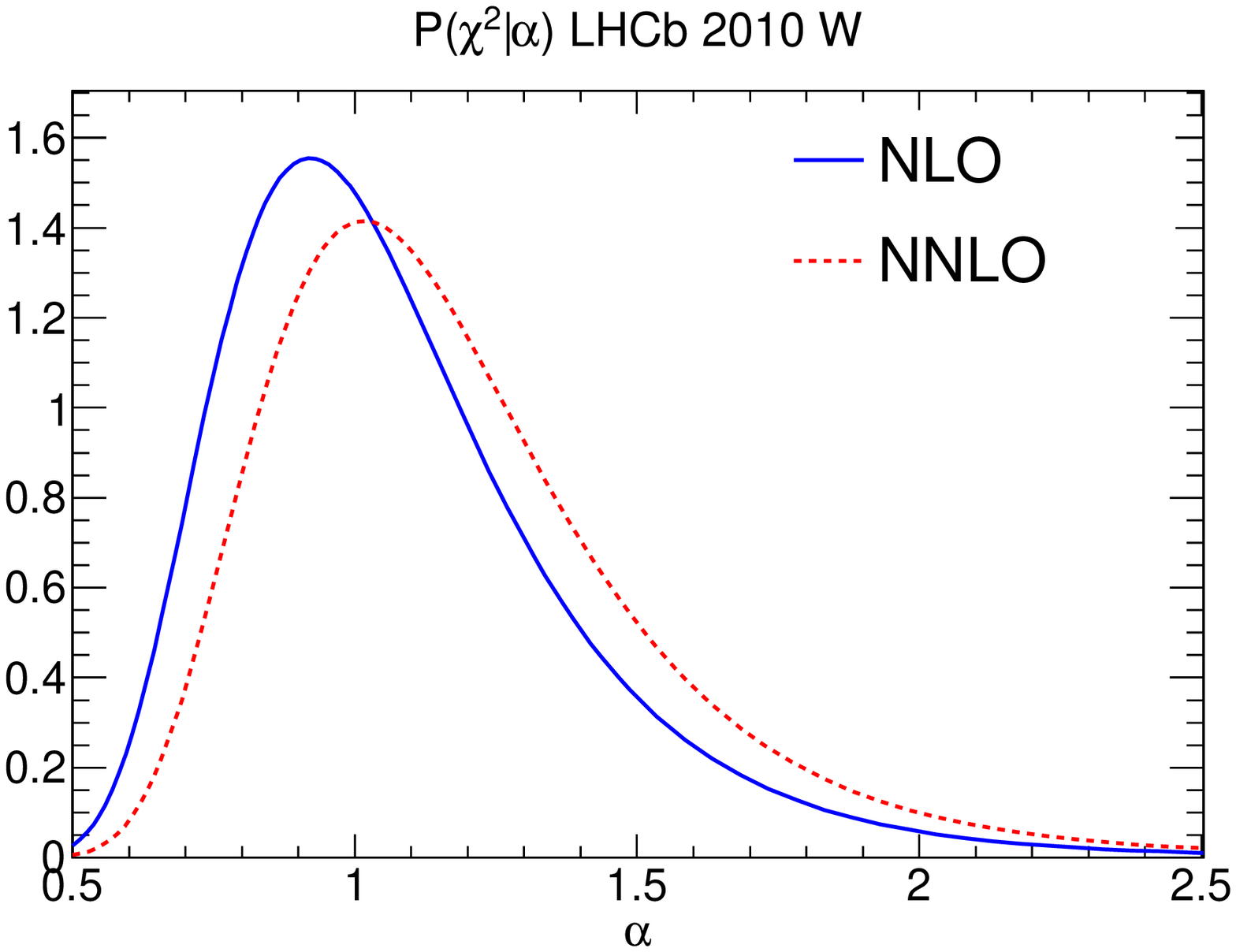}
    \end{center}
    \vskip-0.5cm
    \caption{\small The $P\lp \alpha \rp$ distributions for each
of the LHC experiments included in NNPDF2.3: the ATLAS 2010 jets,
the ATLAS 2010 W,Z data, the CMS 2011 W electron asymmetry and
the LHCb 2010 W data, all  determined using either NLO or
NNLO theory, with $\alpha_s(M_{\rm Z})=0.119$. The mean values of $\alpha$
computed from each of these distributions are given in Tab.~\ref{tab:rw}.}
    \label{fig:palpha}
\end{figure}

We have thus performed the reweighting of a set of $500$
NNPDF2.3 noLHC NLO or NNLO replicas with $\alpha_s(M_{\rm Z})=0.119$ with 
the various LHC data sets. The $\chi^2$
obtained comparing to the data predictions obtained using this 
reweighted set are compared
in Tab.~\ref{tab:estfit2dataset} to those of the standard NNPDF2.3
NLO and NNLO fits. The good agreement between reweighted and refitted
$\chi^2$ values provides evidence for the consistency and
efficiency of the fitting methodology. Furthermore, in
Fig.~\ref{fig:palpha} we show the probability distributions $P(\alpha)$ 
for each of the four LHC data sets added to the fit, at NLO and NNLO, and 
in Tab.~\ref{tab:rw} the values of $\langle\alpha\rangle$ and $N_{\rm eff}$ 
obtained by reweighting at NLO or NNLO first
with each LHC data set individually, and then with all LHC data. The values of
$P(\alpha)$ distributions generally show good consistency of the LHC data 
with the rest, with perhaps a marginal
inconsistency for the ATLAS and CMS gauge boson production. This might
suggest some tension between the flavor decomposition favoured by
low energy and collider data, to which we
will return in Sec.~\ref{sec-reduced}. The values of $N_{\rm eff}$
show that even though the impact of the LHC data is moderate, it is
not negligible: for example, the ATLAS gauge boson production data alone
effectively discard more than two thirds of the starting replicas.
The impact of the LHC data is clearly
seen in Tab.~\ref{tab:estfitsigdataset}: the
uncertainty in the prediction for gauge boson production 
at the LHC is roughly
halved if NNPDF2.3 PDFs rather than NNPDF2.3 noLHC PDFs are used.


\begin{table}
\centering
\begin{tabular}{c||c|c||c|c}
\hline
\multicolumn{5}{c}{NNPDF2.3 noLHC reweighted with LHC data}  \\
\hline 
\hline 
 & \multicolumn{2}{c||}{NLO} & 
\multicolumn{2}{c}{NNLO} \\
\hline 
& $N_{\rm eff}$ & $\la \alpha \ra$ & $N_{\rm eff}$ & $\la \alpha \ra$ \\
ATLAS W/Z & 285  & 1.4 & 134 & 1.6 \\
CMS W e asy & 284 & 1.6  & 290  & 1.6 \\
LHCb W & 492 & 1.1 & 483 & 1.2 \\
ATLAS inclusive jets & 476  &  1.0 & 456  & 0.9 \\
\hline
All LHC data & 338 & 1.1 & 271 & 1.2 \\
\hline 
\end{tabular}
\caption{\small \label{tab:rw} The effective number of replicas
  $N_{\rm eff}$ and average uncertainty rescaling $\langle
  \alpha\rangle$ (defined in Refs.~\cite{Ball:2010gb,Ball:2011gg}) 
obtained by reweighting a set of  $N_{\rm rep}=500$ NNPDF2.3 noLHC NLO
and NNLO replicas with $\alpha_s(M_{\rm Z})=0.119$ with each of the LHC
data sets, and with all of them.  }
\end{table}

\begin{figure}[ht]
    \begin{center}
      \includegraphics[width=0.48\textwidth]{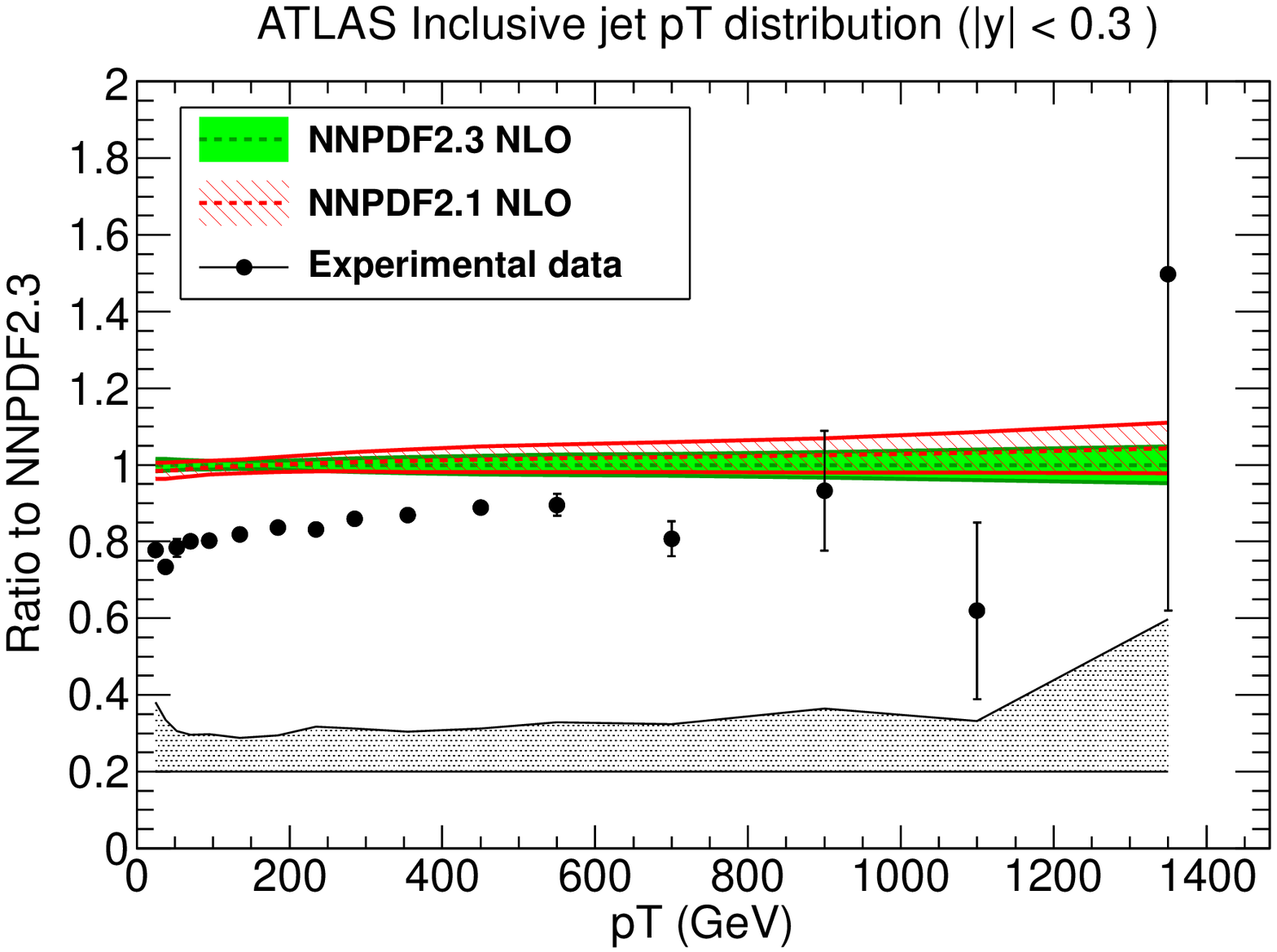}
      \includegraphics[width=0.48\textwidth]{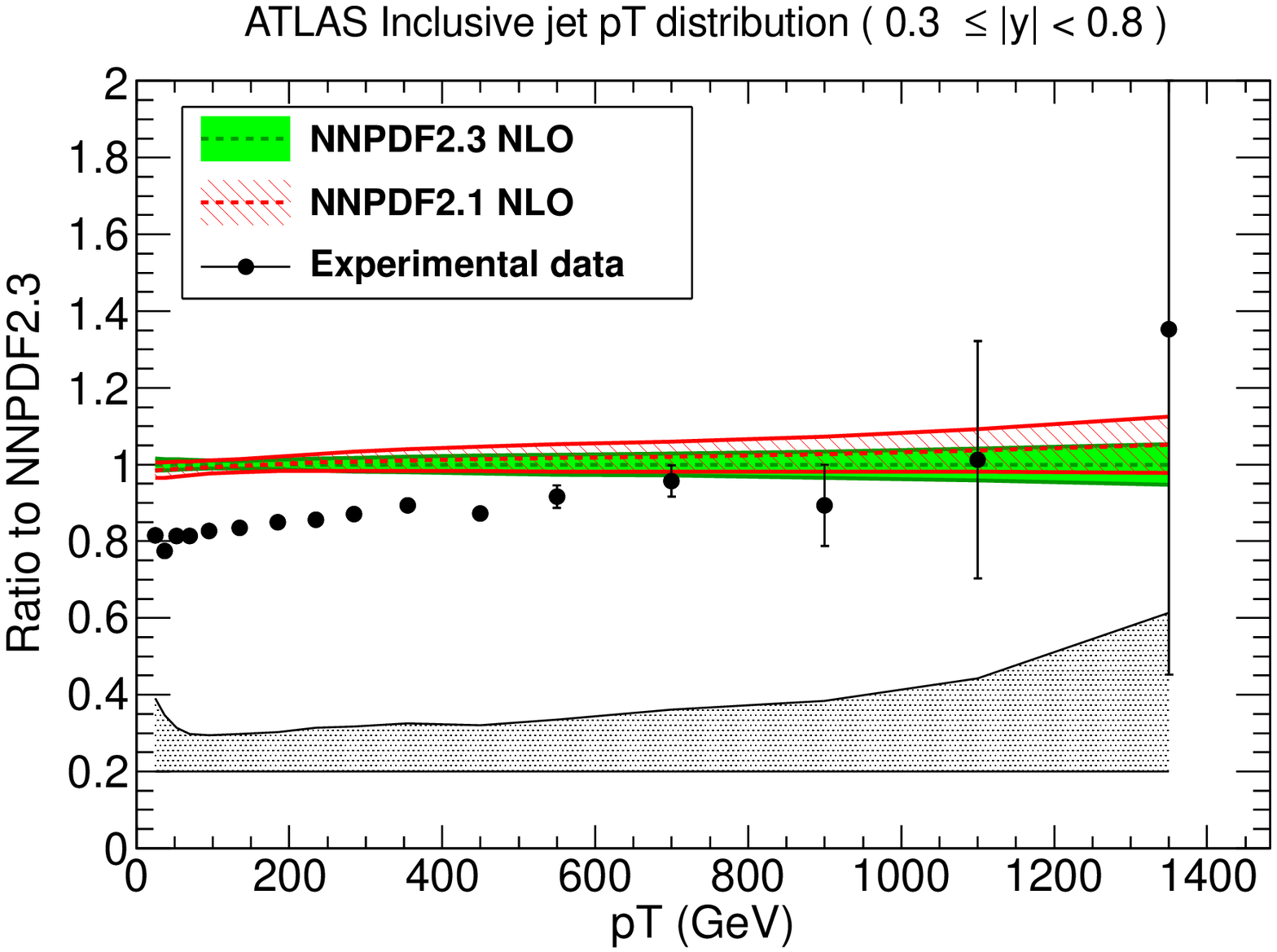}
      \includegraphics[width=0.48\textwidth]{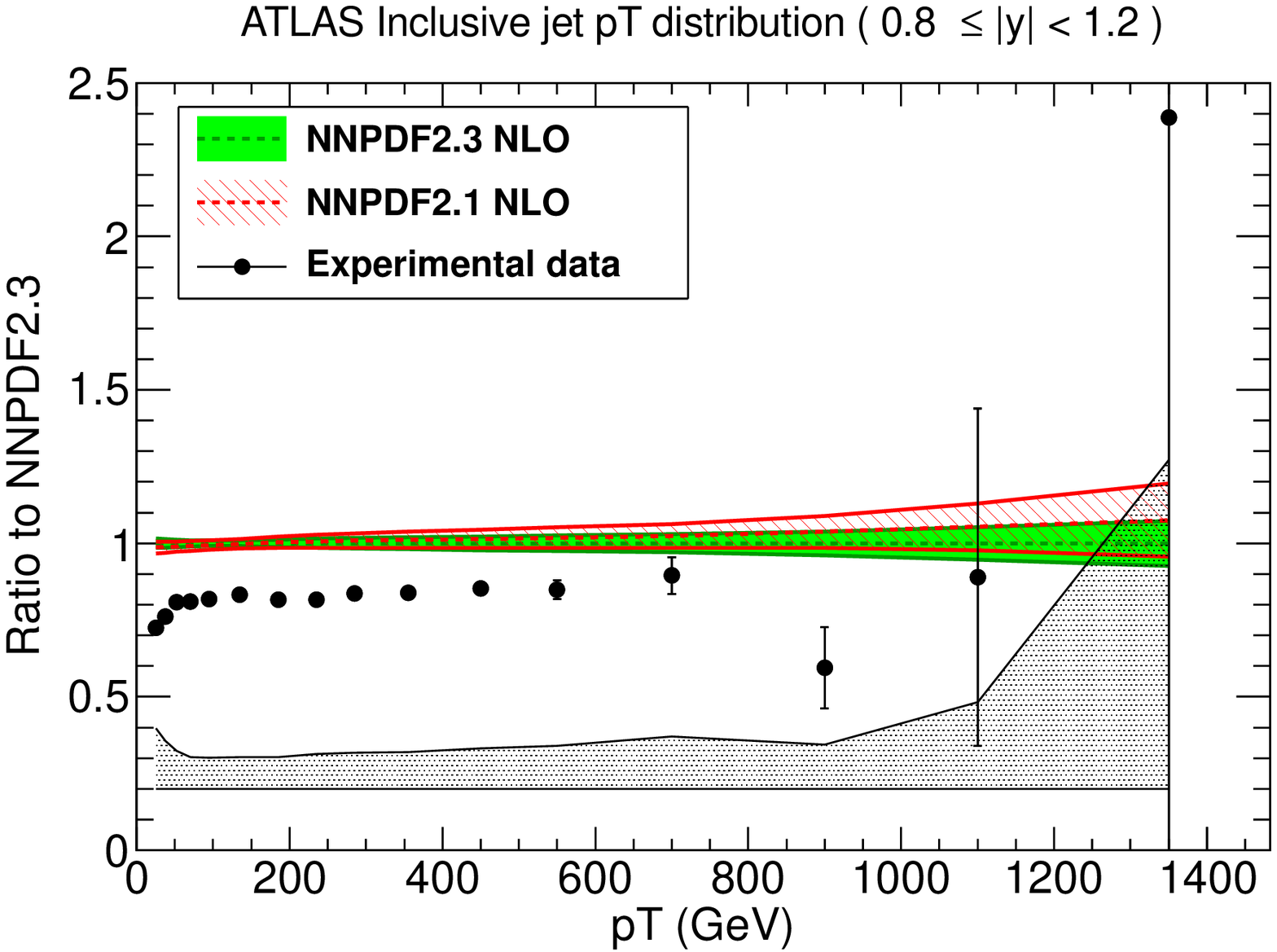}
      \includegraphics[width=0.48\textwidth]{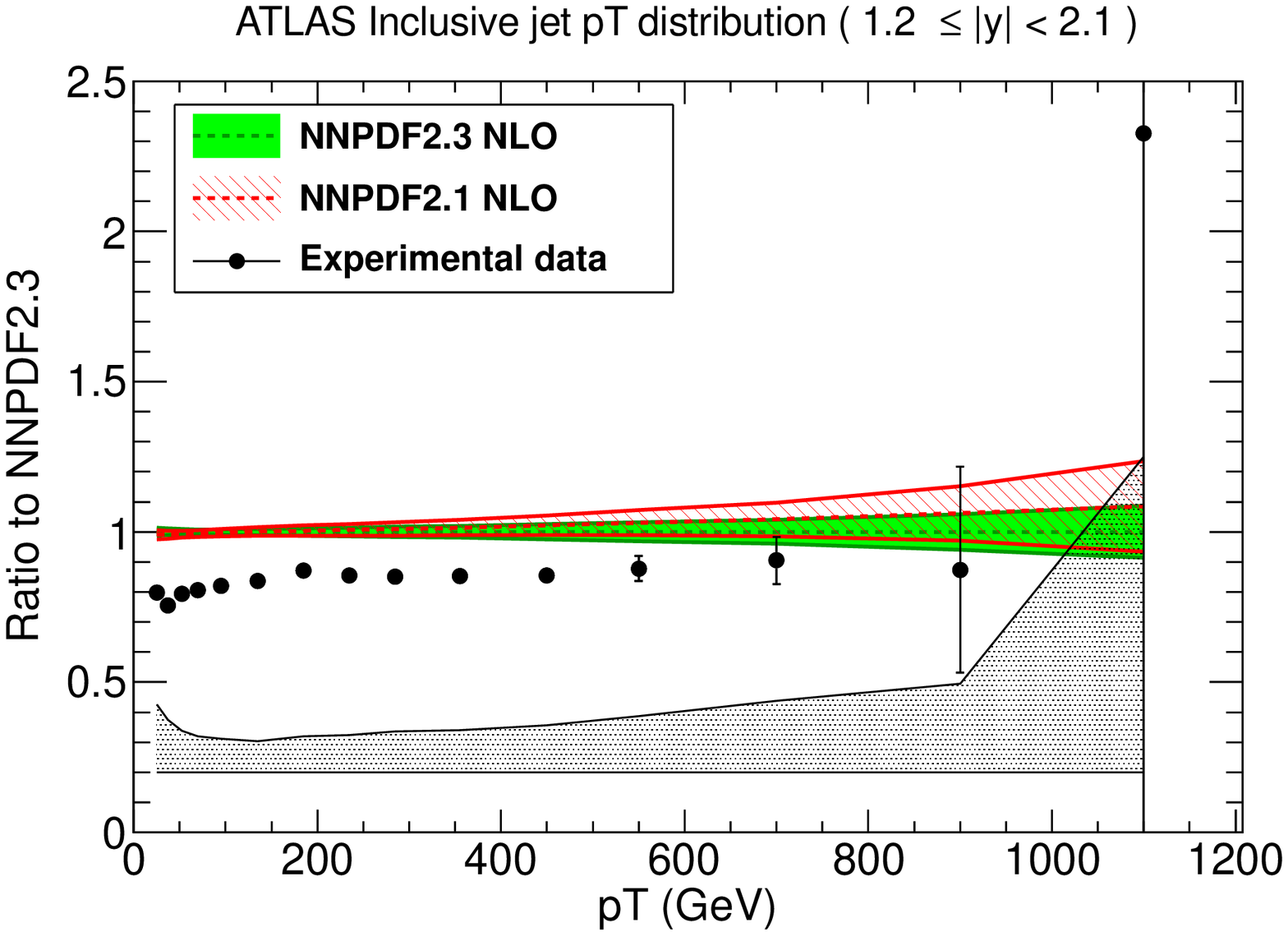}
      \includegraphics[width=0.48\textwidth]{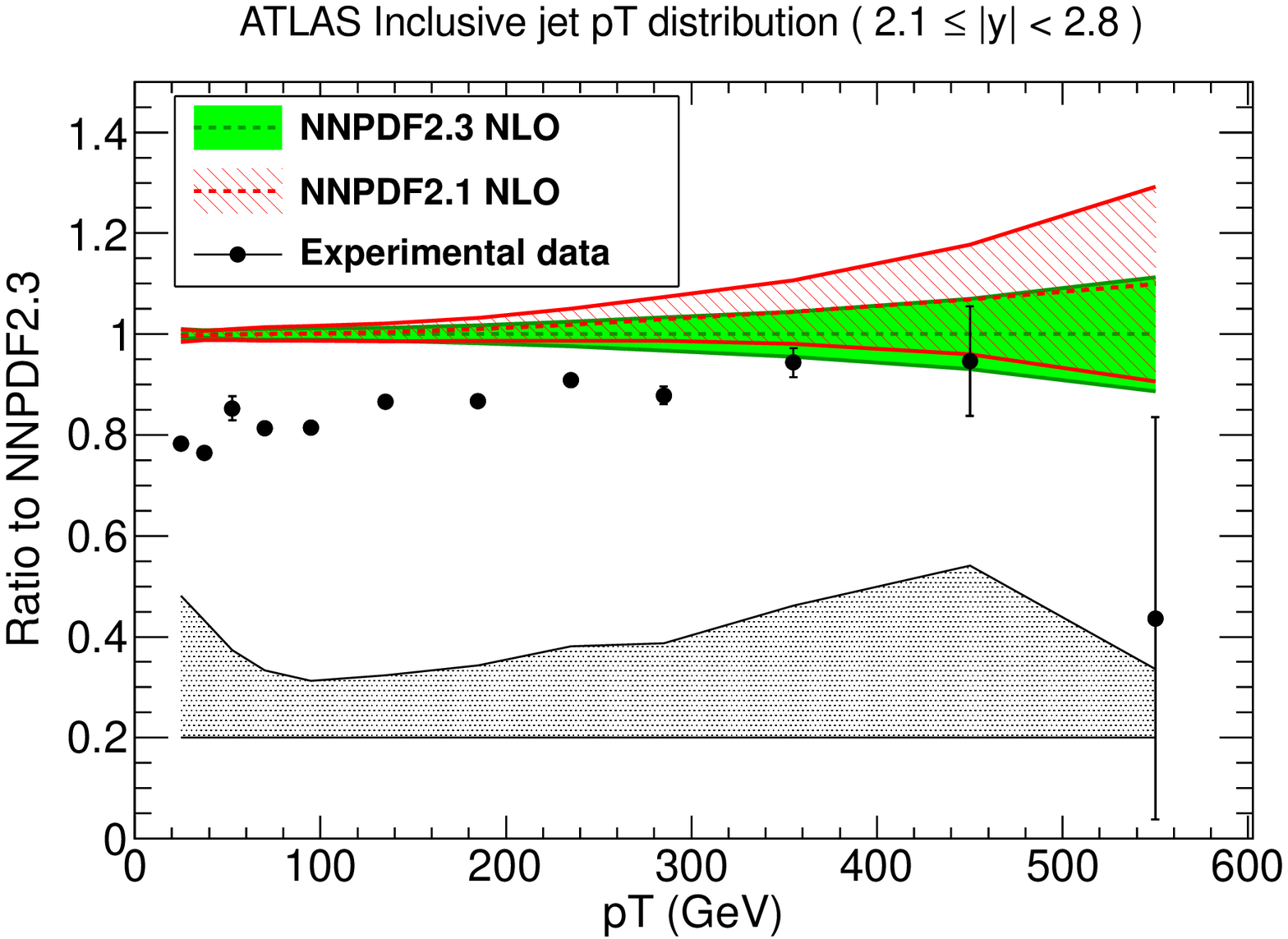}
      \includegraphics[width=0.48\textwidth]{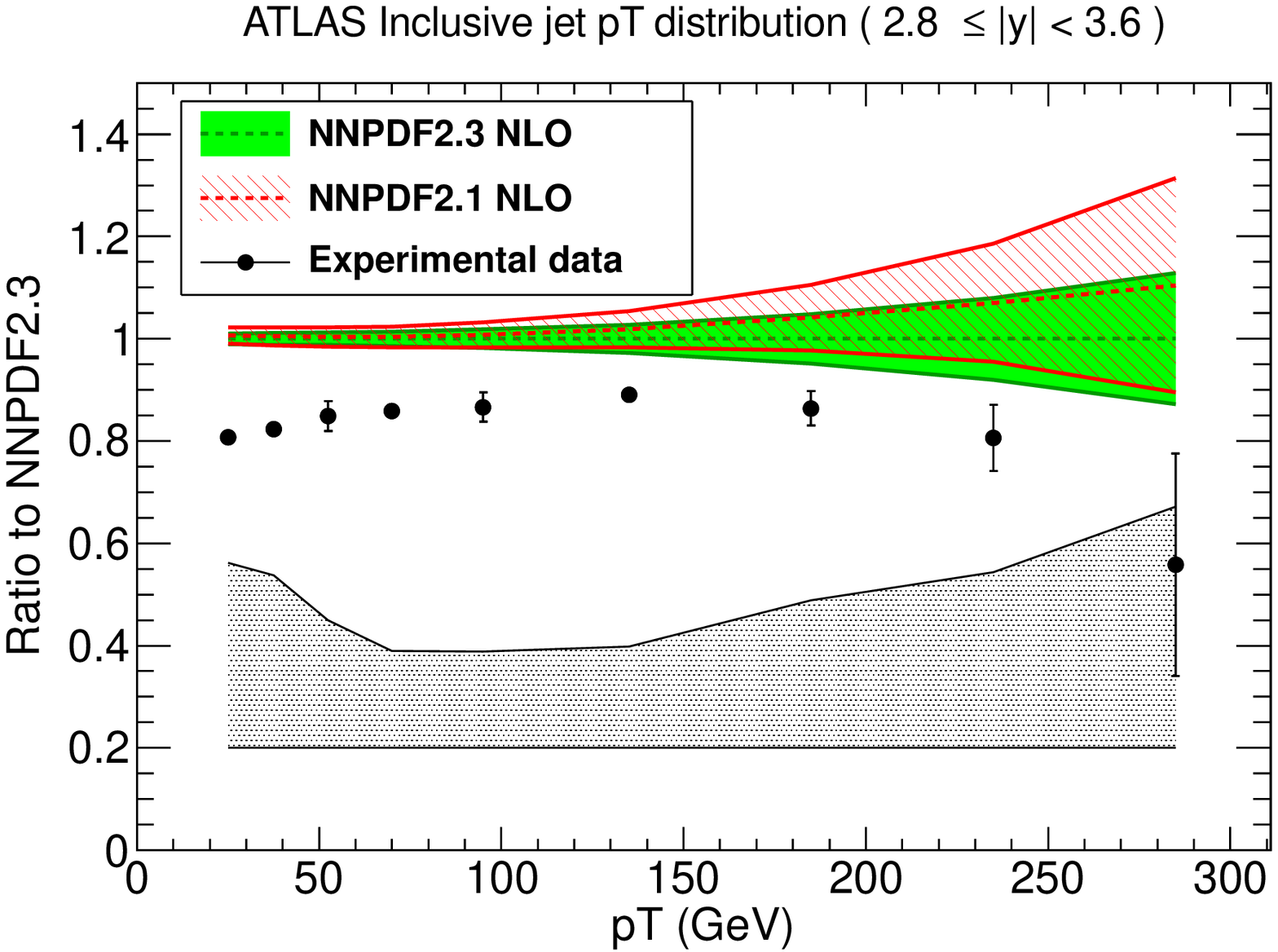}
    \end{center}
    \vskip-0.5cm
    \caption{\small Comparison of the ATLAS inclusive jet data 
with predictions from NNPDF2.1 and NNPDF2.3 NLO PDFs with
$\alpha_s(M_{\rm Z})=0.119$. We show the ratio
of data over theory, normalized to NNPDF2.3, divided into
rapidity bins. The experimental error bars are statistical, while the
(correlated) systematic uncertainty, including normalization errors, 
is shown as a band in the bottom
of each plot.}
    \label{fig:LHCdataplots-jets}
\end{figure}

\begin{figure}[ht]
    \begin{center}
      \includegraphics[width=0.45\textwidth]{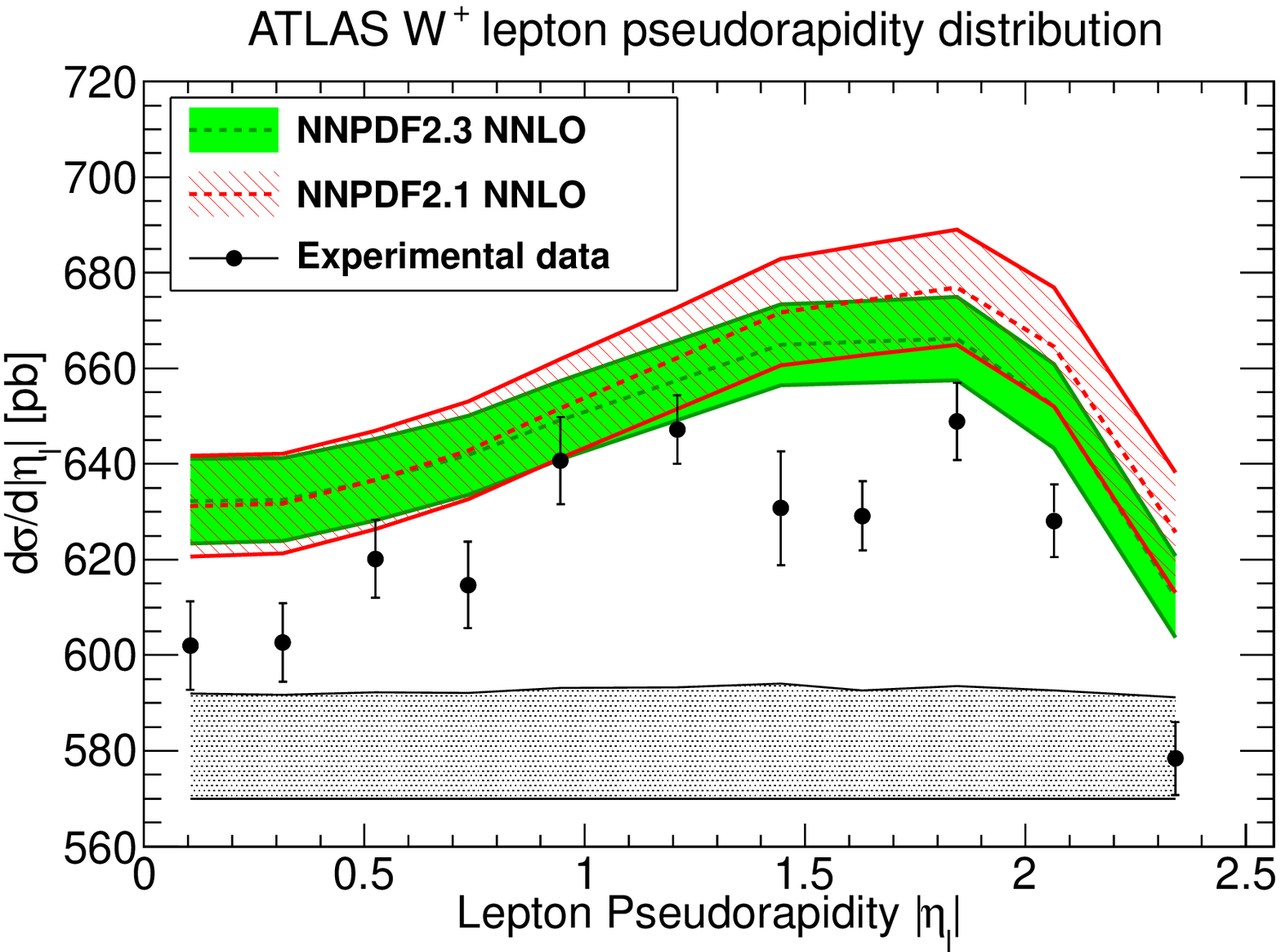}
   \includegraphics[width=0.45\textwidth]{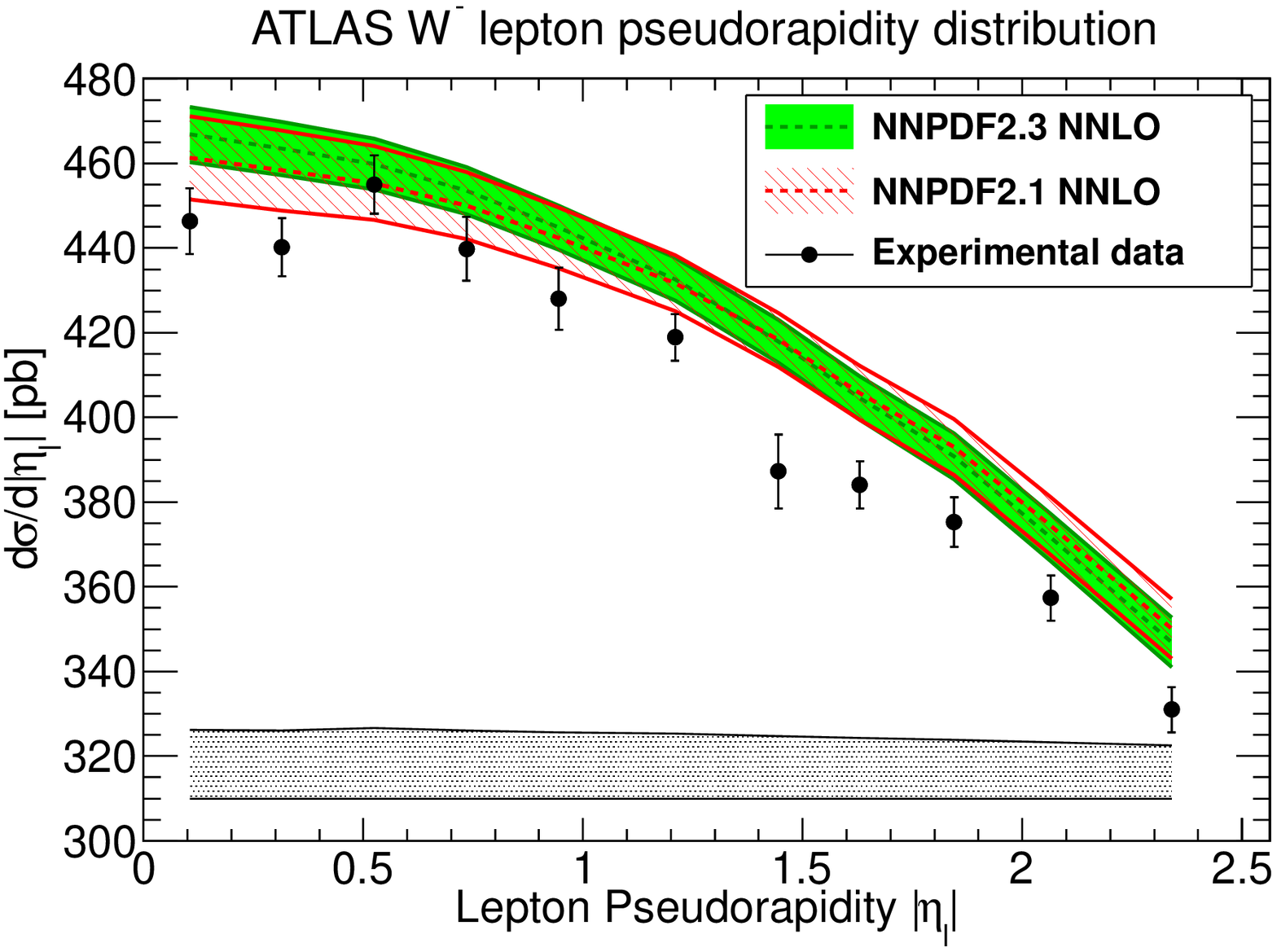}
   \includegraphics[width=0.45\textwidth]{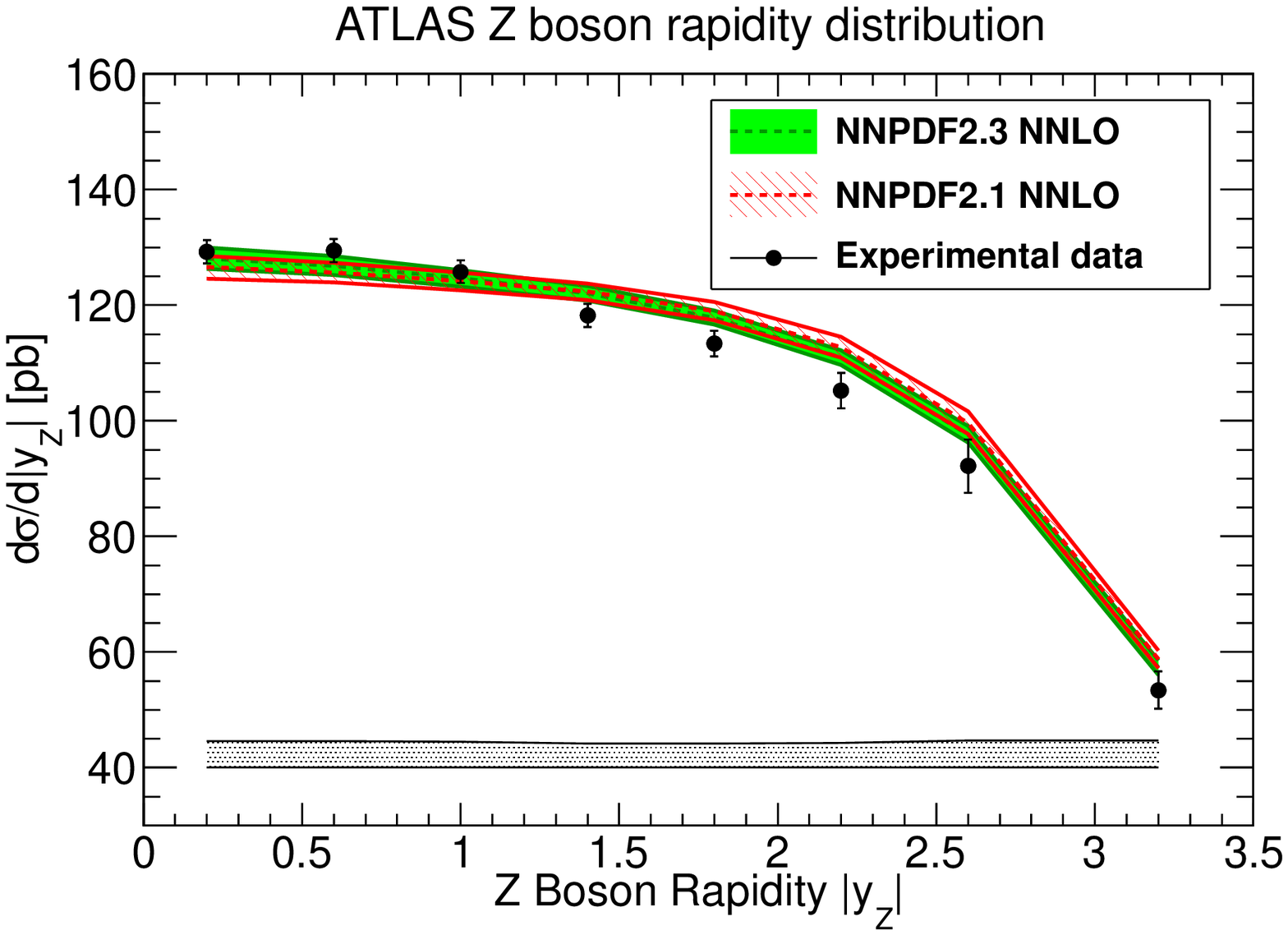}
   \includegraphics[width=0.45\textwidth]{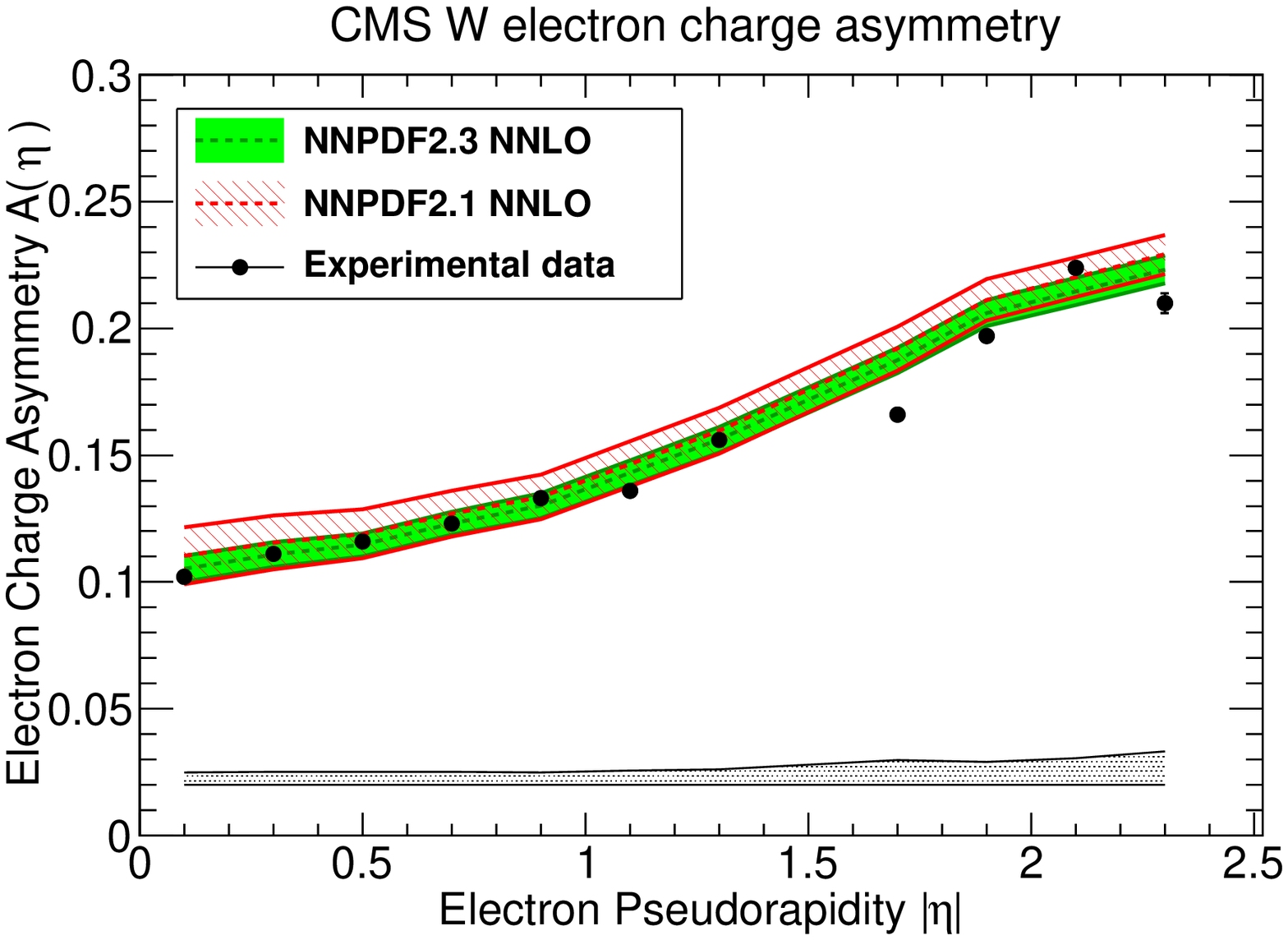}
    \end{center}
    \vskip-0.5cm
    \caption{\small Same as Fig.~\ref{fig:LHCdataplots-jets}, but for  ATLAS and
CMS gauge boson
production data, now using NNLO PDFs.}
    \label{fig:LHCdataplots-ewk}
\end{figure}

\begin{figure}[ht]
    \begin{center}
   \includegraphics[width=0.45\textwidth]{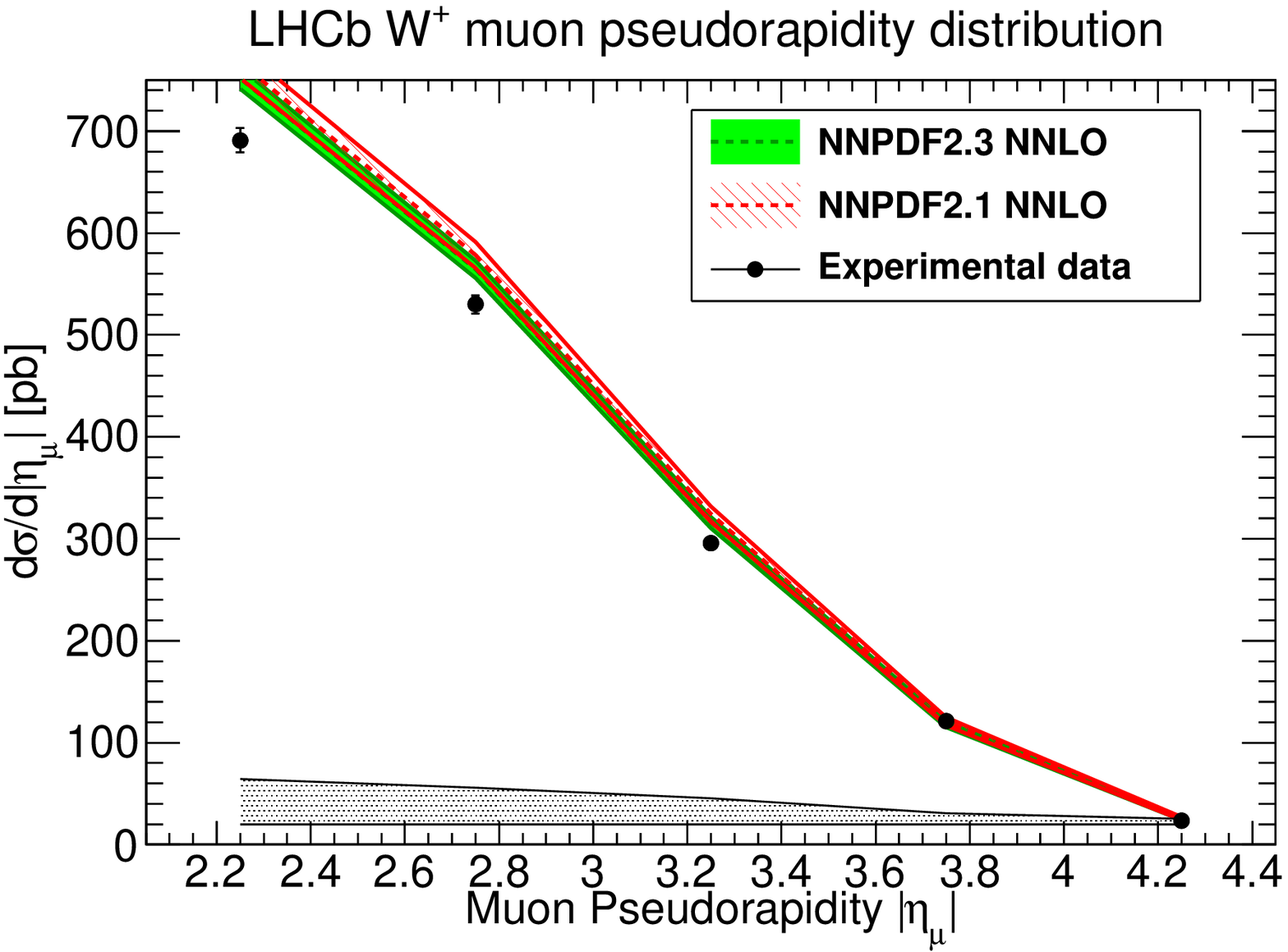}
   \includegraphics[width=0.45\textwidth]{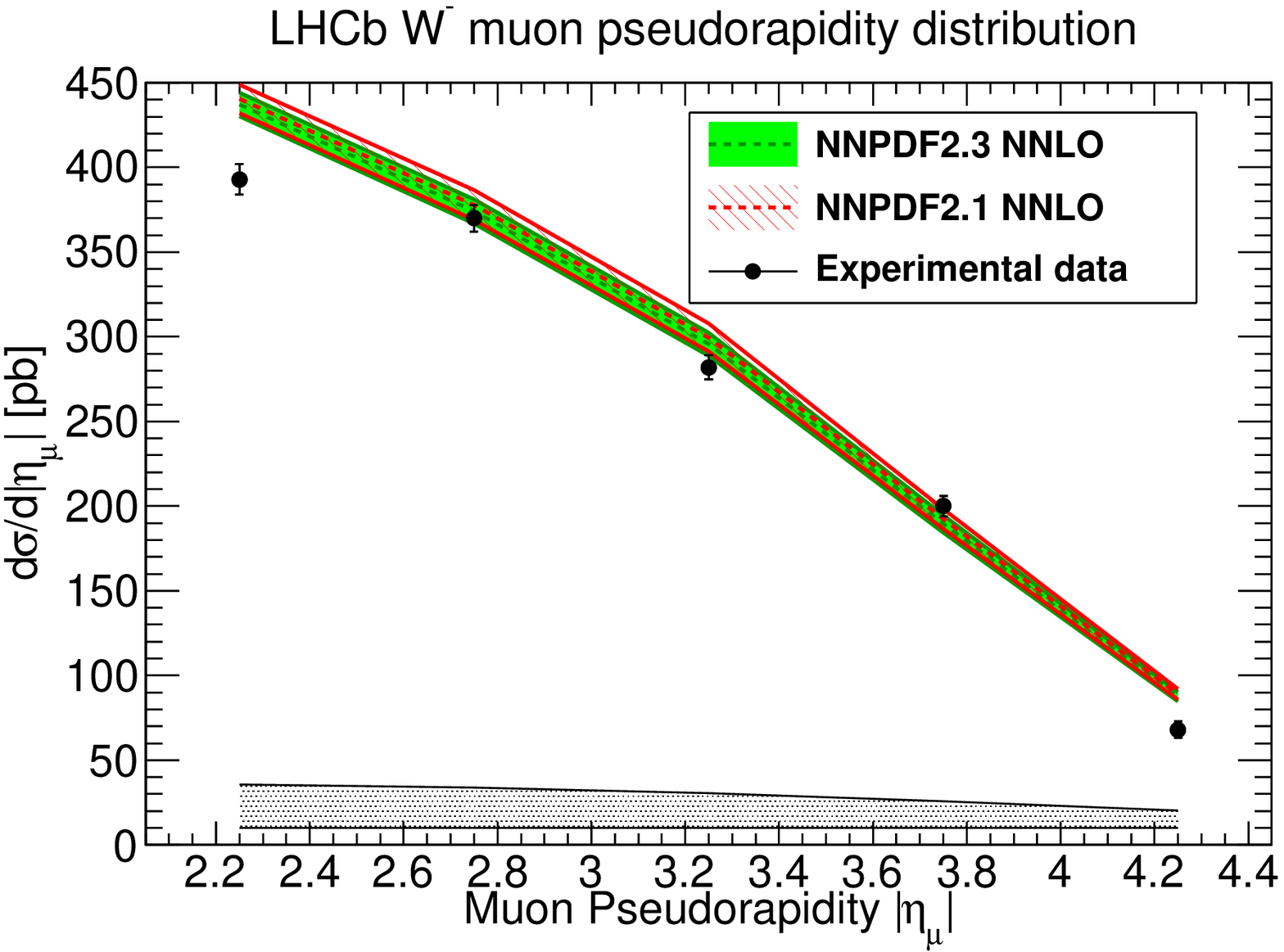}
    \end{center}
    \vskip-0.5cm
    \caption{\small Same as Fig.~\ref{fig:LHCdataplots-jets}, but for
LHCb electroweak
gauge boson production data, now using NNLO PDFs.}
    \label{fig:LHCdataplots-ewk-lhcb}
\end{figure}

The impact of the LHC data can also be seen by comparing the
predictions for the fitted observables before and after their
inclusion in the fit: as shown in Sec.~\ref{sec-comp} the $\chi^2$ values of
Tab.~\ref{tab:estfit2dataset} already show that the fit quality is
acceptable before including the LHC data, and quite good after
including them. This is also seen from  a data-theory comparison:
the predictions from the NNPDF2.1 and NNPDF2.3 $\alpha_s(M_{\rm Z})=0.119$
sets
for the various LHC
observables are compared to the data in
Fig.~\ref{fig:LHCdataplots-jets} (ATLAS jets),
Fig.~\ref{fig:LHCdataplots-ewk} (ATLAS and CMS W,Z production) and
Fig.~\ref{fig:LHCdataplots-ewk-lhcb} (LHCb W production).   For
the LHC electroweak data we show the predictions from the NNLO PDF
sets, while for the ATLAS inclusive jets we show the corresponding NLO
predictions: for jets the results are normalized to the NNPDF2.3 prediction.
Note that much of the uncertainty in the jet data is the totally correlated 
normalization uncertainty, which can shift the entire data set up or
down, so that inspection of the plot can be somewhat misleading and
the $\chi^2$ values of Tab.~\ref{tab:estfit2dataset} should be checked
in order to assess fit quality.

\clearpage

\subsection{Consistency of LHC data with low-energy data}
\label{sec-reduced}

It has been previously
observed~\cite{Ball:2010gb,Lai:2010vv,Thorne:2010kj}  
that there seems to be some tension
between the flavor decomposition favoured by low energy data, and that
obtained from
Tevatron W production data.
The LHC data will eventually solve any such discrepancy: in fact, they
already
shed some light on this issue. To investigate this, as already
mentioned, we have constructed  sets of NLO and NNLO NNPDF2.3 collider
PDFs. The NNPDF2.3 collider PDFs are based on a data set which includes
only the
HERA-I inclusive data, the ZEUS HERA-II data and the H1 and ZEUS
$F_2^c$ charm structure function data, the W and Z production data
from the Tevatron and the LHC, the CDF and D0 Run II inclusive jet
production data and the ATLAS inclusive jet data. This reduces the size of the data set from
about 3500 to about 1200 data points (see Tab.~\ref{tab:sets-numpts}).

\begin{figure}[t]
\begin{center}
\epsfig{width=0.905\textwidth,figure=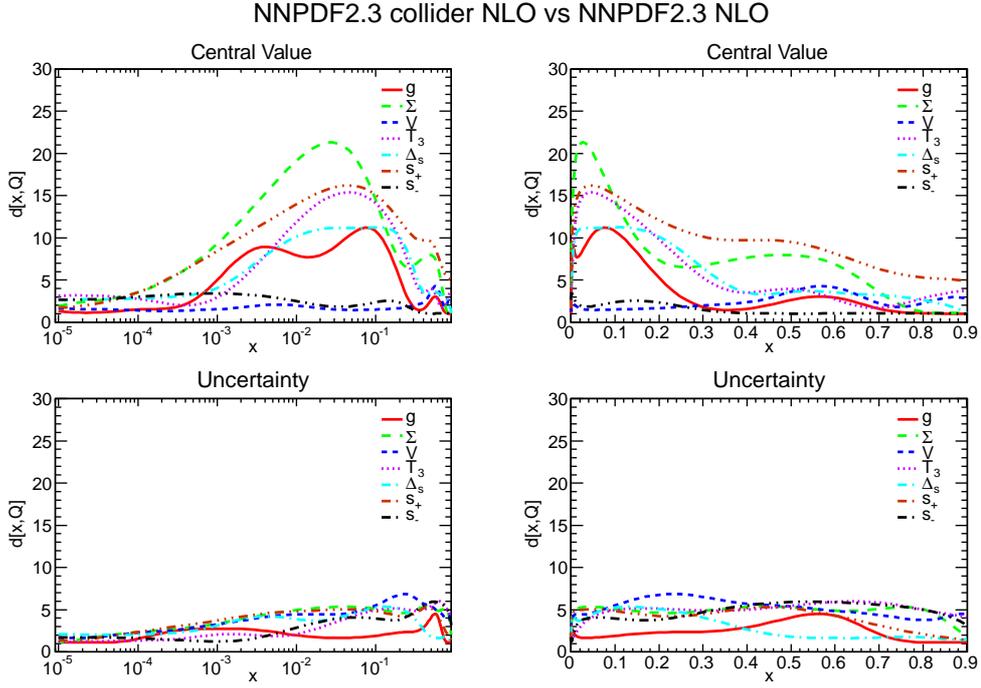}
\caption{ \small Distances between PDFs from the NNPDF2.3 collider and
  NNPDF2.3 NLO sets.
\label{fig:distances-23coll-vs-23-nlo.eps}} 
\end{center}
\end{figure}

\begin{figure}[t]
\begin{center}
\epsfig{width=0.905\textwidth,figure=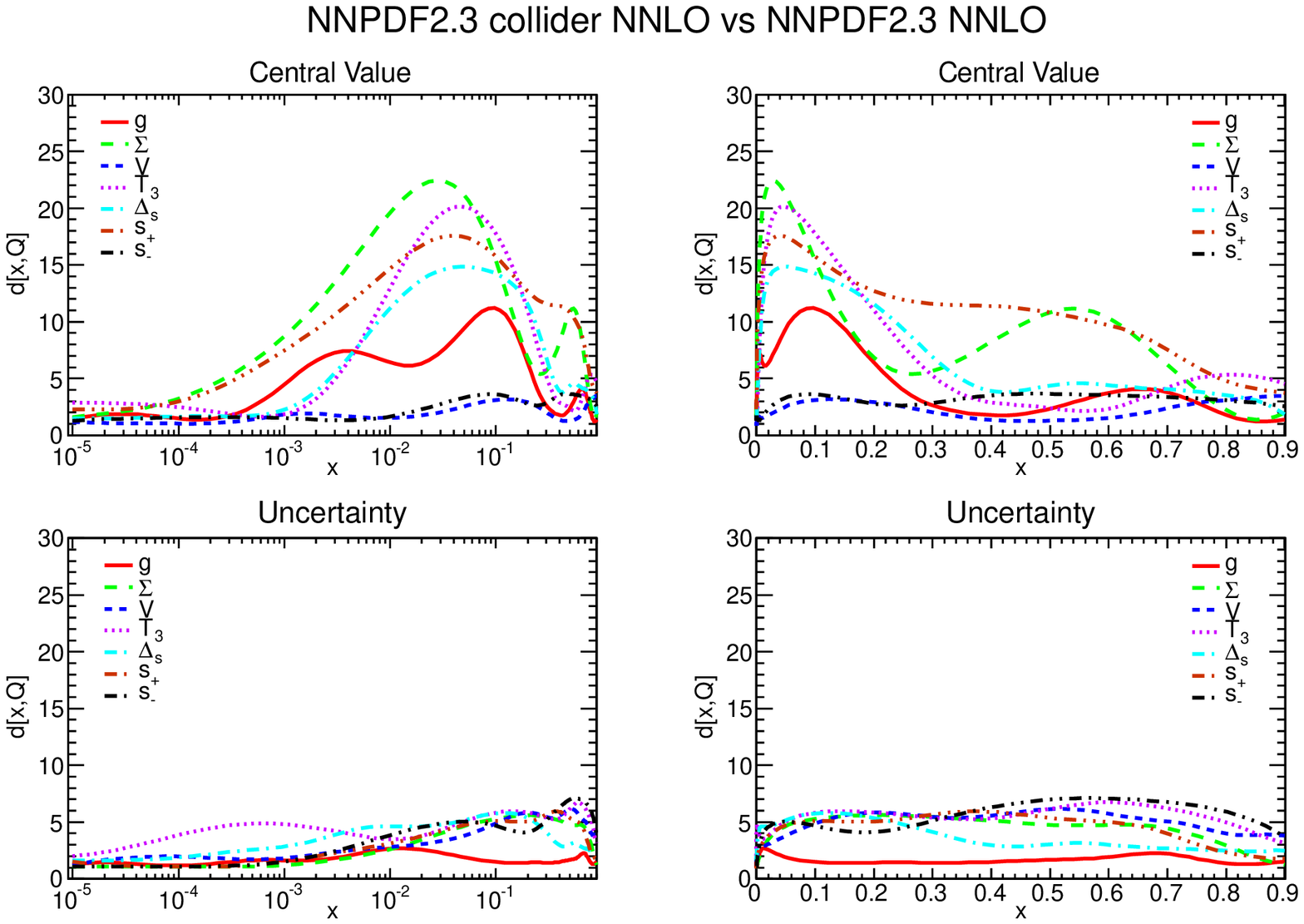}
\caption{ \small Distances between PDFs from the NNPDF2.3 collider and
  NNPDF2.3 NNLO sets.
\label{fig:distances-23coll-vs-23-nnlo.eps}} 
\end{center}
\end{figure}

\begin{figure}[t]
\begin{center}
\epsfig{width=0.905\textwidth,figure=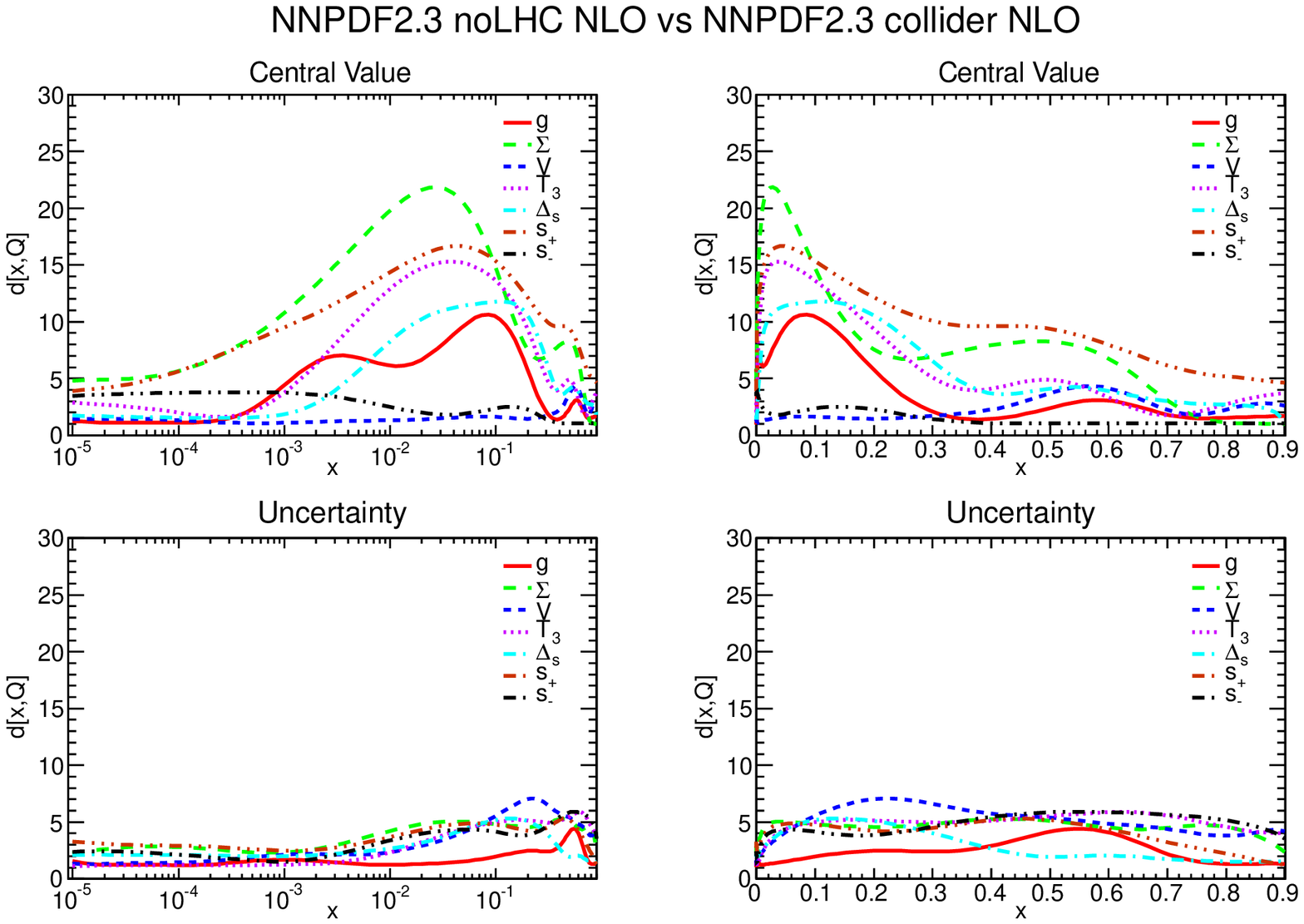}
\caption{ \small 
Distances between PDFs from the NNPDF2.3 collider and
  NNPDF2.3 noLHC NLO sets.
\label{fig:distances-23coll-vs-23nolhc-nlo.eps}} 
\end{center}
\end{figure}

\begin{figure}[t]
\begin{center}
\epsfig{width=0.905\textwidth,figure=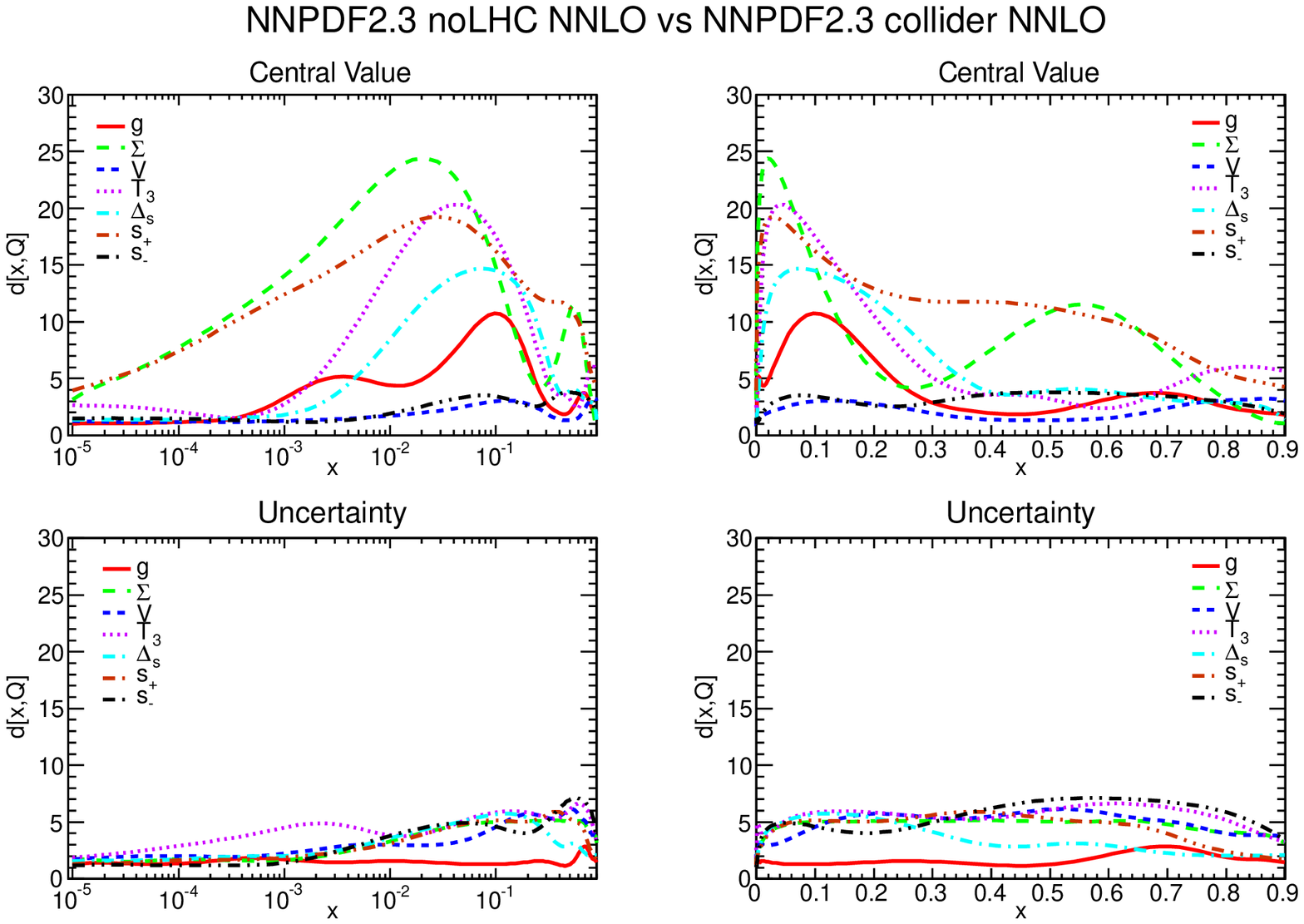}
\caption{ \small Distances between PDFs from the NNPDF2.3 collider and
  NNPDF2.3 noLHC NLO sets.
\label{fig:distances-23coll-vs-23nolhc-nnlo.eps}} 
\end{center}
\end{figure}

The distance between PDFs in the NNPDF2.3 collider and NNPDF2.3
default data sets at NLO and NNLO with $\alpha_s(M_{\rm Z})=0.119$ are shown in
Figs.~\ref{fig:distances-23coll-vs-23-nlo.eps}--\ref{fig:distances-23coll-vs-23-nnlo.eps}
respectively, while   the distances between NNPDF2.3 collider  and
NNPDF2.3 noLHC PDFs are in Figs.~\ref{fig:distances-23coll-vs-23nolhc-nlo.eps}--\ref{fig:distances-23coll-vs-23nolhc-nnlo.eps}.
We note that almost all PDFs change at the one or two sigma level,
both at NLO and NNLO. There are no significant differences between the
pattern of distances 
observed at NLO or NNLO, or when comparing to the standard or
noLHC sets. This in particular suggests that the tension between
collider data and low energy data, if it exists, is only mild: indeed,
if the low energy data were inconsistent with collider data, distances
should be significantly smaller when comparing to the noLHC  set than
to the default set, because the former has much fewer collider data.

To investigate this in more detail, we have   
sampled the distances shown in
Figs.~\ref{fig:distances-23coll-vs-23-nlo.eps}--\ref{fig:distances-23coll-vs-23nolhc-nnlo.eps}
for each PDF at 100 points in $x$, 50 equally spaced on a log scale from
$x=10^{-4}$ to $10^{-1}$, and 50 more  on a linear scale from $0.1$ to
$0.9$. 
For each comparison, we have then produced a histogram of the
distribution of distances. The distance is defined for each $x$ value
as a sum over replicas of normalized square differences of predictions obtained
from a gaussian distribution, and
thus it should follow, for each point and PDF, 
a $\chi^2$ distribution with one degree
of freedom. The combined histogram still follows the same
distribution if correlations are uniform. Indeed only if there is a
large number of points that are correlated to one particular point
will the histogram be distorted. We have verified that this is not the
case by checking that the histograms do not change when redone with a
decreasing number of points, which completely modifies the 
pattern of correlations.

\begin{figure}[t]
\begin{center}
\epsfig{width=0.43\textwidth,figure=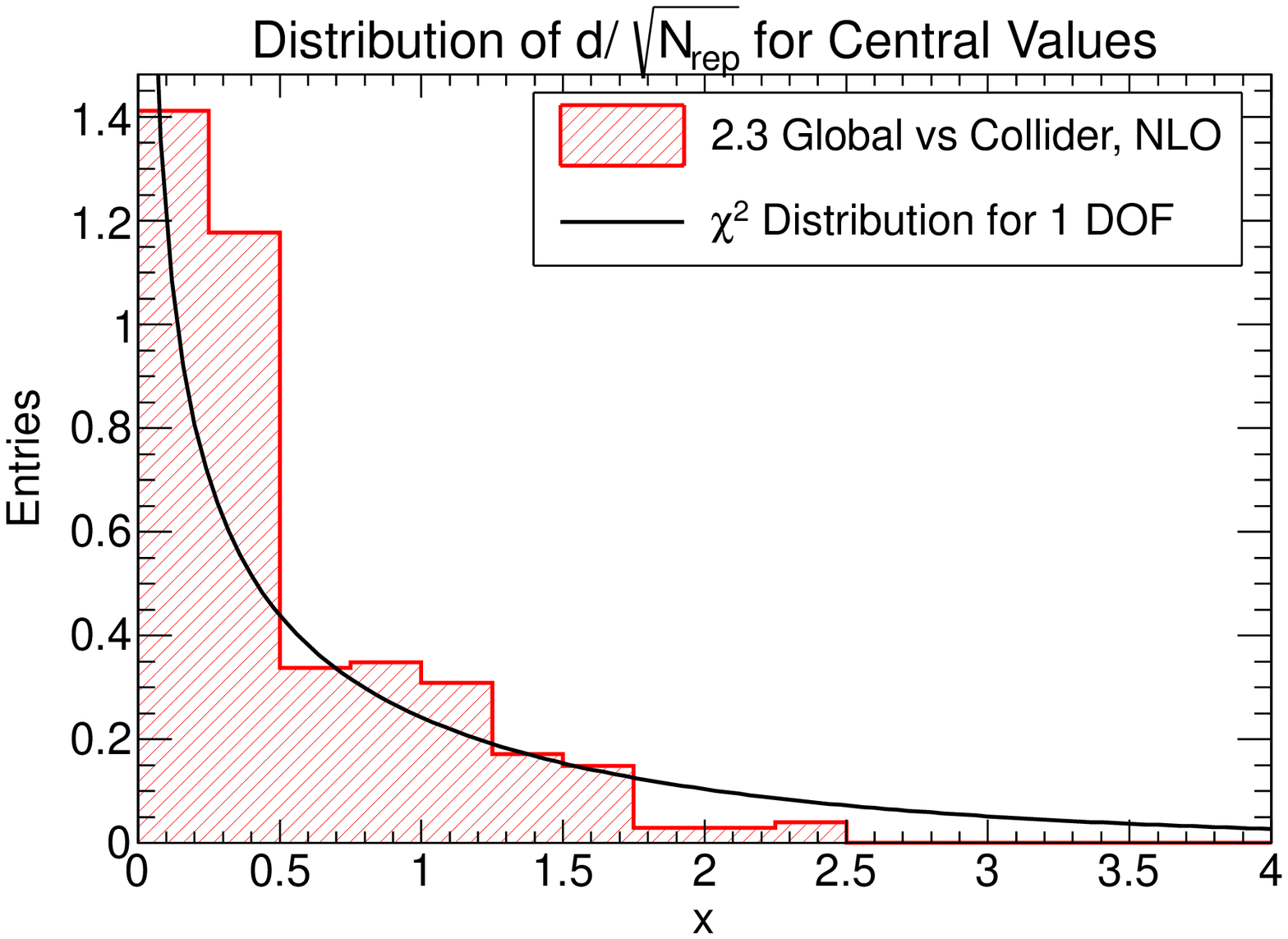 }
\epsfig{width=0.43\textwidth,figure=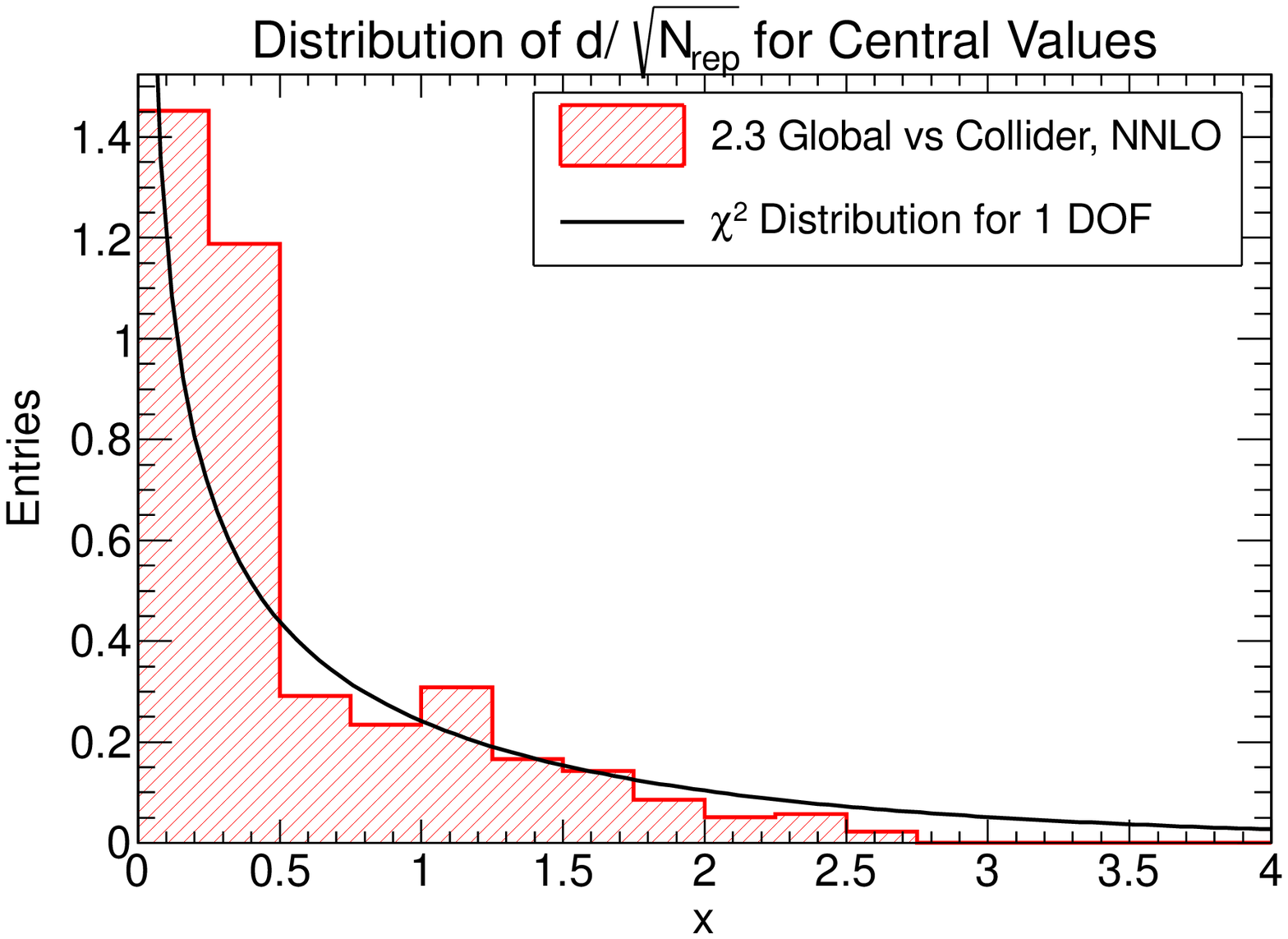 }
\epsfig{width=0.43\textwidth,figure=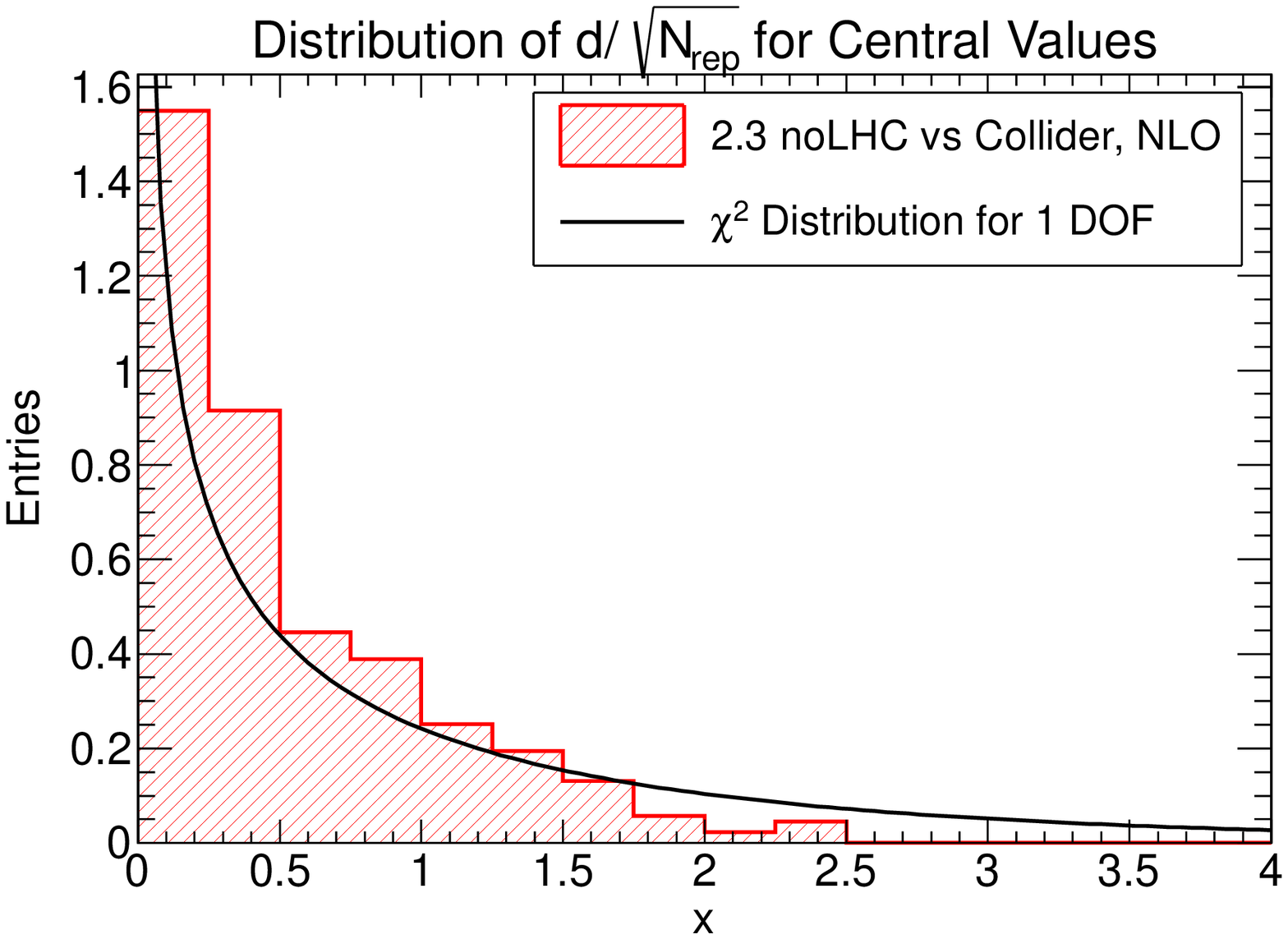 }
\epsfig{width=0.43\textwidth,figure=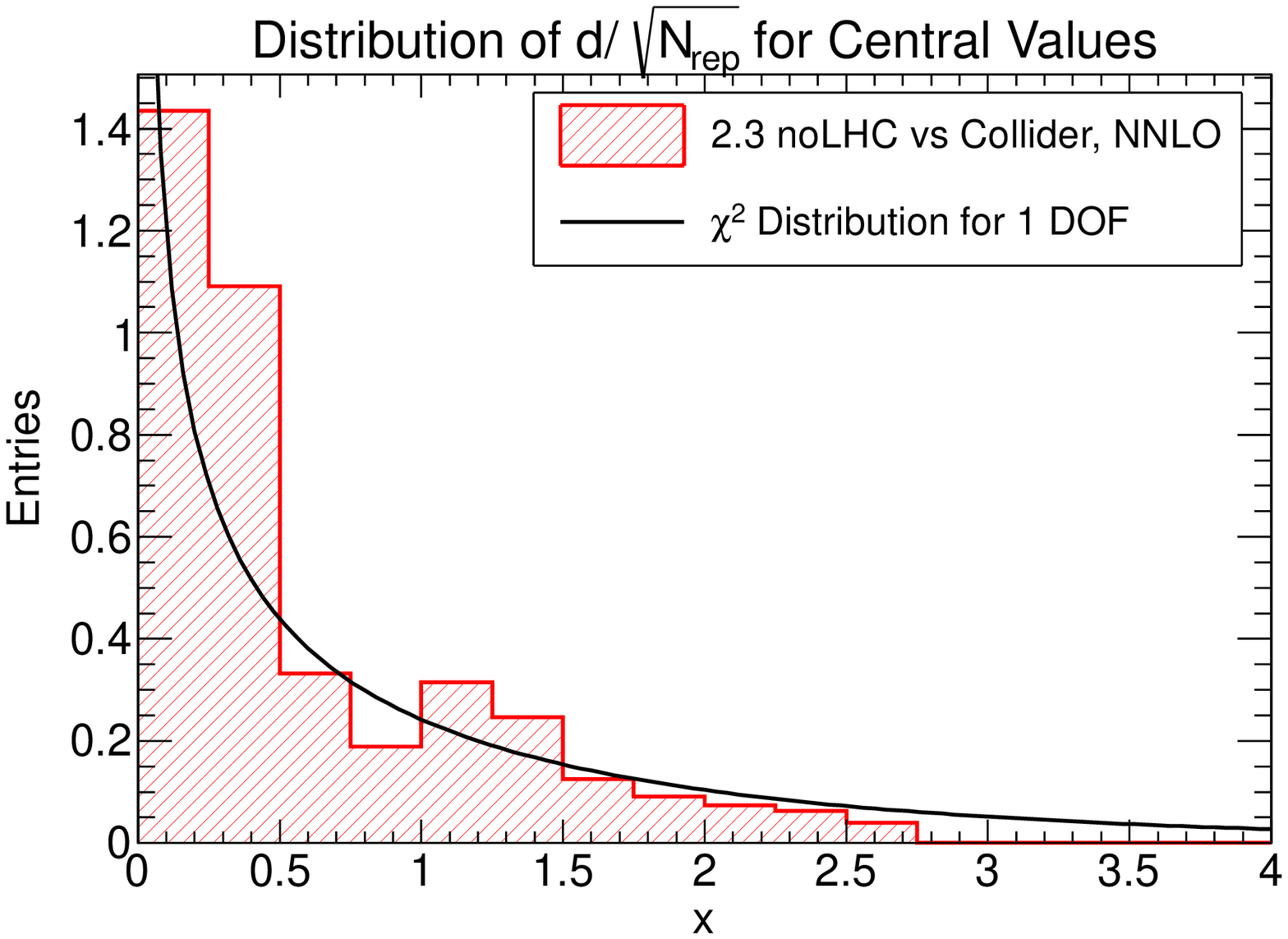 }
\caption{ \small Distribution of 
 distances between PDFs from the NNPDF2.3 collider and
  NNPDF2.3  NLO and NNLO sets (top) and between 
PDFs from the NNPDF2.3 collider and
  NNPDF2.3 noLHC NLO and NNLO sets (bottom). The $\chi^2$
  distribution with one degree of freedom is also shown.
\label{fig:distance-histo-q-23-vs-23coll-nlo.eps}} 
\end{center}
\end{figure}

\begin{figure}[t]
\begin{center}
\epsfig{width=0.43\textwidth,figure=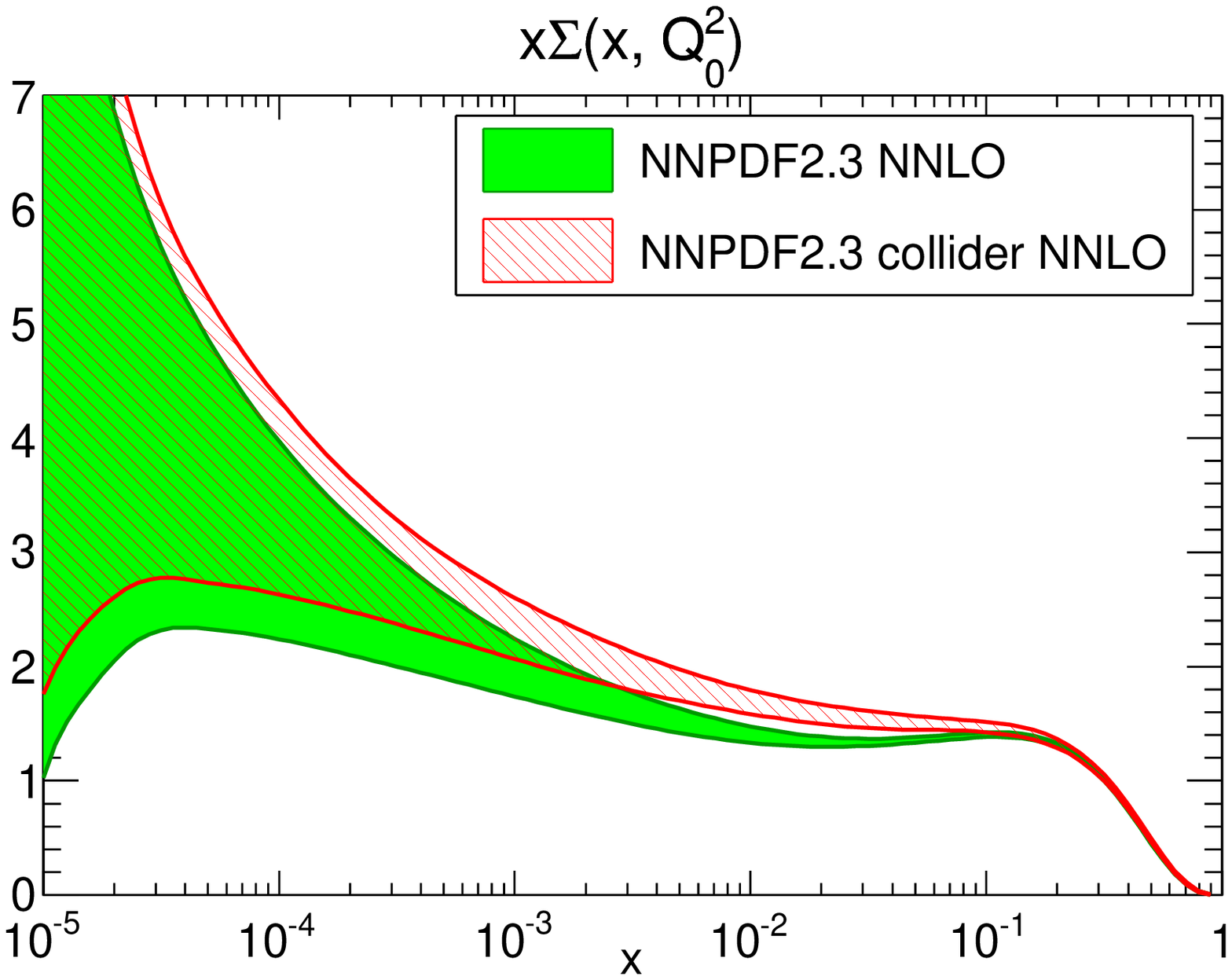}
\epsfig{width=0.43\textwidth,figure=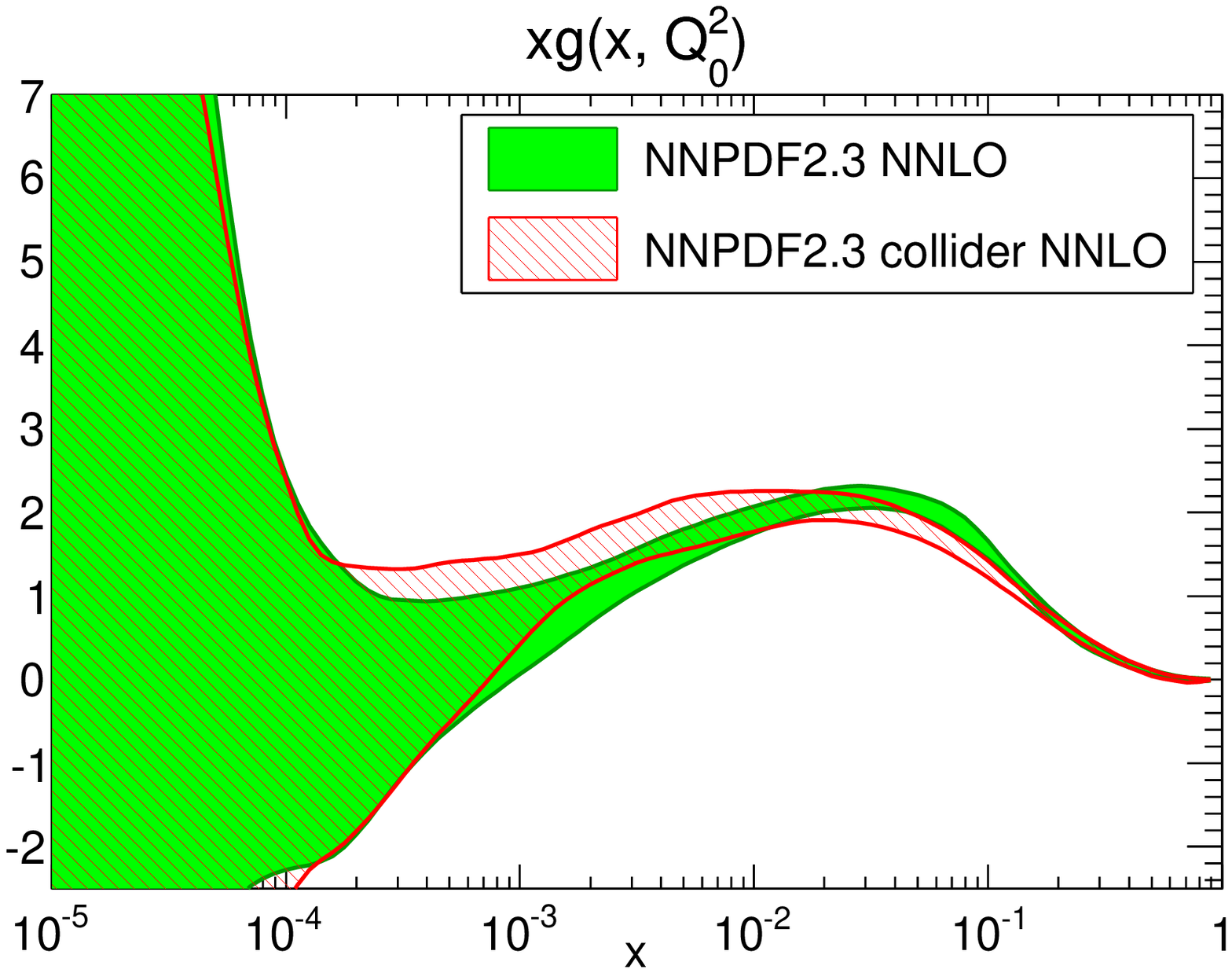}\\
\epsfig{width=0.43\textwidth,figure=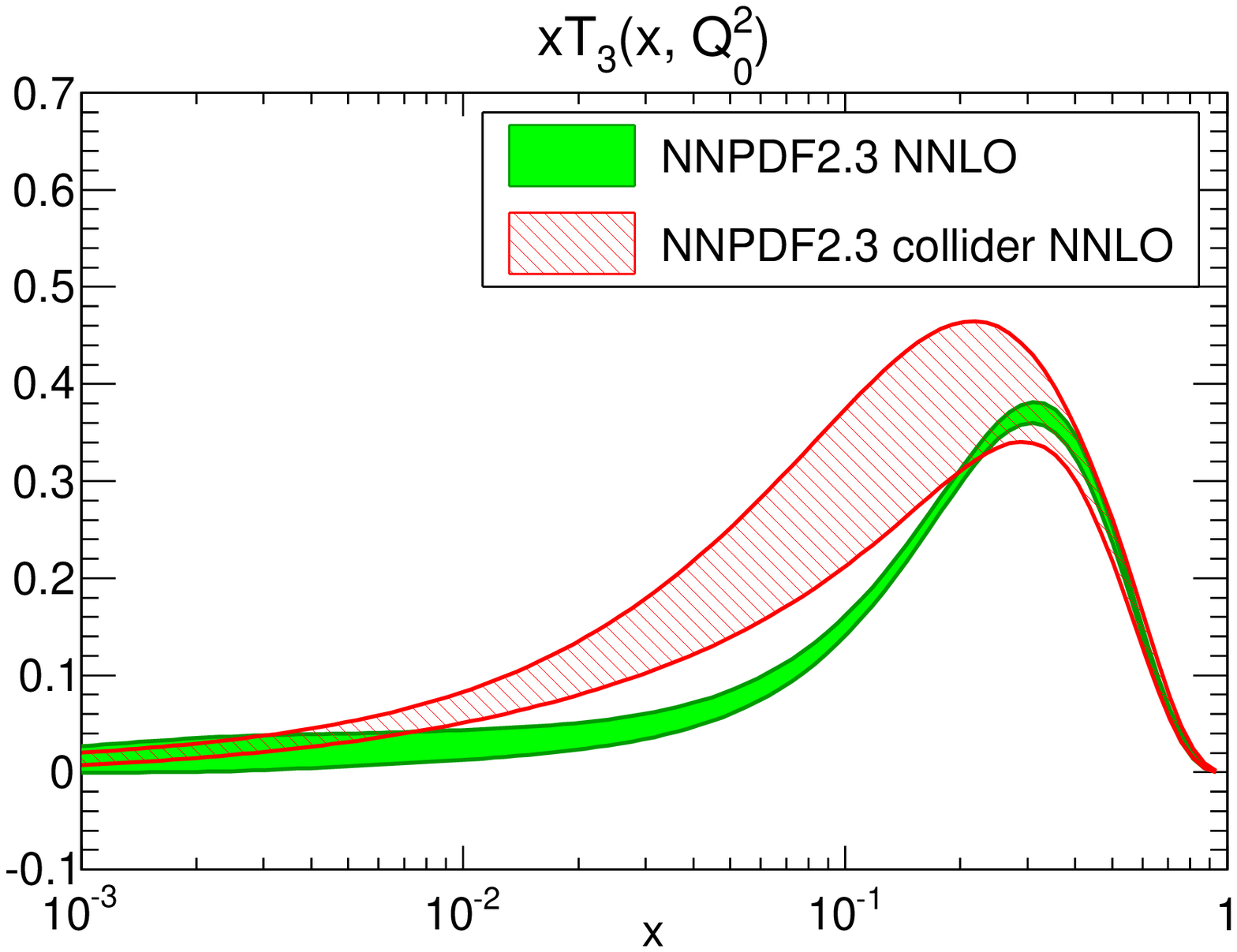}
\epsfig{width=0.43\textwidth,figure=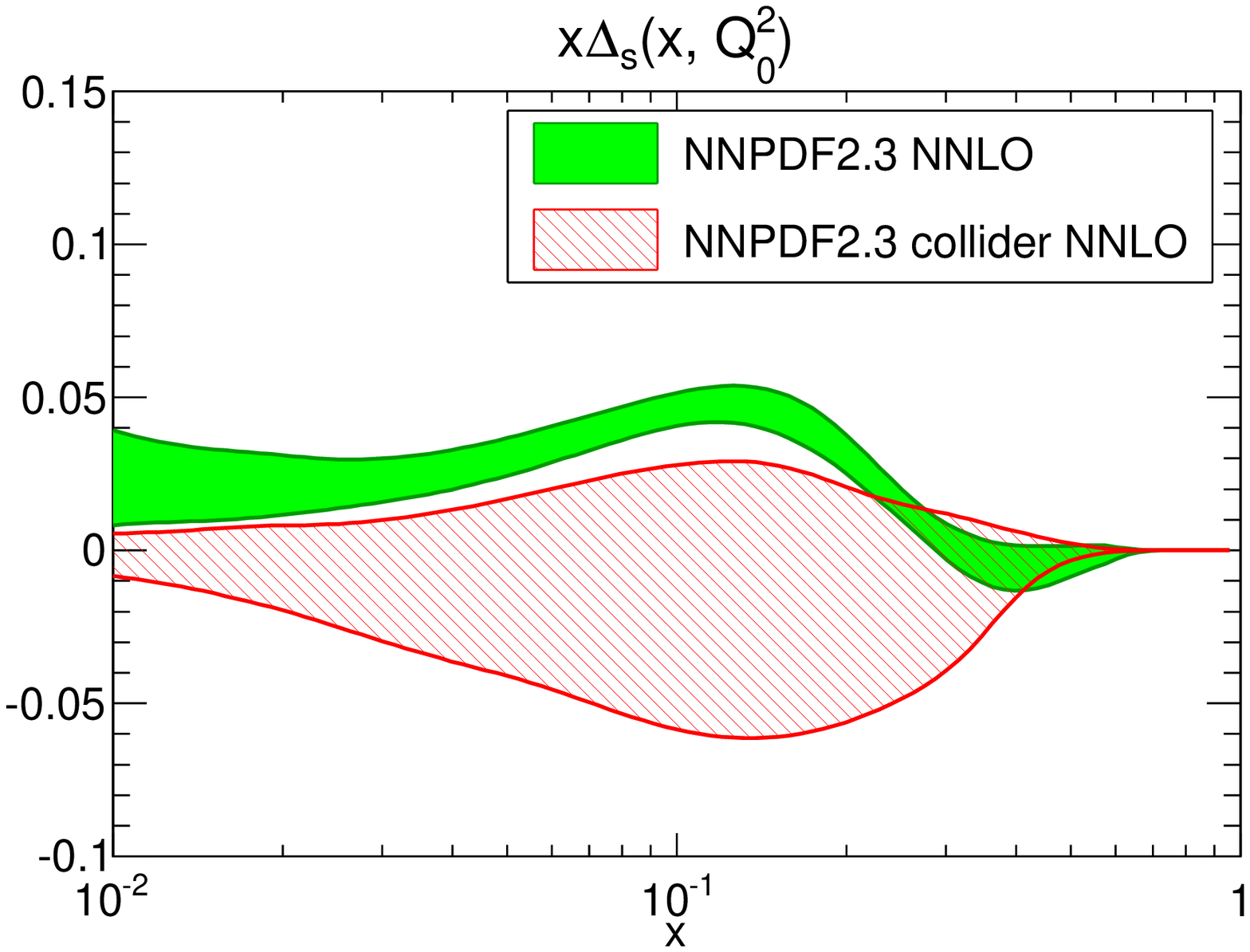}
\caption{ \small Comparison of the singlet, gluon (top), triplet
  $u+\bar u-d-\bar d$ and sea asymmetry $\bar d-\bar u$ at $Q^2=2$~GeV$^2$
from the NNLO NNPDF2.3 collider and NNPDF2.3 sets with $\alpha_s(M_{\rm Z})=0.119$.
\label{fig:xT3_Q_2_log-23-vs-23coll-nnlo.eps}} 
\end{center}
\end{figure}

The normalized histograms are compared to this $\chi^2$ distribution in
Fig.~\ref{fig:distance-histo-q-23-vs-23coll-nlo.eps}: no significant 
difference is apparent between
the four histograms, which are all well consistent with the
theoretical distribution.  This suggests 
that the differences between PDFs
in the collider only and global fits are consistent with a purely
statistical distribution, based on the given PDF uncertainties. 
This means that, the different behaviour of 2.3 and 2.3~collider PDFs
which is seen in Fig.~\ref{fig:xT3_Q_2_log-23-vs-23coll-nnlo.eps}
(e.g. for the triplet at small $x$) is compatible with  a statistical
fluctuation.   

Based on these conclusions, one might be tempted to recommend usage of
the collider PDFs, in that they are free from nuclear corrections and
much less sensitive to possible higher twist corrections, while
retaining the abundant, statistically accurate, 
and theoretically very reliable deep-inelastic data from the HERA
experiments. 
This option
however turns out to be not viable at present, because the collider PDFs
still have rather large statistical uncertainties. This is apparent from
Tab.~\ref{tab:estfitsigdataset}, where it is seen that the
uncertainties of all the observables measured by fixed-target
experiments become unacceptably large when the NNPDF2.3 collider fit
is used. It is also visible in
Fig.~\ref{fig:xT3_Q_2_log-23-vs-23coll-nnlo.eps}, where we compare the
singlet, gluon triplet and sea asymmetry in the collider and default
NNLO fits. This means that a
collider-only fit is presently not viable, though this situation may
change for future collider PDFs, which will include both the final,
yet unpublisjed,
HERA-II combined deep-inelastic data, as well as  future LHC data. 

\clearpage

\subsection{The strangeness fraction  of the proton}
\label{sec-strange}

The ATLAS
collaboration has recently~\cite{Aad:2012sb} presented evidence that
the strange quark distribution at low $Q^2$ and $x$ is rather larger
than hitherto thought, so that, for the specific kinematics probed by
ATLAS ($Q^2=1.9$~GeV$^2$, $x=0.023$) the quark sea is essentially
symmetric. This result, if correct with the stated uncertainty, 
disagrees at the two sigma
level with the result obtained using the NNPDF2.1 set. The ATLAS
analysis is based on combining ATLAS and HERA data, i.e.~on a subset
of the data used to construct the NNPDF2.3 collider fit discussed in 
Sec.~\ref{sec-reduced}: as we have seen there, PDFs based on collider
data only have very large uncertainties, so such a discrepancy seems
surprising. This is especially so given the fact that the strangeness
distributions in the NNPDF2.1  and NNPDF2.3 sets (the latter including
ATLAS data) always agree at the one sigma level, as can be seen from
Fig.~\ref{fig:pdfs-23-vs-21-nnlo-10000.eps}.

In order to check the ATLAS result, we have
produced a version of the
NNPDF2.3 fit based on exactly the same data set, namely only the combined
HERA-I data (HERAI-AV in Tab.~2 of \cite{Ball:2011mu}) and the ATLAS
gauge production  data (ATLAS W$^\pm$ and ATLAS Z in
Tab.~\ref{tab:exp-sets-errors}), with a single value of
$\alpha_s(M_{\rm Z})=0.119$. This PDF set  will be denoted in the
following as  NNPDF2.3 HERA+ATLASWZ.  Following Ref.~\cite{Aad:2012sb}, we define 
the $x$-dependent
 strangeness fraction
\be
\label{eq:rs}
r_s (x,Q^2)=\frac{ s(x,Q^2)+\bar{s}(x,Q^2) }{ 2 \bar{d}(x,Q^2) } \, .
\ee
A perhaps more significant measure of the strangeness content is 
the strangeness momentum fraction normalized to the light sea
momentum fraction~\cite{Lai:2007dq,Alekhin:2008mb,Ball:2009mk,Martin:2009iq}
\be
\label{eq:rsint}
K_s (Q^2)=\frac{\int_0^1 dx x \lp s(x,Q^2)+\bar{s}(x,Q^2)\rp }{ 
\int_0^1 dx x\lp \bar{u}(x,Q^2) + \bar{d}(x,Q^2)\rp }  \, ,
\ee
traditionally~\cite{Abramowicz:1982zr} taken to be
$K_s\approx 0.5$ at scales of a few GeV.

\begin{figure}[ht]
    \begin{center}
      \includegraphics[width=0.49\textwidth]{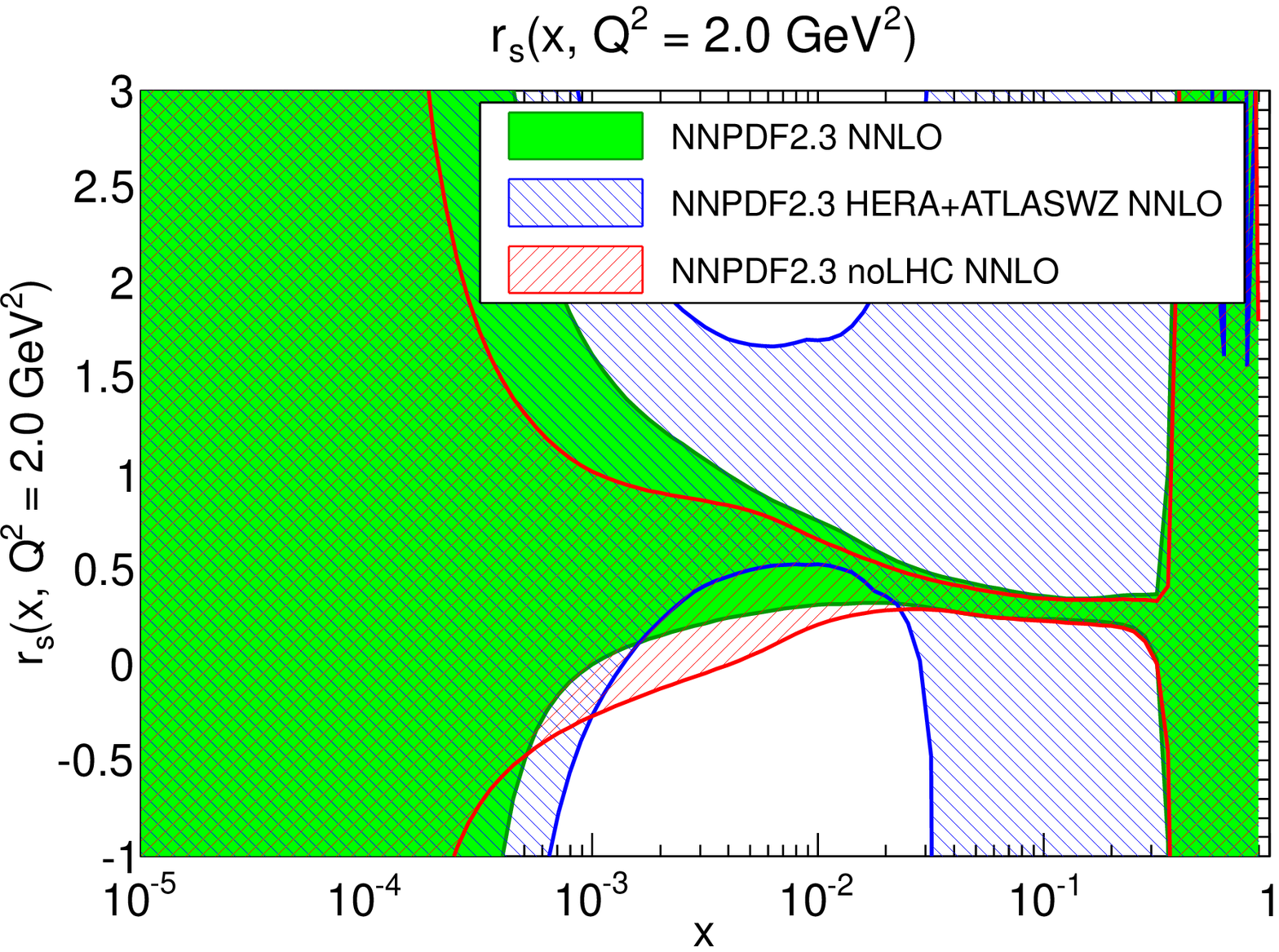}
      \includegraphics[width=0.49\textwidth]{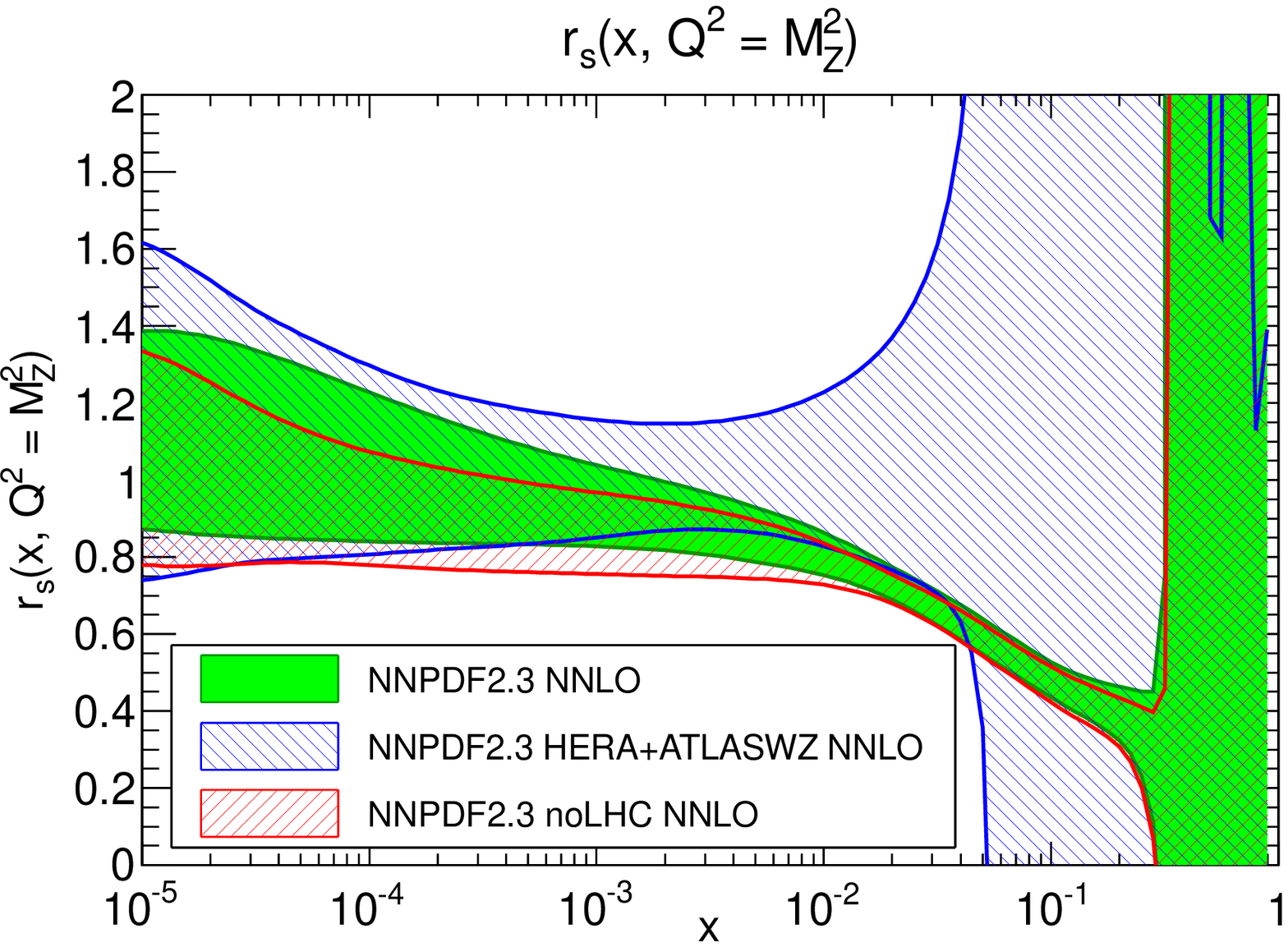}
    \end{center}
    \vskip-0.5cm
    \caption{\small The strangeness fraction $r_s(x,Q^2)$ Eq.~(\ref{eq:rs})
 computed as a function
of $x$ for $Q^2=2$~GeV$^2$ (left) and
$Q^2=M_{\rm Z}^2$~GeV$^2$ (right)
from NNPDF2.3 noLHC,
NNPDF2.3 and NNPDF2.3 HERA+ATLASWZ NNLO PDFs with
$\alpha_s(M_{\rm Z})=0.119$. All error bands are one sigma.}
    \label{fig:rs-xplot}
\end{figure}

The strangeness fraction $r_s(x,Q^2)$ Eq.~(\ref{eq:rs}) was determined in
Ref.~\cite{Aad:2012sb} at the two points in the $(x,Q^2)$ plane
$(x,Q^2)=(0.023,1.9~{\rm GeV}^2)$ and $(x,Q^2)=(0.013,M_{\rm Z}^2)$.
In Fig.~\ref{fig:rs-xplot} we show $r_s(x,Q^2)$ computed as a function
of $x$ for the relevant scales, as obtained using the NNPDF2.3 noLHC,
NNPDF2.3 and NNPDF2.3 HERA+ATLASWZ NNLO PDF sets with
$\alpha_s(M_{\rm Z})=0.119$. It is clear that the strangeness fraction
computed with
NNPDF2.3 PDFs,
especially in the range $10^{-3}\lsim x \lsim 10^{-1}$, is somewhat
larger than the  NNPDF2.3 noLHC one, though they are fully consistent
at the one sigma level for all values of $x$. This means that even
though the ATLAS data do push the
strangeness fraction
towards slightly higher values, the effect is of marginal statistical
significance. 
If the comparison is made with the NNPDF2.3 HERA+ATLASWZ fits,
then uncertainties are so large that results become completely compatible.

\begin{figure}[ht]
    \begin{center}
      \includegraphics[width=0.495\textwidth]{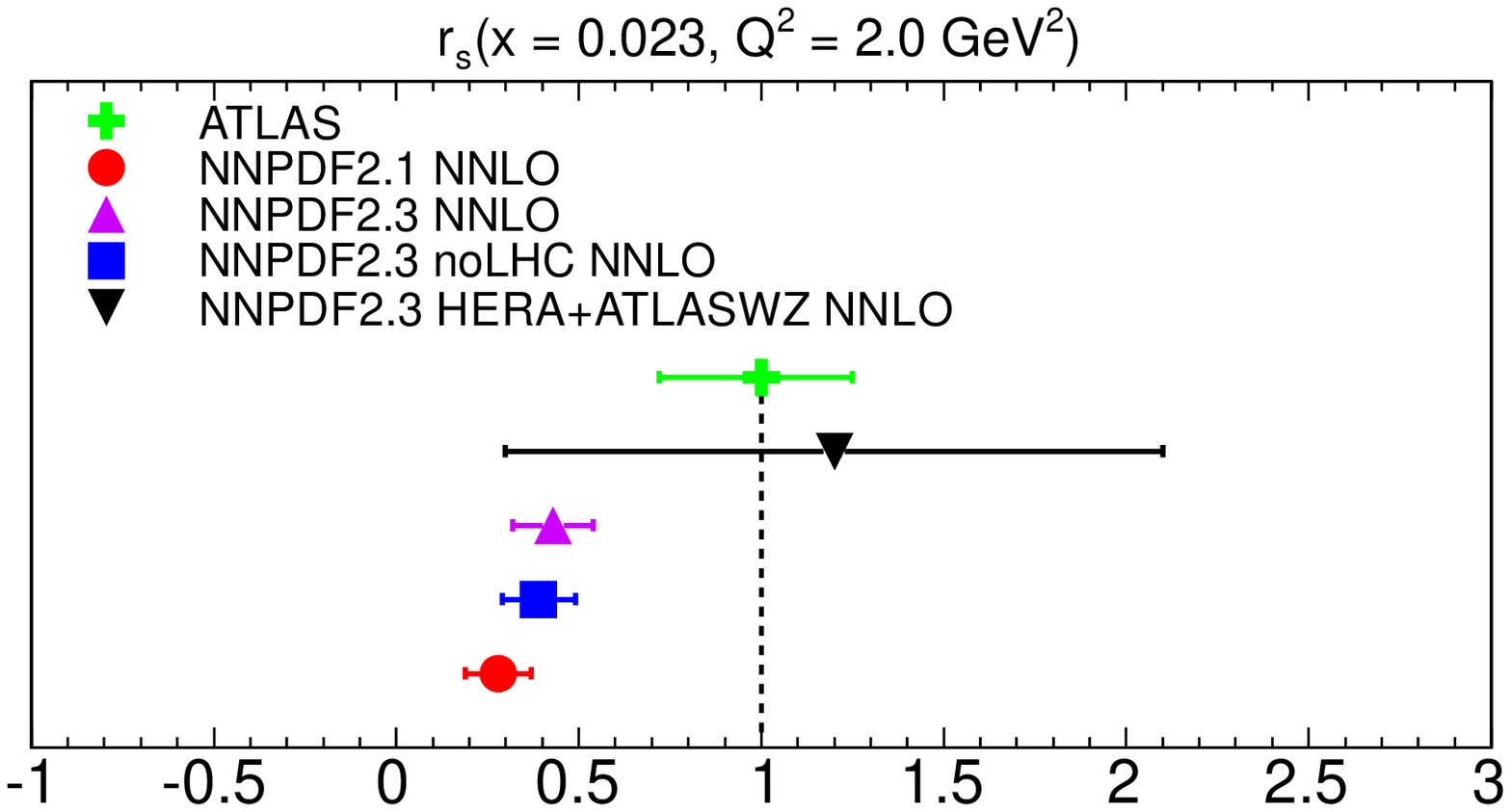}
      \includegraphics[width=0.495\textwidth]{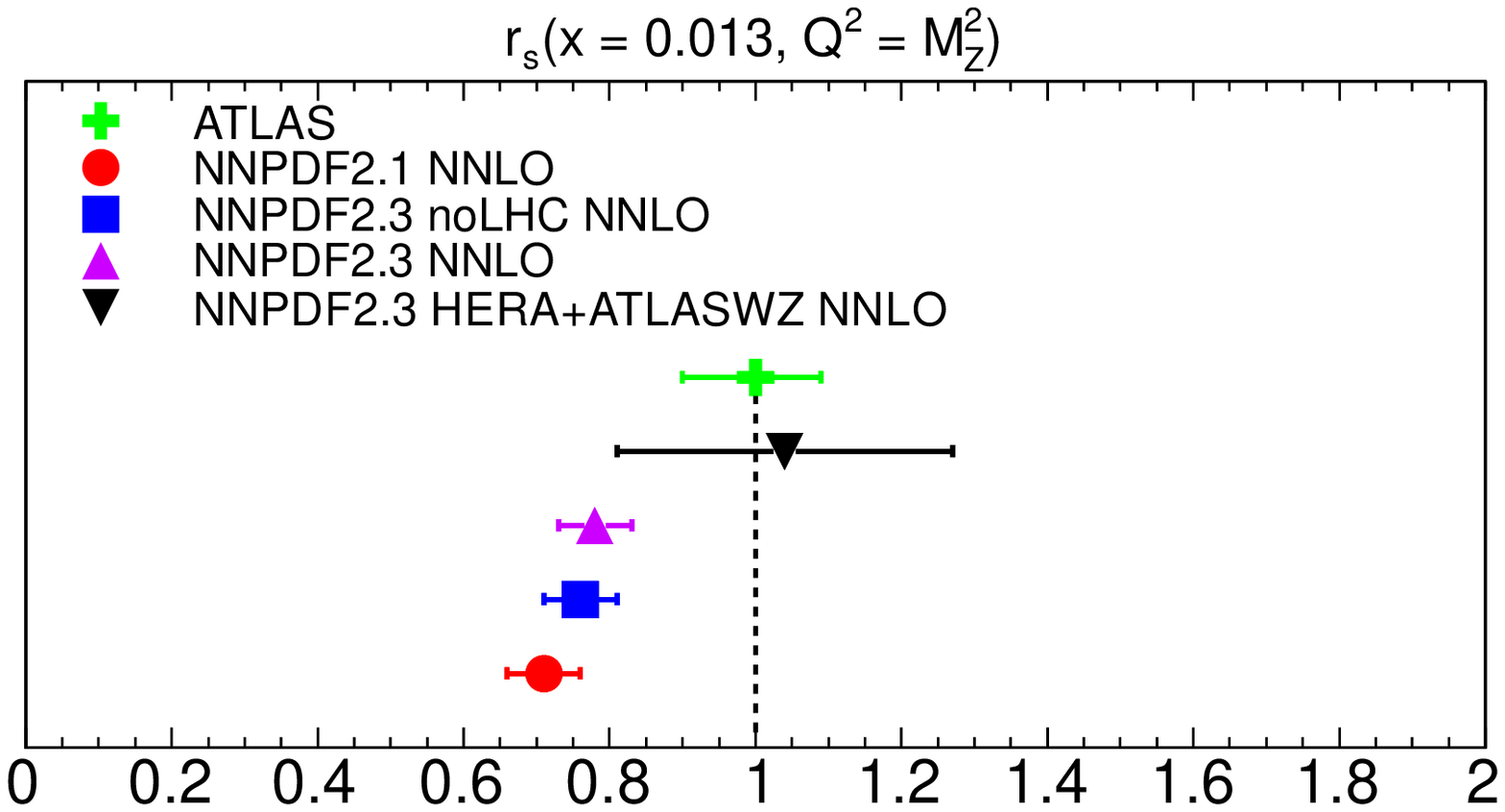}
      \includegraphics[width=0.495\textwidth]{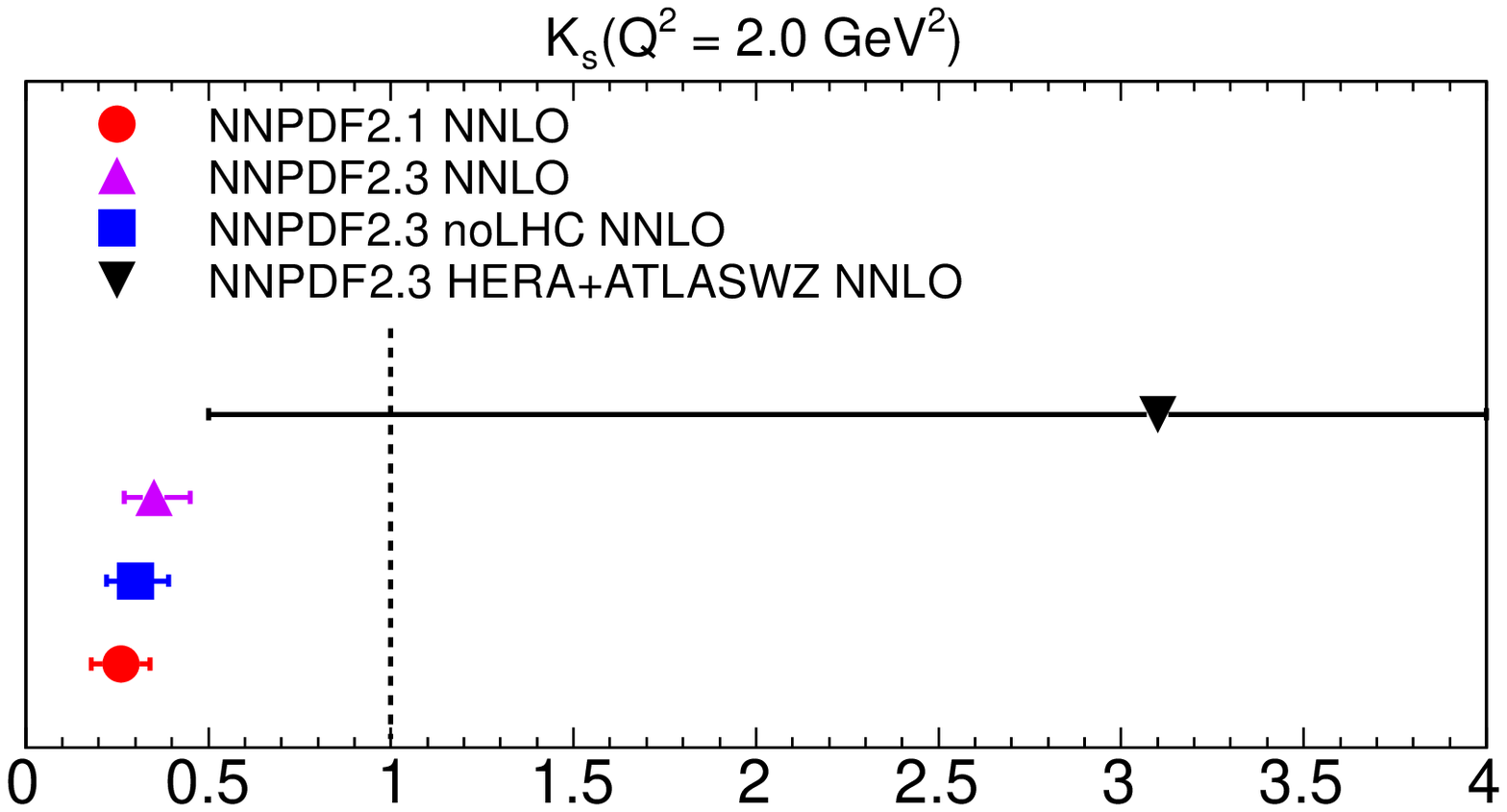}
      \includegraphics[width=0.495\textwidth]{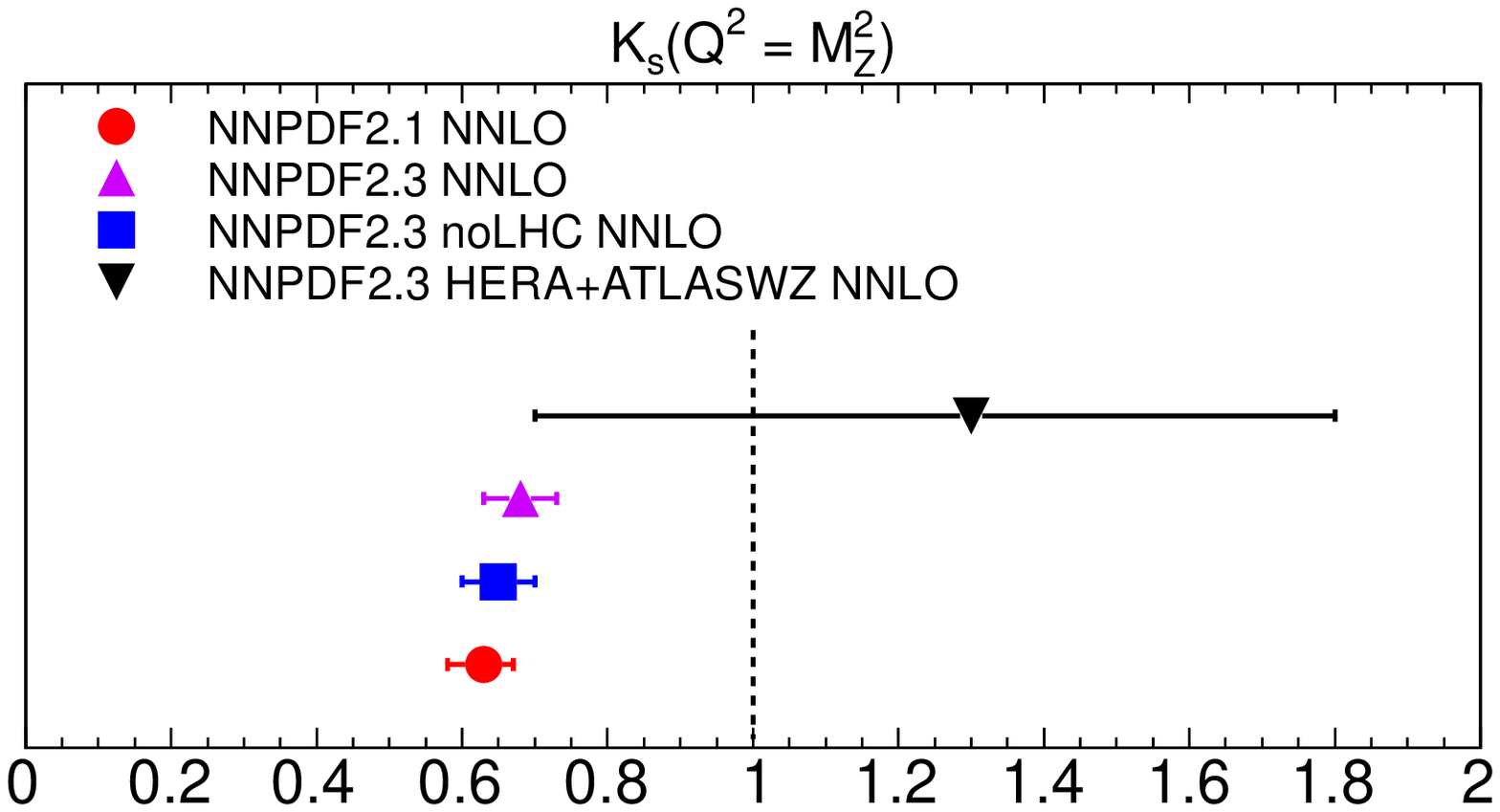}
    \end{center}
    \vskip-0.5cm
    \caption{\small Graphical representation of the results
of Tab.~\ref{tab:rs}. Note the different scale on the $x$ axis. }
    \label{fig:rs-comp}
\end{figure}

The values of the strangeness fraction at the
specific values of $(x,Q^2)$, together with the strangeness
momentum fraction Eq.~(\ref{eq:rsint}) are given
in Tab.~\ref{tab:rs}, and represented graphically in
Fig.~\ref{fig:rs-comp}. The values of the strangeness fraction from 
Ref.~\cite{Aad:2012sb} are also shown (note that in the latter case
the low-scale
value is given at $Q^2=1.9$~GeV$^2$, while for NNPDF it is at $Q^2=2$~GeV$^2$.)

The conclusions are the same as were drawn
from Fig.~\ref{fig:rs-xplot}: the ATLAS data favour a somewhat larger
central value of the strangeness  fraction, which remains however
compatible at the one sigma level with the value obtained without LHC
data (NNPDF2.3 noLHC), and also consistent with the previous,
slightly less
accurate NNPDF2.1 value. However, it seems that the 
``standard''  belief that the strange 
momentum fraction is of order
$K_s\approx\frac{1}{2}$ is still essentially correct.

 If one then attempts a determination based on HERA and ATLAS
data only, then uncertainties are so large that no conclusion can be
drawn: the NNPDF2.3 HERA+ATLASWZ result for $r_s$ has an uncertainty which is
three to four times bigger than that of the ATLAS result of
Ref.~\cite{Aad:2012sb}, and indeed for this PDF set $K_S$ is
essentially undetermined. Because the ATLAS analysis is based on the same
data, and only differs from our analysis in the fitting methodology
(in particular, the use of a rather simple functional form for PDFs),
it appears likely that the uncertainties in the results of 
Ref.~\cite{Aad:2012sb} are significantly underestimated.

\begin{table}[h]
\begin{center}
\begin{tabular}{|c|c|c|}
\hline
\centering PDF Set &  $r_s(0.023,2~\mathrm{GeV}^2)$ & $r_s(0.013,M_{\rm Z}^2)$ \\
\hline
\hline
NNPDF2.1   &   $0.28 \pm 0.09$ & $0.71  \pm 0.05$ \\ 
NNPDF2.3 noLHC    &	$0.39 \pm 0.10$ & $0.76 \pm 0.05$\\
NNPDF2.3   &	$0.43 \pm 0.11$ & $0.78 \pm 0.05$ \\
\hline
NNPDF2.3 HERA+ATLASWZ  	&	$1.2 \pm 0.9$ & $1.04 \pm 0.23$ \\
ATLAS (Ref.~\cite{Aad:2012sb}) 	&	$1.00\,{}^{+0.25}_{-0.28}$ & $1.00\,{}^{+0.09}_{-0.10}$	\\
\hline
\end{tabular}
\end{center}
\begin{center}
\begin{tabular}{|c|c|c|}
\hline
\centering PDF Set &  $K_s(2~\mathrm{GeV}^2)$ & $K_s(M_{\rm Z}^2)$ \\
\hline
\hline
NNPDF2.1  &   $ 0.26^{+0.08}_{-0.08}  $ & $ 0.63^{+0.04}_{-0.05}  $   \\ 
NNPDF2.3 noLHC   & $ 0.30^{+0.09}_{-0.08}  $  & $ 0.65^{+0.05}_{-0.05}  $  \\
NNPDF2.3   &  $ 0.35^{+0.10}_{-0.08}  $ 	  &  $ 0.68^{+0.05}_{-0.05}  $  \\
\hline
NNPDF2.3 HERA+ATLASWZ 	&  $ 3.1^{+0.9}_{-2.6}  $  &   $ 1.3^{+0.5}_{-0.6}  $   \\
\hline
\end{tabular}
\end{center}
\caption{\small \label{tab:rs} 
The strangeness fraction  Eq.~(\ref{eq:rs}) (top table) and
strangeness
momentum fraction Eq.~(\ref{eq:rsint}) (bottom table) determined using
the NNPDF2.1, NNPDF2.3 noLHC, NNPDF2.3 and NNPDF2.3 HERA+ATLASWZ NNLO PDF
sets with $\alpha_s(M_{\rm Z})=0.119$. The values of Ref.~\cite{Aad:2012sb}
are also shown (note that for ATLAS the low-scale
value is given at $Q^2=1.9$~GeV$^2$). The PDF
uncertainties in the strangeness fraction Eq.~(\ref{eq:rs}) are
one sigma errors while in the strangeness
momentum fraction Eq.~(\ref{eq:rsint}) they are 68\% confidence levels.}
\end{table}

\section{Phenomenology}

\label{sec-pheno}

We discuss now some phenomenological implications
of the NNPDF2.3 parton set.
After briefly discussing NNPDF2.3  parton luminosities, we 
use them to compute several LHC reference ``standard candle'' cross-sections. 

\subsection{Parton luminosities}
\label{sec-lumi}

At a hadron collider, all factorized observables depend on parton
distributions through a parton luminosity, which, following Ref.~\cite{Campbell:2006wx}, we define 
 as
\be
\Phi_{ij}\lp M_X^2\rp = \frac{1}{s}\int_{\tau}^1
\frac{dx_1}{x_1} f_i\lp x_1,M_X^2\rp f_j\lp \tau/x_1,M_X^2\rp \ ,
\label{eq:lumdef}
\ee
where $f_i(x,M^2)$ is a PDF and $\tau \equiv M_X^2/s$. The parton
luminosity thus contains all the information on the dependence of
hadronic cross-sections on PDFs.

\begin{figure}[ht!]
\centering
\epsfig{width=0.49\textwidth,figure=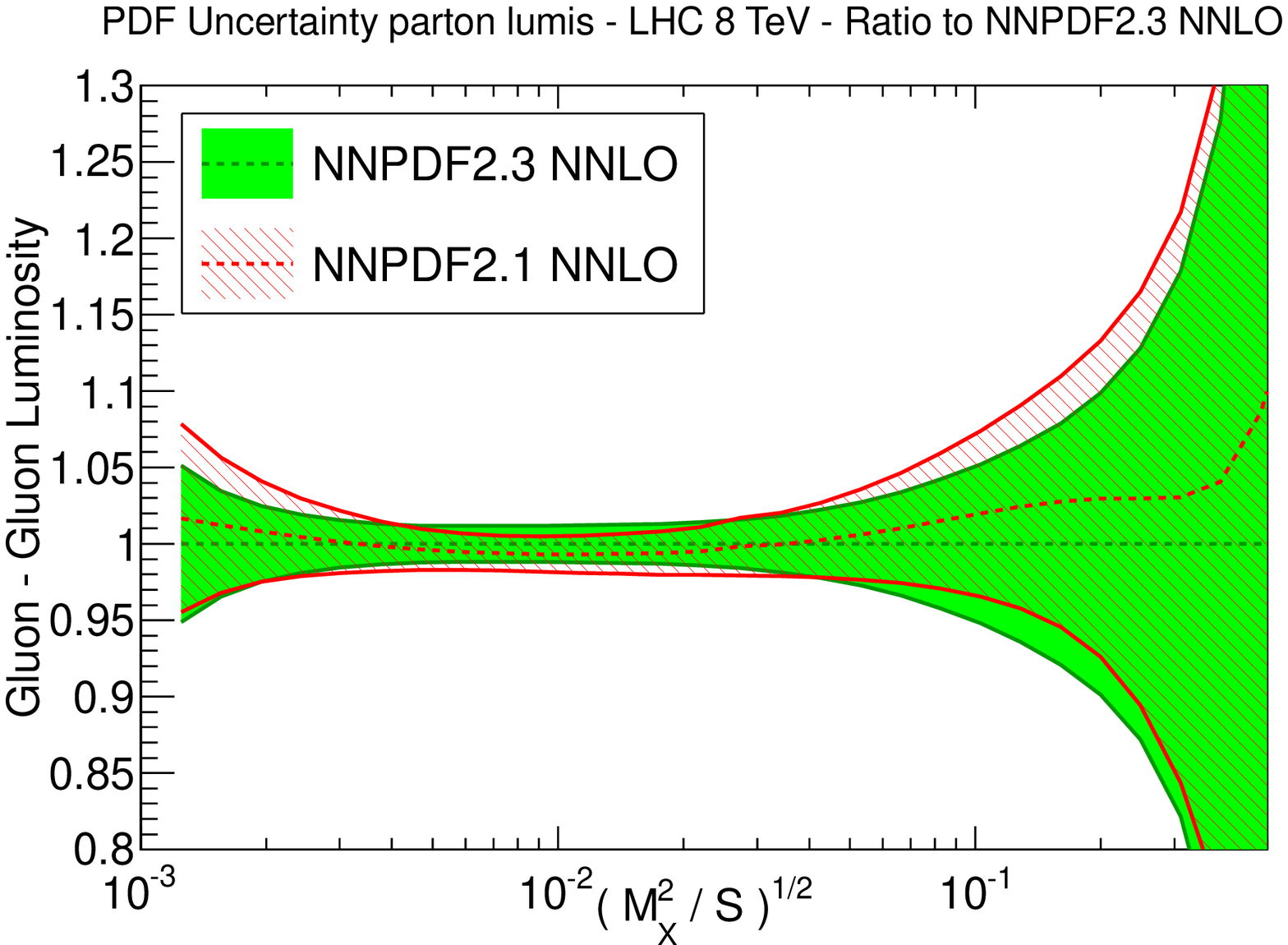}
\epsfig{width=0.49\textwidth,figure=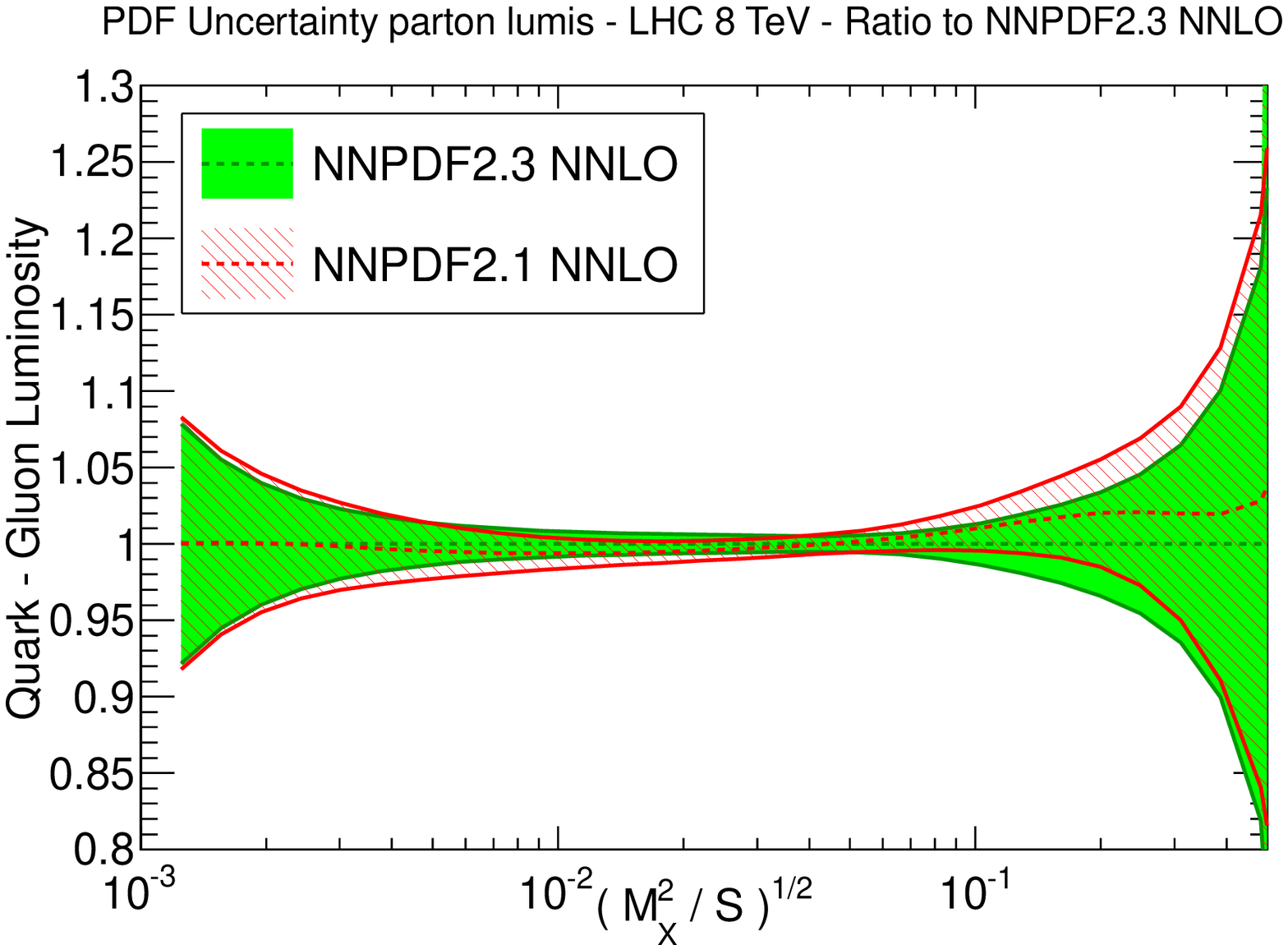}\\
\epsfig{width=0.49\textwidth,figure=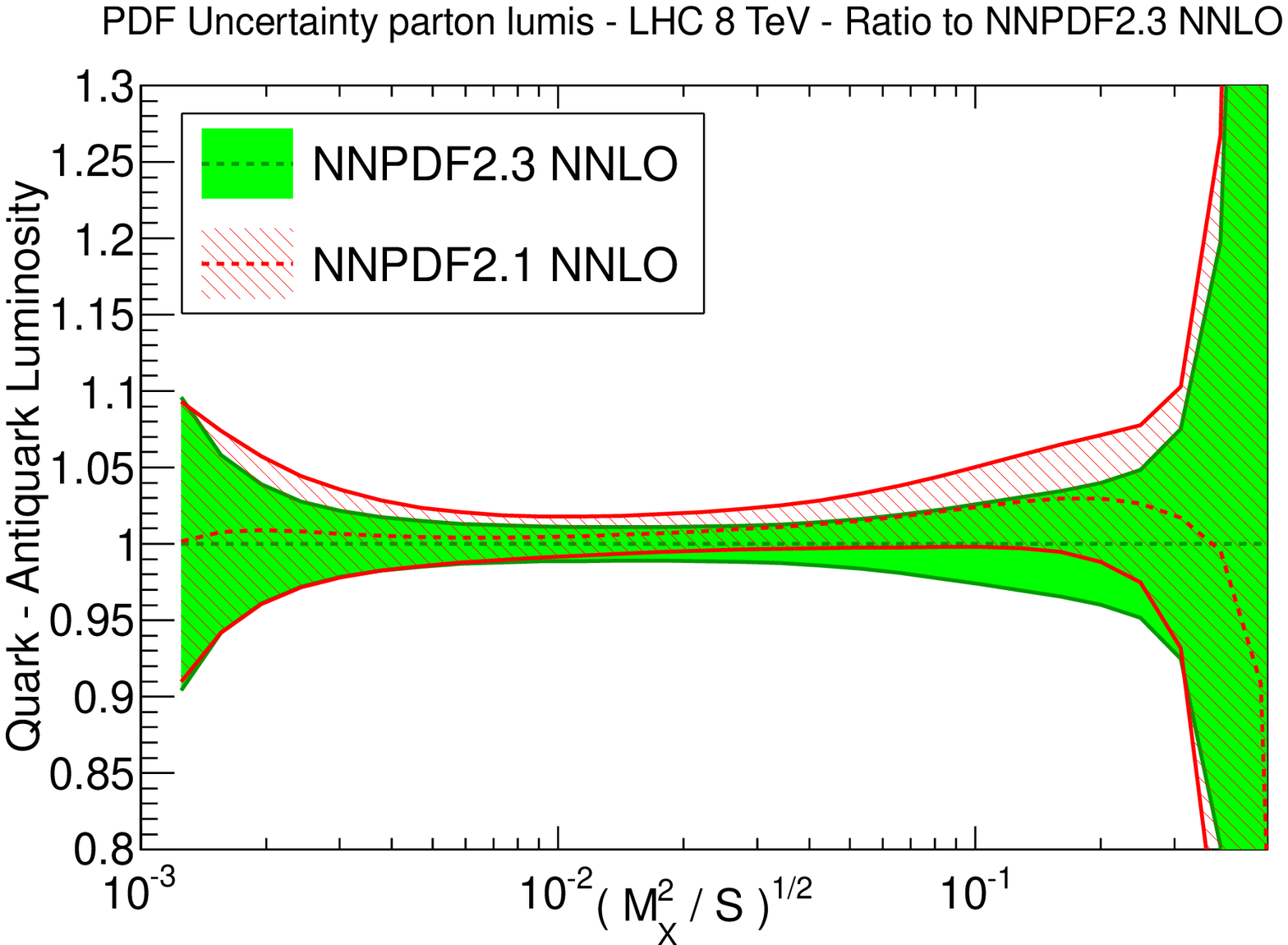}
\epsfig{width=0.49\textwidth,figure=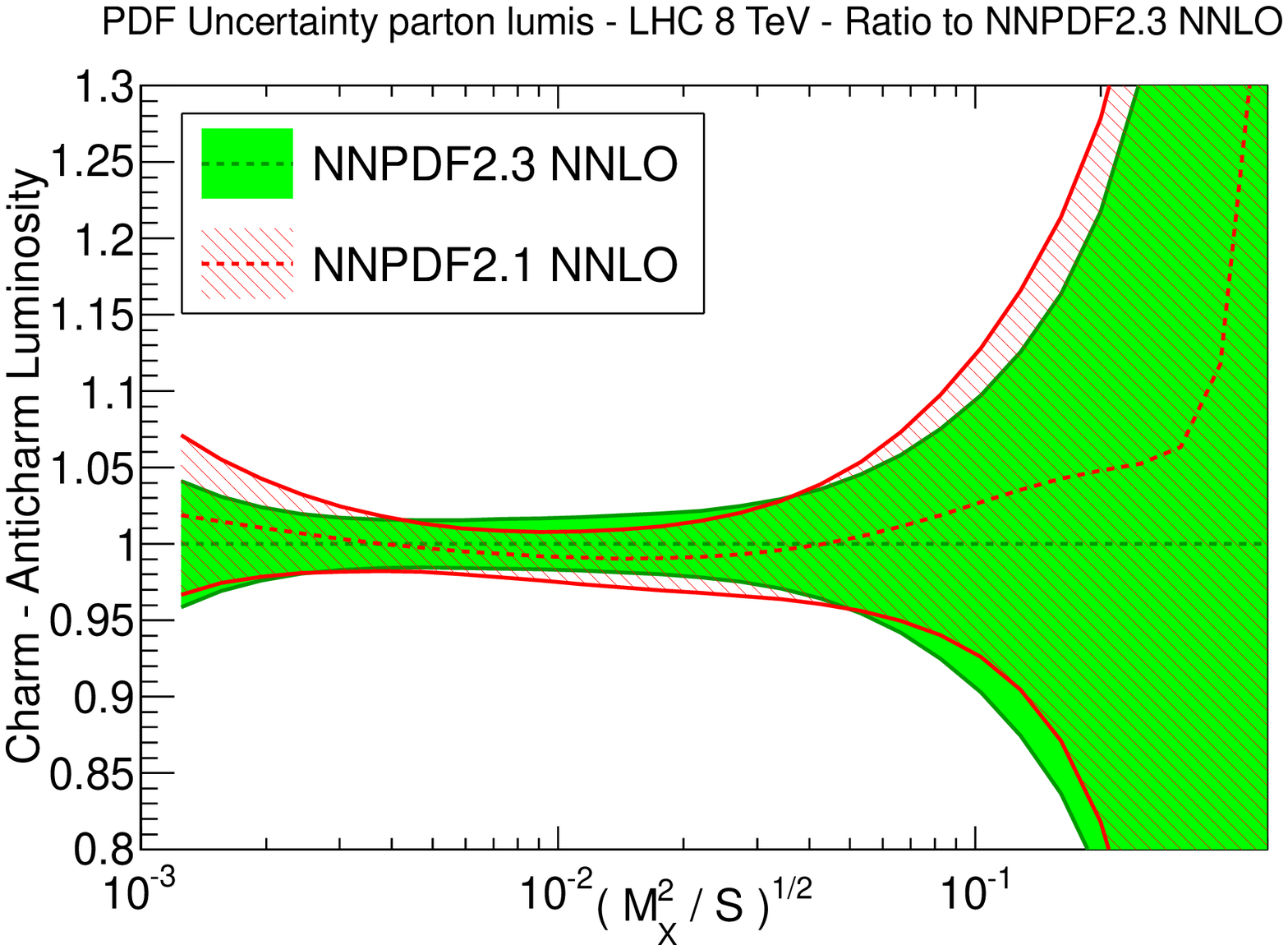}\\
\epsfig{width=0.49\textwidth,figure=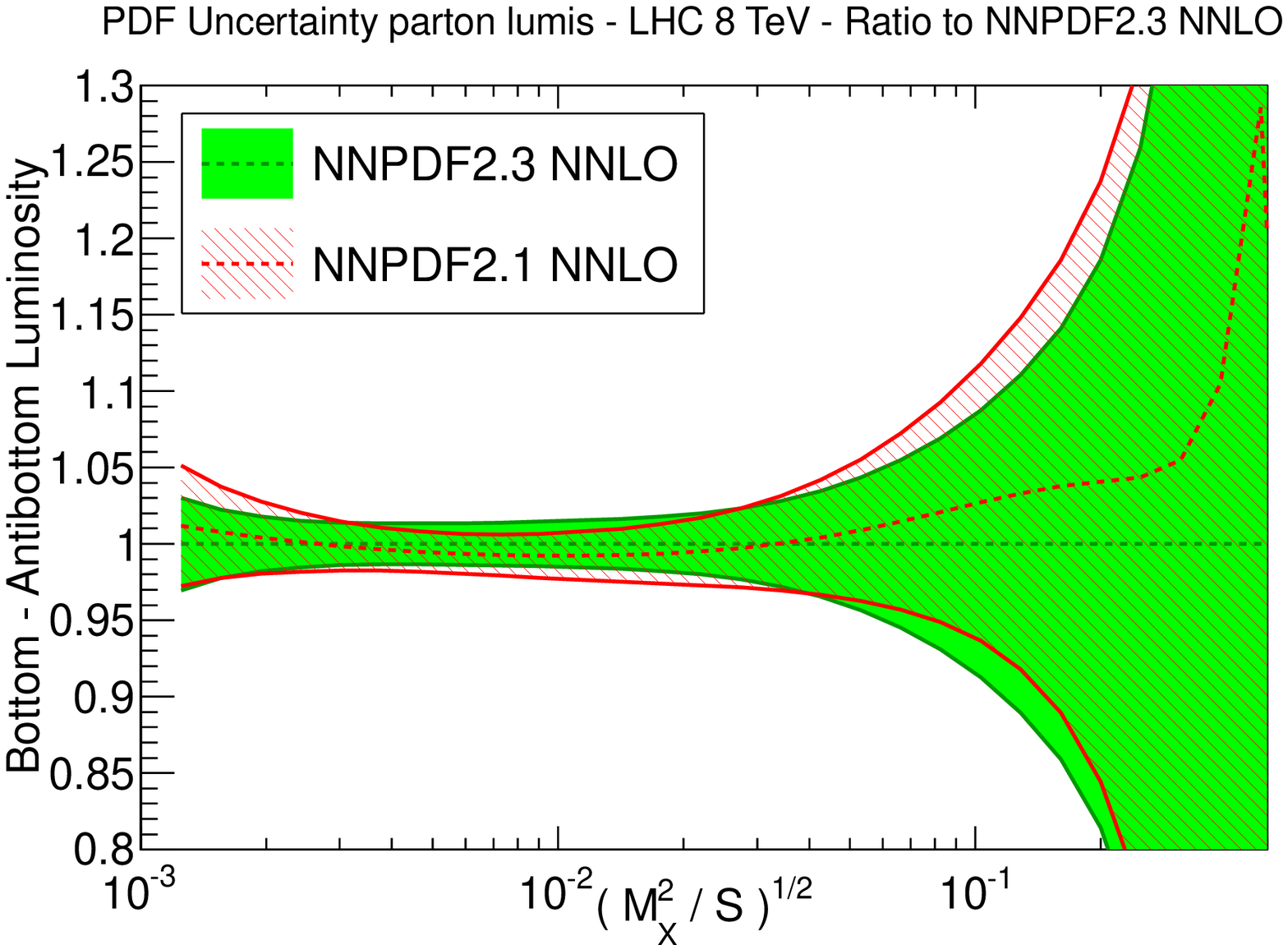}
\epsfig{width=0.49\textwidth,figure=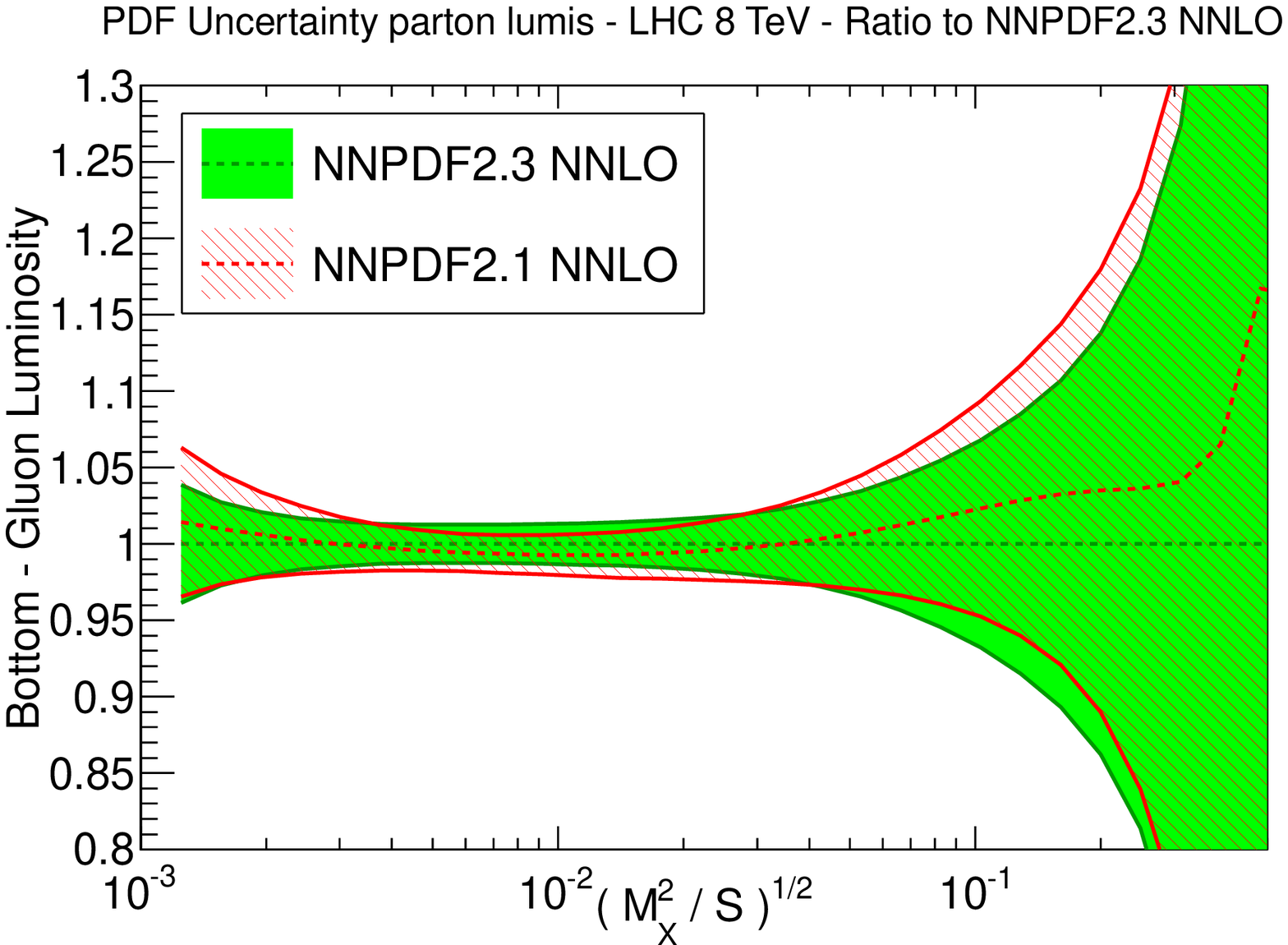}
\caption{\small Comparison of the parton luminosities
 for LHC at 8~TeV, computed using the
NNPDF2.1 and NNPDF2.3 NNLO PDFs, using $N_{\rm rep}=100$ replicas
from both sets. 
From left to right we show $\Phi_{gg}$,  $\Phi_{qg}$, (top)
 $\Phi_{q\bar{q}}$,  $\Phi_{c\bar{c}}$, (middle)  $\Phi_{b\bar{b}}$, 
$\Phi_{bg}$ (bottom).
 All luminosities
are plotted as ratios to the NNPDF2.3 NNLO central value. 
All uncertainties shown are one sigma.
\label{fig_fluxes}}
\end{figure}

Parton luminosities computed for LHC 8 TeV using NNPDF2.1 and NNPDF2.3 PDFs
at NNLO with $\alpha_s(M_{\rm Z})=0.119$  are
compared in Fig.~\ref{fig_fluxes}.
The NLO luminosities are quite similar.
 All the luminosities are very compatible at the one sigma
level. In particular, the 
gluon-gluon luminosity, which is relevant for Higgs production at the
LHC, is quite stable in the region which corresponds to Standard Model 
Higgs
production.
The heavy quark PDFs follow the behaviour of the gluon, from which they
are generated dynamically via perturbative evolution. 
Note that the masses of the heavy quarks
$m_{\rm c}$ and $m_{\rm b}$ are the same in the NLO and NNLO analyses.

In going from NNPDF2.1 to NNPDF2.3, the uncertainty on  the 
gluon-gluon luminosity is  reduced somewhat for larger
final state invariant masses, while the  $q\bar{q}$ luminosity is
somewhat smaller in the same region. As discussed in
Sec.~\ref{sec-comp}, the former effect is due
both to the improved genetic algorithm
minimization and the impact of the ATLAS inclusive jet data, while the latter
is due to the impact of the LHC electroweak
vector boson production data.

\clearpage


\begin{table}[h!]
  \centering
  \footnotesize
  \begin{tabular}{c||c|c||c|c||c|c}
    \hline
    & \multicolumn{2}{c||}{$\sigma({\rm W}^+)$} 
    & \multicolumn{2}{c||}{$\sigma({\rm W}^-)$}
    &  \multicolumn{2}{c}{$\sigma({\rm Z}^0)$}\\
    \hline
    \hline
    & NLO & NNLO & NLO & NNLO & NLO & NNLO \\
    \hline
    NNPDF2.1 & 5.891 $\pm$ 0.133 & 6.198 $\pm$ 0.098 & 
               4.015 $\pm$ 0.091 & 4.213 $\pm$ 0.068 &
               0.928 $\pm$ 0.018 & 0.972 $\pm$ 0.013 \\
    \hline
    NNPDF2.3 & 5.854 $\pm$ 0.076 & 6.122 $\pm$ 0.078 & 
               4.057 $\pm$ 0.053 & 4.202 $\pm$ 0.059 & 
               0.931 $\pm$ 0.011 & 0.968 $\pm$ 0.012 \\
    NNPDF2.3 noLHC & 5.886 $\pm$ 0.082 & 6.198 $\pm$ 0.109 & 
                    4.039 $\pm$ 0.066 & 4.218 $\pm$ 0.072 & 
                    0.931 $\pm$ 0.012 & 0.974 $\pm$ 0.015 \\
    NNPDF2.3 collider & 5.845 $\pm$ 0.104 & 6.127 $\pm$ 0.107 & 
                         4.071 $\pm$ 0.074 & 4.234 $\pm$ 0.068 & 
                         0.953 $\pm$ 0.016 & 1.000 $\pm$ 0.020 \\
    \hline
  \end{tabular}
  \begin{tabular}{c||c|c||c|c}
    \hline
    &  \multicolumn{2}{c||}{$\sigma({\rm W})/\sigma({\rm Z}^0)$}    & 
    \multicolumn{2}{c}{$\sigma({\rm W}^+)/\sigma({\rm W}^-)$} \\
    \hline
    \hline
    &  NLO & NNLO & NLO & NNLO \\
    \hline
    NNPDF2.1    & 10.678 $\pm$ 0.031 & 10.707 $\pm$ 0.034 &  
                   1.467 $\pm$ 0.018 &  1.471 $\pm$ 0.021 \\
    \hline
    NNPDF2.3 & 10.650 $\pm$ 0.022 & 10.669 $\pm$ 0.034 & 
                1.443 $\pm$ 0.009 &  1.457 $\pm$ 0.013 \\
    NNPDF2.3 noLHC & 10.662 $\pm$ 0.025 & 10.692 $\pm$ 0.026 & 
                     1.457 $\pm$ 0.022 &  1.470 $\pm$ 0.020 \\
    NNPDF2.3 collider & 10.403 $\pm$ 0.131 & 10.359 $\pm$ 0.124 & 
                          1.436 $\pm$ 0.014 &  1.447 $\pm$ 0.015 \\
    \hline
  \end{tabular}
  \caption{\label{tab:LHCobs7WZ}  
    \small Total cross-sections for W and Z production
    at the LHC at $\sqrt{s}=7$~TeV. All 
    uncertainties shown are one sigma (in nb). Branching ratios are
    included in the cross-section.}

\vspace{10mm}

  \centering
  \footnotesize
  \begin{tabular}{c||c|c||c|c}
    \hline
    & \multicolumn{2}{c||}{$\sigma({\rm t\bar{t}})$} 
    & \multicolumn{2}{c}{$\sigma({\rm H})$} \\
    \hline
    \hline
    & NLO & NNLO & NLO & NNLO \\
    \hline
    NNPDF2.1 & 160.1 $\pm$ 5.4 & 158.6 $\pm$ 4.4 & 
               11.40 $\pm$ 0.18 & 15.22 $\pm$ 0.22 \\
    \hline
    NNPDF2.3 & 158.3 $\pm$ 4.0 &  157.1 $\pm$ 4.2 & 
               11.46 $\pm$ 0.13 & 15.31 $\pm$ 0.20 \\
    NNPDF2.3noLHC & 158.4 $\pm$ 4.3  & 157.1 $\pm$ 4.7 & 
                    11.48 $\pm$ 0.14 & 15.30 $\pm$ 0.17 \\
    NNPDF2.3--collider & 151.2 $\pm$ 6.1  & 150.0 $\pm$ 5.3 & 
                         10.80 $\pm$ 0.25 & 14.44 $\pm$ 0.27 \\
    \hline
  \end{tabular}
  \caption{\label{tab:LHCobs7tth}  
    \small 
     Total cross-sections for top quark pair production and Higgs production 
    in gluon fusion at the LHC at $\sqrt{s}=7$~TeV (in pb).
    All uncertainties shown are one sigma}
\end{table}


\subsection{Total cross-sections}
\label{sec-tot}

We present now results for several benchmark total cross-sections at
the LHC at NLO and NNLO using NNPDF2.3 PDFs as well as NNPDF2.1 PDFs
for comparison, with
$\alpha_s(M_{\rm Z})=0.119$ at the LHC, and $\sqrt{s}=7$~TeV and
$\sqrt{s}=8$~TeV. 
We  determine the following observables:
\begin{itemize}
\item electroweak gauge boson production total cross-sections and 
${\rm W}^+/{\rm W}^-$ and W/Z cross-section ratios, using the {\tt Vrap}
  code~\cite{Anastasiou:2003ds} with scale $Q^2=M_{\rm V}^2$;
\item top  pair production total cross-section, using
the   {\tt top++} code \cite{Czakon:2011xx} with $Q^2=m_{\rm t}^2$; 
at NNLO the
  approximate cross-sections of Ref.~\cite{Baernreuther:2012ws} are
  used; the settings are the
  default ones of Ref.~\cite{Cacciari:2011hy}, and are the same at NLO
  and NNLO; in particular the
  top quark mass is taken to be $m_{\rm t}=173.3$~GeV;
  use the same settings for the calculations with NLO and NNLO PDFs.
\item Standard Model Higgs boson production cross-sections
with $m_{\rm H}=125$ GeV in the gluon fusion using the 
{\tt iHixs} code~\cite{Anastasiou:2011pi} with $Q=m_{\rm H}$.
\end{itemize}
All uncertainties shown are PDF uncertainties only: in particular they
do not include the uncertainty due to the variation of $\alpha_s$, nor
theoretical uncertainties, in particular the uncertainty due to missing higher
orders, usually estimated by varying renormalization and factorization scales. 

\begin{figure}[ht!]
\centering
\epsfig{width=0.49\textwidth,figure=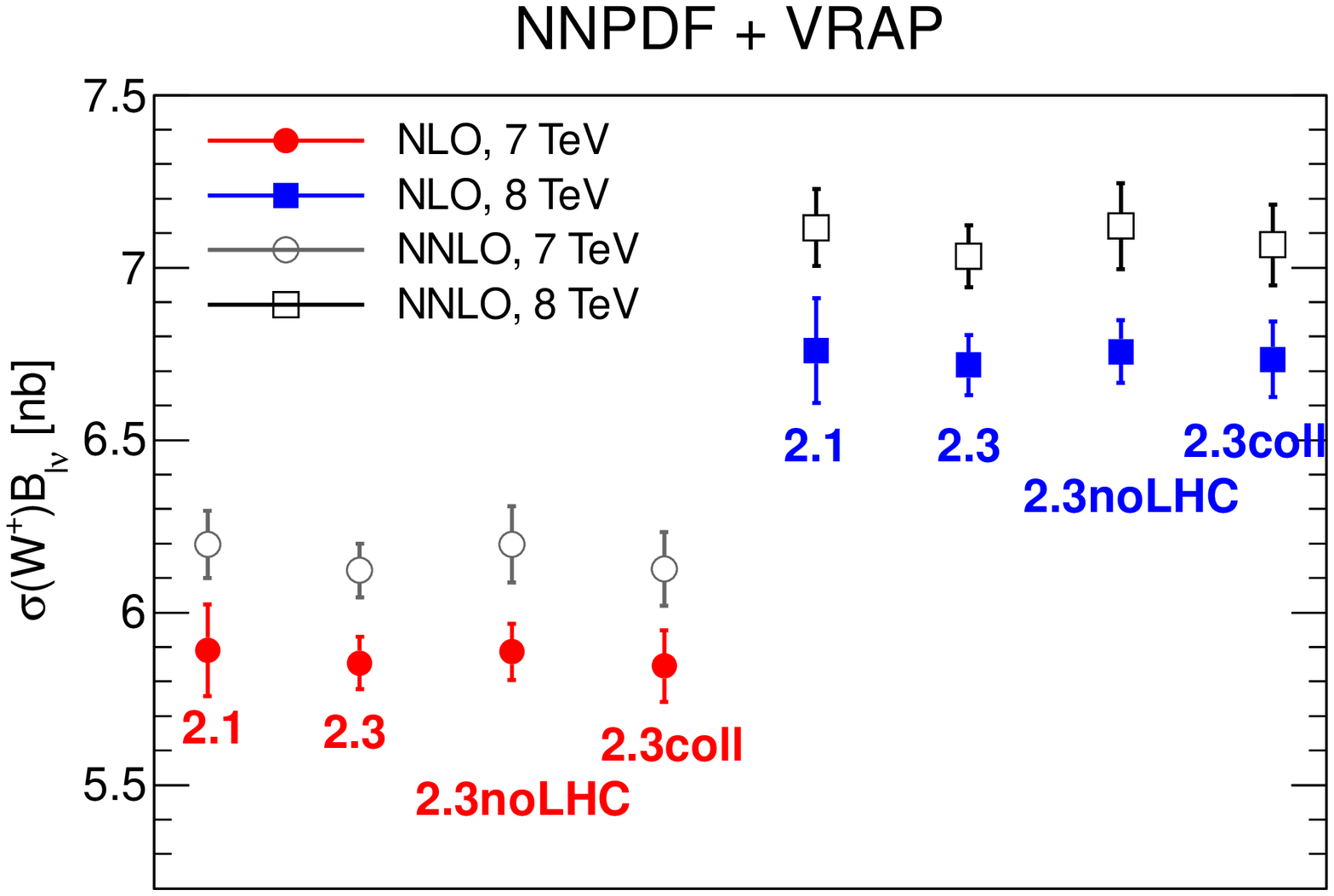}
\epsfig{width=0.49\textwidth,figure=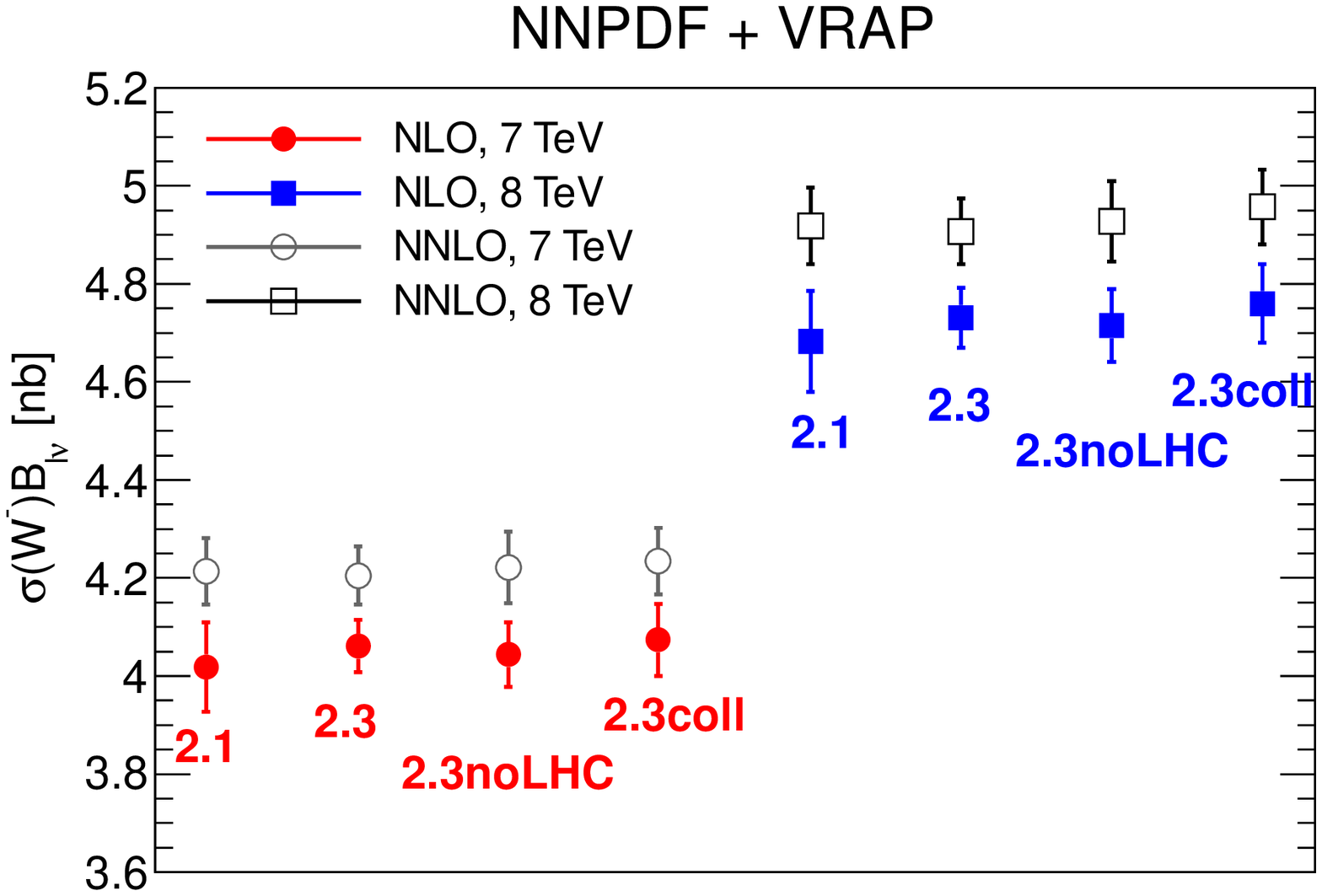}\\
\epsfig{width=0.49\textwidth,figure=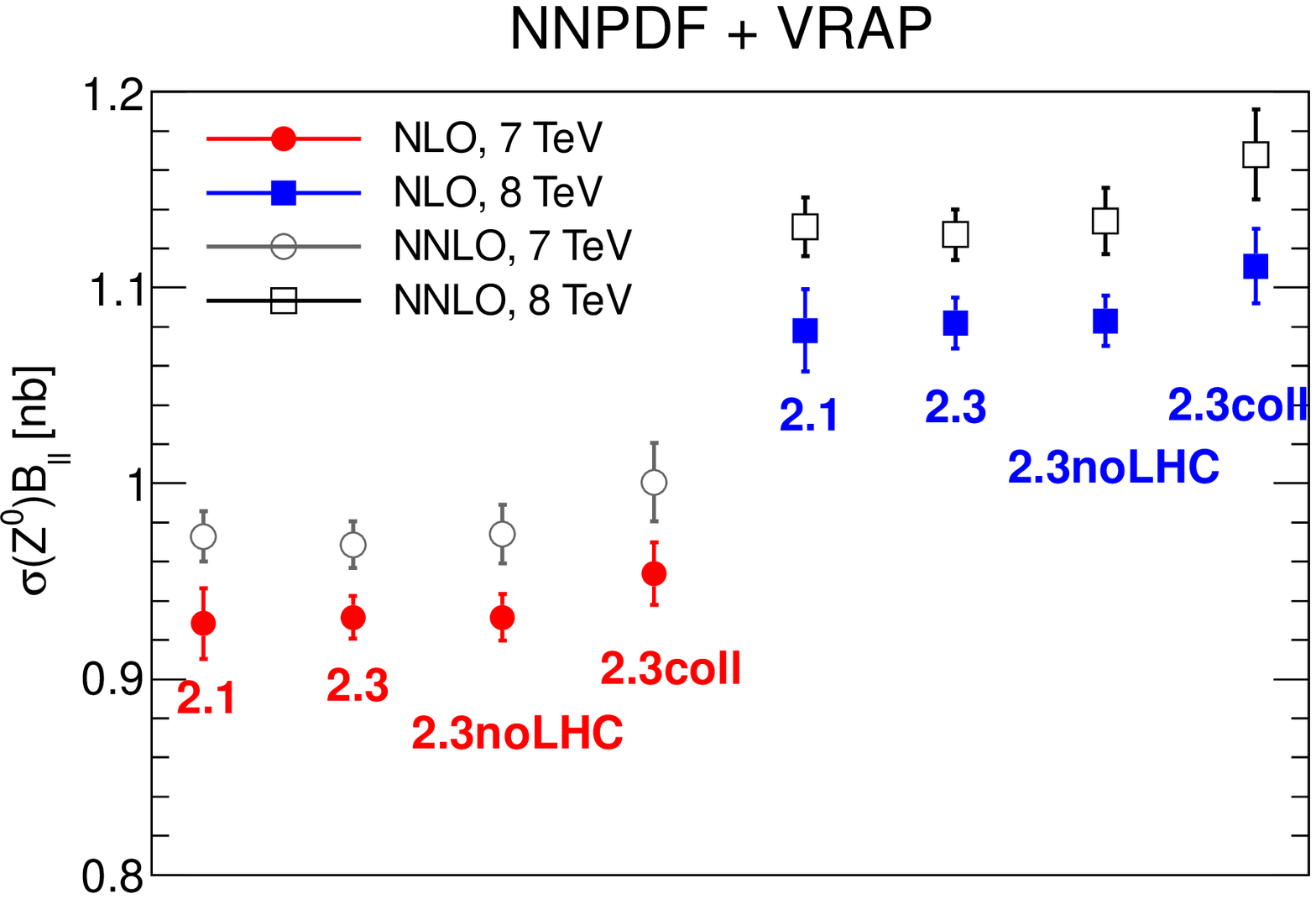}\\
\epsfig{width=0.49\textwidth,figure=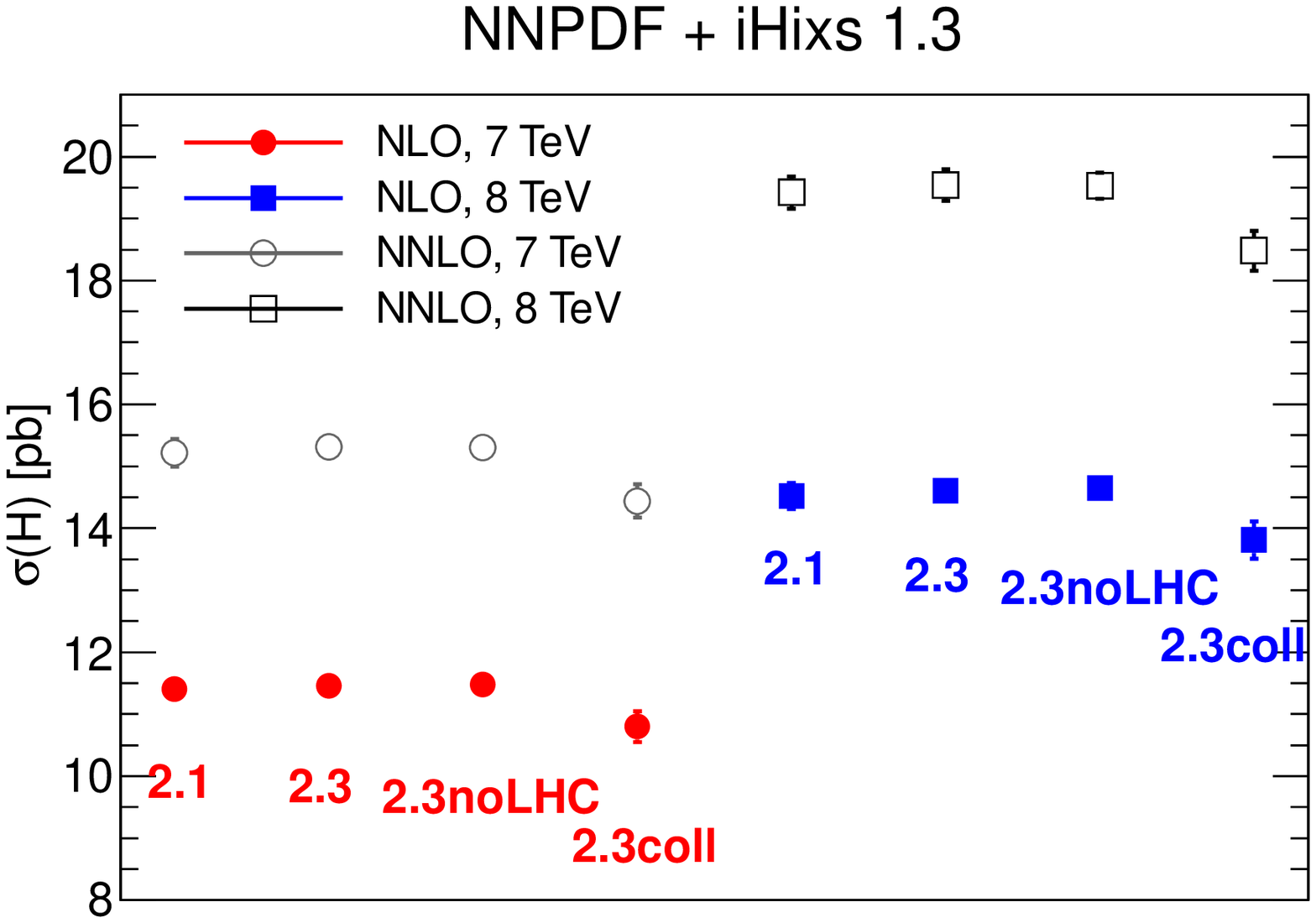}
\epsfig{width=0.49\textwidth,figure=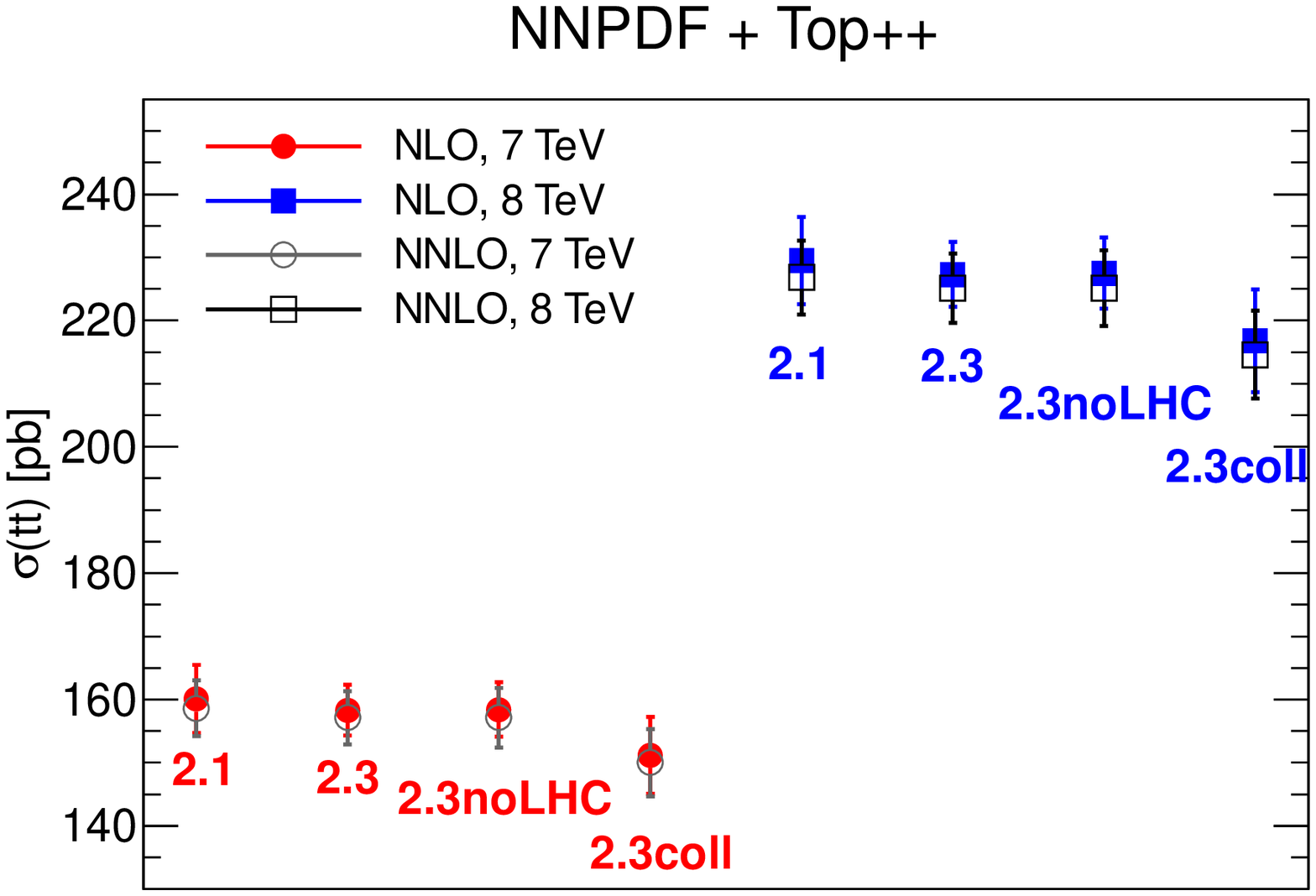}
\caption{\small Graphical representation of the results of
  Tabs.~\ref{tab:LHCobs7WZ}--\ref{tab:LHCobs8tth} 
\label{fig:EWKplots}}
\end{figure}

Results are collected in Tabs.~\ref{tab:LHCobs7WZ} \& \ref{tab:LHCobs7tth} ($\sqrt{s}=7$~TeV)
and in Tabs.~\ref{tab:LHCobs8WZ} \& \ref{tab:LHCobs8tth} ($\sqrt{s}=8$~TeV), and 
represented graphically in Fig.~\ref{fig:EWKplots}.
For all observables it is clear that even though everything is
consistent within uncertainties, the accuracy increases when going
from NNPDF2.1 to NNPDF2.3. Comparison with results obtained using 
the NNPDF2.3 noLHC sets shows that improvement is partly due to the
improved methodology, but the LHC data have a visible impact both on
the central value and uncertainty. Results obtained using the collider
only set are not yet competitive, even for these very inclusive observables,
except for quantities such as the ${\rm W}^+/{\rm W}^-$ ratio which are 
determined primarily by the collider data.


\begin{table}
  \centering
  \footnotesize
  \begin{tabular}{c||c|c||c|c||c|c}
    \hline
    & \multicolumn{2}{c||}{$\sigma({\rm W}^+)$} 
    & \multicolumn{2}{c||}{$\sigma({\rm W}^-)$}
    & \multicolumn{2}{c}{$\sigma({\rm Z}^0)$}\\
    \hline
    \hline
    & NLO & NNLO & NLO & NNLO & NLO & NNLO \\
    \hline
    NNPDF2.1  & 6.759 $\pm$ 0.152 & 7.116 $\pm$ 0.112 & 
                4.683 $\pm$ 0.103 & 4.918 $\pm$ 0.078 & 
                1.078 $\pm$ 0.021 & 1.131 $\pm$ 0.015 \\
    \hline
    NNPDF2.3 & 6.718 $\pm$ 0.087 & 7.033 $\pm$ 0.090 & 
               4.731 $\pm$ 0.061 & 4.907 $\pm$ 0.067 & 
               1.082 $\pm$ 0.013 & 1.127 $\pm$ 0.013 \\
    NNPDF2.3 noLHC & 6.756 $\pm$ 0.090 & 7.120 $\pm$ 0.124 &  
                    4.715 $\pm$ 0.074 & 4.927 $\pm$ 0.082 & 
                    1.083 $\pm$ 0.013 & 1.134 $\pm$ 0.017 \\
    NNPDF2.3 collider & 6.734 $\pm$ 0.110 & 7.066 $\pm$ 0.117 & 
                         4.760 $\pm$ 0.080 & 4.957 $\pm$ 0.076 & 
                         1.111 $\pm$ 0.019 & 1.168 $\pm$ 0.023 \\
    \hline
  \end{tabular}
  \begin{tabular}{c||c|c||c|c}
    \hline
    &  \multicolumn{2}{c||}{$\sigma({\rm W})/\sigma({\rm Z}^0)$}    & 
    \multicolumn{2}{c}{$\sigma({\rm W}^+)/\sigma({\rm W}^-)$} \\
    \hline
    \hline
    &  NLO & NNLO & NLO & NNLO \\
    \hline
    NNPDF2.1    & 10.611 $\pm$ 0.033 & 10.639 $\pm$ 0.035 & 
                   1.443 $\pm$ 0.017 &  1.447 $\pm$ 0.019 \\
    \hline
    NNPDF2.3 & 10.580 $\pm$ 0.023 & 10.598 $\pm$ 0.035 & 
                 1.420 $\pm$ 0.008 &  1.433 $\pm$ 0.012 \\
    NNPDF2.3 noLHC & 10.591 $\pm$ 0.026 & 10.622 $\pm$ 0.028 & 
                     1.433 $\pm$ 0.019 &  1.445 $\pm$ 0.018 \\
    NNPDF2.3 collider & 10.347 $\pm$ 0.128 & 10.299 $\pm$ 0.124 & 
                          1.415 $\pm$ 0.014 &  1.425 $\pm$ 0.013 \\
    \hline
  \end{tabular}
  \caption{\label{tab:LHCobs8WZ}  
    \small Same as Tab.~\ref{tab:LHCobs7WZ}, but for  $\sqrt{s}=8$~TeV.}

\vspace{10mm}

  \centering
  \footnotesize
  \begin{tabular}{c||c|c||c|c}
    \hline
    & \multicolumn{2}{c||}{$\sigma({\rm t\bar{t}})$} 
    & \multicolumn{2}{c}{$\sigma({\rm H})$} \\
    \hline
    \hline
    & NLO & NNLO & NLO & NNLO \\
    \hline
    NNPDF2.1 & 229.5 $\pm$ 6.9 & 226.8 $\pm$ 5.8 & 
               14.52 $\pm$ 0.21 & 19.42 $\pm$ 0.26 \\
    \hline
    NNPDF2.3 & 227.3 $\pm$ 5.1  & 225.1 $\pm$ 5.5  &  
               14.61 $\pm$ 0.16 & 19.54 $\pm$ 0.25 \\
    NNPDF2.3noLHC & 227.5 $\pm$ 5.6 &  225.1 $\pm$ 6.0 & 
                    14.65 $\pm$ 0.17 & 19.53 $\pm$ 0.21 \\
    NNPDF2.3--collider & 216.8 $\pm$ 8.1  & 214.6 $\pm$ 7.0 &  
                         13.81 $\pm$ 0.30 & 18.48 $\pm$ 0.32 \\
    \hline
  \end{tabular}
  \caption{\label{tab:LHCobs8tth}
      \small
Same as Tab.~\ref{tab:LHCobs7tth}, but for  $\sqrt{s}=8$~TeV.}
\end{table}

\section{Conclusions}
\label{sec:conclusions}

The NNPDF2.3 PDF set is the first global PDF set to include
systematically all relevant LHC data, and it is thus arguably the most
accurate PDF set currently available. 
LHC data are likely to play an increasing role in future 
refinements of PDF sets. In particular, the internal 
inconsistencies which were noticed long ago in fixed target DIS 
data~\cite{Forte:2002fg}, and the tensions between fixed target 
and collider data, some of which have been discussed here in 
Sec.~\ref{sec-collider}, suggest that a reliable PDF determination 
should avoid low energy data and data obtained using nuclear targets, 
and should thus be based only on lepton-hadron or hadron-hadron collider 
data. Current collider data are not yet sufficient to give a competitive PDF
determination by themselves (see Sec.~\ref{sec-collider}). However 
this situation is likely to evolve very rapidly thanks to the 
excellent performance of the 
LHC and its experiments, not only as the data on the usual inclusive processes 
become more precise, but also through the incorporation of new processes, 
such as W boson production in association with charm
quarks for strangeness determination~\cite{CMSWc,Stirling:2012vh}. 
Cleaner data are in turn likely to stimulate further 
improvements in the computational and statistical methodologies used 
for their analysis, and in the theoretical framework used to describe them.

\bigskip
\bigskip

All the NNPDF2.3 PDF sets that have been discussed in this work
are available from the NNPDF web site,
\begin{center}
{\bf \url{http://nnpdf.hepforge.org/}~}
\end{center}
and through the LHAPDF interface~\cite{Bourilkov:2006cj}. On the NNPDF
web site a Mathematica interface  is also available, as well as a more
complete selection of PDF plots.

Specifically, the new PDF sets that have been produced in the present analysis and are available in LHAPDF are the following:

\begin{itemize}

\item NNPDF2.3 NLO and NNLO sets of $N_{\rm rep}=100$ replicas,
  provided for all values of $\alpha_s$  
from 0.114 to 0.124 varied in  steps of $\delta\alpha_s=0.001$:\\ 
{\tt NNPDF23\_nlo\_as\_0114.LHgrid},
$\ldots$,  {\tt NNPDF23\_nlo\_as\_0124.LHgrid};\\
{\tt NNPDF23\_nnlo\_as\_0114.LHgrid},
$\ldots$,  {\tt NNPDF23\_nnlo\_as\_0124.LHgrid};\\

\item NNPDF2.3 NLO and NNLO PDF sets based on reduced data sets,
  provided for all values of $\alpha_s$  
from 0.116 to 0.120 in steps of $0.001$:
\\ 
{\tt NNPDF23\_nlo\_noLHC\_as\_0116.LHgrid}, $\ldots$,
{\tt NNPDF23\_nlo\_noLHC\_as\_0120.LHgrid};\\
{\tt NNPDF23\_nnlo\_noLHC\_as\_0116.LHgrid}, $\ldots$,
{\tt NNPDF23\_nnlo\_noLHC\_as\_0120.LHgrid};\\
{\tt NNPDF23\_nlo\_collider\_as\_0116.LHgrid}, $\ldots$,
{\tt NNPDF23\_nlo\_collider\_as\_0120.LHgrid};\\
{\tt NNPDF23\_nnlo\_collider\_as\_0116.LHgrid}, $\ldots$,
{\tt NNPDF23\_nnlo\_collider\_as\_0120.LHgrid};

\item NNPDF2.3 NLO and NNLO PDF sets in the $n_f=4$ and $n_f=5$ schemes
  (number of active flavors only increases up to the given value),
  provided for all values of $\alpha_s$  
from 0.116 to 0.120 in steps of $0.001$:
\\ 
{\tt NNPDF23\_nlo\_FFN\_NF4\_as\_0116.LHgrid}, $\ldots$,
{\tt NNPDF23\_nlo\_FFN\_NF4\_as\_0120.LHgrid};\\
{\tt NNPDF23\_nnlo\_FFN\_NF4\_as\_0116.LHgrid}, $\ldots$,
{\tt NNPDF23\_nnlo\_FFN\_NF4\_as\_0120.LHgrid};\\
{\tt NNPDF23\_nlo\_FFN\_NF5\_as\_0116.LHgrid}, $\ldots$,
{\tt NNPDF23\_nlo\_FFN\_NF5\_as\_0120.LHgrid};\\
{\tt NNPDF23\_nnlo\_FFN\_NF5\_as\_0116.LHgrid}, $\ldots$,
{\tt NNPDF23\_nnlo\_FFN\_NF5\_as\_0120.LHgrid};
\end{itemize}

\bigskip
\bigskip
\begin{center}
\rule{5cm}{.1pt}
\end{center}
\bigskip
\bigskip

{\bf\noindent  Acknowledgments \\}
We are grateful to Tancredi Carli, Eric Feng, Paolo Francavilla, Pavel
Starovoitov and Mark Sutton for discussions about the ATLAS jet
data and the  {\tt APPLgrid} interface. We are grateful to Jeff
Berryhill, Amanda Cooper-Sarkar, 
Uta Klein,  Ronan McNulty, Michele Pioppi,  Voica Radescu, Tara Shears
and Pin Tang for discussions on the ATLAS, CMS and LHCb electroweak
measurements, and to Sasha Glazov for providing the  {\tt APPLgrid}
tables used in the ATLAS W, Z  analysis. We are grateful to
Giancarlo Ferrera for assistance with  {\tt DYNNLO} and Klaus Rabbertz
for help with {\tt FastNLO}. We thank Joey Huston, Pavel Nadolsky,
Robert Thorne, Graeme Watt for many fruitful exchanges and
discussions, and S.~Alekhin for pointing out the error
Eq.~(\ref{eq:bug}). 
J.~R. is supported by a Marie Curie 
Intra--European Fellowship of the European Community's 7th Framework 
Programme under contract number PIEF-GA-2010-272515.
The work of V.~B. was supported by the ERC grant 291377, 
“LHCtheory: Theoretical predictions and analyses of LHC physics: 
advancing the precision frontier”.



\appendix

\include{sec-fk-bench}

\bibliography{nnpdf23}

\end{document}